\documentclass[11pt]{report}
\usepackage[table]{xcolor}
\usepackage[utf8]{inputenc}
\usepackage[english]{babel}
\usepackage{amsmath}
\usepackage{amsfonts}
\usepackage{amssymb}
\usepackage{amsthm}
\usepackage[basic]{complexity}
\usepackage{tikz}
\usepackage{hyphenat}
\usepackage{graphicx}
\usepackage{nicefrac}
\usepackage{booktabs}
\usepackage{dsfont}
\usepackage{bm}
\usepackage{setspace}

\usepackage{libertine}

\usepackage[authoryear]{natbib}

\usepackage{geometry}
\usepackage{tikz}
\usetikzlibrary{automata, positioning, calc, shapes, arrows, fit, patterns, patterns.meta, decorations.pathreplacing}

\usepackage{tabularx}
\renewcommand\tabularxcolumn[1]{m{#1}}

\usepackage[unicode=true, bookmarks=false, breaklinks=true, pdfborder={0 0 1}, colorlinks=true, backref=page]{hyperref}
\hypersetup{linkcolor=my-blue-dark, citecolor=my-blue-dark, filecolor=my-blue, urlcolor=my-blue-dark}

\renewcommand*{\backref}[1]{}
\renewcommand*{\backrefalt}[4]{%
	\ifcase #1 (Not cited)%
	\or        (Cited on page~#2)%
	\else      (Cited on pages~#2)%
	\fi}

\theoremstyle{plain}
\newtheorem{theorem}{Theorem}

\newtheorem{proposition}[theorem]{Proposition}
\newtheorem{definition}[theorem]{Definition}

\usepackage{xcolor}

\definecolor{my-red}{HTML}{C62828}
\definecolor{my-red-light}{HTML}{E57373}
\definecolor{my-red-verylight}{HTML}{FFCDD2}
\definecolor{my-red-dark}{HTML}{B71C1C}

\definecolor{my-pink}{HTML}{EC407A}
\definecolor{my-pink-light}{HTML}{F48FB1}
\definecolor{my-pink-verylight}{HTML}{F8BBD0}
\definecolor{my-pink-dark}{HTML}{880E4F}

\definecolor{my-purple}{HTML}{8E24AA}
\definecolor{my-purple-light}{HTML}{BA68C8}
\definecolor{my-purple-verylight}{HTML}{e5cefc}
\definecolor{my-purple-dark}{HTML}{6A1B9A}

\definecolor{my-indigo}{HTML}{3949AB}
\definecolor{my-indigo-light}{HTML}{7986CB}
\definecolor{my-indigo-verylight}{HTML}{9FA8DA}
\definecolor{my-indigo-dark}{HTML}{1A237E}

\definecolor{my-blue}{HTML}{1E88E5}
\definecolor{my-blue-light}{HTML}{64B5F6}
\definecolor{my-blue-verylight}{HTML}{d3e5ed}
\definecolor{my-blue-veryverylight}{HTML}{f2f9ff}
\definecolor{my-blue-dark}{HTML}{0D47A1}

\definecolor{my-cyan}{HTML}{00BCD4}
\definecolor{my-cyan-light}{HTML}{4DD0E1}
\definecolor{my-cyan-verylight}{HTML}{80DEEA}
\definecolor{my-cyan-dark}{HTML}{0097A7}

\definecolor{my-teal}{HTML}{009688}
\definecolor{my-teal-light}{HTML}{4DB6AC}
\definecolor{my-teal-verylight}{HTML}{B2DFDB}
\definecolor{my-teal-dark}{HTML}{00695C}

\definecolor{my-green}{HTML}{39ac39}
\definecolor{my-green-light}{HTML}{8cd98c}
\definecolor{my-green-verylight}{HTML}{b3e6b3}
\definecolor{my-green-dark}{HTML}{339933}

\definecolor{my-grass}{HTML}{689F38}
\definecolor{my-grass-light}{HTML}{8BC34A}
\definecolor{my-grass-verylight}{HTML}{AED581}
\definecolor{my-grass-dark}{HTML}{33691E}

\definecolor{my-lime}{HTML}{CDDC39}
\definecolor{my-lime-light}{HTML}{DCE775}
\definecolor{my-lime-verylight}{HTML}{E6EE9C}
\definecolor{my-lime-dark}{HTML}{AFB42B}

\definecolor{my-yellow}{HTML}{fffc29}
\definecolor{my-yellow-light}{HTML}{fffd7a}
\definecolor{my-yellow-verylight}{HTML}{fefdbb}
\definecolor{my-yellow-dark}{HTML}{FFD600}

\definecolor{my-orange}{HTML}{FF8F00}
\definecolor{my-orange-light}{HTML}{FFC107}
\definecolor{my-orange-verylight}{HTML}{ffe5a4}
\definecolor{my-orange-verylight}{HTML}{ffe6bb}
\definecolor{my-orange-dark}{HTML}{FF6F00}

\definecolor{my-brown}{HTML}{6D4C41}
\definecolor{my-brown-light}{HTML}{795548}
\definecolor{my-brown-verylight}{HTML}{BCAAA4}
\definecolor{my-brown-dark}{HTML}{3E2723}

\definecolor{my-gray}{HTML}{616161}
\definecolor{my-gray-light}{HTML}{9E9E9E}
\definecolor{my-gray-verylight}{HTML}{f0f0f0}
\definecolor{my-gray-veryverylight}{HTML}{f6f6f6}
\definecolor{my-gray-dark}{HTML}{424242}

\definecolor{my-steel}{HTML}{546E7A}
\definecolor{my-steel-light}{HTML}{78909C}
\definecolor{my-steel-verylight}{HTML}{B0BEC5}
\definecolor{my-steel-dark}{HTML}{37474F}

\definecolor{ColorBlind1}{HTML}{D81B60}
\definecolor{ColorBlind2}{HTML}{1E88E5}
\definecolor{ColorBlind3}{HTML}{FFC107}
\definecolor{ColorBlind4}{HTML}{004D40}

\definecolor{MyOrange}{rgb}{1, 0.5, 0}
\definecolor{MyLightOrange}{rgb}{1, 0.9, 0.7}
\definecolor{MyGrey}{rgb}{0.3, 0.3, 0.3}
\definecolor{MyLightGrey}{rgb}{0.9, 0.9, 0.9}
\definecolor{MyGreen}{rgb}{0, 0.6, 0}
\definecolor{MyBlue}{rgb}{0, 0, 0.6}
\definecolor{MyLightBlue}{rgb}{0.7, 0.8, 1}
\definecolor{MyRed}{rgb}{0.7, 0, 0}
\definecolor{MyLightRed}{rgb}{1, 0.7, 0.7}
\definecolor{MyLightYellow}{HTML}{f9f6ec}

\usepackage{framed}

\renewenvironment{leftbar}[1][\hsize]
{%
	\MakeFramed{\hsize#1\advance\hsize-\width\FrameRestore}%
}
{\endMakeFramed}

\usepackage{xparse}
\NewDocumentEnvironment{illustration}{o}{%
	\begin{leftbar}\noindent{\bfseries Illustration\IfNoValueTF{#1}{}{ \normalfont(#1)\bfseries}.}%
}
	{\end{leftbar}}

\usepackage{pifont}
\newcommand{\cmark}{{\color{MyGreen}\ding{51}}}

\newcommand{\xmark}{{\color{MyRed}\ding{55}}}

\usepackage{eurosym}
\usepackage{textcomp}

\newcommand{\tuple}[1]{\left\langle #1 \right\rangle}

\renewcommand{\phi}{\varphi}

\newcommand{\Rplus}{\ensuremath{\mathbb{R}_{\geq 0}}}

\DeclareMathOperator*{\argmax}{arg\,max}
\DeclareMathOperator*{\argmin}{arg\,min}

\newcommand{\agentSet}{\mathcal{N}}
\newcommand{\projSet}{\mathcal{P}}
\newcommand{\allocSet}{\textsc{Feas}}

\newcommand{\profile}{\boldsymbol{A}}

\newcommand{\bigO}{\mathcal{O}}
\newcommand{\complexP}{\ensuremath{\mathtt{P}}}
\newcommand{\complexNP}{\ensuremath{\mathtt{NP}}}
\newcommand{\complexcoNP}{\ensuremath{\mathtt{coNP}}}
\newcommand{\complexFPT}{\ensuremath{\mathtt{FPT}}}

\newcommand{\pbRule}{{\normalfont\textsc{R}}}
\newcommand{\maxWelCard}{{\normalfont\textsc{MaxCard}}}
\newcommand{\maxWelCost}{{\normalfont\textsc{MaxCost}}}
\newcommand{\greedy}{{\normalfont\textsc{Greed}}}
\newcommand{\ordinalGreedy}[1]{\ensuremath{{\normalfont\textsc{Greed}}_{#1}}}
\newcommand{\greWelCard}{{\normalfont\textsc{GreedCard}}}
\newcommand{\greWelCost}{{\normalfont\textsc{GreedCost}}}
\newcommand{\mes}{{\normalfont\textsc{MES}}}
\newcommand{\mesSat}[1]{\mes{}{\normalfont[#1]}}
\newcommand{\seqPhragmen}{{\normalfont\textsc{SeqPhrag}}}
\newcommand{\maximinSupport}{{\normalfont\textsc{MaximinSupp}}}

\newcommand{\appScore}{\mathit{app}}
\newcommand{\satisfaction}{\mathit{sat}}

\newcommand{\cardSatisfaction}{\mathit{sat}^{\mathit{card}}}
\newcommand{\costSatisfaction}{\mathit{sat}^{\mathit{cost}}}

\newcommand{\share}{\mathit{share}}

\newcommand{\utilSW}{\textsc{Util-SW}}
\newcommand{\nashSW}{\textsc{Nash-SW}}
\newcommand{\egalSW}{\textsc{Egal-SW}}
\newcommand{\ccSW}{\textsc{CC-SW}}

\usepackage{todonotes}

\begin{document}
	\begin{titlepage}
		\null
		
		\vspace{3em}
		
		\begin{center}
			\huge \bfseries The (Computational) Social Choice Take on Indivisible Participatory Budgeting
		\end{center}
	
		\vspace{2.5em}
		
		\begin{center}
			\large
			\begin{tabular}{ccc}
				Simon Rey$^1$ & Felicia Schmidt$^2$ & Jan Maly$^{2,3}$ \\[0.8em]
				\texttt{s.j.rey@uva.nl} &\texttt{felicia.schmidt@wu.ac.at} & \texttt{jan.maly@wu.ac.at} \\[0.8em]
\small $^1$Institute for Logic, Language  & \small $^2$Institute for Data, Process &\small $^3$Institute of Logic   \\
\small and Computation (ILLC) & \small and Knowledge Management  & \small and Computation\\
\small University of Amsterdam & \small WU Vienna University & \small TU Wien\\
& \small   of Economics and Business  & 
			\end{tabular}
		\end{center}
	
		\vspace{2em}
		
		\begin{center}
			Last compilation: \today
		\end{center}
	
		\vspace{2em}
	
		\begin{center}
			{\bfseries \large Abstract}\\[1em]
			
			\parbox{0.9\linewidth}{
				In this survey, we review the literature investigating participatory budgeting as a social choice problem. Participatory Budgeting (PB) is a democratic tool aiming at making budgeting decisions in a more democratic manner. Specifically, citizens are asked to vote on how to allocate a given amount of money to a set of projects that can potentially be funded. From a social choice perspective, it corresponds then to the problem of aggregating opinions about which projects should be funded, into a budget allocation satisfying a budget constraint. This problem has received substantial attention in recent years and the literature is growing at a fast pace. In this survey, we present the most important research directions from the literature, each time presenting a large set of representative results. We only focus on the \emph{indivisible} case, that is, PB problems in which projects can either be fully funded or not at all.
				
				\qquad The aim of the survey is to present a comprehensive overview of the state of the research on PB. We aim at providing both a general overview of the main research questions that are being investigated, and formal and unified definitions of the most important technical concepts from the literature.
			}
		\end{center}
	\end{titlepage}
	
	\null\vfill
	
	\begin{center}
		{\bfseries \large Disclaimer}\\[1em]
		
		\parbox{0.8\linewidth}{
			A survey can never be exhaustive since the state of the research keeps changing. This document is intended to be a living document that gets updated every now and then as the literature grows. If you feel that some papers are not presented correctly, or simply missing, feel free to contact us. We will be more than happy to correct it and to acknowledge your contribution.
		}
	\end{center}
	
	\vfill
	
	\begin{center}		
		{\bfseries \large Acknowledgements}\\[1em]
		
		\parbox{0.8\linewidth}{
			We would like to thank all who helped writing and improving this survey. Throughout the process, we have received very valuable feedback from Edith Elkind, Ulle Endriss, Piotr Faliszewski, Martin Lackner, Jannik Peters and Dominik Peters. We are very grateful to all of you.
		}
	\end{center}

	\vfill\null
	
	\tableofcontents
	
	\chapter{Introduction}
	
	
	Participatory budgeting (PB) is an \emph{innovative democratic institution} through which citizens are involved in the decision process of the allocation of public funds \citep{WNT21}. It has originally been developed by politicians in the city of Porto Alegre, Brazil, where it was first implemented in 1989. After years of dictatorship, the hope of the local politicians was to establish Brazil's new representative democracy by increasing the number of administrative mechanisms involving citizens \citep{Aber00}. After this initial success, PB spread to other Brazilian cities, and not long after was used worldwide \citep{Oliv17, Dias18, DEJ19}.
	
	In this document, we survey the scientific literature on dedicated to PB from a (computational) social choice perspective. Before delving into the details, let us mention some generalities, both about PB and social choice theory.
	
	\section{Participatory Budgeting Processes}
	\label{sec:Def_PB}
	
	With the rise of PB processes throughout the world, more and more diverse mechanisms have been implemented under the name of PB, making it hard to provide a clear definition of what actually is a PB process. Instead of providing a direct definition, political scientists usually prefer to characterise PB processes through the properties they satisfy. Following this idea, \citet{SHR08} present five criteria making any budgeting process involving non-elected citizens a PB process:
	\begin{itemize}
		\item It should be about the allocation of scarce resources;
		\item It should involve a public institution (city, district) with an elected body and power over administration and allocation of resources;
		\item It has to be repeated over the years;
		\item It has to allow for public deliberation phases;
		\item It should implement some mechanisms enforcing accountability on the result.
	\end{itemize}
	
	\noindent The above list informs us about the organisational aspects of a PB process. Complementing this approach, \citet{Wamp12} identifies five\footnote{Note that \citet{Wamp12} only discusses \emph{four} core principles. Social inclusion was only added at a later stage \citep[see, \textit{e.g.},][]{WNT21}.} \emph{core} principles that need to be implemented (at least in part) by any PB processes in order to generate social change:
	\begin{itemize}
		\item \textbf{Voice}: citizens are offered a chance to voice their opinions and ideas;
		\item \textbf{Vote}: by voting in the PB process, citizens actively take part in state-sanctioned decision-making processes;
		\item \textbf{Social justice}: areas that are more in need are targeted to achieve a better redistribution of resources;
		\item \textbf{Social inclusion}: traditionally marginalised groups are offered additional opportunities to be represented;
		\item \textbf{Oversight}: citizens are involved at every step of the process to organise it, monitor the implementation of the projects,  and so on.
	\end{itemize}
	
	These two sets of principles and criteria only offer us a general overview of the key components of any PB process, but do not touch on how PB processes are actually implemented. It is once again hard to provide a general description of how a PB process is organised given the multiplicity of actual implementations. Still, researchers have been able to single out several key steps that almost all PB processes follow \citep{Wamp00, Caba04, Shah07}. We present them below.
	
	\begin{itemize}
		\item Regular meetings are held by the municipality to discuss potential projects that could be funded using the available budget. Typically, these projects are proposed by the citizens. 
		\item A shortlist of potential projects is decided upon, usually, by collecting all proposals that are feasible and fit the requirements of the PB process. Additionally, the cost of each possible project is determined, either by experts from the municipality or by the citizens who submitted the project.
		\item Citizens vote on the shortlisted projects to determine which of them will be funded, given the budget constraint.
		\item The municipality reports back to the citizens on the advancement of the actual realisation of the selected projects.
	\end{itemize}
	
	This description of a typical PB process is the one we adopt for this survey. Specifically, when referring to a PB process, we will have in mind a mechanism implementing these steps. Note that we will mainly focus on the voting stage, \textit{i.e.}, the third bullet point of the above.
	
	As should be clear by now, this description only fits most PB processes, but not all. For instance, the above is phrased as if the organising entity was a municipality. This is the typical case; however, the scale of the process can vary significantly: from schools,\footnote{See \href{https://www.participatorybudgeting.org/pb-at-ps139/}{participatorybudgeting.org/pb-at-ps139} for an example of PB within a primary school.} or housing communities,\footnote{See the example of social housing in Scotland for instance: \href{https://sharedfuturecic.org.uk/participatory-budgeting-within-social-housing-ideas-for-better-engaging-with-tenants-and-residents-groups/}{sharedfuturecic.org.uk/participatory-budgeting-within-social-housing-ideas-for-better-engaging-with-tenants-and-residents-groups}.} to neighbourhoods of a city,\footnote{\textit{E.g.}, Amsterdam organises individual PB processes for each district \citep{amspb}.} and even to subnational entities.\footnote{PB processes were organised at the scale of a regional department in Peru \citep{Shah07}.}
	
	It is also interesting to note that not all the processes include a voting stage. Indeed, some PB processes are organised as a simple deliberative mechanism throughout which the set of projects to implement is determined meeting after meeting. This was typically the case for some of the first PB processes implemented in Brazil \citep{Caba04}. Given that we focus on the voting stage of PB processes, we will assume that all PB processes are organised around a voting stage.
	
	\section{A Social Choice Problem}
	
	As should be clear from the typical structure outlined above, several steps of a PB process involve the citizens' participation: first when submitting proposals, and second when voting on the shortlisted projects. In both cases, individual preferences are taken into account to reach a collective decision. This perspective on PB makes it a typical social choice problem. Indeed, social choice theory is the science of collective decision making based on individual preferences \citep{ASS02, ASS11, BCELP16}.
	
	A burgeoning literature on PB has emerged from the (computational) social choice community, focusing predominantly on the voting stage. The aim of this survey is to present the main findings coming from this line of research. It complements the first survey of \citet{AzSh21} which was written before the sharp increase of publications on the topic and only covers the literature until 2019. 
	
	In contrast to the survey of \citet{AzSh21}, we only focus on indivisible participatory budgeting, also called \emph{discrete} PB \citep{AzSh21}, that is, the special case of PB where projects can either be fully funded or not all (projects cannot be fractionally implemented). Within this framework, we present what we believe to be the most important concepts and results. We aim to provide a comprehensive set of definitions and to unify concepts and notations that appeared in different publications. 
	
	\section{Outline of this Survey}
	
	We will first present the basic model and our notations (Chapter~\ref{chap:Preliminaries}). Then we turn to the different ballot formats that have been proposed for PB (Chapter~\ref{chap:Ballot_Design}). Once the design of the ballots will be clarified, we will discuss rules for aggregating said ballots (Chapter~\ref{chap:PB_Rules}). We will then present how to asses the quality of these rules in terms of fairness (Chapter~\ref{chap:Fairness}) and other axiomatic properties (Chapter~\ref{chap:Axiomatic}). After that, we will look at the algorithmic aspects of PB (Chapter~\ref{chap:Algorithmic}). Having discussed the standard model for PB, we will then present variations and extensions of the standard model that have been introduced (Chapter~\ref{chap:Extensions}). We will finally provide interesting pointers to go beyond what we presented in the survey, be it related frameworks or actual implementation of PB in practice (Chapter~\ref{chap:Beyond}). We will conclude this survey by mentioning what we consider to be the most important directions for future work (Chapter~\ref{chap:Conclusion}).

	\chapter{Preliminaries}
	\label{chap:Preliminaries}
	
	Almost the entire computational social choice literature focuses on the voting stage of PB. The only exception we are aware of is the work of \citet{REH21}. The voting stage will also be the main focus of this survey. In the following we introduce the standard model of the voting stage of PB processes. We then try to clarify different related concepts: preferences, utilities, satisfaction and ballots.
	
	\section{The Standard Model of PB}
	\label{subsec:Standard_Model}
	
	The voting stage of a PB process is represented as a tuple of three elements  $I = \tuple{\projSet, c, b}$ called an \emph{instance} where $\projSet = \{p_1, \ldots, p_m\}$ is the \emph{set of projects}; $c: \projSet \rightarrow \mathbb{R}_{> 0}$ is the \emph{cost function}, associating every project $p \in \projSet$ with its cost $c(p) \in \mathbb{R}_{> 0}$; and $b \in \mathbb{R}_{> 0}$ is the \emph{budget limit}. We assume that all projects are feasible, \textit{i.e.}, that $c(p)$ for all $p \in \projSet$. For any subset of projects $P \subseteq \projSet$, we denote by $c(P)$ its total cost $\sum_{p \in P} c(p)$. 
	An instance $I = \tuple{\projSet, c, b}$ is said to have \emph{unit costs} if for every project $p \in \projSet$, we have $c(p) = 1$ and $b \in \mathbb{N}_{> 0}$. These instances are especially interesting because they correspond to multi-winner elections \citep{LaSk23}.
	
	Note here that we make the common assumption\footnote{\citet{MSRE22} is the only paper considering negative costs, which only is relevant since they also consider multi-dimensional costs.} that both the costs and the budget limit have to be positive.
	Importantly, we also consider them to be real---and not rational---numbers. For most results this assumption has no impact and tends to simplify the exposition, however, when it comes to computational complexity it is safer to assume that costs and budget limits are in $\mathbb{Q}$ to avoid having to deal with the representation of real numbers \citep[see][for more details on the topic]{BCSS98}.
	
	Let $\agentSet = \{1, \ldots, n\}$ be the \emph{set of voters} involved in the PB process, these are the citizens participating in the process, not the officials/organisers. When facing an instance $I = \tuple{\projSet, c, b}$, they are asked to submit their preferences over the projects in $\projSet$. They do so by submitting a ballot whose format is determined by the rules of the process. Several \emph{ballot formats} have been considered for PB, as we shall see later. For now, let us denote by $A_i$ the ballot that voter $i \in \agentSet$ is submitting. The vector $\profile = (A_1, \ldots, A_n)$ of the ballots of the voters is called a \emph{profile}. Note that we use the terms voters and agents interchangeably, purely for stylistic reasons.\footnote{The reader may get bored to always read the same terminology all the time.}
	
	The outcome of the voting stage $I = \tuple{\projSet, c, b}$ is a \emph{budget allocation} $\pi \subseteq \projSet$ such that $c(\pi) \leq b$. We will denote by $\allocSet(I)$ the set of all \emph{feasible} budget allocations for instance~$I$, defined as $\allocSet(I) = \{\pi \subseteq \projSet \mid c(\pi) \leq b\}$. A budget allocation $\pi \in \allocSet(I)$ is \emph{exhaustive} if there is no $p \in \projSet \setminus \pi$ such that $c(\pi \cup \{p\}) \leq b$.
	
	Budget allocations are determined using PB rules. A \emph{PB rule} $\pbRule$ is a function taking as input an instance $I$ and a profile $\profile$ and returning a set of feasible budget allocations $\pbRule(I, \profile) \subseteq \allocSet(I)$. PB rules that always return a single budget allocation are called \emph{resolute}. For simplicity, we will denote the output $\{\pi\}$ of a resolute PB rule by just $\pi$. PB rules that are not resolute are called \emph{irresolute}, they thus potentially return several \emph{tied} budget allocations. Unless explicitly stated, we will assume rules to be \emph{resolute}. At a few occasions we will discuss randomised PB rules, which are rules that return for any instance $I$ and profile $\profile$, not a budget allocation, but a probability distribution over $\allocSet(I)$.
	
	In the coming chapters, and particularly in chapters~\ref{chap:Fairness} and~\ref{chap:Axiomatic}, we will introduce several properties of budget allocations. To avoid unnecessary definitions, we will use the exact same properties for rules. For a given property $\mathcal{X}$ of a budget allocation, we say that a rule $\pbRule$ satisfies $\mathcal{X}$ if for every instance $I$ and profile $\profile$, the outcome of the (resolute) rule $\pbRule(I, \profile)$ satisfies~$\mathcal{X}$. When needed, we will explicitly specify how properties of budget allocations are lifted to \emph{irresolute} rules.
	
	\section{The Voters: Preferences, Utility, Satisfaction and Ballots}
	\label{subsec:Voters_Pref_Utility}
	
	Going through the literature on PB, and more generally on computational social choice, it appears that the terms preferences, utility, satisfaction, and ballots are used in a somewhat interchangeable fashion. In the following we suggest exact definitions for each of those, hoping that it will help to clarify and unify the use of these terms.
	
	One distinction that seems important to us is that of the private and public information of the voters. The information submitted by the voters, their \emph{ballots}, is the only information that is publicly available, especially to the decision maker. In no case can the ballots be assumed to represent the internal preference model of the voters. Hopefully, the ballots reflect some aspects of the preferences of the voters, but they cannot be claimed to capture them entirely. This observation is based on the following two main arguments. First, we know that almost none of the rules we are studying prevent voters from rationally behaving strategically, so there is no reason to assume their ballots to be truthful \citep{Gibb73, Satt75, DiLi07, Meir18, Peters18}. Second, even if voters try to vote truthfully, it is debatable whether they would be able to produce a ballot that faithfully represents their true internal preferences due to their \emph{bounded rationality} \citep{Dhillon02,BDST11}. It is therefore questionable to assume that a voter's ballot represents their true preferences, even if voters behave truthfully.
	
	We thus urge researchers to always clarify the assumption they are making about the voters, about their internal state and about how they cast their ballots. To help with that, we present below what we believe to be the best way to use this terminology.
	
	\begin{itemize}
		\item \textbf{Preferences}: The preferences are private information accessible only to the voters themselves, reflecting their views on the possible outcomes of the decision making scenario. Remember from the above that this information may not be accessible in full to the voters (notably because of bounded rationality). In economic theory, it is usually assumed that preferences take the form of \emph{weak} or \emph{incomplete rankings} over the different outcomes \citep{Lewin96}, though other representations of the preferences can be argued for \citep[see, \textit{e.g.},][]{hansson01}. Note that the term ``preferences'' sometimes indicates that the preferences are \emph{ordinal}, \textit{i.e.}, they are based on rankings of the outcomes.
		
		\item \textbf{Utility}: The utility of a voter is a specific type of preferences for which every outcome can be mapped to a specific numerical value. These preferences are sometimes referred to as \emph{cardinal preferences}.
		
		\item \textbf{Satisfaction}: The satisfaction of a voter is often used synonymously with their utility. In computational social choice, it is also often used when ballots do not allow agents to report their full utility functions (because of the limited expressiveness of the ballots). In this case, it represents an approximation of the utility of a voter that would be compatible with the ballot submitted. We shall see a concrete example later in this survey. We claim that it is important to always be clear that such satisfaction functions can at most be \emph{proxies} to the utilities of the agents, and in no case their actual level of satisfaction or utility (even if the ballots would allow voters to submit their full preferences). In the following, we use satisfaction as meaning ``the satisfaction that the decision maker is assuming the voter enjoys''.
		
		\item \textbf{Ballots}: The ballot of an agent is the information they submitted. This information is formatted in the format required by the type of ballots that is being used. Let us emphasise once again that a ballot is the sole information submitted by the (potentially strategically-behaving) voter and not necessarily a representation of their private information.
	\end{itemize}

	In the following we will adopt these definitions and try to use the terms accordingly.
	
	\chapter{Ballot Design}
	\label{chap:Ballot_Design}
	
	Ballot design is an important part of the research on PB. Indeed, the outcome space being combinatorial in nature, the design of the ballots is critical to achieve a good balance between the amount of information elicited and the practical usability of said ballot. To get the maximum amount of information, we would want to offer the possibility for the agents to submit their preferences over all possible budget allocations. These could take the forms of orderings over $\allocSet(I)$, or utility functions associating a score to every feasible budget allocation $\pi \in \allocSet(I)$. This approach clearly cannot be implemented in real life as the size of $\allocSet(I)$ is exponential in the number of projects, which in itself might already be quite large (in 2023 there were 138 projects in the municipal Warsaw PB process\footnote{See the data hosted on \href{http://pabulib.org}{pabulib.org} \citep{SST20} and the specific Warsaw 2023 file: \href{http://pabulib.org/media/files/poland_warszawa_2023_.pb}{poland\_warszawa\_2023\_.pb.}}).
	
	Several ballot formats have then been designed in the pursuit of the best trade-off between the amount of information that is elicited and the usability of the ballot. All of these formats are \emph{project-based} ballots, \textit{i.e.}, the information collected concerns the projects and not the feasible budget allocations. This is mainly because the set of all the feasible budget allocations can be huge. In what follows, we distinguish between cardinal ballots (Section \ref{subsec:Cardinal_Ballots})---that associate a score to each project---and ordinal ballots (Section \ref{subsec:Ordinal_Ballots})---that require agents to rank the projects. We will conclude this section by comparing the different formats (Section \ref{subsec:Comparison_Ballots}) and discussing how to define satisfaction functions based on different ballots (Section \ref{subsec:Satisfaction}).
	
	To get an overview of the different ballot formats that have been introduced and the papers studying them, we present a classification of the papers we have reviewed in Table \ref{tab:summary_paperPerBallots}, based on the ballot format they are considering.
	
	\begin{table}
		\begin{center}
			\begin{tabularx}{\linewidth}{cX}
				\toprule
				\multicolumn{2}{c}{\textbf{Cardinal Ballots}} \\
				\midrule
				
				\textbf{Generic} & 
				\citet{BNPS21} ---
				\citet{CLM22} ---
				\citet{LCG22} ---
				\citet{FBG23} ---
				\citet{FSTW19} ---
				\citet{HKPP21} ---
				\citet{JMW20} ---
				\citet{Laru21}$^\star$ ---
				\citet{LCG22} ---
				\citet{MSW22} ---
				\citet{MSWW22} ---
				\citet{PKL21} ---
				\citet{PPS21}
				\\ \midrule
				
				\textbf{Approval} & 
				\citet{AzGa21} ---
				\citet{AGPSV22} ---
				\citet{ALT18} ---
				\citet{BBH21} ---
				\citet{BBL22} ---
				\citet{BBS20} ---
				\citet{BFLMP23} ---
				\citet{JST20} ---
				\citet{JSTZ21} ---
				\citet{LMR21} ---
				\citet{LCG22} ---
				\citet{MREL23} ---
				\citet{MSRE22} ---
				\citet{REH20} ---
				\citet{REH21} ---
				\citet{SBY22} ---
				\citet{TaFa19}
				\\ \midrule
				
				\textbf{$t$-Approval} &
				\citet{FBG23}
				\\ \midrule
				
				\textbf{Knapsack} & 
				\citet{BNPS21} ---
				\citet{FBG23} ---
				\citet{GKSA19}
				\\ \midrule
				
				\textbf{$t$-Threshold} & 
				\citet{BNPS21} ---
				\citet{FBG23}
				\\ \midrule
				
				\textbf{Cumulative} & 
				\citet{SSST20} \\
				
				\bottomrule \addlinespace[\belowrulesep]
				
				\multicolumn{2}{c}{\textbf{Ordinal Ballots}} \\
				\midrule
				
				\textbf{Strict Orders} & 
				\citet{LuBo11} ---
				\citet{PPS21}
				\\ \midrule
				
				\textbf{Weak Orders} & 
				\citet{AzLe21} ---
				\citet{Laru21}$^\star$
				\\ \midrule
				
				\begin{tabular}{@{}c@{}}
					\textbf{Value-for} \\ \textbf{Money}
				\end{tabular} & 
				\citet{BNPS21} ---
				\citet{GKSA19} ---
				\citet{FBG23}
				\\ \midrule
				
				\textbf{Value} & 
				\citet{BNPS21} ---
				\citet{FBG23}
				\\ \bottomrule
			\end{tabularx}
		\end{center}
		\footnotesize
		$^\star$\citet{Laru21} considers that the agents submit weak and complete rankings over the projects, that are then converted into cardinal scores (via positional scoring functions) for the aggregation.
		
		\caption{Papers studying indivisible PB organised by the type of ballots they consider. We categorised the papers based on the main ballot format used in their study, not necessarily based on \emph{all} the format mentioned in the paper.}
		\label{tab:summary_paperPerBallots}
	\end{table}
	
	\section{Cardinal Ballots}
	\label{subsec:Cardinal_Ballots}
	
	Let us start with cardinal ballots. Loosely speaking, when these ballots are used, agents are asked to submit a score for all projects. Additional constraints are sometimes imposed on the scores. Note that we refer to this ballot format as \emph{cardinal ballots} and not utility functions or cardinal preferences as they are usually called, in line with our discussion in Section \ref{subsec:Voters_Pref_Utility}.
	
	Formally, a \emph{cardinal ballot} $A_i: \projSet \rightarrow \mathbb{R}_{\geq 0}$ for agent $i \in \agentSet$ is a mapping from projects to a non-negative score. Note that in our definition cardinal ballots associate scores to projects and not budget allocations. Of course the definition can easily be adapted to allow voters to submit scores over budget allocations, but since there are almost no papers \citep*[the only potential exception being][]{JST20} working with cardinal ballots over budget allocations, we decided to keep the simpler definition. Importantly, when it comes to computational complexity where the representation of real numbers can be a problem, it can be safer to assume that cardinal ballots map to rational numbers (in $\mathbb{Q}$).
	
	A common assumption \citep*[see, \textit{e.g.},][]{PPS21} is that the score of a budget allocation for an agent is simply the sum of the scores of the projects it contains. We call this the \emph{additivity} assumption.
	
	Even though cardinal ballots can be used as is for PB, several important variations have been introduced that we discuss below.
	
	\begin{illustration}[Cardinal Ballots]
		Consider a PB election with 5 projects (taken from the 2019 PB process Buurtbudget Kleine Wereld in Amsterdam). When asked to submit cardinal ballots over the projects, the voters could be presented the following interface.
		\begin{center}
			\begin{tikzpicture}
				\node[fill=my-gray-veryverylight, draw, rounded corners=5pt] at (0, 0) {
					\begin{tikzpicture}
						\node[anchor = east] at (0, 0) {Sports boot camp for single mothers};
						\node[anchor = east] at (0, -0.8) {Bench on a plank};
						\node[anchor = east] at (0, -1.6) {Cooking classes for children};
						\node[anchor = east] at (0, -2.4) {Theater performance with young people};
						\node[anchor = east] at (0, -3.2) {Dutch bootcamps};
						
						\node[anchor = east] at (2.2, 0) {\euro\;840};
						\node[anchor = east] at (2.2, -0.8) {\euro\;10\,000};
						\node[anchor = east] at (2.2, -1.6) {\euro\;1\,000};
						\node[anchor = east] at (2.2, -2.4) {\euro\;1\,500};
						\node[anchor = east] at (2.2, -3.2) {\euro\;10\,000};
						
						\fill[my-gray-light, rounded corners=1.5pt] (4, 0.05) rectangle (6, -0.05);
						\fill[my-green-light] (4, 0) circle (3pt);
						\node[anchor=west] at (6.2, 0) {\small 0};
						
						\fill[my-gray-light, rounded corners=1.5pt] (4, -0.75) rectangle (6, -0.85);
						\fill[my-green-light] (5, -0.8) circle (3pt);
						\node[anchor=west] at (6.2, -0.8) {\small 50};
						
						\fill[my-gray-light, rounded corners=1.5pt] (4, -1.55) rectangle (6, -1.65);
						\fill[my-green-light] (5.5, -1.6) circle (3pt);
						\node[anchor=west] at (6.2, -1.6) {\small 75};
						
						\fill[my-gray-light, rounded corners=1.5pt] (4, -2.35) rectangle (6, -2.45);
						\fill[my-green-light] (4.2, -2.4) circle (3pt);
						\node[anchor=west] at (6.2, -2.4) {\small 10};
						
						\fill[my-gray-light, rounded corners=1.5pt] (4, -3.15) rectangle (6, -3.25);
						\fill[my-green-light] (5.24, -3.2) circle (3pt);
						\node[anchor=west] at (6.2, -3.2) {\small 62};
					\end{tikzpicture}
				};
			\end{tikzpicture}
		\end{center}
	\end{illustration}
	
	\subsection{Approval Ballots}
	
	When using approval ballots, agents are asked to submit a subset of projects they approve of. We represent approval ballots as cardinal ballots by requiring the score of each project to either be~$0$ or~$1$. For agent $i \in \agentSet$, their approval ballot $A_i: \projSet \rightarrow \{0, 1\}$ is a mapping from $\projSet$ to $\{0, 1\}$, where for any $p \in \projSet$, $A_i(p) = 1$ indicates that agent $i$ approves of project $p$, and $A_i(p) = 0$ that $i$ does not approve of $p$. We will sometimes call voter $i \in \agentSet$ a \emph{supporter} of project $p \in \projSet$ whenever $A_i(p) = 1$.
	
	It is important to state that approval ballots are the most widely used ballot format in real life PB processes. At the same time, and potentially for that exact reason, it is also the most studied format in the literature (see Table~\ref{tab:summary_paperPerBallots}).
	
	One of the main drawbacks of approval ballots is that they are semantically weak: not much information is communicated. In particular, it is unclear what an agent intends to communicate when not approving a project (setting $A_i(p) = 0$ for project $p$). It is notably ambiguous whether this case should be treated as stating a rejection of the project, or simply stating an indifference status regarding the project. One way of circumventing this issue is to enforce additional constraints on the ballots that allow us to interpret them more accurately.
	
	\begin{illustration}[Approval Ballots]
		Consider a PB election with 5 projects (taken from the 2019 PB process in Warszawa, Ursynów). When asked to submit approval ballots over the projects, the voters could be presented the following interface.
		\begin{center}
			\begin{tikzpicture}
				\node[fill=my-gray-veryverylight, draw, rounded corners=5pt] at (0, 0) {
					\begin{tikzpicture}
						\node[anchor = east] at (0, 0) {Ground trampolines in parks};
						\node[anchor = east] at (0, -0.8) {Real Christmas illumination};
						\node[anchor = east] at (0, -1.6) {Sunday healthy food market};
						\node[anchor = east] at (0, -2.4) {Butterfly park};
						\node[anchor = east] at (0, -3.2) {Camera in the bird's nest};
						
						\node[anchor = east] at (2.2, 0) {z\l\;175\,000};
						\node[anchor = east] at (2.2, -0.8) {z\l\;149\,000};
						\node[anchor = east] at (2.2, -1.6) {z\l\;39\,000};
						\node[anchor = east] at (2.2, -2.4) {z\l\;257\,030};
						\node[anchor = east] at (2.2, -3.2) {z\l\;3\,900};
						
						\draw[rounded corners=0] (3, 0.2) rectangle (3.4, -0.2);
						\node at (3.2, 0) {\cmark};
						\draw[rounded corners=0] (3, -0.6) rectangle (3.4, -1);
						\draw[rounded corners=0] (3, -1.4) rectangle (3.4, -1.8);
						\draw[rounded corners=0] (3, -2.2) rectangle (3.4, -2.6);
						\node at (3.2, -2.4) {\cmark};
						\draw[rounded corners=0] (3, -3) rectangle (3.4, -3.4);
						\node at (3.2, -3.2) {\cmark};
					\end{tikzpicture}
				};
			\end{tikzpicture}
		\end{center}
	\end{illustration}
	
	\subsection{Semantically Enriched Approval Ballots}
	
	As explained above, the semantics of approval ballots is not well defined. This leads to various problems and has prompted researchers to introduce some additional constraints on the approval ballots to correct this.
	
	In practice, it is often the case that voters can only approve of a limited number of projects. When asked for \emph{$t$-approval ballots}, agents can only approve up to $t \in \mathbb{N}_{> 0}$ different projects. This is formalised by imposing $\sum_{p \in \projSet}A_i(p) \leq t$ for the ballot $A_i$ of each agent $i \in \agentSet$. We can therefore gain some understanding of the not approved projects: they are not part of the top-$t$ projects of the voter (assuming that voters can actually order the projects based on their preferences).
	
	One important variation of approval ballots, both in theoretical terms and because of its actual usage, is the \emph{knapsack ballot} \citep{GKSA19}. A knapsack ballot is an approval ballot with the additional constraint that the total cost of the approved projects cannot exceed the budget limit $b$. Formally speaking, it is an approval ballot $A_i$ such that $c(\{p \in \projSet \mid A_i(p) = 1\}) \leq b$. Phrasing it differently, when submitting knapsack ballots, agents are asked to provide their most preferred feasible budget allocation. In this sense, knapsack ballots have a clear meaning that can be used to make potentially better decisions.
	
	Another semantically enriched variation of approval ballot are \emph{$t$-threshold approval ballots}  \citep{BNPS21, FBG23}. Here, agents are assumed to have private additive utility functions that they are aware of, and they are asked to submit an approval ballot, approving of a project if and only if it provides them with utility at least $t \in \mathbb{R}$.
	
	\begin{illustration}[Knapsack Ballots]
		Consider a PB election with 5 projects (taken from the 2020 PB process in Katowice, Osiedle Witosa) and a budget of z\l\;535\,658. When asked to submit knapsack ballots over the projects, the voters could be presented the following interface.
		\begin{center}
			\begin{tikzpicture}
				\node[fill=my-gray-veryverylight, draw, rounded corners=5pt] at (0, 0) {
					\begin{tikzpicture}
						\node[anchor = east] at (0, 0) {Improvement of sports and recreational infrastructure};
						\node[anchor = east] at (0, -0.8) {Green acoustic screen};
						\node[anchor = east] at (0, -1.6) {Benches with their own roof};
						\node[anchor = east] at (0, -2.4) {District bike};
						\node[anchor = east] at (0, -3.2) {Asphalt sidewalk};
						
						\node[anchor = east] at (2.2, 0) {z\l\;231,000};
						\node[anchor = east] at (2.2, -0.8) {z\l\;264\,364};
						\node[anchor = east] at (2.2, -1.6) {z\l\;83\,556};
						\node[anchor = east] at (2.2, -2.4) {z\l\;200\,000};
						\node[anchor = east] at (2.2, -3.2) {z\l\;140\,000};
						
						\draw[rounded corners=0] (3, 0.2) rectangle (3.4, -0.2);
						\draw[rounded corners=0] (3, -0.6) rectangle (3.4, -1);
						\draw[rounded corners=0] (3, -1.4) rectangle (3.4, -1.8);
						\node at (3.2, -1.6) {\cmark};
						\draw[rounded corners=0] (3, -2.2) rectangle (3.4, -2.6);
						\node at (3.2, -2.4) {\cmark};
						\draw[rounded corners=0] (3, -3) rectangle (3.4, -3.4);
						\node at (3.2, -3.2) {\cmark};
						
						\fill[my-gray-light, rounded corners=2pt] (-8.7, -3.8) rectangle (0.6, -4.2);
						\fill[my-green-light, rounded corners=2pt] (-8.7, -3.8) rectangle (-1.346, -4.2);
						\node[anchor = west] at (0.8, -4) {423\,556/535\,658 z\l\;};
						\node[anchor=west] at (-8.65, -4) {\scriptsize Budget used};
					\end{tikzpicture}
				};
			\end{tikzpicture}
		\end{center}
	\end{illustration}

	\subsection{Cumulative Ballots}
	
	When using cumulative ballots \citep{SSST20}, agents are asked to distribute a certain amount of money (usually $\nicefrac{b}{n}$, \textit{i.e.}, their share of the budget) over all the projects. Formally, a cumulative ballot $A_i$ is a cardinal ballot such that $\sum_{p \in \projSet} A_i(p) \leq 1$. The idea behind cumulative ballots is that agents control some share of the budget and indicate how they would want to use that share.
	
	Note that one could also assume that $A_i(p)$ represents the fraction of the budget limit $b$ that voter $i$ believes should be allocated to project $p$ (in total). This interpretation, however, does not fit with the assumption that projects are indivisible.
	
	\begin{illustration}[Cumulative Ballots]
		Consider a PB election with 5 projects (taken from the 2019 PB process Oost Begroot IJburg Zeeburgereiland in Amsterdam). When asked to submit cumulative ballots over the projects, the voters could be presented the following interface.
		\begin{center}
			\begin{tikzpicture}
				\node[fill=my-gray-veryverylight, draw, rounded corners=5pt] at (0, 0) {
					\begin{tikzpicture}
						\node[anchor = east] at (0, 0) {Discover and create green oases};
						\node[anchor = east] at (0, -0.8) {The wild bee back on IJburg!};
						\node[anchor = east] at (0, -1.6) {Circular sculpture route};
						\node[anchor = east] at (0, -2.4) {Cake workshop for kids};
						\node[anchor = east] at (0, -3.2) {Urban music event};
						
						\node[anchor = east] at (2.2, 0) {\euro\;10\,305};
						\node[anchor = east] at (2.2, -0.8) {\euro\;40\,000};
						\node[anchor = east] at (2.2, -1.6) {\euro\;12\,000};
						\node[anchor = east] at (2.2, -2.4) {\euro\;6\,190};
						\node[anchor = east] at (2.2, -3.2) {\euro\;47\,779};
						
						\fill[my-gray-light, rounded corners=1.5pt] (4, 0.05) rectangle (6, -0.05);
						\fill[my-green-light] (4, 0) circle (3pt);
						\node[anchor=west] at (6.2, 0) {\small 0};
						
						\fill[my-gray-light, rounded corners=1.5pt] (4, -0.75) rectangle (6, -0.85);
						\fill[my-green-light] (5.24, -0.8) circle (3pt);
						\node[anchor=west] at (6.2, -0.8) {\small 62};
						
						\fill[my-gray-light, rounded corners=1.5pt] (4, -1.55) rectangle (6, -1.65);
						\fill[my-green-light] (4.2, -1.6) circle (3pt);
						\node[anchor=west] at (6.2, -1.6) {\small 10};
						
						\fill[my-gray-light, rounded corners=1.5pt] (4, -2.35) rectangle (6, -2.45);
						\fill[my-green-light] (4.3, -2.4) circle (3pt);
						\node[anchor=west] at (6.2, -2.4) {\small 15};
						
						\fill[my-gray-light, rounded corners=1.5pt] (4, -3.15) rectangle (6, -3.25);
						\fill[my-green-light] (4.1, -3.2) circle (3pt);
						\node[anchor=west] at (6.2, -3.2) {\small 5};
						
						\fill[my-gray-light] (-5.2, -3.8) rectangle (5.3, -4.2);
						\fill[my-green-light] (-5.2, -3.8) rectangle (4.46, -4.2);
						\node[anchor = west] at (5.5, -4) {92/100};
						\node[anchor=west] at (-5.2, -4) {\scriptsize Points used};
					\end{tikzpicture}
				};
			\end{tikzpicture}
		\end{center}
	\end{illustration}
	
	\section{Ordinal Ballots}
	\label{subsec:Ordinal_Ballots}
	
	The second main category of ballots that have been studied for PB are ordinal ballots. In this context, the ballot of an agent is an ordering over the projects. Formally, agent $i$'s ballot $A_i$ is a strict linear order over $\projSet$. We will typically denote it by $\succ_i$ where for two projects $p, p' \in \projSet$, $p \succ_i p'$ indicates that agent $i$ prefers $p$ over $p'$.
	
	Ordinal ballots can be used as is for aggregation purposes. However, because projects have different costs, the exact semantics of the ordering is not always clear. Several specific ways of ranking the projects have thus been proposed.
	
	When submitting \emph{ranking-by-value ballots} \citep{BNPS21}, agents are assumed to provide a strict total order over the projects such that a project $p$ is ranked above another project $p'$ if and only $p$ is preferred to $p'$.
	
	Similarly, \emph{ranking by value-for-money ballots} \citep{GKSA19} requires agents to provide ranking of the projects based on their value for money. Note that this is only well-defined when agents are assumed to have private utility functions that they are aware of.
	
	We have only mentioned \emph{strict} rankings above, but weak rankings have also been considered \citep{AzLe21}. A weak ranking will typically be denoted by $\succsim$ with $\succ$ being the strict part of the ranking and $\sim$ the indifference relation, defined as $p \succ p'$ if $p \succsim p'$ but not $p' \succsim p$; and $p \sim p'$ if $p \succsim p'$ and $p' \succsim p$, for any two projects $p$ and $p'$. Of course rankings by value or value for money can be considered either as strict or weak rankings.
	
	Finally, it is worth mentioning that in practice voters are only asked to submit incomplete ordinal ballots, typically ranking a small number of projects. This type of ballot was favoured by voters in a study carried out by \citet{YHP+24} which we will discuss in more detail in Section~\ref{subsec:experimental_comparison_ballots}. We are, however, not aware of any theoretical work studying this ballot format, which are often called $t$-top-truncated ballots \citep{BFLR12} and which \citet{YHP+24} called \emph{choose $t$ and rank}.
	
	\begin{illustration}[Ordinal Ballots]
		Consider a PB election with 5 projects (taken from the 2019 PB process in Kraków, Bieżanów-Prokocim). When asked to submit ordinal ballots over the projects, the voters could be presented the following interface, accompanied by a suitable prompt depending on the type of rankings that is expected (by value, ...).
		\begin{center}
			\begin{tikzpicture}
				\node[fill=my-gray-veryverylight, draw, rounded corners=5pt] at (0, 0) {
					\begin{tikzpicture}
						\node[anchor = east] at (0, 0) {Renovation of sidewalks and revitalization};
						\node[anchor = east] at (0, -0.8) {Safe pedestrian crossing};
						\node[anchor = east] at (0, -1.6) {Outdoor gym};
						\node[anchor = east] at (0, -2.4) {Construction of a playground with a safe surface};
						\node[anchor = east] at (0, -3.2) {Self-defense and assertiveness training for girls};
						
						\node[anchor = east] at (1.8, 0) {z\l\;80\,000};
						\node[anchor = east] at (1.8, -0.8) {z\l\;30\,000};
						\node[anchor = east] at (1.8, -1.6) {z\l\;58\,000};
						\node[anchor = east] at (1.8, -2.4) {z\l\;80\,000};
						\node[anchor = east] at (1.8, -3.2) {z\l\;10\,440};
						
						\foreach \x [count = \y] in {0, -0.8, -1.6, -2.4, -3.2}{
							\node[draw] at (3.8, \x) {\y};
							\ifnum\pdfstrcmp{\y}{1}=0 \else
								\node[draw, fill=my-gray-light, rounded corners=3pt, inner sep=2pt] at (3, \x) {
									\begin{tikzpicture}
										\fill[my-gray-dark] (0, 0) -- (0.8, 0) -- (0.4, 0.4) -- cycle;
									\end{tikzpicture}
								};
							\fi
							\ifnum\pdfstrcmp{\y}{5}=0 \else
								\node[draw, fill=my-gray-light, rounded corners=3pt, inner sep=2pt] at (4.6, \x) {
									\begin{tikzpicture}[inner sep=0, outer sep=0]
										\fill[my-gray-dark] (0, 0) -- (0.8, 0) -- (0.4, -0.4) -- cycle;
									\end{tikzpicture}
								};
							\fi
						}
					\end{tikzpicture}
				};
			\end{tikzpicture}
		\end{center}
	\end{illustration}
	
	\section{Comparison of Ballot Formats}
	\label{subsec:Comparison_Ballots}
	
	Comparing the merits of different ballot formats is not an easy task. Two approaches have been explored in the literature focusing either on theoretical or empirical results.
	
	\begin{table}[t]
		\begin{center}
			\resizebox{\linewidth}{!}{
				\begin{tabular}{ccccc}
					\toprule
					& \multicolumn{2}{c}{\textbf{Deterministic}} & \multicolumn{2}{c}{\textbf{Randomised}} \\
					& \multicolumn{2}{c}{\textbf{Distortion}} & \multicolumn{2}{c}{\textbf{Distortion}} \\
					\textbf{Bound}
					& Lower & Upper & Lower & Upper \\
					\midrule
					\textbf{Knapsack} & $\Omega(\nicefrac{2^m}{\sqrt{m}})$ & $\bigO(m \cdot 2^m)$ & $\Omega(m)$ & $m$ \\
					\textbf{Rankings by Value} & $\Omega(m^2)$ & $\bigO(m^2)$ & $\Omega(\sqrt{m})$ & $\bigO(\sqrt{m} \cdot \log(m))$ \\
					\textbf{Rankings by Value-for-Money} & \multicolumn{2}{c}{Unbounded} & $\Omega(\sqrt{m})$ & $\bigO(\sqrt{m} \cdot \log(m))$ \\
					\textbf{Det. $t$-Threshold Approval}$^\star$ & \multicolumn{2}{c}{Unbounded} & $\Omega(\sqrt{m})$ & $m$ \\
					\textbf{Rand. $t$-Threshold Approval}$^\star$ & \multicolumn{2}{c}{---} & $\Omega(\nicefrac{\log(m)}{\log\log(m)})$ & $\bigO(\log^2(m))$ \\
					\bottomrule
				\end{tabular}
			}
		\end{center}
		\footnotesize
		$^\star$For $t$-threshold approval ballots, \citet{BNPS21} distinguish between two cases. In the deterministic case (Det.) the threshold $t$ is chosen arbitrarily by the decision maker once for all the agents. In the randomised (Rand.) case, for each agent, the threshold $t$ is sampled at random from a given distribution. Note that this distinction makes little sense in the deterministic case.
		
		\caption{Summary of the results on the distortion of some of the ballot formats obtained by \citet{BNPS21}. The deterministic distortion corresponds to the situation where only deterministic PB rules are considered. In the randomised distortion setting, randomised PB rules are also considered.}
		\label{tab:summary_distortionBallots}
	\end{table}
	
	\subsection{Comparison via Distortion}
	
	One way to compare different ballot formats is via the \emph{distortion} \citep{PrRo06} they induce. It is a measure of the amount of information communicated by a ballot format for the purpose of identifying a budget allocation that maximises utilitarian social welfare. Specifically, under the assumption that agents have cardinal preferences, the distortion of a ballot format measures the ratio between the maximum social welfare achievable in the knowledge of the full preferences of the agents, to the maximal social welfare achievable when agents submit their ballots according to the specific format. 
	
	\citet{BNPS21} provide a complete analysis of the distortion induced by four of the ballot formats we introduced: knapsack and $t$-threshold approval ballots, rankings by value and rankings by value for money. Table \ref{tab:summary_distortionBallots} presents their findings. Note that they also complemented their theoretical approach with an empirical one on real-life data. Their findings suggest that approval ballots, and more specifically knapsack ballots, may not be the best ballot format when it comes to PB.\footnote{The intuition as to why knapsack ballots do not behave well with respect to distortion is that in the worst case, when all projects cost exactly the budget limit $b$, knapsack ballots only elicit the favourite project of each agent, and it is well understood that this information alone is not enough to make a high-quality decision.}

	\begin{figure}
		\centering
		\begin{tikzpicture}
			
			\node[anchor = north east] at (0, 0) {
				\begin{tikzpicture}
					\node[anchor = east] at (0, 0.175) {Rank Value-for-Money};
					\node[anchor = east] at (0, -0.375) {Cardinal Ballot};
					\node[anchor = east] at (0, -0.925) {$t$-Threshold Approval};
					\node[anchor = east] at (0, -1.475) {Rank Value};
					\node[anchor = east] at (0, -2.025) {Knapsack Ballot};
					\node[anchor = east] at (0, -2.575) {$t$-Approval};
				\end{tikzpicture}
			};
			
			\fill[my-gray-verylight] (-4.24, -0.645) rectangle (10.25, -0.655);
			\fill[my-gray-verylight] (-4.24, -1.195) rectangle (10.25, -1.205);
			\fill[my-gray-verylight] (-4.24, -1.745) rectangle (10.25, -1.755);
			\fill[my-gray-verylight] (-4.24, -2.295) rectangle (10.25, -2.305);
			\fill[my-gray-verylight] (-4.24, -2.845) rectangle (10.25, -2.855);
			
			\node[anchor = north west, label={[align = center]below:Voting Time\\in Seconds}] at (0, 0) {
				\begin{tikzpicture}
					\fill[ColorBlind1] (0, 0) rectangle (3, 0.35);
					\node[anchor = east, white] at (3, 0.175) {\footnotesize 780};
					\fill[ColorBlind2] (0, -0.55) rectangle (1.75, -0.2);
					\node[anchor = east, white] at (1.75, -0.375) {\footnotesize 454};
					\fill[ColorBlind3] (0, -1.1) rectangle (1.70, -0.75);
					\node[anchor = east, white] at (1.70, -0.925) {\footnotesize 443};
					\fill[my-grass-dark] (0, -1.65) rectangle (1.66, -1.3);
					\node[anchor = east, white] at (1.66, -1.475) {\footnotesize 431};
					\fill[my-purple-light] (0, -2.2) rectangle (1.48, -1.85);
					\node[anchor = east, white] at (1.48, -2.025) {\footnotesize 386};
					\fill[my-cyan-light] (0, -2.75) rectangle (1.33, -2.4);
					\node[anchor = east, white] at (1.33, -2.575) {\footnotesize 345};
				\end{tikzpicture}
			};
			
			\node[anchor = north west, label={[align = center]below:Reported\\Ease of Use}] at (3.5, 0) {
				\begin{tikzpicture}
					\fill[ColorBlind1] (0, 0) rectangle (2.4, 0.35);
					\node[anchor = east, white] at (2.4, 0.175) {\footnotesize 3.36};
					\fill[ColorBlind2] (0, -0.55) rectangle (2.91, -0.2);
					\node[anchor = east, white] at (2.91, -0.375) {\footnotesize 4.07};
					\fill[ColorBlind3] (0, -1.1) rectangle (2.77, -0.75);
					\node[anchor = east, white] at (2.77, -0.925) {\footnotesize 3.88};
					\fill[my-grass-dark] (0, -1.65) rectangle (2.81, -1.3);
					\node[anchor = east, white] at (2.81, -1.475) {\footnotesize 3.94};
					\fill[my-purple-light] (0, -2.2) rectangle (2.86, -1.85);
					\node[anchor = east, white] at (2.86, -2.025) {\footnotesize 4.00};
					\fill[my-cyan-light] (0, -2.75) rectangle (3, -2.4);
					\node[anchor = east, white] at (3, -2.575) {\footnotesize 4.2};
				\end{tikzpicture}
			};

			\node[anchor = north west, label={[align = center]below:Reported\\Expressiveness}] at (7, 0) {
				\begin{tikzpicture}
					\fill[ColorBlind1] (0, 0) rectangle (2.65, 0.35);
					\node[anchor = east, white] at (2.65, 0.175) {\footnotesize 3.72};
					\fill[ColorBlind2] (0, -0.55) rectangle (2.91, -0.2);
					\node[anchor = east, white] at (2.91, -0.375) {\footnotesize 4.08};
					\fill[ColorBlind3] (0, -1.1) rectangle (2.89, -0.75);
					\node[anchor = east, white] at (2.89, -0.925) {\footnotesize 4.05};
					\fill[my-grass-dark] (0, -1.65) rectangle (2.94, -1.3);
					\node[anchor = east, white] at (2.94, -1.475) {\footnotesize 4.12};
					\fill[my-purple-light] (0, -2.2) rectangle (2.96, -1.85);
					\node[anchor = east, white] at (2.96, -2.025) {\footnotesize 4.15};
					\fill[my-cyan-light] (0, -2.75) rectangle (3, -2.4);
					\node[anchor = east, white] at (3, -2.575) {\footnotesize 4.21};
				\end{tikzpicture}
			};
		\end{tikzpicture}
		
		\caption{Some of the experimental findings of \citet{FBG23} comparing different ballot formats. The voting time column indicates the time in seconds it took participants to submit their opinion for each ballot format. The reported ease of use and expressiveness columns represents the average value reported by the participants about the ease of use and the expressiveness of each ballot format, on a scale from 1 to 5 (the higher the better). The figures have been reproduced with the authorisation of the authors, using the data available in the GitHub repository \href{https://github.com/rfire01/Participatory-Budgeting-Experiment}{github.com/rfire01/Participatory-Budgeting-Experiment}.}
		\label{fig:experimental_comparison_ballots}
	\end{figure}
	
	\subsection{Comparison via Real-Life Experiments}\label{subsec:experimental_comparison_ballots}
	
	Another approach to compare ballot formats for PB is to run experiments with human participants that use different formats. This is the approach that \citet{FBG23}, \citet{YHP+24} and \citet{GG24} followed. \citet{FBG23} recruited 1800 participants on Amazon Mechanical Turk who were then asked to cast their ballot in a format which was selected from a set of 6 for a specific PB instance (selected from a set of 4 instances). For each participant, the time they needed to vote was measured. Additionally, they asked the participants to self-report on the ease of use of the different formats. 
	
	Some of the findings from \citet{FBG23} are presented in Figure~\ref{fig:experimental_comparison_ballots}. They studied the following ballot formats: generic cardinal ballots, 5-approval ballots, knapsack ballots, 10-threshold approval ballots, rankings by value and rankings by value for money. Summarising, all the ballot formats they study require a similar amount of time for the participants to cast, except for ranking by value-for-money for which participants take significantly longer. The results are the same for the self-reported measures. Notably, for all measures $t$-approval ballots outperform all the other ballot formats, though not by a large margin.

	\citet{YHP+24} had 180 participants, all of them students, take part in a fictional PB setting. The goal, among other things, was to understand how ballot design influences the perceived ease and fairness of the vote. The participants voted on the same projects using the following ballot types: approval ballots, 5-approval ballots, 5-top-truncated ballots, cumulative ballots with 5 points, cumulative ballots with 10 points, and the distribution of 10 points across 5 approved projects. Notably, participants' votes were generally consistent across ballot types. Participants where then asked to mark the easiness and expressiveness of each ballot type as well as rank them by recommendation. The recommendation of a ballot type was statistically highly influenced by the perceived expressiveness of the ballot, while the ease of vote seemed less influential. This, in combination with a perceived lack of clarity in approval ballots, can give an explanation as to why 5-top-truncated ballots were a clear favourite.

	\citet{GG24}, using data from the Stanford Participatory Budgeting Platform\footnote{\href{https://pbstanford.org/}{pbstanford.org}}, showed that while ballot complexity increases the time voters spend on their ballot, it is not correlated to a higher degree of vote abandonment. Furthermore, they investigated how the chosen ballot design influences the cost of projects in the winning outcome, concluding that for most elections, using $t$-approval ballots led to significantly more expensive projects in the outcome than using knapsack ballots.

	\section{Ballot-Based Satisfaction}
	\label{subsec:Satisfaction}
	
	Before we consider how to use the ballots to determine budget allocations through PB rules, let us discuss how to model satisfaction based on the different ballot formats we have introduced. Many of the concepts that we will introduce in the rest of this paper rely on measures of satisfaction.
	
	\subsection{Generic Cardinal Ballots}
	
	When asked for cardinal ballots, voters are asked to report their satisfaction level for each project. There is thus no need to consider anything other than the ballot, at least as long as we are under the \emph{additivity} assumption. This means the satisfaction of a voter is the sum of the score they submitted for the projects that have been selected.
	
	\subsection{Approval Ballots}
	
	When it comes to approval ballots, there is no obvious way to define a measure of the satisfaction of a voter. \citet{BFLMP23} introduced the concept of \emph{approval-based satisfaction functions}, which are functions translating a budget allocation into a satisfaction level for the agents, given their approval ballots. Let us provide their definition.
	
	\begin{definition}[Approval-Based Satisfaction Functions]
		Given an instance $I = \tuple{\projSet, c, b}$ and a profile $\profile$, an \emph{(approval-based) satisfaction function} is a mapping $\satisfaction: 2^\projSet \rightarrow \mathbb{R}_{\geq 0}$ satisfying the following two conditions:
		\begin{itemize}
			\item $\satisfaction(P) \geq \satisfaction(P')$ for all $P, P' \subseteq \projSet$ such that $P \supseteq P'$: the satisfaction is inclusion-monotonic;
			\item $\satisfaction(P) = 0$ if and only if $P = \emptyset$: the satisfaction is zero only for the empty set.
		\end{itemize}
		
		\noindent The satisfaction of agent $i \in \agentSet$ for a budget allocation $\pi \in \allocSet(I)$ is defined as:
		\[\satisfaction_i(\pi) = \satisfaction(\{p \in \pi \mid A_i(p) = 1\}).\]
	\end{definition}
	
	\noindent Note that in contrast to the case of cardinal ballots, satisfaction functions are not generally assumed to be additive. However, we will sometimes make this assumption, \textit{i.e.}, requiring that $\satisfaction(P) = \sum_{p \in P} \satisfaction(\{p\})$ for any $P \subseteq \projSet$.
	
	One might wonder what the difference between an approval profile together with a satisfaction function $\satisfaction$, and a cardinal profile is. Assuming $\satisfaction$ is additive, an approval profile with a satisfaction function is a special case of a cardinal profile in which every agent approving a project $p$ has the same satisfaction for $p$. This is a natural assumption, given the limited information about the voters' preferences. However, some authors have proposed to model the satisfaction of voters in a way that also takes additional information into account, for example the non-approved projects in the winning bundle. This cannot be modelled with a satisfaction function as defined by \citet{BFLMP23}. See the discussion below for more details.
	
	\medskip
	
	Several satisfaction functions have been introduced in the literature, we define them below.
	
	\begin{itemize}
		\item \textbf{Cardinality Satisfaction Function} \citep{TaFa19}: measures the satisfaction of the voters as the number of selected and approved projects:
		\[\cardSatisfaction(P) = |P|.\]
		\item \textbf{Cost Satisfaction Function} \citep{TaFa19}: measures the satisfaction of the voters as the cost of the selected and approved projects:
		\[\costSatisfaction(P) = c(P).\]
		Note that with indivisible projects, this is equivalent to the \emph{overlap satisfaction function} of \citet{GKSA19}.
		\item \textbf{Chamberlin-Courant Satisfaction Function} \citep{TaFa19}: measures the satisfaction of the voters as being 1 if at least one approved project was selected, and 0 otherwise:
		\[\satisfaction^{CC}(P) = \mathds{1}_{P \neq \emptyset}.\]
		\item \textbf{Share} \citep{LMR21}: measures the resources the decision maker used to satisfy the voters:
		\[\satisfaction^{\share}(P) = \sum_{p \in P} \frac{c(p)}{|\{i \in \agentSet \mid A_i(p) = 1\}|}.\]
		It is important to keep in mind that the share has not been introduced as a satisfaction function but can still be interpreted as one (while being cautious as to how to use it).
		\item \textbf{Square Root and Log Satisfaction Functions} \citep{BFLMP23}: measures the satisfaction of the voters as (marginally) diminishing when the cost of a project increases:
		\[\satisfaction^{\sum\sqrt{~}}(P) = \sum_{p \in P} \sqrt{c(p)} \qquad\qquad  \satisfaction^{\sum\ln}(P) = \sum_{p \in P} \ln(1 + c(p)).\]
		Note that we could also consider satisfaction functions that implement global marginal diminishing satisfaction:
		\[\satisfaction^{\sqrt{~}}(P) = \sqrt{c(P)}  \qquad\qquad  \satisfaction^{\ln}(P) = \ln(1 + c(P)).\]
		
	\end{itemize}

	\begin{illustration}[Approval-Based Satisfaction Functions]
		Consider the instance $I$ and the approval ballot $A$ summarized in the following table.
		\begin{center}
			\begin{tabular}{rcccccc}
				\toprule
				& $p_1$ & $p_2$ & $p_3$ & $p_4$ & $p_5$ & $p_6$ \\
				\midrule
				Cost & 1 & 2 & 4 & 10 & 20 & 25 \\
				\midrule
				$A$ & \cmark & \cmark & \cmark & \cmark & \cmark & \cmark \\
				\midrule
				\multicolumn{7}{c}{$b = 30$} \\
				\bottomrule
			\end{tabular}
		\end{center}
		Below, we present the relative positions of some budget allocations based on several satisfaction functions, on an axis going from the smallest achievable satisfaction to the highest. We are considering three (exhaustive) budget allocations: ${\color{ColorBlind1}\pi} = \{p_4, p_5\}$, ${\color{ColorBlind2}\gamma} = \{p_1, p_2, p_6\}$, and ${\color{ColorBlind4}\delta} = \{p_1, p_2, p_3, p_4\}$.
		\begin{center}
			\begin{tikzpicture}
				\node[anchor=east] at (0, 0) {
					\begin{tikzpicture}
						\node[anchor=east] at (-0.5, 0) {$\cardSatisfaction$};
						\fill[my-gray-light, rounded corners=1.5pt] (0, 0.05) rectangle (5, -0.05);
						\fill[ColorBlind1] (2.5, 0) circle (5pt);
						\node[anchor = center] at (2.5, 0) {\footnotesize \color{white} $\pi$};
						\fill[ColorBlind2] (3.75, 0) circle (5pt);
						\node[anchor = center] at (3.75, 0) {\footnotesize \color{white} $\gamma$};
						\fill[ColorBlind4] (5, 0) circle (5pt);
						\node[anchor = center] at (5, 0) {\footnotesize \color{white} $\delta$};
					\end{tikzpicture}
				};
				
				\node[anchor=east] at (0, -0.8) {
					\begin{tikzpicture}
						\node[anchor=east] at (-0.5, 0) {$\costSatisfaction$};
						\fill[my-gray-light, rounded corners=1.5pt] (0, 0.05) rectangle (5, -0.05);
						\fill[ColorBlind4] (2.83333, 0) circle (5pt);
						\node[anchor = center] at (2.83333, 0) {\footnotesize \color{white} $\delta$};
						\fill[ColorBlind2] (4.6666, 0) circle (5pt);
						\node[anchor = center] at (4.6666, 0) {\footnotesize \color{white} $\gamma$};
						\fill[ColorBlind1] (5, 0) circle (5pt);
						\node[anchor = center] at (5, 0) {\footnotesize \color{white} $\pi$};
					\end{tikzpicture}
				};
				
				\node[anchor=east] at (0, -1.6) {
					\begin{tikzpicture}
						\node[anchor=east] at (-0.5, 0) {$\satisfaction^{\sqrt{~}}$};
						\fill[my-gray-light, rounded corners=1.5pt] (0, 0.05) rectangle (5, -0.05);
						\fill[ColorBlind4] (3.763863, 0) circle (5pt);
						\node[anchor = center] at (3.763863, 0) {\footnotesize \color{white} $\delta$};
						\fill[ColorBlind2] (4.830458, 0) circle (5pt);
						\node[anchor = center] at (4.830458, 0) {\footnotesize \color{white} $\gamma$};
						\fill[ColorBlind1] (5, 0) circle (5pt);
						\node[anchor = center] at (5, 0) {\footnotesize \color{white} $\pi$};
					\end{tikzpicture}
				};
				
				\node[anchor=east] at (0, -2.4) {
					\begin{tikzpicture}
						\node[anchor=east] at (-0.5, 0) {$\satisfaction^{\sum\sqrt{~}}$};
						\fill[my-gray-light, rounded corners=1.5pt] (0, 0.05) rectangle (5, -0.05);
						\fill[ColorBlind2] (4.855784, 0) circle (5pt);
						\node[anchor = center] at (4.855784, 0) {\footnotesize \color{white} $\gamma$};
						\fill[ColorBlind4] (4.96206, 0) circle (5pt);
						\node[anchor = center] at (4.96206, 0) {\footnotesize \color{white} $\delta$};
						\fill[ColorBlind1] (5, 0) circle (5pt);
						\node[anchor = center] at (5, 0) {\footnotesize \color{white} $\pi$};
					\end{tikzpicture}
				};
				
				\node[anchor=east] at (0, -3.2) {
					\begin{tikzpicture}
						\node[anchor=east] at (-0.5, 0) {$\satisfaction^{\ln}$};
						\fill[my-gray-light, rounded corners=1.5pt] (0, 0.05) rectangle (5, -0.05);
						\fill[ColorBlind4] (4.208477, 0) circle (5pt);
						\node[anchor = center] at (4.208477, 0) {\footnotesize \color{white} $\delta$};
						\fill[ColorBlind2] (4.9028951, 0) circle (5pt);
						\node[anchor = center] at (4.9028951, 0) {\footnotesize \color{white} $\gamma$};
						\fill[ColorBlind1] (5, 0) circle (5pt);
						\node[anchor = center] at (5, 0) {\footnotesize \color{white} $\pi$};
					\end{tikzpicture}
				};
				
				\node[anchor=east] at (0, -4) {
					\begin{tikzpicture}
						\node[anchor=east] at (-0.5, 0) {$\satisfaction^{\sum\ln}$};
						\fill[my-gray-light, rounded corners=1.5pt] (0, 0.05) rectangle (5, -0.05);
						\fill[ColorBlind2] (4.354005, 0) circle (5pt);
						\node[anchor = center] at (4.354005, 0) {\footnotesize \color{white} $\gamma$};
						\fill[ColorBlind1] (4.692473, 0) circle (5pt);
						\node[anchor = center] at (4.692473, 0) {\footnotesize \color{white} $\pi$};
						\fill[ColorBlind4] (5, 0) circle (5pt);
						\node[anchor = center] at (5, 0) {\footnotesize \color{white} $\delta$};
					\end{tikzpicture}
				};
			\end{tikzpicture}
		\end{center}
		It is interesting to see how the ordering of the budget allocations, and the magnitude of the difference in satisfaction  varies from one satisfaction function to the other.
	\end{illustration}

	Both the cardinality and the cost satisfaction are quite standard within the literature, even though they can easily be criticised: there is no good reason to assume that the satisfaction of an agent is the same for two projects, one being very expensive while the other being particularly cheap; though it is also not sensible to assume a perfect correlation between the cost of a project, and the satisfaction voters derive from it.
	
	In general, all the above apply seamlessly to all approval-like ballots ($t$-approval, knapsack, $t$-threshold...). Some satisfaction functions are however more meaningful with some ballots than with others. In particular, additional satisfaction functions could be interesting to study when using semantically richer approval ballots.
	
	It is also worth noting that \citet{BFLMP23} presented results that apply to whole classes of satisfaction functions, and not just functions from the list above. Some of these results will be presented later (notably in sections~\ref{subsec:Justified_Representation} and~\ref{subsec:Priceability}).
	
	\medskip
	
	Finally, note that a satisfaction function as defined by \citet{BFLMP23} only depends on the projects in the winning bundle that the voter approved. Therefore, these functions cannot capture satisfaction functions that also depend on the non-approved projects in the winning bundle or on the approved projects that have not been funded. 
	
	\citet{GKSA19} introduce a measure of the dissatisfaction of the voters in terms of the $L_1$ distance between a given budget allocation and their ballot. This cannot be modelled by approval-based satisfaction (even though they introduce it in a framework with knapsack ballots) as the satisfaction of a voters depends on projects that are outside of the selected and approved ones. It is important to keep in mind however that the authors deem this measure of satisfaction to be of very limited relevance when the projects are indivisible.
	
	\citet{LMR21} proposed the notion of \emph{relative satisfaction}, which normalises the satisfaction of a voter by the maximum satisfaction achievable:
	\[\mathit{relsat}_{\satisfaction}(P) = \frac{\satisfaction(P)}{\max \{\satisfaction(P') \mid P' \in \allocSet(I) \text{ and } A_i(p) = 1, \forall p \in P'\}},\]
	where $\satisfaction$ is any satisfaction function. \citet{LMR21} only considered relative satisfaction associated with the cost satisfaction function $\costSatisfaction$. This can also not be modelled as an approval-based satisfaction function, as it depends on the full approval ballot of the voter.

	\subsection{Ordinal Ballots}
	
	To measure satisfaction with ordinal ballots, one can associate each project in the ordering with a given satisfaction level. This is usually done through \emph{positional scoring functions} that associate each project with a score that only depends on the position of the project in the ranking. That is the approach followed by \citet{Laru21} for instance.
	
	Satisfaction with ordinal ballots can also be defined in more general terms (not simply mapping projects to scores). For instance, \citet{AzLe21} compare sets of projects according to the cost of the projects ranked above a certain threshold, where the threshold is context-dependent. Note that this assumption is never explicitly stated and that this reflects our understanding of their definitions.

	\chapter{Participatory Budgeting Rules}
	\label{chap:PB_Rules}
	
	We have seen many ways of collecting the opinion of the voters. The next natural step is thus to use that information to select ``good'' budget allocations. This is done through the use of PB rules. In this section we will present the main rules that have been introduced in the literature. Note that in what follows, and in almost the entirety of the paper, we will mainly focus on cardinal and approval ballots.
	
	Our exposition will start with welfare-maximising rules (Section \ref{subsec:Welfare_Rules}). Thereafter, we will discuss three rules based on the idea of finding budget allocations that spread the cost of the selected projects nicely among the voters: the sequential Phragmén rule (Section \ref{subsec:Sequential_Phragmen}), the maximin support rule (Section \ref{subsec:MaximinSupport}), and the method of equal shares (Section \ref{subsec:MES}). A brief overview of the other rules that have been introduced in PB will conclude this part of our survey (Section \ref{subsec:Other_Rules}).
	
	\section{Welfare-Based Rules}
	\label{subsec:Welfare_Rules}
	
	In a purely utilitarian view, agents are assumed to have cardinal preferences over budget allocations and the aim is to select a budget allocation that maximises the overall utility of the agents. That is, utilitarian rules aim to achieve high utilitarian social welfare, where the utilitarian social welfare---which we denote by \utilSW{}---is defined as follows: for a given instance $I = \tuple{\projSet, c, b}$, budget allocation $\pi \in \allocSet(I)$ and a utility function $\mu_i: 2^\projSet \to \mathbb{R}_{\geq 0}$ for every agent $i \in \agentSet$:
	\[\utilSW(I, (\mu_i)_{i \in \agentSet}, \pi) = \sum_{i \in \agentSet} \mu_i(\pi).\]
	Here, $\mu_i(\pi)$ denotes the utility of agent $i$ for allocation $\pi$. As already mentioned, the decision maker does not have access to the utility of the agents, so welfare-maximising rules have to be defined in terms of the assumed satisfaction of an agent given their ballot.
	
	\subsection{Exact Welfare Maximisation}
	
	We start with rules that select outcomes maximising the social welfare.
	
	\subsubsection{Welfare Maximisers with Cardinal Ballots}
	
	When using cardinal ballots we usually assume that the satisfaction of an agent is equivalent to their cardinal ballot. Therefore the above definition directly induces a PB rule if, in a slight abuse of notation, we equate the ballot of a voter with their utility: for a given $I$ and $\profile$, select the budget allocation that maximises \utilSW:\footnote{Note that even though the signature of the functions may look the same, there is a clear conceptual difference between the social welfare defined with utility functions, and \utilSW{} for cardinal ballots: the former uses private information of the voters, while the latter is only defined with respect to public information provided by the voters.}
	\[\utilSW(I, \profile, \pi) = \sum_{i \in \agentSet} \sum_{p \in \pi} A_i(p).\]
	This measures the total satisfaction of the voters (assuming additivity for the cardinal ballots).
	
	Given an instance $I$ and a profile $\profile$, selecting budget allocations $\pi \in \allocSet(I)$ that maximise $\utilSW(I, \profile, \pi)$ defines a rule, the utilitarian welfare maximising rule for cardinal ballots.
	
	\begin{illustration}[Utilitarian Welfare Maximiser for Cardinal Ballots]
		Consider the instance $I$ and the profile $\profile$ of cardinal ballots as depicted in the table below.
		\begin{center}
			\begin{tabular}{rcccccc}
				\toprule
				& $p_1$ & $p_2$ & $p_3$ & $p_4$ & $p_5$ & $p_6$ \\
				\midrule
				Cost & 2 & 3 & 3 & 10 & 15 & 23 \\
				\midrule
				$A_1$ & 7 & 31 & 43 & 17 & 2 & 4 \\
				$A_2$ & 17 & 37 & 35 & 14 & 17 & 8 \\
				$A_3$ & 17 & 44 & 8 & 5 & 32 & 16 \\
				$A_4$ & 25 & 35 & 14 & 19 & 25 & 40 \\
				\midrule
				Total scores & 66 & 147 & 100 & 55 & 76 & 68\\
				\midrule
				\multicolumn{7}{c}{$b = 25$} \\
				\bottomrule
			\end{tabular}
		\end{center}
		Given the profile described above, and especially the ``Total scores'' line in the table above, we can check that the budget allocation that maximises the utilitarian social welfare with cardinal ballots is $\pi = \{p_1, p_2, p_3, p_5\}$. This budget allocation achieves a social welfare of $83 + 106 + 101 + 99 = 389$, where the each component of the sum corresponds to an agent.
	\end{illustration}
	
	\subsubsection{Welfare Maximisers with Approval Ballots}
	
	We now turn to the case of approval ballots. In Section \ref{subsec:Satisfaction}, we have introduced so-called satisfaction functions to measure the satisfaction of the voters when using approval ballots. The definition of the utilitarian social welfare can then be parametrised by a satisfaction function. Given a satisfaction function $\satisfaction$, the utilitarian social welfare of a budget allocation $\pi \in \allocSet(I)$ given an instance $I = \tuple{\projSet, c, b}$, profile $\profile$ of approval ballots is defined as:
	\[\utilSW[\satisfaction](I, \profile, \pi) = \sum_{i \in \agentSet} \satisfaction_i(\pi).\]
	Remember that $\satisfaction_i(\pi) = \satisfaction(\{p \in \pi \mid A_i(p) = 1\})$.
	
	The utilitarian welfare maximiser for approval ballots given a satisfaction function $\satisfaction$ thus selects the budget allocations maximising the above quantity.
		
	\begin{illustration}[Utilitarian Welfare Maximiser for Approval Ballots]
		Consider the instance $I$ and the profile $\profile$ of approval ballots as depicted in the table below.
		\begin{center}
			\begin{tabular}{rcccccc}
				\toprule
				& $p_1$ & $p_2$ & $p_3$ & $p_4$ & $p_5$ & $p_6$ \\
				\midrule
				Cost & 4 & 6 & 7 & 10 & 17 & 19 \\
				\midrule
				$A_1$ & \cmark & \xmark & \xmark & \xmark & \cmark & \xmark \\
				$A_2$ & \cmark & \xmark & \xmark & \cmark & \xmark & \xmark \\
				$A_3$ & \xmark & \cmark & \xmark & \xmark & \xmark & \xmark \\
				$A_4$ & \xmark & \cmark & \cmark & \xmark & \cmark & \cmark \\
				$A_5$ & \cmark & \xmark & \cmark & \cmark & \cmark & \xmark \\
				\midrule
				\multicolumn{7}{c}{$b = 25$} \\
				\bottomrule
			\end{tabular}
		\end{center}
		We present below the (irresolute) outcomes of the utilitarian welfare maximiser on $I$ and $\profile$ for different satisfaction functions. We indicate the detail of the social welfare for each agent (only for the first budget allocation when several are returned).
		\begin{center}
			\renewcommand{\arraystretch}{1.2}
			\begin{tabular}{ccc}
				\toprule
				& \textbf{Irresolute outcome} & \textbf{Achieved social welfare} \\
				\midrule
				$\cardSatisfaction$ & \begin{tabular}{@{}c@{}}$\{p_1, p_2, p_3\}$, $\{p_1, p_2, p_4\}$, \\ and $\{p_1, p_3, p_4\}$\end{tabular} & $1 + 1 + 1 + 2 + 2 = 7$ \\
				$\costSatisfaction$ & $\{p_3, p_5\}$ & $17 + 0 + 0 + 24 + 24 = 65$ \\
				$\satisfaction^{\ln}$ & $\{p_1, p_2, p_4\}$ & $\ln(5) + \ln(15) + \ln(7) + \ln(7) + \ln(15) \approx 10.92$ \\
				$\satisfaction^{\sqrt{~}}$ & $\{p_2, p_5\}$ & $\sqrt{17} + \sqrt{0} + \sqrt{6} + \sqrt{23} + \sqrt{17} \approx 15.49$ \\
				$\satisfaction^{\share}$ & $\{p_2, p_6\}$ & $0 + 0 + 3 + 22 + 0 = 25$ \\
				$\satisfaction^{CC}$ & \begin{tabular}{@{}c@{}}$\{p_1, p_2\}$, $\{p_1, p_2, p_3\}$, \\ and $\{p_1, p_2, p_4\}$\end{tabular}  & $1 + 1 + 1 + 1 + 1 = 5$ \\
				$\satisfaction^{\sum\ln}$ & $\{p_1, p_3, p_4\}$ & $\ln(5) \times 3 + \ln(8) \times 2 + \ln(11) \times 2 \approx 13.78$ \\
				$\satisfaction^{\sum\sqrt{~}}$ & $\{p_1, p_5\}$ & $\sqrt{4} \times 3 + \sqrt{17} \times 3 \approx 18.37$ \\
				\bottomrule
			\end{tabular}
		\end{center}
		Interestingly, all irresolute outcomes are different here, illustrating once more how crucial the choice of the satisfaction function is. By breaking ties in favour of the budget allocation with the smallest cardinality, this would also hold for resolute outcomes.
	\end{illustration}
	
	\medskip
	
	Among the first PB rules to have been introduced in the literature are two utilitarian welfare maximisation rules \citep{TaFa19}. They make use of the cardinality and cost satisfaction functions. 
	
	The \emph{cardinality welfare maximising rule} \maxWelCard{} is defined for any instance $I$ and approval profile $\profile$ as:
	\begin{align*}
		\maxWelCard(I, \profile) & = \argmax_{\pi \in \allocSet(I)} \utilSW\left[\cardSatisfaction\right](I, \profile, \pi) \\
		& = \argmax_{\pi \in \allocSet(I)} \sum_{i \in \agentSet} |\{p \in \pi \mid A_i(p) = 1\}|.
	\end{align*}
	
	Similarly, the \emph{cost welfare maximising rule } \maxWelCost{} is defined for any instance $I$ and approval profile $\profile$ as:
	\begin{align*}
		\maxWelCost(I, \profile) & = \argmax_{\pi \in \allocSet(I)} \utilSW\left[\costSatisfaction\right](I, \profile, \pi) \\
		& = \argmax_{\pi \in \allocSet(I)} \sum_{i \in \agentSet} c(\{p \in \pi \mid A_i(p) = 1\}).
	\end{align*}

	These definitions give rise to \emph{irresolute rules}. Remember that we want to work with resolute rules in this survey. They can be obtained by using some fixed tie-breaking mechanism among all budget allocations maximising $\utilSW$.
	
	\medskip
	
	Interestingly, these two rules can be reinterpreted in terms of \emph{approval score}: given an instance $I$ and a profile $\profile$ of approval ballots, the approval score of a project $p \in \projSet$ in $\profile$, denoted by $\appScore(p, \profile)$, is defined as $\appScore(p, \profile) = |\{i \in \agentSet \mid A_i(p) = 1\}|$. For any $I$ and $\profile$, we then have:
	\begin{align*}
		\maxWelCard(I, \profile) & = \argmax_{\pi \in \allocSet(I)} \sum_{p \in \pi} \appScore(p, \profile), \\
		\maxWelCost(I, \profile) & = \argmax_{\pi \in \allocSet(I)} \sum_{p \in \pi} \appScore(p, \profile) \cdot c(p).
	\end{align*}
	These two formulation will prove useful when drawing parallel with the knapsack problem \citep{KPP04}.
	
	\subsection{Approximate Welfare Maximisation}
	
	As we will see later (Section \ref{subsec:AlgorithmicSocialWelfare}), it is in general computationally difficult to compute the welfare maximisers, at least at a theoretical level.\footnote{In practice, there are rather efficient techniques for solving knapsack problems that can be used for additive satisfaction functions. More involved algorithms also exist for non-additive satisfaction functions.} For this reason, greedy approximations of the utilitarian social welfare have also been considered (though not all of them are approximation algorithms in the formal sense).
	
	\subsubsection{General Greedy Scheme}
	
	Let us first introduce a general greedy scheme for selecting projects that will be at the core of the greedy rules introduced later. This was first introduced in the context of PB by \citet{TaFa19} under the name ``proportional greedy scheme''.
	
	\begin{definition}[Greedy Scheme]
		\label{def:GreedyScheme}
		Consider an instance $I = \tuple{\projSet, c, b}$, a profile $\profile$ and a function $f$ mapping instances, profiles and subsets of projects $P \subseteq \projSet$ to a score $f(I,\profile, P) \in \mathbb{R}_{\geq 0}$. The \emph{greedy scheme} $\greedy(I, \profile, f)$ constructs a budget allocation $\pi$---initially empty---iteratively. In each round, let $P^\star \subseteq \projSet \setminus \pi$ consists of all projects $p^\star$ such that $c(\pi) + c(p^\star) \leq b$, and:
		\[p^\star \in \argmax_{p \in \projSet \setminus \pi} \frac{f(I, \profile, \pi \cup \{p\}) - f(I, \profile, \pi)}{c(p)}.\]
		If $P^\star$ is empty, the procedure stops and $\pi$ is the outcome of $\greedy(I, \profile, f)$. Otherwise, a project from $P^\star$ is selected and added to $\pi$ and a new round starts.
	\end{definition}
	
	\noindent The outcome of $\greedy(I, \profile, f)$ needs not be unique since $P^\star$ may contain more than one element for some rounds. To make the outcome unique (\textit{i.e.,} for $\greedy(I, \profile, f)$ to be resolute), one needs to break ties between the projects in $P^\star$ at each round.
	
	To approximate the welfare maximisers, one then simply needs to define the function $f$ as the suitable notion of utilitarian social welfare. For instance, for a given profile $\profile$ of cardinal ballots, the greedy approximation of the utilitarian welfare maximiser would return the outcome of $\greedy(I, \profile, \utilSW)$. When considering approval ballots, the rule would be defined as $\greedy(I, \profile, \utilSW[\satisfaction])$ for a given satisfaction function $\satisfaction$.
	
	\begin{illustration}[Greedy Approximation of the Utilitarian Welfare Maximiser]
		Consider the instance $I$ and the profile $\profile$ of approval ballots as depicted in the table below.
		\begin{center}
			\begin{tabular}{rcccc}
				\toprule
				& $p_1$ & $p_2$ & $p_3$ & $p_4$ \\
				\midrule
				Cost & 1 & 3 & 7 & 9 \\
				\midrule
				$A_1$ & \xmark & \cmark & \xmark & \cmark \\
				$A_2$ & \xmark & \xmark & \cmark & \cmark \\
				$A_3$ & \cmark & \xmark & \cmark & \xmark \\
				\midrule
				\multicolumn{5}{c}{$b = 10$} \\
				\bottomrule
			\end{tabular}
		\end{center}
		We illustrate the outcome of $\greedy(I, \profile, \utilSW[\satisfaction^{CC}])$, $\greedy(I, \profile, \utilSW[\satisfaction^{\ln}])$ and $\greedy(I, \profile, \utilSW[\satisfaction^{\sqrt{~}}])$ below. All the rules terminate in two rounds. For each round and each project $p$ we present the increase in satisfaction divided by the cost of $p$ that selecting $p$ would incur. Projects in orange are the ones selected. We also indicate the social welfare achieved by the greedy solution, and the highest achievable social welfare.
		\begin{center}
			\resizebox{\linewidth}{!}{
				\renewcommand{\arraystretch}{1.2}
				\begin{tabular}{rccccc}
					\toprule
					& \multicolumn{2}{c}{Cost-Relative Marginal Satisfaction} & Irresolute & \multicolumn{2}{c}{Social Welfare} \\
					& Round 1 & Round 2 & Outcome & Achieved & Maximal \\
					\midrule
					$\satisfaction^{CC}$ & \renewcommand{\arraystretch}{1}\begin{tabular}{@{}c@{}}{\color{MyOrange}$p_1$: $\nicefrac{1}{1}$} \\$p_2$: $\nicefrac{1}{3}$\\$p_3$: $\nicefrac{2}{7}$\\$p_4$: $\nicefrac{2}{9}$\end{tabular} & \renewcommand{\arraystretch}{1}\begin{tabular}{@{}c@{}}{\color{MyOrange}$p_2$: $\nicefrac{1}{3}$} \\ $p_3$: $\nicefrac{2}{7}$\\$p_4$: $\nicefrac{1}{9}$\end{tabular} & $\{p_1, p_2\}$ & 2 & 3 \\
					\midrule
					$\satisfaction^{\ln}$ & \renewcommand{\arraystretch}{1}\begin{tabular}{@{}c@{}}{\color{MyOrange}$p_1$: $\nicefrac{\ln(2)}{1} \approx 0.69$} \\$p_2$: $\nicefrac{\ln(4)}{3} \approx 0.46$ \\ $p_3$: $\nicefrac{2 \times \ln(8)}{7} \approx 0.59$\\ $p_4$: $\nicefrac{2 \times \ln(10)}{9} \approx 0.51$\end{tabular} & \renewcommand{\arraystretch}{1}\begin{tabular}{@{}c@{}}$p_2$: $\nicefrac{\ln(4) - \ln(2)}{3} \approx 0.23$ \\ {\color{MyOrange}$p_3$: $\nicefrac{\ln(8) + \ln(9) - \ln(2)}{7} \approx 0.5119$}\\ $p_4$: $\nicefrac{2 \times \ln(10)}{9} \approx 0.5117$\end{tabular} & $\{p_1, p_3\}$ & \renewcommand{\arraystretch}{1}\begin{tabular}{@{}c@{}}$\ln(8) + \ln(9)$ \\ $\approx 4.28$\end{tabular} & \renewcommand{\arraystretch}{1}\begin{tabular}{@{}c@{}}$\ln(4) + 2 \times \ln(8)$ \\ $\approx 5.55$\end{tabular} \\
					\midrule
					$\satisfaction^{\sqrt{~}}$ & \renewcommand{\arraystretch}{1}\begin{tabular}{@{}c@{}}{\color{MyOrange}$p_1$: $\nicefrac{\sqrt{1}}{1} = 1$} \\$p_2$: $\nicefrac{\sqrt{3}}{3} \approx 0.57$ \\ $p_3$: $\nicefrac{2 \times \sqrt{7}}{7} \approx 0.76$\\ $p_4$: $\nicefrac{2 \times \sqrt{9}}{9} \approx 0.66$\end{tabular} & \renewcommand{\arraystretch}{1}\begin{tabular}{@{}c@{}} $p_2$: $\nicefrac{\sqrt{3} - \sqrt{1}}{3} \approx 0.24$ \\ $p_3$: $\nicefrac{\sqrt{7} + \sqrt{8} - \sqrt{1}}{7} \approx 0.64$\\ {\color{MyOrange}$p_4$: $\nicefrac{2 \times \sqrt{9}}{9} \approx 0.66$}\end{tabular} & $\{p_1, p_4\}$ & \renewcommand{\arraystretch}{1}\begin{tabular}{@{}c@{}}$\sqrt{9} + \sqrt{9} + \sqrt{1}$\\ $= 7$\end{tabular} & \renewcommand{\arraystretch}{1}\begin{tabular}{@{}c@{}}$\sqrt{4} + \sqrt{7} + \sqrt{7}$ \\ $\approx 7.29$\end{tabular} \\
					\bottomrule
				\end{tabular}
			}
		\end{center}
	\end{illustration}
	
	\subsubsection{Greedy Scheme with Additive Satisfaction}
	
	The above formulation of the greedy scheme is particularly interesting when drawing connections between \utilSW{} welfare with \emph{additive satisfaction} and the knapsack problems. Indeed, the prolific literature on the knapsack problem \citep{KPP04} teaches us that our greedy scheme approximates its welfare objective within a factor 2 when used with additive satisfaction \citep*[Chapter 2]{KPP04}.\footnote{Note that for the factor 2 approximation to be formally correct, one needs to either take the outcome of the rules as we defined them, or the most valuable item, whichever has the highest score.}
	
	\medskip
	
	Interestingly, when the satisfaction is additive, we can reformulate the greedy scheme in terms of selecting the projects according to a given ordering of the projects. We introduce below this ordinal variant of the greedy scheme.
	
	\begin{definition}[Ordinal Greedy Scheme]
		Consider an instance $I = \tuple{\projSet, c, b}$ and a strict ordering $\rhd$ over $\projSet$. The \emph{ordinal greedy scheme} $\ordinalGreedy{\rhd}(I)$ is a procedure selecting a budget allocation $\pi$ iteratively as follows. The budget allocation $\pi$ is initially empty. Projects are considered in the order defined by $\rhd$.
		When considering project $p$ for current budget allocation $\pi$, $p$ is selected (added to $\pi$) if and only $c(\pi \cup \{p\}) \leq b$.
		If there is a next project according to $\rhd$, it is considered; otherwise $\pi$ is the output of $\greedy(I, \rhd)$.
	\end{definition}
	
	\noindent Note that, as opposed to $\greedy$, the outcome of $\ordinalGreedy{\rhd}$ is always unique.
	
	\greedy{} and \ordinalGreedy{\rhd} are connected in the following way. Consider an instance $I$, a profile $\profile$ and a function $f$ as described in Definition~\ref{def:GreedyScheme}. We say that an ordering of the projects $\rhd$ is compatible with $f$ when for every projects $p, p' \in \projSet$, we have $p \rhd p'$ if and only if we have $\nicefrac{f(I, \profile, \{p\})}{c(p)} \geq \nicefrac{f(I, \profile, \{p'\})}{c(p')}$. The following holds:
	\[\greedy(I, \profile, f) =  \{\ordinalGreedy{\rhd} \mid \rhd \text{ is compatible with } f\}.\]
	
	\medskip
	
	To conclude this section, we focus on two particular rules that have received a lot of attention both in the literature and in practice, namely the greedy approximation of $\utilSW[\cardSatisfaction]$ and of $\utilSW[\costSatisfaction]$ (defined for approval ballots). We denote these two rules \greWelCard{} and \greWelCost{} respectively. These were initially studied by \citet{TaFa19}, and are now quite standard in the literature.
	
	Since both $\cardSatisfaction$ and $\costSatisfaction$ are additive, we can interpret \greWelCard{} and \greWelCost{} using the ordinal greedy scheme. \greWelCard{} would then select projects in order of their approval score divided by their cost. More interestingly, \greWelCost{} would select projects in order of their approval score times cost divided by their cost, \textit{i.e.}, just their approval score.
	
	A final important fact to keep in mind is that the greedy cost welfare rule, \greWelCost{}, is actually the rule that is the most widely used in practice. It is usually described through its ordinal interpretation.
	
	\begin{illustration}[Ordinal Greedy Scheme]
		Consider the instance $I$ and the profile $\profile$ of approval ballots as depicted in the table below.
		\begin{center}
			\begin{tabular}{rcccccc}
				\toprule
				& $p_1$ & $p_2$ & $p_3$ & $p_4$ & $p_5$ & $p_6$ \\
				\midrule
				Cost & 1 & 2 & 5 & 6 & 11 & 12 \\
				\midrule
				$A_1$ & \xmark & \xmark & \cmark & \xmark & \xmark & \cmark \\
				$A_2$ & \xmark & \cmark & \xmark & \xmark & \cmark & \cmark \\
				$A_3$ & \cmark & \xmark & \xmark & \cmark & \xmark & \cmark \\
				$A_4$ & \xmark & \cmark & \xmark & \cmark & \cmark & \xmark \\
				\midrule
				\multicolumn{7}{c}{$b = 20$} \\
				\bottomrule
			\end{tabular}
		\end{center}
	
		\noindent For $I$ and $\profile$, we consider the greedy approximation of the utilitarian welfare when used with the satisfaction functions $\cardSatisfaction$, $\costSatisfaction$, $\satisfaction^{\sum\ln}$, and $\satisfaction^{\sum\sqrt{~}}$. Since all satisfaction functions are additive we consider here the ordinal interpretation of the rules. For each satisfaction function we present the weak ordering of the projects used to define the greedy scheme (where sets in the description of the order denote indifference classes), the corresponding irresolute outcome, and the social welfare achieved---that can be compared to the maximum achievable social welfare. Observe that we defined the rules to be resolute, but for added generality we present the irresolute version of the rule here. To achieve a resolute outcome, we can apply an arbitrary tie-breaking function to the weak orderings presented below.
		
		\begin{center}
			\begin{tabular}{ccccc}
				\toprule
				& Corresponding & Irresolute & \multicolumn{2}{c}{Social Welfare} \\
				& Weak Ordering & Outcome & Achieved & Maximum \\
				\midrule
				$\cardSatisfaction$ & $\{p_1, p_2\} \rhd p_4 \rhd p_6 \rhd p_3 \rhd p_5$ & $\{p_1, p_2, p_3, p_4\}$ & 6 & 7 \\
				$\costSatisfaction$ & $p_6 \rhd \{p_2, p_4, p_5\} \rhd \{p_1, p_3\}$ & $\{p_2, p_4, p_6\}$ & 52 & 52 \\
				$\satisfaction^{\sum\ln}$ & $p_2 \rhd p_1 \rhd p_4 \rhd p_6 \rhd p_5 \rhd p_3$ & $\{p_1, p_2, p_4, p_5\}$ & $\approx 11.75$ & $\approx 13.78$ \\
				$\satisfaction^{\sum\sqrt{~}}$ & $p_2 \rhd p_1 \rhd p_6 \rhd p_4 \rhd p_5 \rhd p_3$ & $\{p_1, p_2, p_3, p_6\}$ & $\approx 16.46$ & $\approx 18.12$ \\
				\bottomrule
			\end{tabular}
		\end{center}
		Note that we did not include $\satisfaction^\share$ in our analysis since for that satisfaction function the cost-relative marginal satisfaction is always 1 for all projects (that have at least one supporter). The corresponding greedy rule thus always returns all exhaustive budget allocations.
	\end{illustration}
	
	\subsection{Other Welfare-Based Rules}
	
	On top of the four rules we defined above, \citet{TaFa19} introduce five extra rules. They additionally consider welfare defined in terms of $\satisfaction^{CC}$ (see Section \ref{subsec:Satisfaction}), and another greedy scheme to approximate the maximum social welfare. \citet{BBS20} complemented the work of \citet{TaFa19}, showing that two of their rules are actually equivalent, and introducing another greedy scheme (hybrid greedy rules).
	
	Another measure of social welfare was studied by \citet{SBY22} in the context of PB with approval ballots: maximin social welfare---which we call egalitarian social welfare in Section \ref{subsec:AlgorithmicSocialWelfare}---that measures the welfare of a society as the satisfaction of its least satisfied member. \citet{SBY22} consider the maximisation of the egalitarian social welfare as a PB rule, studying its computation and its axiomatic properties.
	
	Our focus was mainly on approval ballots, though a similar approach has been followed for cardinal ones. \citet{FSTW19} study utilitarian and Chamberlin-Courant social welfare (that aims at finding \emph{diverse} knapsacks in their terminology) with cardinal ballots. They also study the maximisation of the Nash social welfare, defined as the product of the satisfaction of the agents (once again defined formally in Section \ref{subsec:AlgorithmicSocialWelfare}). Their motivation is more algorithmic, however, and they don't necessarily aim to devise PB rules.

	Finally coming to ordinal ballots, \citet{Laru21} studies welfare-maximising rules with weak ordinal ballots where positional scoring functions are used to measure the satisfaction of a voter (thus obtaining something equivalent to cardinal ballots). Within this framework, \citet{Laru21} defines greedy approximations of the utilitarian social welfare, and one greedy approximation for Chamberlin-Courant social welfare (there called Rawlsian social welfare) that aims at providing every agent with at least one satisfactory project (see Section \ref{subsec:AlgorithmicSocialWelfare} for a formal definition).

	\section{The Sequential Phragmén Rule}
	\label{subsec:Sequential_Phragmen}
	
	We now leave the world of rules based on measures of social welfare and turn to other kinds of rules. The first one we present is the \emph{sequential Phragmén rule}, an adaptation of a rule introduced at the end of the 19th century by the Swedish mathematician Lars Edvard Phragmén \citep{Jans16}. This rule aims to provide \emph{proportional representation}, which will be studied in more detail in Chapter~\ref{chap:Fairness}.
	
	This rule can only be applied with approval ballots. It was formally studied in the multi-winner literature by \citet{BFJL17}, and has then been adapted for the PB setting by \citet{LCG22}.
	
	\begin{definition}[Sequential Phragmén, Continuous Formulation]
		Given an instance $I$ and a profile $\profile$ of approval ballots, the \emph{Sequential Phragmén rule}, \seqPhragmen, constructs budget allocations using the following continuous process.
		
		Voters receive money in a virtual currency. They all start with a budget of~0 and that budget continuously increases as time passes. At time~$t$, a voter will have received $t$ money. For any time $t$, let $P^\star_t$ be the set of projects $p \in \projSet$ for which the supporters of $p$ altogether have more than $c(p)$ money available. As soon as, for a given $t$, $P^\star_t$ is non-empty, if there exists a $p \in P^\star_t$ such that $c(\pi \cup \{p\}) > b$, the process stops; otherwise one project from $P^\star_t$ is selected, the budget of its supporters is set to 0, and the process resumes.
	\end{definition}

	\noindent Breaking the ties among the projects in any $P^\star_t$ in the above definition will lead to a resolute rule. In the irresolute variant, one would consider all possible ways of breaking such ties.
	
	The termination condition we stated above can be surprising at first sight. It is needed for the rule to satisfy priceability, which however comes at the cost of exhaustiveness (see Sections \ref{subsec:Priceability} and \ref{subsec:Exhaustiveness}).\footnote{Note that phrasing the termination condition as it is here also implies that none of the results rely on the way ties are being broken. If one were to use the stopping condition ``the rule stops as soon as it would select a project leading to a violation of the budget constraint'', priceability would only be satisfied when ties are broken in favour of the most expensive project.}
	
	\medskip

	The sequential Phragmén rule can also be formalised in a discrete fashion where the \emph{loads} of the voters are to be balanced. These two formulations are equivalent. We provide below the second formulation \citep*[see, \textit{e.g.},][]{BFLMP23}. 
	
	\begin{definition}[Sequential Phragmén, Discrete Formulation]
		Given an instance $I$ and a profile $\profile$ of approval ballots, the \emph{sequential Phragmén rule}, \seqPhragmen, constructs a budget allocation $\pi$,
		initially empty, iteratively as follows.
		A load $\ell_i: 2^\projSet \rightarrow \Rplus$, is associated with every agent $i \in \agentSet$, initialised
		as $\ell_i(\emptyset) = 0$ for all $i \in \agentSet$.
		Given $\pi$, the new maximum load for selecting project $p \in \projSet \setminus \pi$ is defined as:
		\[\ell^\star(\pi, p) = \frac{c(p) + \sum_{i \in \agentSet} A_i(p) \cdot  \ell_i(\pi)}{|\{i \in
			\agentSet \mid A_i(p) = 1\}|}.\]
		At a given round with current budget allocation $\pi$, let $P^\star \subseteq \projSet$ be such that:
		\[P^\star = \argmin_{p \in \projSet \setminus \pi} \ell^\star(\pi, p).\]
		If there exists $p \in P^\star$ such that $c(\pi \cup \{p\}) > b$, sequential Phragmén terminates and outputs $\pi$.
		Otherwise, a project $p \in P^\star$ is selected and the agents' loads are
		updated: If $A_i(p) = 0$, then $\ell_i(\pi \cup \{p\}) = \ell_i(\pi)$, and otherwise $\ell_i(\pi \cup
		\{p\}) = \ell^\star(\pi, p)$. Then, $\pi$ is updated to $\pi \cup \{p\}$ and a new round starts.
	\end{definition}
	
	\noindent As before, to obtain a resolute rule one needs to break the ties among the projects in any $P^\star$. The irresolute variant is obtained by considering all possible ways of breaking such ties.
	
	\begin{illustration}[Sequential Phragmén]
		Consider the instance $I$ and the profile $\profile$ of approval ballots as depicted in the table below.
		\begin{center}
			\begin{tabular}{rccccc}
				\toprule
				& $p_1$ & $p_2$ & $p_3$ & $p_4$ & $p_5$ \\
				\midrule
				Cost & 4 & 6 & 8 & 10 & 15 \\
				\midrule
				$A_1$ & \cmark & \xmark & \xmark & \cmark & \cmark \\
				$A_2$ & \xmark & \xmark & \xmark & \xmark & \cmark \\
				$A_3$ & \cmark & \cmark & \xmark & \cmark & \cmark \\
				$A_4$ & \cmark & \cmark & \cmark & \cmark & \cmark \\
				\midrule
				\multicolumn{6}{c}{$b = 10$} \\
				\bottomrule
			\end{tabular}
		\end{center}
		Let us consider a run of \seqPhragmen{} on $I$ and $\profile$. We will first use the continuous formulation. In the illustration below, each bar represents, for different timestamps, the amount of money collected by the supporters of the projects.
		
		\begin{center}
			\resizebox{\linewidth}{!}{
				\begin{tikzpicture}
					\node[anchor=south east] at (-0.5, 0.4) {Time};
					\node[anchor=east] at (-0.5, 0) {$p_1$};
					\node[anchor=east] at (-0.5, -0.6) {$p_2$};
					\node[anchor=east] at (-0.5, -1.2) {$p_3$};
					\node[anchor=east] at (-0.5, -1.8) {$p_4$};
					\node[anchor=east] at (-0.5, -2.4) {$p_5$};
					
					\node[anchor=south] at (1.5, 0.4) {$t_1 =  \nicefrac{4}{3}$};
					
					\fill[my-gray-light, rounded corners=3pt] (0, 0.15) rectangle (3, -0.15);
					\fill[my-green-light, rounded corners=3pt] (0, 0.15) rectangle (3, -0.15);
					\node[anchor=east, white] at (3, 0) {\scriptsize $4$};
					
					\fill[my-gray-light, rounded corners=3pt] (0, -0.45) rectangle (3, -0.75);
					\fill[my-green-light, rounded corners=3pt] (0, -0.45) rectangle (1.33, -0.75);
					\node[anchor=east, white] at (1.33, -0.6) {\scriptsize $\approx 2.67$};
					
					\fill[my-gray-light, rounded corners=3pt] (0, -1.05) rectangle (3, -1.35);
					\fill[my-green-light, rounded corners=3pt] (0, -1.05) rectangle (0.5, -1.35);
					\node[anchor=west, white] at (0.5, -1.2) {\scriptsize $\approx 1.33$};
					
					\fill[my-gray-light, rounded corners=3pt] (0, -1.65) rectangle (3, -1.95);
					\fill[my-green-light, rounded corners=3pt] (0, -1.65) rectangle (1.2, -1.95);
					\node[anchor=east, white] at (1.2, -1.8) {\scriptsize $4$};
					
					\fill[my-gray-light, rounded corners=3pt] (0, -2.25) rectangle (3, -2.55);
					\fill[my-green-light, rounded corners=3pt] (0, -2.25) rectangle (1.066, -2.55);
					\node[anchor=east, white] at (1.066, -2.4) {\scriptsize $\approx 5.33$};
					
					\node[anchor=north] at (1.5, -2.8) {$p_1$ is selected};
					
					\node[anchor=south] at (5, 0.4) {$t_2 = \nicefrac{4}{3} + 0.01$};
					
					\fill[my-gray-light, rounded corners=3pt] (3.5, -0.45) rectangle (6.5, -0.75);
					\node[anchor=west, white] at (3.5, -0.6) {\scriptsize $0.02$};
					
					\fill[my-gray-light, rounded corners=3pt] (3.5, -1.05) rectangle (6.5, -1.35);
					\node[anchor=west, white] at (3.5, -1.2) {\scriptsize $0.01$};
				
					\fill[my-gray-light, rounded corners=3pt] (3.5, -1.65) rectangle (6.5, -1.95);
					\node[anchor=west, white] at (3.5, -1.8) {\scriptsize $0.03$};
					
					\fill[my-gray-light, rounded corners=3pt] (3.5, -2.25) rectangle (6.5, -2.55);
					\fill[my-green-light, rounded corners=3pt] (3.5, -2.25) rectangle (3.7686, -2.55);
					\node[anchor=west, white] at (3.7686, -2.4) {\scriptsize $\approx 1.34$};
					
					\node[anchor=north, align=center] at (5, -2.8) {Supporters of $p_1$ have\\ been reinitialised};
					
					\node[anchor=south] at (8.5, 0.4) {$t_3 = \nicefrac{4}{3} + 3$};
					
					\fill[my-gray-light, rounded corners=3pt] (7, -0.45) rectangle (10, -0.75);
					\fill[my-green-light, rounded corners=3pt] (7, -0.45) rectangle (10, -0.75);
					\node[anchor=east, white] at (10, -0.6) {\scriptsize $6$};
					
					\fill[my-gray-light, rounded corners=3pt] (7, -1.05) rectangle (10, -1.35);
					\fill[my-green-light, rounded corners=3pt] (7, -1.05) rectangle (8.125, -1.35);
					\node[anchor=east, white] at (8.125, -1.2) {\scriptsize $3$};
					
					\fill[my-gray-light, rounded corners=3pt] (7, -1.65) rectangle (10, -1.95);
					\fill[my-green-light, rounded corners=3pt] (7, -1.65) rectangle (9.7, -1.95);
					\node[anchor=east, white] at (9.7, -1.8) {\scriptsize $9$};
					
					\fill[my-gray-light, rounded corners=3pt] (7, -2.25) rectangle (10, -2.55);
					\fill[my-green-light, rounded corners=3pt] (7, -2.25) rectangle (9.6666, -2.55);
					\node[anchor=east, white] at (9.6666, -2.4) {\scriptsize $\approx 13.33$};
					
					\node[anchor=north, align=center] at (8.5, -2.8) {$p_2$ is selected};
					
					\node[anchor=south] at (12, 0.4) {$t_3 = \nicefrac{4}{3} + 3 + \nicefrac{9}{4}$};
					
					\fill[my-gray-light, rounded corners=3pt] (10.5, -1.05) rectangle (13.5, -1.35);
					\fill[my-green-light, rounded corners=3pt] (10.5, -1.05) rectangle (11.34375, -1.35);
					\node[anchor=east, white] at (11.34375, -1.2) {\scriptsize $2.25$};
					
					\fill[my-gray-light, rounded corners=3pt] (10.5, -1.65) rectangle (13.5, -1.95);
					\fill[my-green-light, rounded corners=3pt] (10.5, -1.65) rectangle (13.425, -1.95);
					\node[anchor=east, white] at (13.425, -1.8) {\scriptsize $9.75$};
					
					\fill[my-gray-light, rounded corners=3pt] (10.5, -2.25) rectangle (13.5, -2.55);
					\fill[my-green-light, rounded corners=3pt] (10.5, -2.25) rectangle (13.5, -2.55);
					\node[anchor=east, white] at (13.5, -2.4) {\scriptsize $15$};
					
					\node[anchor=north, align=center] at (12, -2.8) {$p_5$ is too expensive\\ \seqPhragmen{} stops};
				\end{tikzpicture}
			}
		\end{center}
		
		One can check that the formulation in terms of load would lead to the same result. In the table below, each row corresponds to a round. In each round we indicate for each project $p$ the new maximum load it would incur. We use a different colour to indicate the smallest such value per round.
		
		\begin{center}
			\renewcommand{\arraystretch}{1.2}
			\resizebox{\linewidth}{!}{
				\begin{tabular}{cccccc}
					\toprule
					& $p_1$ & $p_2$ & $p_3$ & $p_4$ & $p_5$ \\
					Cost & 4 & 6 & 8 & 10 & 15 \\
					\midrule
					$\ell^\star(\emptyset, p)$ & {\color{my-orange-dark}$\nicefrac{4}{3} \approx 1.33$} & $\nicefrac{6}{2} = 3$ & $8$ & $\nicefrac{10}{3} \approx 9.33$ & $\nicefrac{15}{4} = 3.75$ \\
					$\ell^\star(\{p_1\}, p)$ & --- & {\color{my-orange-dark}$\frac{6 + 2 \times \nicefrac{4}{3}}{2} \approx 4.33$} & $8 + \nicefrac{4}{3} \approx 9.33$ & $\frac{10 + 3 \times \nicefrac{4}{3}}{3} \approx 4.67$ & $\frac{15 + 4 \times \nicefrac{4}{3}}{4} \approx 5.08$ \\
					$\ell^\star(\{p_1, p_2\}, p)$ & --- & --- & $8 + \nicefrac{13}{3} \approx 12.33$ & $\frac{10 + \nicefrac{4}{3} + 2 \times \nicefrac{13}{3}}{3} \approx 6.67$ & {\color{my-orange-dark}$\frac{15 + 2 \times \nicefrac{4}{3} + 2 \times \nicefrac{13}{3}}{4} \approx 5.58$} \\
					\midrule
					Selected & \cmark & \cmark & \xmark & \xmark & \xmark \\
					\bottomrule
				\end{tabular}
			}
		\end{center}
	\end{illustration}
	
	\section{The Maximin Support Rule}
	\label{subsec:MaximinSupport}
	
	We can also adapt the definition of sequential Phragmén slightly and allow a redistribution of the loads in each round. This leads to the definition of the \emph{maximin support rule}. This rule was first introduced as a multi-winner voting rule by \citet{SFFB16, SFFB22}. Its adaptation to the PB setting is due to \citet{ALT18}. Note that they named it sequential Phragmén though they actually define a generalisation of the maximin support rule from multi-winner voting.\footnote{This was observed by \citet{BFLMP23}.}
	
	The maximin support rule is defined for approval ballots as follows.
	
	\begin{definition}[Maximin Support Rule]
		Given an instance $I$ and a profile $\profile$ of approval ballots, the \emph{maximin support rule}, \maximinSupport, constructs a budget allocation $\pi$,
		initially empty, iteratively as follows.
		
		Given $I$, $\profile$ and a subset of projects $P \subseteq \projSet$, a \emph{load distribution} $\boldsymbol{\ell} = (\ell_i)_{i \in \agentSet}$ for $P$ is a collection of functions $\ell_i: P \rightarrow \mathbb{R}_{\geq 0}$ for every agent $i \in \agentSet$ such that $\sum_{i \in \agentSet} \ell_i(p) = c(p)$ for all projects $p \in P$ and $\ell_i(p) = 0$ for all agents $i \in \agentSet$ and projects $p \in \projSet$ such that $A_i(p) = 0$. Omitting $I$ and $\profile$, we denote by $\mathcal{L}(P)$ the set of all the load distributions for $P \subseteq \projSet$.
		
		At a given round with current budget allocation $\pi$, let $P^\star \subseteq \projSet$ be such that:
		\[P^\star = \argmin_{p \in \projSet \setminus \pi} \max_{\substack{\boldsymbol{\ell} \in \mathcal{L}(\pi \cup \{p\}) \\ i \in \agentSet}} \sum_{p' \in \pi \cup \{p\}} \ell_i(p').\]
		If there exists $p \in P^\star$ such that $c(\pi \cup \{p\}) > b$, the maximin support rule terminates and outputs $\pi$.
		Otherwise, a project $p \in P^\star$ is selected ($\pi$ is updated to $\pi \cup \{p\}$) and a new round begins.
	\end{definition}
	
	\noindent Once again, to obtain a resolute rule one needs to break the ties among the projects in any $P^\star$. The irresolute variant is obtained by considering all possible ways of breaking such ties.
	
	Note that in their definition, \citet{ALT18} provide a linear program to efficiently compute the optimum load distribution in each round.
	
	Interestingly, we know that with unit instances (in the multi-winner voting setting) \maximinSupport{} provides approximation guarantees (to the optimum load distribution ) that \seqPhragmen{} does not \citep{CeSt21}. The corresponding result for PB is not known. Overall, further investigation of this rule is needed.
	
	\section{The Method of Equal Shares}
	\label{subsec:MES}
	
	The next rule we introduce is called the \emph{method of equal shares} (formerly known as Rule X). It is similar to \seqPhragmen{} (in its continuous formulation) or \maximinSupport{} except that agents receive all their money from the outset.
	
	This rule has been introduced for PB by \citet{PPS21}\footnote{Note that for all reference to \citet{PPS21} we advise the reader to consider the extended version, updated in November 2022 and available at \href{https://arxiv.org/abs/2008.13276}{arxiv.org/abs/2008.13276}.} for generic cardinal ballots, based on the version for multi-winner voting introduced by \citet{PeSk20}. We provide their definition below.\footnote{For a description of the Method of Equal Shares aimed at non-experts, see \href{https://equalshares.net/}{equalshares.net}, a website maintained by Dominik Peters.}
	
	\begin{definition}[Method of Equal Shares for Cardinal Ballots]
		Given an instance $I=\tuple{\projSet, c, b}$ and a profile $\profile$ of cardinal ballots, the \emph{method of equal shares}, \mes{}, constructs a budget allocation $\pi$, initially empty,
		iteratively as follows.
		
		A load $\ell_i: 2^\projSet \rightarrow \Rplus$, is associated with every agent $i \in \agentSet$, initialised as
		$\ell_i(\emptyset) = 0$ for all $i \in \agentSet$. The load represents how much virtual money the agents have spent.
		
		Given $\pi$ and a scalar $\alpha \geq 0$, the contribution of agent $i \in \agentSet$ for project
		$p \in \projSet \setminus \pi$ is defined by:
		\[\gamma_i(\pi, \alpha, p) = \min\left(\nicefrac{b}{n} - \ell_i(\pi), \alpha \cdot A_i(p)\right).\]
		This is the amount $i$ would pay to buy project $p$ for a given $\alpha$. Note that the above means that agents are initially provided $\nicefrac{b}{n}$ units of the virtual currency.
		
		Given a budget allocation $\pi$, a project $p \in \projSet \setminus \pi$ is said to be $\alpha$-affordable, for $\alpha \geq 0$, if
		\[\sum_{i \in \agentSet} \gamma_i(\pi, \alpha, p) \geq c(p).\]
		A project is thus $\alpha$-affordable if, for the given $\alpha$, all the agents can contribute enough to afford $p$.
		
		At a given round with current budget allocation $\pi$, if no project is $\alpha$-affordable for any $\alpha$,
		\mes{} terminates.
		
		Otherwise, it selects a project $p \in \projSet \setminus \pi$ that is $\alpha^\star$-affordable where $\alpha^\star$
		is the smallest $\alpha$ such that one project is $\alpha$-affordable.
		The agents' loads are then updated: $\ell_i(\pi \cup \{p\}) = \ell_i(\pi) + \gamma_i(\pi, \alpha, p)$. Then, $\pi$ is updated to $\pi \cup \{p\}$ and a new round starts.
	\end{definition}
	
	\noindent The above definition gives rise to a resolute rule (when ties among $\alpha^\star$-affordable projects are broken arbitrarily). For the irresolute variant of the rule, one would need to consider all ways to break the ties between $\alpha^\star$-affordable projects at each round.
	
	We observe that \mes{} does not necessarily spend the whole budget, \textit{i.e.}, it is not \emph{exhaustive} (see Definition~\ref{def:exhaustive}). Indeed, it is possible that no project is $\alpha$-affordable for any $\alpha$, in which case \mes{} returns the empty set. For this reason, in practice \mes{} nearly always needs to be combined with a completion method. We discuss this point in more detail in Section~\ref{subsec:Exhaustiveness}.
	
	\medskip

	The definition of \mes{} can easily be adapted for approval ballots. Note that since approval ballots are just restricted cases of cardinal ones, \mes{} can actually be used as is. However, since the cardinal ballots are meant to represent the voters' utilities for the projects, it is more natural to parametrize \mes{} for approval ballots by a satisfaction function \citep{BFLMP23}.
	
	\begin{definition}[Method of Equal Shares for Approval Ballots]
		Given an instance $I=\tuple{\projSet, c, b}$ and a profile $\profile$ of approval ballots, the \emph{method of equal shares for the satisfaction function $\satisfaction$},  \mesSat{$\satisfaction$}, constructs a budget allocation $\pi$, initially empty,
		iteratively as follows.
		
		A load $\ell_i: 2^\projSet \rightarrow \Rplus$, is associated with every agent $i \in \agentSet$, initialised as
		$\ell_i(\emptyset) = 0$ for all $i \in \agentSet$. The load represents how much virtual money the agents have spent.
		
		Given $\pi$ and a scalar $\alpha \geq 0$, the contribution of agent $i \in \agentSet$ for project
		$p \in \projSet \setminus \pi$ is defined by:
		\[\gamma_i(\pi, \alpha, p) = A_i(p) \cdot \min\left(\nicefrac{b}{n} - \ell_i(\pi), \alpha \cdot \satisfaction(\{p\})\right).\]
		This is the amount $i$ would pay to buy project $p$ for a given $\alpha$. Importantly, $i$ only contributes to $p$ if $A_i(p) = 1$, \textit{i.e.}, if $i$ approves of $p$. Note that the above means that agents are initially provided $\nicefrac{b}{n}$ units of the virtual currency.
		
		Given a budget allocation $\pi$, a project $p \in \projSet \setminus \pi$ is said to be $\alpha$-affordable, for $\alpha \geq 0$, if
		\[\sum_{i \in \agentSet} \gamma_i(\pi, \alpha, p) \geq c(p).\]
		A project is thus $\alpha$-affordable if, for the given $\alpha$, all the agents can contribute enough to afford $p$.
		
		At a given round with current budget allocation $\pi$, if no project is $\alpha$-affordable for any $\alpha$,
		\mesSat{$\satisfaction$} terminates.
		
		Otherwise, it selects a project $p \in \projSet \setminus \pi$ that is $\alpha^\star$-affordable where $\alpha^\star$
		is the smallest $\alpha$ such that one project is $\alpha$-affordable. 
		The agents' loads are then updated: $\ell_i(\pi \cup \{p\}) = \ell_i(\pi) + \gamma_i(\pi, \alpha, p)$. Then, $\pi$ is updated to $\pi \cup \{p\}$ and a new round starts.
	\end{definition}
	
	\noindent Notice that in the above, $\satisfaction$ is only ever used on singletons. Notably, this implies that even if $\satisfaction$ is not additive, \mesSat{$\satisfaction$} is still well-defined (but not necessarily meaningful).
	
	\begin{illustration}[Method of Equal Shares with Approval Ballots]
		Consider the instance $I$ and the profile $\profile$ of approval ballots as depicted in the table below.
		\begin{center}
			\begin{tabular}{rccccc}
				\toprule
				& $p_1$ & $p_2$ & $p_3$ & $p_4$ & $p_5$\\
				\midrule
				Cost & 1 & 1 & 3 & 6 & 9 \\
				\midrule
				$A_1$ & \xmark & \cmark & \xmark & \cmark & \cmark \\
				$A_2$ & \cmark & \cmark & \cmark & \xmark & \cmark \\
				$A_3$ & \xmark & \cmark & \xmark & \xmark & \cmark \\
				$A_4$ & \xmark & \xmark & \cmark & \xmark & \cmark \\
				\midrule
				\multicolumn{6}{c}{$b = 10$} \\
				\bottomrule
			\end{tabular}
		\end{center}
		We illustrate below the different steps that \mes{} would perform on $I$ and $\profile$. For each round, we present different values that were used but not named in the definition of the procedure, namely:
		\begin{itemize}
			\item \textbf{Available Contributions}: a vector indicating for each agent their potential contribution for the round, \textit{i.e.}, $A_i(p) \times \nicefrac{b}{n} - \ell_i(\pi)$;
			\item \textbf{Affordable}: a Boolean indicating whether the supporters of $p$ have enough potential contribution to afford the project, \textit{i.e.}, $\mathds{1}_{A_i(p) \times \nicefrac{b}{n} - \ell_i(\pi)}$; projects that are not affordable need not be considered further;
			\item \textbf{Full price}: the minimum maximum price that a supporter would contribute, \textit{i.e.}, $\max_{i \in \agentSet} \gamma_i(\pi, \alpha^\star, p)$ where $\alpha^\star$ is the smallest $\alpha$ for which $p$ is $\alpha$-affordable (though the value of $\alpha^\star$ is not needed to compute this value);
			\item \textbf{Full price per satisfaction}: the price per unit of satisfaction paid by supporters contributing the full price, \textit{i.e.}, $\alpha^\star$, the smallest $\alpha$ for which $p$ is $\alpha$-affordable.
		\end{itemize}
		
		Let us start with \mesSat{$\costSatisfaction$}. In the following table, each segment of rows corresponds (after the first one) to a round.
		\begin{center}
			\resizebox{\linewidth}{!}{
				\begin{tabular}{rccccc}
					\toprule
					& $p_1$ & $p_2$ & $p_3$ & $p_4$ & $p_5$ \\
					Cost & 1 & 1 & 3 & 6 & 9 \\
					Satisfaction & 1 & 1 & 3 & 6 & 9 \\
					\midrule
					Available Contribs. & $0, 2.5, 0, 0$ & $2.5, 2.5, 2.5, 0$ & $0, 2.5, 0, 2.5$ & $2.5, 0, 0, 0$ & $2.5, 2.5, 2.5, 2.5$ \\
					Affordable & \cmark & \cmark & \cmark & \xmark & \cmark \\
					Full price & $1$ & $\nicefrac{1}{3}$ & $1.5$ & --- & $\nicefrac{9}{4}$ \\
					Full price per sat. & $1$ & $\nicefrac{1}{3}$ & $0.5$ & --- & {\color{my-orange-dark}$0.25$} \\
					\midrule
					Available Contribs. & $0, \nicefrac{1}{4}, 0, 0$ & $\nicefrac{1}{4}, \nicefrac{1}{4}, \nicefrac{1}{4}, 0$ & $0, \nicefrac{1}{4}, 0, \nicefrac{1}{4}$ & --- & --- \\
					Affordable & \xmark & \xmark & \xmark & --- & --- \\
					\bottomrule
				\end{tabular}
			}
		\end{center}
		The (irresolute) outcome is thus $\{p_5\}$.
		
		Let us now consider \mesSat{$\cardSatisfaction$}. Note that in this case, the full price and the full price per satisfaction are the same.
		\begin{center}
			\resizebox{\linewidth}{!}{
				\begin{tabular}{rccccc}
					\toprule
					& $p_1$ & $p_2$ & $p_3$ & $p_4$ & $p_5$ \\
					Cost & 1 & 1 & 3 & 6 & 9 \\
					Satisfaction & 1 & 1 & 1 & 1 & 1 \\
					\midrule
					Available Contribs. & $0, 2.5, 0, 0$ & $2.5, 2.5, 2.5, 0$ & $0, 2.5, 0, 2.5$ & $2.5, 0, 0, 0$ & $2.5, 2.5, 2.5, 2.5$ \\
					Affordable & \cmark & \cmark & \cmark & \xmark & \cmark \\
					Full price (per sat.) & $1$ & {\color{my-orange-dark}$\nicefrac{1}{3}$} & $1.5$ & --- & $\nicefrac{9}{4}$ \\
					\midrule
					Available Contribs. & $0, \nicefrac{13}{6}, 0, 0$ & --- & $0, \nicefrac{13}{6}, 0, \nicefrac{13}{6}$ & --- & $\nicefrac{13}{6}, \nicefrac{13}{6}, \nicefrac{13}{6}, 2.5$ \\
					Affordable & \cmark & --- & \cmark & --- & \cmark \\
					Full price (per sat.) & {\color{my-orange-dark}$1$} & --- & $1.5$ & --- & $2.5$ \\
					\midrule
					Available Contribs. & --- & --- & $0, \nicefrac{7}{6}, 0, \nicefrac{13}{6}$ & --- & $\nicefrac{13}{6}, \nicefrac{7}{6}, \nicefrac{13}{6}, 2.5$ \\
					Affordable & --- & --- & \cmark & --- & \xmark \\
					Full price (per sat.) & --- & --- & {\color{my-orange-dark}$\nicefrac{11}{6}$} & --- & --- \\
					\bottomrule
				\end{tabular}
			}
		\end{center}
		The (irresolute) outcome is thus $\{p_1, p_2, p_3\}$.
	\end{illustration}

	In the case of cardinal ballots, the method is slightly more involved as not all voters have the same satisfaction for a given a project. Thus, the full price per satisfaction is potentially different for each agent. 

	\section{Other Rules for Participatory Budgeting}
	\label{subsec:Other_Rules}
	
	We have introduced what we believe to be the most prominent rules in the literature for PB. These are obviously not the only ones that have been defined. We briefly comment on other rules.
	
	In the multi-winner literature, Thiele methods play an important role \citep{LaSk23}. It can thus be surprising that this is not the case in the PB setting. It turns out that these rules do not behave as nicely in PB as they did in multi-winner voting. In particular, \emph{Proportional Approval Voting} (PAV), that provides interesting proportionality guarantees in multi-winner voting \citep{ABCEFW17}, no longer enjoys them in the non unit-cost setting as shown by \citet{PPS21} and \citet{LCG22}.
	
	Among the other rules that have been defined, \citet{SSST20} propose an adaptation of the multi-winner variant of the \emph{Single Transferable Votes} rule (STV) in the PB setting with cumulative ballots.
	
	When considering ordinal ballots, \citet{AzLe21} introduced the \emph{expanding-approvals rule} for PB. \citet{PPS21} proposed an ordinal version of \mes{}, showing that it is an expanding-approvals rule.\footnote{This part is only available in the extended version, available at \href{https://arxiv.org/abs/2008.13276}{arxiv.org/abs/2008.13276}.} For weak rankings, \citet{SrNa22} presented two different classes of rules and studied their algorithmic and axiomatic properties, either by translating them into approval votes or by focusing on fairness issues utilising the minimal amount a project needs to be funded with to satisfy a voter.

	Finally, focusing on the distribution of money between districts, \citet{BTS24} propose to find a PB outcome that aids in a socially just budget distribution with a single-leader-multiple-follower game.
	
	\chapter{Fairness in Indivisible Participatory Budgeting}
	\label{chap:Fairness}	
	
	All the ingredients are now in place: We have introduced the basic setting (Chapter \ref{chap:Preliminaries}), discussed how the voters can submit their opinion (Chapter \ref{chap:Ballot_Design}), and finally how to take these opinions into account to select budget allocations (Chapter \ref{chap:PB_Rules}). However, defining the rules is only the first step, we still need to assess their respective merits.
	
	Throughout this section we will study different PB rules in terms of their fairness guarantees. This represents the largest share of the literature devoted to the analysis of PB rules and has proved to be a rich and fruitful research direction.
	
	This section is mainly organised around the different types of fairness requirements that have been introduced. We will start with the concepts revolving around justified representation (Section \ref{subsec:Justified_Representation}), which will naturally lead us to the idea of the core (Section \ref{subsec:Core}). We will then discuss the idea of priceability (Section \ref{subsec:Priceability}). Broadening our perspective, our next focus will be fairness in ordinal PB (Section \ref{subsec:Proportionality_Ordinal}), and more generally all the other notions of fairness that have been introduced (Section \ref{subsec:Other_Proportionality_Axioms}). Fairness in extended models of PB will then be discussed (Section \ref{subsec:Fairness_Extended}). We will conclude by attempting to unify everything by drawing taxonomies linking the different requirements to each other (Section \ref{subsec:Taxonomies_Proportionality}).
	
	\section{Extended and Proportional Justified Representation}
	\label{subsec:Justified_Representation}
	
	The main part of the research on fairness in PB focuses on adapting the well studied concept of \emph{justified representation} from the multi-winner voting literature to PB \citep{ABCEFW17, AEHLSS18, BFKN19, PeSk20, LaSk23}. The idea behind justified representation is that groups of agents that are large enough and similar enough, the so-called \emph{cohesive groups}, deserve to be satisfied with a fraction of the outcome that is proportional to their size.
	
	In the following we will define the most important concepts related to justified representation and present the main results from the literature. Figures~\ref{fig:Taxonomy_Cardinal_Ballots} and~\ref{fig:Taxonomy_Approval_Ballots} summarise (among others) the results presented in this section.
	
	\subsection{Justified Representation with Cardinal Ballots}
	
	We first consider the general case of cardinal ballots. A special focus on approval ballots will come later.
	
	\subsubsection{Cohesive Groups for Cardinal Ballots}
	
	Let us start with the definition of cohesive groups. We follow the definition of \citet{PPS21}.
	
	\begin{definition}[$(\alpha, P)$-Cohesive Groups]
		\label{def:Cohesive_CardinalBallots}
		Given an instance $I = \tuple{\projSet, c, b}$ and a profile $\profile$ of cardinal ballots, a non-empty group of agents $N \subseteq \agentSet$ is said to be \emph{$(\alpha, P)$-cohesive}, for a function $\alpha: \projSet \rightarrow \mathbb{R}_{\geq 0}$ and a set of projects $P \subseteq \projSet$, if the following two conditions are satisfied:
		\begin{itemize}
			\item $\alpha(p) \leq A_i(p)$ for all $i \in N$ and $p \in P$, that is, $\alpha$ is lower-bounding the score of the agents in $N$;
			\item $\frac{|N|}{n} \cdot b \geq c(P)$, that is, $N$'s share of the budget is enough to afford $P$.
		\end{itemize}
	\end{definition}
	
	\begin{definition}[$(\alpha, P)$-Cohesive Groups]
		\label{def:Cohesive_CardinalBallots}
		Given an instance $I = \tuple{\projSet, c, b}$ and a profile $\profile$ of cardinal ballots, a non-empty group of agents $N \subseteq \agentSet$ is said to be \emph{$(c, P)$-cohesive}, for a given claim $c \in \mathbb{R}_{> 0}$ and a set of projects $P \subseteq \projSet$, if the following two conditions are satisfied:
		\begin{itemize}
			\item there exists a function $\alpha: P \rightarrow \mathbb{R}_{> 0}$ such that $c = \sum_{p \in P}\alpha(p)$ and $\alpha(p) \leq A_i(p)$ for all $i \in N$ and $p \in P$, that is, $c$ is a claim that all agents can agree on for the projects in $P$;
			\item $\frac{|N|}{n} \cdot b \geq c(P)$, that is, $N$'s share of the budget is enough to afford $P$.
		\end{itemize}
	\end{definition}
	
	\noindent Overall, for any $(\alpha, P)$-cohesive group of agents $N \subseteq \agentSet$, it should be that $N$ is $(i)$ large enough to afford the projects in $P$, and, $(ii)$ cohesive enough to ``deserve'' the satisfaction they receive from the projects in $P$, measured by $\alpha$.
	
	We will make use of a specific collection of functions $\alpha$ denoted by $\alpha^{\min}$ and defined for any profile $\profile$ and subset of agents $N \subseteq \agentSet$ as:
	\[\alpha_{N, \profile}^{\min}(p) = \min_{i \in N} A_i(p), \text{ for all } p \in \projSet.\]
	This function simply takes the minimum score submitted by any agent in $N$ for project $p$.
	Note that for every group of agents $N \subseteq \agentSet$ and subset of projects $P \subseteq \projSet$, if $\nicefrac{|N|}{n} \cdot b \geq c(P)$ then $N$ is $(\alpha_{N, \profile}^{\min}, P)$-cohesive.
	
	\begin{illustration}[Cohesive Groups with Cardinal Ballots]
		Consider the following instance $I$ and profile $\profile$ of cardinal ballots (that actually correspond to cumulative ballots in which 8 points are to be distributed).
		\begin{center}
			\begin{tabular}{ccccc}
				\toprule
				& $p_1$ & $p_2$ & $p_3$ & $p_4$ \\
				Cost & 1 & 1 & 2 & 6 \\
				\midrule
				$A_1$ & 2 & 3 & 3 & 0 \\
				$A_2$ & 0 & 4 & 3 & 1 \\
				$A_3$ & 1 & 1 & 2 & 4 \\
				\midrule
				\multicolumn{5}{c}{$b = 6$} \\
				\bottomrule
			\end{tabular}
		\end{center}
		For $I$ and $\profile$, there are 26 cohesive groups if we exclude the groups $N$ for which $\alpha_{N, \profile}^{\min}(p) = 0$ for some $p$. We will describe some of them for the reader to understand the concept.
		
		First, note that $\nicefrac{b}{n} = 2$. Thus, for every agent $i \in \agentSet$, the group $\{i\}$ is $(\alpha_{\{i\}, \profile}^{\min}, \{p_1, p_2\})$-cohesive for any $\alpha$ such that $\alpha(p_1) \leq A_i(p_1)$ and $\alpha(p_2) \leq A_i(p_2)$. The same is true for the set of projects $\{p_1\}$, $\{p_2\}$ and $\{p_3\}$. Overall agent 1 would ``deserve a satisfaction of at least 5, agent 2 of at least 4, and agent 3 of at least 2 (the notion of deserving is introduced formally in the following sections).
		
		Consider now the group $N = \{1, 3\}$. It is $(\alpha, \{p_1, p_2, p_3\})$-cohesive for any $\alpha$ such that $\alpha(p_1) \leq 1$, $\alpha(p_2) \leq 1$ and $\alpha(p_3) \leq 2$. The group would thus deserve a satisfaction of at least 4. Note that the group $\{1, 2, 3\}$ is $(\alpha, \{p_4\})$-cohesive but only if $\alpha(p_4) = 0$, \textit{i.e.}, a null claim for the group satisfaction. We leave it to the reader to list the rest of the cohesive groups.
		
		An important thing to keep in mind here is that even though cohesive groups are cohesive for (potentially) infinitely many $\alpha$, only $\alpha^{\min}$ matters, as it defines the strongest claim in terms of group satisfaction.
		
		\medskip
		
		We can try to present the cohesive groups in a graphical manner. We first represent the instance as follows where the ballots are in rows and the projects in columns. The height of a row is $\nicefrac{b}{n}$ and the width of a column is the cost of a project. A cell is coloured if the corresponding voter has assigned a non-zero score to the corresponding project. The value in a coloured cell indicates the score assigned to the project by the voter.
		
		\begin{center}
			\scalebox{0.8}{
				\begin{tikzpicture}
					\foreach \x in {0, 1, 1.5, 2, 5} {
						\draw[dashed, my-gray-light] (\x, -1) -- (\x, 3.5);
					}
					
					\foreach \x in {0, 1, 2, 3} {
						\draw[dashed, my-gray-light] (-0.9, \x) -- (6.1, \x);
					}
				
					\draw[fill=ColorBlind1!20!white] (0, 0) rectangle (1, 1);
					\node[anchor=west, ColorBlind1] at (0, 0.5) {3};
					\draw[fill=ColorBlind2!20!white] (0, 1) rectangle (1, 2);
					\node[anchor=west, ColorBlind2] at (0, 1.5) {3};
					\draw[fill=ColorBlind3!20!white] (0, 2) rectangle (1, 3);
					\node[anchor=west, ColorBlind3] at (0, 2.5) {2};
					
					\draw[fill=ColorBlind1!20!white] (1, 0) rectangle (1.5, 1);
					\node[anchor=west, ColorBlind1] at (1, 0.5) {3};
					\draw[fill=ColorBlind2!20!white] (1, 1) rectangle (1.5, 2);
					\node[anchor=west, ColorBlind2] at (1, 1.5) {4};
					\draw[fill=ColorBlind3!20!white] (1, 2) rectangle (1.5, 3);
					\node[anchor=west, ColorBlind3] at (1, 2.5) {1};
					
					\draw[fill=ColorBlind1!20!white] (1.5, 0.0) rectangle (2, 1);
					\node[anchor=west, ColorBlind1] at (1.5, 0.5) {2};
					\draw[fill=ColorBlind3!20!white] (1.5, 2) rectangle (2, 3);
					\node[anchor=west, ColorBlind3] at (1.5, 2.5) {1};
					
					\draw[fill=ColorBlind2!20!white] (2, 1) rectangle (5, 2);
					\node[anchor=west, ColorBlind2] at (2, 1.5) {1};
					\draw[fill=ColorBlind3!20!white] (2, 2) rectangle (5, 3);
					\node[anchor=west, ColorBlind3] at (2, 2.5) {4};
					
					\draw [decorate, decoration={brace, amplitude=10pt, mirror, raise=4pt}, yshift=0pt] (0, 0) -- (1, 0);
					\node[align=center] at (0.5, 3.3) {$p_3$};
					\node[align=center] at (0.5, -0.8) {$2$};
					\draw [decorate, decoration={brace, amplitude=10pt, mirror, raise=4pt}, yshift=0pt] (1, 0) -- (1.5, 0);
					\node at (1.25, 3.3) {$p_2$};
					\node at (1.25, -0.8) {$1$};
					\draw [decorate, decoration={brace, amplitude=10pt, mirror, raise=4pt}, yshift=0pt] (1.5, 0) -- (2, 0);
					\node at (1.75, 3.3) {$p_1$};
					\node at (1.75, -0.8) {$1$};
					\draw [decorate, decoration={brace, amplitude=10pt, mirror, raise=4pt}, yshift=0pt] (2, 0) -- (5, 0);
					\node at (3.5, 3.3) {$p_4$};
					\node at (3.5, -0.8) {$6$};
					
					\draw [decorate, decoration={brace, amplitude=10pt, mirror, raise=4pt}, yshift=0pt] (5, 0) -- (5, 1);
					\node[ColorBlind1] at (-0.5, 0.5) {$A_1$};
					\node at (5.8, 0.5) {$2$};
					\draw [decorate, decoration={brace, amplitude=10pt, mirror, raise=4pt}, yshift=0pt] (5, 1) -- (5, 2);
					\node[ColorBlind2] at (-0.5, 1.5) {$A_2$};
					\node at (5.8, 1.5) {$2$};
					\draw [decorate, decoration={brace, amplitude=10pt, mirror, raise=4pt}, yshift=0pt] (5, 2) -- (5, 3);
					\node[ColorBlind3] at (-0.5, 2.5) {$A_3$};
					\node at (5.8, 2.5) {$2$};
				\end{tikzpicture}
			}
		\end{center}
		
		In this representation, a group of voters $N$ is $(\alpha_{N, \profile}^{\min}, P)$-cohesive if by putting side by side all the columns of the projects in $P$ and the rows of the agents in $N$ the following two conditions are met:
		\begin{itemize}
			\item The resulting shape is a fully coloured rectangle (agents approve of all the projects in $P$);
			\item The resulting rectangle is taller than it is wide (the cost of $P$ is no more than $N$'s share of the budget).
		\end{itemize}
		Moreover, the value of $\alpha_{N, \profile}^{\min}$ for each project is the minimum of the value appearing in a column. The cohesive groups can thus be represented as:
		
		\begin{center}
			
			\scalebox{0.7}{
				\begin{tikzpicture}
					\node[fill=my-blue-verylight, rounded corners=5pt, anchor=north] at (-0.25, 0) {
						\begin{tikzpicture}[rounded corners=0, anchor=center]
							\draw[fill=ColorBlind1!20!white] (0, 0) rectangle (1, 1);
							\node[anchor=west, ColorBlind1] at (0, 0.5) {3};
							\node at (0.5, 1.3) {$p_3$};
							\node[ColorBlind1] at (-0.5, 0.5) {$A_1$};
							\node at (-0.6, -0.4) {$\alpha_{N, \profile}^{\min}$};
							\node at (0.5, -0.4) {3};
						\end{tikzpicture}
					};
					
					\node[fill=my-blue-verylight, rounded corners=5pt, anchor=north] at (2.5, 0) {
						\begin{tikzpicture}[rounded corners=0, anchor=center]
							\draw[fill=ColorBlind2!20!white] (0, 0) rectangle (1, 1);
							\node[anchor=west, ColorBlind2] at (0, 0.5) {3};
							\node at (0.5, 1.3) {$p_3$};
							\node[ColorBlind2] at (-0.5, 0.5) {$A_2$};
							\node at (-0.6, -0.4) {$\alpha_{N, \profile}^{\min}$};
							\node at (0.5, -0.4) {3};
						\end{tikzpicture}
					};
					
					\node[fill=my-blue-verylight, rounded corners=5pt, anchor=north] at (5.25, 0) {
						\begin{tikzpicture}[rounded corners=0, anchor=center]
							\draw[fill=ColorBlind3!20!white] (0, 0) rectangle (1, 1);
							\node[anchor=west, ColorBlind3] at (0, 0.5) {2};
							\node at (0.5, 1.3) {$p_3$};
							\node[ColorBlind3] at (-0.5, 0.5) {$A_3$};
							\node at (-0.6, -0.4) {$\alpha_{N, \profile}^{\min}$};
							\node at (0.5, -0.4) {2};
						\end{tikzpicture}
					};
					
					\node[fill=my-blue-verylight, rounded corners=5pt, anchor=north] at (8, 0) {
						\begin{tikzpicture}[rounded corners=0, anchor=center]
							\draw[fill=ColorBlind1!20!white] (0, 0) rectangle (1, 1);
							\node[anchor=west, ColorBlind1] at (0, 0.5) {3};
							\draw[fill=ColorBlind2!20!white] (0, 1) rectangle (1, 2);
							\node[anchor=west, ColorBlind2] at (0, 1.5) {3};
							\node at (0.5, 2.3) {$p_3$};
							\node[ColorBlind1] at (-0.5, 0.5) {$A_1$};
							\node[ColorBlind2] at (-0.5, 1.5) {$A_2$};
							\node at (-0.6, -0.4) {$\alpha_{N, \profile}^{\min}$};
							\node at (0.5, -0.4) {3};
						\end{tikzpicture}
					};
					
					\node[fill=my-blue-verylight, rounded corners=5pt, anchor=north] at (10.75, 0) {
						\begin{tikzpicture}[rounded corners=0, anchor=center]
							\draw[fill=ColorBlind1!20!white] (0, 0) rectangle (1, 1);
							\node[anchor=west, ColorBlind1] at (0, 0.5) {3};
							\draw[fill=ColorBlind3!20!white] (0, 1) rectangle (1, 2);
							\node[anchor=west, ColorBlind3] at (0, 1.5) {2};
							\node at (0.5, 2.3) {$p_3$};
							\node[ColorBlind1] at (-0.5, 0.5) {$A_1$};
							\node[ColorBlind3] at (-0.5, 1.5) {$A_3$};
							\node at (-0.6, -0.4) {$\alpha_{N, \profile}^{\min}$};
							\node at (0.5, -0.4) {2};
						\end{tikzpicture}
					};
					
					\node[fill=my-blue-verylight, rounded corners=5pt, anchor=north] at (13.5, 0) {
						\begin{tikzpicture}[rounded corners=0, anchor=center]
							\draw[fill=ColorBlind2!20!white] (0, 0) rectangle (1, 1);
							\node[anchor=west, ColorBlind2] at (0, 0.5) {3};
							\draw[fill=ColorBlind3!20!white] (0, 1) rectangle (1, 2);
							\node[anchor=west, ColorBlind3] at (0, 1.5) {2};
							\node at (0.5, 2.3) {$p_3$};
							\node[ColorBlind2] at (-0.5, 0.5) {$A_2$};
							\node[ColorBlind3] at (-0.5, 1.5) {$A_3$};
							\node at (-0.6, -0.4) {$\alpha_{N, \profile}^{\min}$};
							\node at (0.5, -0.4) {2};
						\end{tikzpicture}
					};
					
					\node[fill=my-blue-verylight, rounded corners=5pt, anchor=north] at (16.25, 0) {
						\begin{tikzpicture}[rounded corners=0, anchor=center]
							\draw[fill=ColorBlind1!20!white] (0, 0) rectangle (1, 1);
							\node[anchor=west, ColorBlind1] at (0, 0.5) {3};
							\draw[fill=ColorBlind2!20!white] (0, 1) rectangle (1, 2);
							\node[anchor=west, ColorBlind2] at (0, 1.5) {3};
							\draw[fill=ColorBlind3!20!white] (0, 2) rectangle (1, 3);
							\node[anchor=west, ColorBlind3] at (0, 2.5) {2};
							\node at (0.5, 3.3) {$p_3$};
							\node[ColorBlind1] at (-0.5, 0.5) {$A_1$};
							\node[ColorBlind2] at (-0.5, 1.5) {$A_2$};
							\node[ColorBlind3] at (-0.5, 2.5) {$A_3$};
							\node at (-0.6, -0.4) {$\alpha_{N, \profile}^{\min}$};
							\node at (0.5, -0.4) {2};
						\end{tikzpicture}
					};
					
					\node[fill=my-blue-verylight, rounded corners=5pt, anchor=north] at (-0.5, -2.75) {
						\begin{tikzpicture}[rounded corners=0, anchor=center]
							\draw[fill=ColorBlind1!20!white] (0, 0) rectangle (0.5, 1);
							\node[anchor=west, ColorBlind1] at (0, 0.5) {3};
							\draw[fill=ColorBlind2!20!white] (0, 1) rectangle (0.5, 2);
							\node[anchor=west, ColorBlind2] at (0, 1.5) {4};
							\node at (0.25, 2.3) {$p_2$};
							\node[ColorBlind1] at (-0.5, 0.5) {$A_1$};
							\node[ColorBlind2] at (-0.5, 1.5) {$A_2$};
							\node at (-0.6, -0.4) {$\alpha_{N, \profile}^{\min}$};
							\node at (0.25, -0.4) {3};
						\end{tikzpicture}
					};
					
					\node[fill=my-blue-verylight, rounded corners=5pt, anchor=north] at (2.5, -2.75) {
						\begin{tikzpicture}[rounded corners=0, anchor=center]
							\draw[fill=ColorBlind1!20!white] (0, 0) rectangle (0.5, 1);
							\node[anchor=west, ColorBlind1] at (0, 0.5) {3};
							\draw[fill=ColorBlind3!20!white] (0, 1) rectangle (0.5, 2);
							\node[anchor=west, ColorBlind3] at (0, 1.5) {1};
							\node at (0.25, 2.3) {$p_2$};
							\node[ColorBlind1] at (-0.5, 0.5) {$A_1$};
							\node[ColorBlind3] at (-0.5, 1.5) {$A_3$};
							\node at (-0.6, -0.4) {$\alpha_{N, \profile}^{\min}$};
							\node at (0.25, -0.4) {1};
						\end{tikzpicture}
					};
					
					\node[fill=my-blue-verylight, rounded corners=5pt, anchor=north] at (5.25, -2.75) {
						\begin{tikzpicture}[rounded corners=0, anchor=center]
							\draw[fill=ColorBlind2!20!white] (0, 0) rectangle (0.5, 1);
							\node[anchor=west, ColorBlind2] at (0, 0.5) {4};
							\draw[fill=ColorBlind3!20!white] (0, 1) rectangle (0.5, 2);
							\node[anchor=west, ColorBlind3] at (0, 1.5) {1};
							\node at (0.25, 2.3) {$p_2$};
							\node[ColorBlind2] at (-0.5, 0.5) {$A_2$};
							\node[ColorBlind3] at (-0.5, 1.5) {$A_3$};
							\node at (-0.6, -0.4) {$\alpha_{N, \profile}^{\min}$};
							\node at (0.25, -0.4) {1};
						\end{tikzpicture}
					};
					
					\node[fill=my-blue-verylight, rounded corners=5pt, anchor=north] at (8, -3.75) {
						\begin{tikzpicture}[rounded corners=0, anchor=center]
							\draw[fill=ColorBlind1!20!white] (0, 0) rectangle (0.5, 1);
							\node[anchor=west, ColorBlind1] at (0, 0.5) {3};
							\node at (0.25, 1.3) {$p_2$};
							\node[ColorBlind1] at (-0.5, 0.5) {$A_1$};
							\node at (-0.6, -0.4) {$\alpha_{N, \profile}^{\min}$};
							\node at (0.25, -0.4) {3};
						\end{tikzpicture}
					};
					
					\node[fill=my-blue-verylight, rounded corners=5pt, anchor=north] at (10.75, -3.75) {
						\begin{tikzpicture}[rounded corners=0, anchor=center]
							\draw[fill=ColorBlind2!20!white] (0, 0) rectangle (0.5, 1);
							\node[anchor=west, ColorBlind2] at (0, 0.5) {4};
							\node at (0.25, 1.3) {$p_2$};
							\node[ColorBlind2] at (-0.5, 0.5) {$A_2$};
							\node at (-0.6, -0.4) {$\alpha_{N, \profile}^{\min}$};
							\node at (0.25, -0.4) {4};
						\end{tikzpicture}
					};
					
					\node[fill=my-blue-verylight, rounded corners=5pt, anchor=north] at (13.5, -3.75) {
						\begin{tikzpicture}[rounded corners=0, anchor=center]
							\draw[fill=ColorBlind3!20!white] (0, 0) rectangle (0.5, 1);
							\node[anchor=west, ColorBlind3] at (0, 0.5) {1};
							\node at (0.25, 1.3) {$p_2$};
							\node[ColorBlind3] at (-0.5, 0.5) {$A_3$};
							\node at (-0.6, -0.4) {$\alpha_{N, \profile}^{\min}$};
							\node at (0.25, -0.4) {1};
						\end{tikzpicture}
					};
					
					\node[fill=my-blue-verylight, rounded corners=5pt, anchor=north] at (16.5, -5.125) {
						\begin{tikzpicture}[rounded corners=0, anchor=center]
							\draw[fill=ColorBlind1!20!white] (0, 0) rectangle (0.5, 1);
							\node[anchor=west, ColorBlind1] at (0, 0.5) {3};
							\draw[fill=ColorBlind2!20!white] (0, 1) rectangle (0.5, 2);
							\node[anchor=west, ColorBlind2] at (0, 1.5) {4};
							\draw[fill=ColorBlind3!20!white] (0, 2) rectangle (0.5, 3);
							\node[anchor=west, ColorBlind3] at (0, 2.5) {1};
							\node at (0.25, 3.3) {$p_2$};
							\node[ColorBlind1] at (-0.5, 0.5) {$A_1$};
							\node[ColorBlind2] at (-0.5, 1.5) {$A_2$};
							\node[ColorBlind3] at (-0.5, 2.5) {$A_3$};
							\node at (-0.6, -0.4) {$\alpha_{N, \profile}^{\min}$};
							\node at (0.25, -0.4) {1};
						\end{tikzpicture}
					};
					
					\node[fill=my-blue-verylight, rounded corners=5pt, anchor=north] at (-0.5, -6.5) {
						\begin{tikzpicture}[rounded corners=0, anchor=center]
							\draw[fill=ColorBlind1!20!white] (0, 0) rectangle (0.5, 1);
							\node[anchor=west, ColorBlind1] at (0, 0.5) {2};
							\node at (0.25, 1.3) {$p_1$};
							\node[ColorBlind1] at (-0.5, 0.5) {$A_1$};
							\node at (-0.6, -0.4) {$\alpha_{N, \profile}^{\min}$};
							\node at (0.25, -0.4) {2};
						\end{tikzpicture}
					};
					
					\node[fill=my-blue-verylight, rounded corners=5pt, anchor=north] at (2.5, -6.5) {
						\begin{tikzpicture}[rounded corners=0, anchor=center]
							\draw[fill=ColorBlind3!20!white] (0, 0) rectangle (0.5, 1);
							\node[anchor=west, ColorBlind3] at (0, 0.5) {1};
							\node at (0.25, 1.3) {$p_1$};
							\node[ColorBlind3] at (-0.5, 0.5) {$A_3$};
							\node at (-0.6, -0.4) {$\alpha_{N, \profile}^{\min}$};
							\node at (0.25, -0.4) {1};
						\end{tikzpicture}
					};
					
					\node[fill=my-blue-verylight, rounded corners=5pt, anchor=north] at (5.25, -6.5) {
						\begin{tikzpicture}[rounded corners=0, anchor=center]
							\draw[fill=ColorBlind1!20!white] (0, 0) rectangle (0.5, 1);
							\node[anchor=west, ColorBlind1] at (0, 0.5) {2};
							\draw[fill=ColorBlind3!20!white] (0, 1) rectangle (0.5, 2);
							\node[anchor=west, ColorBlind3] at (0, 1.5) {1};
							\node at (0.25, 2.3) {$p_1$};
							\node[ColorBlind1] at (-0.5, 0.5) {$A_1$};
							\node[ColorBlind3] at (-0.5, 1.5) {$A_3$};
							\node at (-0.6, -0.4) {$\alpha_{N, \profile}^{\min}$};
							\node at (0.25, -0.4) {1};
						\end{tikzpicture}
					};
					
					\node[fill=my-blue-verylight, rounded corners=5pt, anchor=north] at (8, -6.5) {
						\begin{tikzpicture}[rounded corners=0, anchor=center]
							\draw[fill=ColorBlind1!20!white] (0, 0) rectangle (0.5, 1);
							\node[anchor=west, ColorBlind1] at (0, 0.5) {3};
							\draw[fill=ColorBlind1!20!white] (0.5, 0) rectangle (1, 1);
							\node[anchor=west, ColorBlind1] at (0.5, 0.5) {2};
							\node at (0.25, 1.3) {$p_2$};
							\node at (0.75, 1.3) {$p_1$};
							\node[ColorBlind1] at (-0.5, 0.5) {$A_1$};
							\node at (-0.6, -0.4) {$\alpha_{N, \profile}^{\min}$};
							\node at (0.25, -0.4) {3};
							\node at (0.75, -0.4) {2};
						\end{tikzpicture}
					};
					
					\node[fill=my-blue-verylight, rounded corners=5pt, anchor=north] at (10.75, -6.5) {
						\begin{tikzpicture}[rounded corners=0, anchor=center]
							\draw[fill=ColorBlind3!20!white] (0, 0) rectangle (0.5, 1);
							\node[anchor=west, ColorBlind3] at (0, 0.5) {1};
							\draw[fill=ColorBlind3!20!white] (0.5, 0) rectangle (1, 1);
							\node[anchor=west, ColorBlind3] at (0.5, 0.5) {1};
							\node at (0.25, 1.3) {$p_2$};
							\node at (0.75, 1.3) {$p_1$};
							\node[ColorBlind3] at (-0.5, 0.5) {$A_3$};
							\node at (-0.6, -0.4) {$\alpha_{N, \profile}^{\min}$};
							\node at (0.25, -0.4) {1};
							\node at (0.75, -0.4) {1};
						\end{tikzpicture}
					};
					
					\node[fill=my-blue-verylight, rounded corners=5pt, anchor=north] at (13.5, -6.5) {
						\begin{tikzpicture}[rounded corners=0, anchor=center]
							\draw[fill=ColorBlind1!20!white] (0, 0) rectangle (0.5, 1);
							\node[anchor=west, ColorBlind1] at (0, 0.5) {3};
							\draw[fill=ColorBlind3!20!white] (0, 1) rectangle (0.5, 2);
							\node[anchor=west, ColorBlind3] at (0, 1.5) {1};
							\draw[fill=ColorBlind1!20!white] (0.5, 0) rectangle (1, 1);
							\node[anchor=west, ColorBlind1] at (0.5, 0.5) {2};
							\draw[fill=ColorBlind3!20!white] (0.5, 1) rectangle (1, 2);
							\node[anchor=west, ColorBlind3] at (0.5, 1.5) {1};
							\node at (0.25, 2.3) {$p_2$};
							\node at (0.75, 2.3) {$p_1$};
							\node[ColorBlind1] at (-0.5, 0.5) {$A_1$};
							\node[ColorBlind3] at (-0.5, 1.5) {$A_3$};
							\node at (-0.6, -0.4) {$\alpha_{N, \profile}^{\min}$};
							\node at (0.25, -0.4) {1};
							\node at (0.75, -0.4) {1};
						\end{tikzpicture}
					};
					
					\node[fill=my-blue-verylight, rounded corners=5pt, anchor=north] at (0, -9.25) {
						\begin{tikzpicture}[rounded corners=0, anchor=center]
							\draw[fill=ColorBlind1!20!white] (0, 0) rectangle (1, 1);
							\node[anchor=west, ColorBlind1] at (0, 0.5) {3};
							\draw[fill=ColorBlind1!20!white] (1, 0) rectangle (1.5, 1);
							\node[anchor=west, ColorBlind1] at (1, 0.5) {3};
							\draw[fill=ColorBlind2!20!white] (0, 1) rectangle (1, 2);
							\node[anchor=west, ColorBlind2] at (0, 1.5) {3};
							\draw[fill=ColorBlind2!20!white] (1, 1) rectangle (1.5, 2);
							\node[anchor=west, ColorBlind2] at (1, 1.5) {4};
							\draw[fill=ColorBlind3!20!white] (0, 2) rectangle (1, 3);
							\node[anchor=west, ColorBlind3] at (0, 2.5) {2};
							\draw[fill=ColorBlind3!20!white] (1, 2) rectangle (1.5, 3);
							\node[anchor=west, ColorBlind3] at (1, 2.5) {1};
							\node at (0.5, 3.3) {$p_3$};
							\node at (1.25, 3.3) {$p_2$};
							\node[ColorBlind1] at (-0.5, 0.5) {$A_1$};
							\node[ColorBlind2] at (-0.5, 1.5) {$A_2$};
							\node[ColorBlind3] at (-0.5, 2.5) {$A_3$};
							\node at (-0.6, -0.4) {$\alpha_{N, \profile}^{\min}$};
							\node at (0.5, -0.4) {2};
							\node at (1.25, -0.4) {1};
						\end{tikzpicture}
					};
					
					\node[fill=my-blue-verylight, rounded corners=5pt, anchor=north] at (3.1, -10.25) {
						\begin{tikzpicture}[rounded corners=0, anchor=center]
							\draw[fill=ColorBlind1!20!white] (0, 0) rectangle (1, 1);
							\node[anchor=west, ColorBlind1] at (0, 0.5) {3};
							\draw[fill=ColorBlind1!20!white] (1, 0) rectangle (1.5, 1);
							\node[anchor=west, ColorBlind1] at (1, 0.5) {3};
							\draw[fill=ColorBlind2!20!white] (0, 1) rectangle (1, 2);
							\node[anchor=west, ColorBlind2] at (0, 1.5) {3};
							\draw[fill=ColorBlind2!20!white] (1, 1) rectangle (1.5, 2);
							\node[anchor=west, ColorBlind2] at (1, 1.5) {4};
							\node at (0.5, 2.3) {$p_3$};
							\node at (1.25, 2.3) {$p_2$};
							\node[ColorBlind1] at (-0.5, 0.5) {$A_1$};
							\node[ColorBlind2] at (-0.5, 1.5) {$A_2$};
							\node at (-0.6, -0.4) {$\alpha_{N, \profile}^{\min}$};
							\node at (0.5, -0.4) {3};
							\node at (1.25, -0.4) {4};
						\end{tikzpicture}
					};
					
					\node[fill=my-blue-verylight, rounded corners=5pt, anchor=north] at (6.2, -10.25) {
						\begin{tikzpicture}[rounded corners=0, anchor=center]
							\draw[fill=ColorBlind1!20!white] (0, 0) rectangle (1, 1);
							\node[anchor=west, ColorBlind1] at (0, 0.5) {3};
							\draw[fill=ColorBlind1!20!white] (1, 0) rectangle (1.5, 1);
							\node[anchor=west, ColorBlind1] at (1, 0.5) {3};
							\draw[fill=ColorBlind3!20!white] (0, 1) rectangle (1, 2);
							\node[anchor=west, ColorBlind3] at (0, 1.5) {2};
							\draw[fill=ColorBlind3!20!white] (1, 1) rectangle (1.5, 2);
							\node[anchor=west, ColorBlind3] at (1, 1.5) {1};
							\node at (0.5, 2.3) {$p_3$};
							\node at (1.25, 2.3) {$p_2$};
							\node[ColorBlind1] at (-0.5, 0.5) {$A_1$};
							\node[ColorBlind3] at (-0.5, 1.5) {$A_3$};
							\node at (-0.6, -0.4) {$\alpha_{N, \profile}^{\min}$};
							\node at (0.5, -0.4) {2};
							\node at (1.25, -0.4) {1};
						\end{tikzpicture}
					};
					
					\node[fill=my-blue-verylight, rounded corners=5pt, anchor=north] at (9.3, -10.25) {
						\begin{tikzpicture}[rounded corners=0, anchor=center]
							\draw[fill=ColorBlind2!20!white] (0, 0) rectangle (1, 1);
							\node[anchor=west, ColorBlind2] at (0, 0.5) {3};
							\draw[fill=ColorBlind2!20!white] (1, 0) rectangle (1.5, 1);
							\node[anchor=west, ColorBlind2] at (1, 0.5) {4};
							\draw[fill=ColorBlind3!20!white] (0, 1) rectangle (1, 2);
							\node[anchor=west, ColorBlind3] at (0, 1.5) {2};
							\draw[fill=ColorBlind3!20!white] (1, 1) rectangle (1.5, 2);
							\node[anchor=west, ColorBlind3] at (1, 1.5) {1};
							\node at (0.5, 2.3) {$p_3$};
							\node at (1.25, 2.3) {$p_2$};
							\node[ColorBlind2] at (-0.5, 0.5) {$A_2$};
							\node[ColorBlind3] at (-0.5, 1.5) {$A_3$};
							\node at (-0.6, -0.4) {$\alpha_{N, \profile}^{\min}$};
							\node at (0.5, -0.4) {2};
							\node at (1.25, -0.4) {1};
						\end{tikzpicture}
					};

					\node[fill=my-blue-verylight, rounded corners=5pt, anchor=north] at (12.4, -10.25) {
						\begin{tikzpicture}[rounded corners=0, anchor=center]
							\draw[fill=ColorBlind1!20!white] (0, 0) rectangle (1, 1);
							\node[anchor=west, ColorBlind1] at (0, 0.5) {3};
							\draw[fill=ColorBlind1!20!white] (1, 0) rectangle (1.5, 1);
							\node[anchor=west, ColorBlind1] at (1, 0.5) {2};
							\draw[fill=ColorBlind3!20!white] (0, 1) rectangle (1, 2);
							\node[anchor=west, ColorBlind3] at (0, 1.5) {2};
							\draw[fill=ColorBlind3!20!white] (1, 1) rectangle (1.5, 2);
							\node[anchor=west, ColorBlind3] at (1, 1.5) {1};
							\node at (0.5, 2.3) {$p_3$};
							\node at (1.25, 2.3) {$p_1$};
							\node[ColorBlind1] at (-0.5, 0.5) {$A_1$};
							\node[ColorBlind3] at (-0.5, 1.5) {$A_3$};
							\node at (-0.6, -0.4) {$\alpha_{N, \profile}^{\min}$};
							\node at (0.5, -0.4) {2};
							\node at (1.25, -0.4) {1};
						\end{tikzpicture}
					};
					
					\node[fill=my-blue-verylight, rounded corners=5pt, anchor=north] at (15.75, -10.25) {
						\begin{tikzpicture}[rounded corners=0, anchor=center]
							\draw[fill=ColorBlind1!20!white] (0, 0) rectangle (1, 1);
							\node[anchor=west, ColorBlind1] at (0, 0.5) {3};
							\draw[fill=ColorBlind1!20!white] (1, 0) rectangle (1.5, 1);
							\node[anchor=west, ColorBlind1] at (1, 0.5) {3};
							\draw[fill=ColorBlind1!20!white] (1.5, 0) rectangle (2, 1);
							\node[anchor=west, ColorBlind1] at (1.5, 0.5) {2};
							\draw[fill=ColorBlind3!20!white] (0, 1) rectangle (1, 2);
							\node[anchor=west, ColorBlind3] at (0, 1.5) {2};
							\draw[fill=ColorBlind3!20!white] (1, 1) rectangle (1.5, 2);
							\node[anchor=west, ColorBlind3] at (1, 1.5) {1};
							\draw[fill=ColorBlind3!20!white] (1.5, 1) rectangle (2, 2);
							\node[anchor=west, ColorBlind3] at (1.5, 1.5) {1};
							\node at (0.5, 2.3) {$p_3$};
							\node at (1.25, 2.3) {$p_2$};
							\node at (1.75, 2.3) {$p_1$};
							\node[ColorBlind1] at (-0.5, 0.5) {$A_1$};
							\node[ColorBlind3] at (-0.5, 1.5) {$A_3$};
							\node at (-0.6, -0.4) {$\alpha_{N, \profile}^{\min}$};
							\node at (0.5, -0.4) {2};
							\node at (1.25, -0.4) {1};
							\node at (1.75, -0.4) {1};
						\end{tikzpicture}
					};
				\end{tikzpicture}
			}
		\end{center}
	\end{illustration}
	
	\subsubsection{Extended Justified Representation for Cardinal Ballots}
	
	We want to ensure that cohesive groups receive what they deserve. But what exactly do cohesive groups deserve? Consider an $(\alpha, P)$-cohesive group $N$. Since agents in $N$ control enough share of the budget to afford $P$, the most natural idea would be to guarantee all agents in $N$ at least as much satisfaction as they all agree $P$ would offer them (captured by $\alpha$). This idea is captured by the following axiom: \emph{strong extended justified representation}.\footnote{Note here that we slightly differ from the definition of \citet{PPS21}. Indeed, in the definition of Strong-EJR (and EJR) they consider any $(\alpha, P)$-cohesive group while we only use a specific $\alpha$, namely $\alpha^{\min}$. The two definitions are however equivalent and we believe this one to be clearer since it requires one less universal quantification.}
	
	\begin{definition}[Strong Extended Justified Representation]
		Given an instance $I = \tuple{\projSet, c, b}$ a profile $\profile$ of cardinal ballots, a budget allocation $\pi \in \allocSet(I)$ is said to satisfy \emph{strong extended justified representation} (Strong-EJR) if for all $P \subseteq \projSet$, all $(\alpha_{N, \profile}^{\min}, P)$-cohesive groups $N$, and all $i^\star \in N$, we have:
		\[\sum_{p \in \pi} A_{i^\star}(p) \geq \sum_{p \in P} \min_{i \in N} A_i(p).\]
	\end{definition}
	
	\noindent Remember that when using cardinal ballots, we made the assumption that the satisfaction of an agent behaves additively, so the left-hand-side of the inequality above represents the agent's satisfaction. It is also important to keep in mind that the right-hand-side of the inequality corresponds to the definition of $\alpha_{N, \profile}^{\min}$.
	
	Even though Strong-EJR is quite appealing (or at least somewhat natural), it is unsatisfiable in general. This was already observed in multi-winner voting \citep{ABCEFW17}.
	
	Given this impossibility, the focus is usually put on (simple) \emph{extended justified representation} \citep{ABCEFW17,PPS21}. This is a weakening of Strong-EJR requiring only one member of each cohesive group to reach the satisfaction threshold. We thus switch one quantifier from a universal one to an existential one in the definition.
	
	\begin{definition}[Extended Justified Representation]
		Given an instance $I = \tuple{\projSet, c, b}$ a profile $\profile$ of cardinal ballots, a budget allocation $\pi \in \allocSet(I)$ is said to satisfy \emph{extended justified representation} (EJR) if for all $P \subseteq \projSet$ and all $(\alpha_{N, \profile}^{\min}, P)$-cohesive groups $N$, there exists $i^\star \in N$ such that:
		\[\sum_{p \in \pi} A_{i^\star}(p) \geq \sum_{p \in P} \min_{i \in N} A_i(p).\]
	\end{definition}
	
	The first thing to note is that EJR does not suffer the same drawback as Strong-EJR: it can always be satisfied.
	
	\begin{theorem}[\citealt{PPS21}]
		\label{thm:EJR_AlwaysExist}
		For every instance $I$, there exists a budget allocation $\pi \in \allocSet(I)$ that satisfies EJR.
	\end{theorem}
	
	\noindent What \citet{PPS21} actually prove is that a \emph{greedy cohesive rule}\footnote{The idea behind a greedy cohesive rule is to consider all the cohesive groups, and to greedily select sets of projects $P$ for which there is a suitable $(\alpha, P)$-cohesive group with ``maximum'' $\alpha$. This is a general scheme for procedures as the notion of ``suitable cohesive group'' differs depending on the goal. Such procedures have notably been defined and studied by \citet{ALT18}, \citet{PPS21} and \citet{MREL23} in the context of PB.} always returns a feasible budget allocation that satisfies EJR (it even satisfies \emph{full justified representation}, see Section~\ref{subsubsec:Full_Justified_Representation}). This rule is interesting at a theoretical level but is quite artificial and thus not really appealing at a practical level. One of its main drawbacks is that it runs in exponential time. This, however, seems to be unavoidable to satisfy EJR.
	
	\begin{theorem}[\citealt{PPS21}]
		\label{thm:EJR_HardToCompute}
		There is no strongly polynomial time algorithm that computes, given an instance $I$ and a profile $\profile$ of cardinal ballots, a budget allocation $\pi \in \allocSet(I)$ satisfying EJR unless $\complexP = \complexNP$, even if there is only one voter.
	\end{theorem}

	\noindent Interestingly, the hardness proof shows that the running time of an algorithm finding an EJR budget allocation has to be exponential in $\log(b)$, while the greedy cohesive rule mentioned above runs in time exponential in $n$, the number of voters. Closing this gap is an interesting open problem.

	Let us quickly mention another computational result: checking whether a given budget allocation satisfies EJR is a \complexcoNP{}-complete problem. This is because it was already the case in the unit-cost setting with approval ballots \citep{ABCEFW17}.

	\medskip
	
	In the hope of achieving polynomial-time computability, a relaxation of EJR has been introduced: EJR \emph{up to one project} (EJR-1). It relaxes EJR in the following way: for each cohesive group $N$, it can be the case that no agent in $N$ enjoys enough satisfaction but at least one agent would reach the desired level of satisfaction if we were to select an additional project. This concept can be interpreted as requiring that the satisfaction can only be at most one project away from the objective.
	
	\begin{definition}[Extended Justified Representation up to One Project]
		Given an instance $I = \tuple{\projSet, c, b}$ and a profile $\profile$ of cardinal ballots, a budget allocation $\pi \in \allocSet(I)$ is said to satisfy \emph{extended justified representation up to one project} (EJR-1) if for all $P \subseteq \projSet$ and all $(\alpha_{N, \profile}^{\min}, P)$-cohesive groups $N$, there exists $i^\star \in N$ such that either:
		\[\sum_{p \in \pi} A_{i^\star}(p) \geq \sum_{p \in P} \min_{i \in N} A_i(p),\]
		or for some $p^\star \in P \setminus \pi$ we have:
		\[A_{i^\star}(p^\star) + \sum_{p \in \pi } A_{i^\star}(p) > \sum_{p \in P} \min_{i \in N} A_i(p).\]
	\end{definition}
	
	\noindent One might be surprised by the strict inequality in the definition above. It is  there for technical reasons: It ensures that EJR and EJR-1 coincide in the unit cost setting when used with approval ballots.\footnote{EJR and EJR-1 do not coincide in the unit cost setting with generic cardinal ballots as presented by \citet{PPS21} in Footnote~8 of the ArXiv version.} It also has interesting consequences in terms of the fairness properties that EJR-1 implies.\footnote{Notably, having a strict inequality ensures that EJR-1 implies a property that could be called basic proportionality, which requires that if for a group of agents $N$ for which there is exists $P \subseteq \projSet$ such that, first, $\nicefrac{|N|}{n} \cdot b \geq c(P)$, second for all $i \in N$, $A_i(p) > 0$ if and only if $p \in P$, and third, $A_i(p) = A_{i'}(p)$ for all $i, i' \in N$ and $p \in P$, then $P$ must be selected. This is not the case if EJR-1 is defined with a weak inequality.}
	
	One of the main results from \citet{PPS21} is that \mes{} does satisfy EJR-1. Given that \mes{} runs in strongly polynomial-time, this shows that a budget allocation satisfying EJR-1 can always be computed in polynomial time.
	
	\begin{theorem}[\citealt{PPS21}]
		For every instance $I$ and profile $\profile$ of cardinal ballots, $\mes(I, \profile)$ satisfies EJR-1.
	\end{theorem}

	\subsubsection{Proportional Justified Representation for Cardinal Ballots}
	
	Going down the list of weakenings of Strong-EJR, we have now reached proportional justified representation (PJR) \citep{SELFFVS17}. While EJR required at least one member of each cohesive group to enjoy the required satisfaction, PJR requires the group altogether---acting as a single agent---to achieve the required satisfaction. We provide below the definition from \citet{LCG22} who defined it for PB with cardinal ballots.

	\begin{definition}[Proportional Justified Representation]
		Given an instance $I = \tuple{\projSet, c, b}$ and a profile $\profile$ of cardinal ballots, a budget allocation $\pi \in \allocSet(I)$ is said to satisfy \emph{proportional justified representation} (PJR) if for all $P \subseteq \projSet$ and all $(\alpha_{N, \profile}^{\min}, P)$-cohesive groups $N$ we have:
		\[\sum_{p \in \pi} \max_{i \in N} A_i(p) \geq \sum_{p \in P} \min_{i \in N} A_i(p).\]
	\end{definition}
	
	It should be quite obvious that any budget allocation satisfying EJR also satisfies PJR. From this, we can derive that theorems~\ref{thm:EJR_AlwaysExist} and~\ref{thm:EJR_HardToCompute} also apply to PJR. Specifically, we know that \emph{(i)} for every instance, there exists a feasible budget allocation that satisfies PJR, and \emph{(ii)} there exists no polynomial-time algorithm computing PJR budget allocations unless $\complexP = \complexNP$. To see why the second point holds, observe that PJR and EJR coincide when there is only a single agent and that Theorem \ref{thm:EJR_HardToCompute} holds for one-agent profiles.
	
	Interestingly, the problem of checking whether a budget allocation satisfies PJR or not is still a \complexcoNP{}-complete problem (remember that this was already the case for EJR), and that already on unit-cost instances with approval ballots \citep{AEHLSS18}.
	
	\medskip
	
	\citet{LCG22} show that a PB adaption of the PAV rule fails to satisfy PJR. This might come as a surprise since PAV satisfies EJR in the case of multi-winner voting elections.

	\subsubsection{Illustration of Justified Representation Concepts with Cardinal Ballots}
	
	We conclude the section on cardinal ballots by illustrating the concepts we have introduced on an example.
	
	\begin{illustration}[Justified Representation Concepts for Cardinal Ballots]
		We repeat the instance $I$ and profile $\profile$ we presented when illustrating cohesive groups. We will review the different concepts we introduced above.
		\begin{center}
			\begin{tabular}{ccccc}
				\toprule
				& $p_1$ & $p_2$ & $p_3$ & $p_4$ \\
				Cost & 1 & 1 & 2 & 6 \\
				\midrule
				$A_1$ & 2 & 3 & 3 & 0 \\
				$A_2$ & 0 & 4 & 3 & 1 \\
				$A_3$ & 1 & 1 & 2 & 4 \\
				\midrule
				\multicolumn{5}{c}{$b = 6$} \\
				\bottomrule
			\end{tabular}
		\end{center}
		Let us start with Strong-EJR. We claim that $\{p_1, p_2, p_3\}$ is the only budget allocation satisfying Strong-EJR for $I$ and $\profile$. Since $\{1, 3\}$ is $(\alpha, \{p_1, p_2, p_3\})$-cohesive for $\alpha$ such that $\alpha(p_1) = \alpha(p_2) = 1$ and $\alpha(p_3) = 2$, agent 3 deserves a satisfaction of 4 (according to Strong-EJR). This can only be achieved if $p_1$, $p_2$ and $p_3$ are selected. This implies that $p_4$ cannot be selected. Since for all agents the satisfaction for $\{p_1, p_2, p_3\}$ is higher than that of $\{p_4\}$, the satisfaction for all cohesive groups meets the requirements of Strong-EJR.
		
		We move to EJR now. It is clear that $\{p_1, p_2, p_3\}$ satisfies EJR since it satisfies Strong-EJR. We claim that $\{p_2, p_3\}$ also does. Indeed, with the budget allocation $\{p_2, p_3\}$, agent 1 has satisfaction 6 and agent 2 satisfaction 7. These are, for both agents, more than the maximum requirement they get from cohesiveness. Thus, every cohesive group containing either 1 or 2 will have at least one agent with enough satisfaction, thus being satisfied in the sense of EJR. Finally, the cohesive groups involving agent 3 alone lead to satisfaction requirements of 2 for agent 3, less than the achieved satisfaction of 3 for $\{p_2, p_3\}$. EJR is thus satisfied. Note that $\{p_2, p_3\}$ fails Strong-EJR as agent only has satisfaction 3 and deserved satisfaction 4 according to Strong-EJR.
		
		Next in line is PJR. We claim that $\{p_1, p_2\}$ satisfies PJR (on top of $\{p_2, p_3\}$ and $\{p_1, p_2, p_3\}$ which satisfy EJR and thus PJR). The reason $\{p_1, p_2\}$ fails EJR is that $N = \{1, 2\}$ is $(\alpha_{N, \profile}^{\min}, \{p_2, p_3\})$-cohesive and thus either one of them needs to have satisfaction at least 6. This is not a problem with PJR, as the requirement is that satisfaction 6 should be achieved for the group $\{1, 2\}$ as if it was a single agent. In the PJR perspective, the group satisfaction of $\{1, 2\}$ for $\{p_1, p_2\}$ is $2 + 4 = 6$. The satisfaction requirement is thus met. One can then check that this is true for all other cohesive groups.
		
		We conclude by considering EJR-1. In addition to $\{p_1, p_2, p_3\}$ and $\{p_2, p_3\}$ that already satisfy EJR, the budget allocations $\{p_2\}, \{p_3\}, \{p_1, p_3\}$ and $\{p_1, p_2\}$ satisfy EJR-1. This is clear since by a single project to each of the budget allocations, we reach either $\{p_1, p_2, p_3\}$ or $\{p_2, p_3\}$ that are known to satisfy EJR. Note that the argument we present here adds the same project for all agents while the definition of EJR-1 allows us to add different projects to boost the satisfaction of different agents. The simplicity of the election we consider here allows us to have such a succinct argument. It is also interesting to check that $\{p_4\}$ fails EJR-1. Indeed with $\{p_4\}$ agent 1 has satisfaction 0 and agent 2 satisfaction 1. Whichever project is added either for agent 1 or 2 will never make one of them reach satisfaction 6 which is required by EJR. EJR-1 is thus not satisfied.
		
		Overall, the situation for $I$ and $\profile$ is as described in the following table. Note that the fact that the budget allocations satisfaying EJR-1 form a subset of the ones satisfying PJR is just a coincidence (PJR does not imply EJR-1).
		
		\begin{center}
			\begin{tabular}{rl}
				\toprule
				Fairness Criteria & Budget Allocations Satisfying the Fairness Criteria\\
				\midrule
				Strong-EJR & $\{p_1, p_2, p_3\}$ \\
				EJR & $\{p_1, p_2, p_3\}, \{p_2, p_3\}$ \\
				PJR & $\{p_1, p_2, p_3\}, \{p_2, p_3\}, \{p_1, p_2\}$ \\
				EJR-1 & $\{p_1, p_2, p_3\}, \{p_2, p_3\}, \{p_1, p_2\}, \{p_2\}, \{p_3\}, \{p_1, p_3\}, \{p_1, p_2\}$ \\
				\bottomrule
			\end{tabular}
		\end{center}
	\end{illustration}
	
	We are now ready to discuss approval ballots.
	
	\subsection{Justified Representation with Approval Ballots}
	
	All we presented above for cardinal ballots also applies in the case of approval ballots. However, since approval ballots are a special case of cardinal ballots, the definitions can be simplified and stronger results can be obtained.
	
	\subsubsection{Cohesive Groups for Approval Ballots}
	
	With cardinal ballots we had to introduce the $\alpha$-parameter to the definition of a cohesive group, since agents could assign different scores to the projects. This is not necessary with approval ballots.
	
	\begin{definition}[$P$-cohesive groups]
		\label{def:Cohesive_ApprovalBallots}
		Given an instance $I = \tuple{\projSet, c, b}$ and a profile $\profile$ of approval ballots, a non-empty group of agents  $N \subseteq \agentSet$ is said to be \emph{$P$-cohesive}, for a set of projects $P \subseteq \projSet$, if the following two conditions are satisfied:
		\begin{itemize}
			\item for all $p \in P$ and $i \in N$, $A_i(p) = 1$, that is, every agent in $N$ approves all projects in $P$;
			\item $\frac{|N|}{n} \cdot b \geq c(P)$, that is, $N$'s share of the budget is enough to afford $P$.
		\end{itemize}
	\end{definition}
	
	Remember the interpretation we had of cohesive groups: these are groups of agents that deserve some satisfaction in the final outcome. When using approval ballots, we will use \emph{(approval-based) satisfaction functions} as introduced in Section \ref{subsec:Satisfaction} as measures of satisfaction.
	
	\begin{illustration}[Cohesive Groups for Approval Ballots]
		Consider the instance $I$ and profile $\profile$ of approval ballots as described in the table below.
		\begin{center}
			\begin{tabular}{cccccc}
				\toprule
				& $p_1$ & $p_2$ & $p_3$ & $p_4$ & $p_5$ \\
				Cost & 1 & 3 & 8 & 9 & 10\\
				\midrule
				$A_1$ & \cmark & \xmark & \cmark & \cmark & \xmark \\
				$A_2$ & \cmark & \cmark & \xmark & \cmark & \cmark \\
				$A_3$ & \cmark & \xmark & \xmark & \cmark & \xmark \\
				\midrule
				\multicolumn{6}{c}{$b = 12$} \\
				\bottomrule
			\end{tabular}
		\end{center}
		There are 11 cohesive groups in this instance. We describe them in the following.
		
		First, each agent represents $\nicefrac{1}{3}$ of the population, and thus controls 4 units of money on their own. Since the cost of $p_1$ is less than 4 and all agents approve of $p_1$, every non-empty subset of agents $N \subseteq \agentSet$ is $\{p_1\}$-cohesive. That makes up for 7 cohesive groups.
		
		In addition, agent 2 controls 4 unit of money which is less than the cost of $p_1$ and $p_2$. Since both projects are approved by agent 2, the group $\{2\}$ is $\{p_2\}$-cohesive and also $\{p_1, p_2\}$-cohesive.
		
		Finally, it should be obvious that $\agentSet$ is $\{p_4\}$-cohesive and also $\{p_1, p_4\}$-cohesive since all agents approve of both $p_1$ and $p_4$.
		
		\medskip
		
		We can present the cohesive groups in a graphical manner.
		
		\begin{center}
			\scalebox{0.8}{
				\begin{tikzpicture}
					\foreach \x in {0, 3.375, 3.75, 4.875, 8.625, 11.625} {
						\draw[dashed, my-gray-light] (\x, -1.5) -- (\x, 5.25);
					}
					
					\foreach \x in {0, 1.5, 3, 4.5} {
						\draw[dashed, my-gray-light] (-1.3, \x) -- (12.925, \x);
					}
					
					\draw[fill=ColorBlind1!20!white] (0, 0) rectangle (3.375, 1.5);
					\draw[fill=ColorBlind2!20!white] (0, 1.5) rectangle (3.375, 3);
					\draw[fill=ColorBlind3!20!white] (0, 3) rectangle (3.375, 4.5);
					
					\draw[fill=ColorBlind1!20!white] (3.375, 0) rectangle (3.75, 1.5);
					\draw[fill=ColorBlind2!20!white] (3.375, 1.5) rectangle (3.75, 3);
					\draw[fill=ColorBlind3!20!white] (3.375, 3) rectangle (3.75, 4.5);
					
					\draw[fill=ColorBlind2!20!white] (3.75, 1.5) rectangle (4.875, 3);
					\draw[fill=ColorBlind2!20!white] (4.875, 1.5) rectangle (8.625, 3);
					
					\draw[fill=ColorBlind1!20!white] (8.625, 0) rectangle (11.625, 1.5);
					
					\draw [decorate, decoration={brace, amplitude=10pt, mirror, raise=4pt}, yshift=0pt] (0, 0) -- (3.375, 0);
					\node[align=center] at (1.6875, 5) {$p_4$};
					\node[align=center] at (1.6875, -0.8) {$9$};
					\draw [decorate, decoration={brace, amplitude=10pt, mirror, raise=4pt}, yshift=0pt] (3.375, 0) -- (3.75, 0);
					\node at (3.5625, 5) {$p_1$};
					\node at (3.5625, -0.8) {$1$};
					\draw [decorate, decoration={brace, amplitude=10pt, mirror, raise=4pt}, yshift=0pt] (3.75, 0) -- (4.875, 0);
					\node at (4.3125, 5) {$p_2$};
					\node at (4.3125, -0.8) {$3$};
					\draw [decorate, decoration={brace, amplitude=10pt, mirror, raise=4pt}, yshift=0pt] (4.875, 0) -- (8.625, 0);
					\node at (6.75, 5) {$p_5$};
					\node at (6.75, -0.8) {$10$};
					\draw [decorate, decoration={brace, amplitude=10pt, mirror, raise=4pt}, yshift=0pt] (8.625, 0) -- (11.625, 0);
					\node at (10.125, 5) {$p_3$};
					\node at (10.125, -0.8) {$8$};
					
					\draw [decorate, decoration={brace, amplitude=10pt, mirror, raise=4pt}, yshift=0pt] (11.625, 0) -- (11.625, 1.5);
					\node[ColorBlind1] at (-0.9, 0.75) {$A_1$};
					\node at (12.525, 0.75) {$4$};
					\draw [decorate, decoration={brace, amplitude=10pt, mirror, raise=4pt}, yshift=0pt] (11.625, 1.5) -- (11.625, 3);
					\node[ColorBlind2] at (-0.9, 2.25) {$A_2$};
					\node at (12.525, 2.25) {$4$};
					\draw [decorate, decoration={brace, amplitude=10pt, mirror, raise=4pt}, yshift=0pt] (11.625, 3) -- (11.625, 4.5);
					\node[ColorBlind3] at (-0.9, 3.75) {$A_3$};
					\node at (12.525, 3.75) {$4$};
				\end{tikzpicture}
			}
		\end{center}
		In this representation, a group of voters $N$ is $P$-cohesive if by putting side by side all the columns of the projects in $P$ and the rows of the agents in $N$ the following two conditions are met:
		\begin{itemize}
			\item The resulting shape is a fully coloured rectangle (agents approve of all the projects in $P$);
			\item The resulting rectangle is taller than it is wide (the cost of $P$ is no more than $N$'s share of the budget).
		\end{itemize}
		
		The 11 cohesive groups for $I$ and $\profile$ are thus:
		\begin{center}
			\scalebox{0.8}{
				\begin{tikzpicture}
					\node[fill=my-blue-verylight, rounded corners=5pt] at (0, 0) {
						\begin{tikzpicture}[rounded corners=0]
							\draw[fill=ColorBlind1!20!white] (3.375, 0) rectangle (3.75, 1.5);
							\node at (3.5625, 1.8) {$p_1$};
							\node[ColorBlind1] at (3, 0.75) {$A_1$};
						\end{tikzpicture}
					};
					
					\node[fill=my-blue-verylight, rounded corners=5pt] at (2, 0) {
						\begin{tikzpicture}[rounded corners=0]
							\draw[fill=ColorBlind2!20!white] (3.375, 0) rectangle (3.75, 1.5);
							\node at (3.5625, 1.8) {$p_1$};
							\node[ColorBlind2] at (3, 0.75) {$A_2$};
						\end{tikzpicture}
					};
				
					\node[fill=my-blue-verylight, rounded corners=5pt] at (4, 0) {
						\begin{tikzpicture}[rounded corners=0]
							\draw[fill=ColorBlind3!20!white] (3.375, 0) rectangle (3.75, 1.5);
							\node at (3.5625, 1.8) {$p_1$};
							\node[ColorBlind3] at (3, 0.75) {$A_3$};
						\end{tikzpicture}
					};
					
					\node[fill=my-blue-verylight, rounded corners=5pt] at (6, 0.75) {
						\begin{tikzpicture}[rounded corners=0]
							\draw[fill=ColorBlind1!20!white] (3.375, 0) rectangle (3.75, 1.5);
							\draw[fill=ColorBlind2!20!white] (3.375, 1.5) rectangle (3.75, 3);
							\node at (3.5625, 3.3) {$p_1$};
							\node[ColorBlind1] at (3, 0.75) {$A_1$};
							\node[ColorBlind2] at (3, 2.25) {$A_2$};
						\end{tikzpicture}
					};
					
					\node[fill=my-blue-verylight, rounded corners=5pt] at (8, 0.75) {
						\begin{tikzpicture}[rounded corners=0]
							\draw[fill=ColorBlind1!20!white] (3.375, 0) rectangle (3.75, 1.5);
							\draw[fill=ColorBlind3!20!white] (3.375, 1.5) rectangle (3.75, 3);
							\node at (3.5625, 3.3) {$p_1$};
							\node[ColorBlind1] at (3, 0.75) {$A_1$};
							\node[ColorBlind2] at (3, 2.25) {$A_3$};
						\end{tikzpicture}
					};
					
					\node[fill=my-blue-verylight, rounded corners=5pt] at (10, 0.75) {
						\begin{tikzpicture}[rounded corners=0]
							\draw[fill=ColorBlind2!20!white] (3.375, 1.5) rectangle (3.75, 3);
							\draw[fill=ColorBlind3!20!white] (3.375, 3) rectangle (3.75, 4.5);
							\node at (3.5625, 4.8) {$p_1$};
							\node[ColorBlind2] at (3, 2.25) {$A_2$};
							\node[ColorBlind3] at (3, 3.75) {$A_3$};
						\end{tikzpicture}
					};
					
					\node[fill=my-blue-verylight, rounded corners=5pt] at (12, 0) {
						\begin{tikzpicture}[rounded corners=0]
							\draw[fill=ColorBlind1!20!white] (3.375, 0) rectangle (3.75, 1.5);
							\draw[fill=ColorBlind2!20!white] (3.375, 1.5) rectangle (3.75, 3);
							\draw[fill=ColorBlind3!20!white] (3.375, 3) rectangle (3.75, 4.5);
							\node at (3.5625, 4.8) {$p_1$};
							\node[ColorBlind1] at (3, 0.75) {$A_1$};
							\node[ColorBlind2] at (3, 2.25) {$A_2$};
							\node[ColorBlind3] at (3, 3.75) {$A_3$};
						\end{tikzpicture}
					};
					
					\node[fill=my-blue-verylight, rounded corners=5pt] at (1.685, -4) {
						\begin{tikzpicture}[rounded corners=0]
							\draw[fill=ColorBlind1!20!white] (0, 0) rectangle (3.375, 1.5);
							\draw[fill=ColorBlind2!20!white] (0, 1.5) rectangle (3.375, 3);
							\draw[fill=ColorBlind3!20!white] (0, 3) rectangle (3.375, 4.5);
							\draw[fill=ColorBlind1!20!white] (3.375, 0) rectangle (3.75, 1.5);
							\draw[fill=ColorBlind2!20!white] (3.375, 1.5) rectangle (3.75, 3);
							\draw[fill=ColorBlind3!20!white] (3.375, 3) rectangle (3.75, 4.5);
							\node at (1.6875, 4.8) {$p_4$};
							\node at (3.5625, 4.8) {$p_1$};
							\node[ColorBlind1] at (-0.375, 0.75) {$A_1$};
							\node[ColorBlind2] at (-0.375, 2.25) {$A_2$};
							\node[ColorBlind3] at (-0.375, 3.75) {$A_3$};
						\end{tikzpicture}
					};
					
					\node[fill=my-blue-verylight, rounded corners=5pt] at (6.55, -4) {
						\begin{tikzpicture}[rounded corners=0]
							\draw[fill=ColorBlind1!20!white] (0, 0) rectangle (3.375, 1.5);
							\draw[fill=ColorBlind2!20!white] (0, 1.5) rectangle (3.375, 3);
							\draw[fill=ColorBlind3!20!white] (0, 3) rectangle (3.375, 4.5);
							\node at (1.6875, 4.8) {$p_4$};
							\node[ColorBlind1] at (-0.375, 0.75) {$A_1$};
							\node[ColorBlind2] at (-0.375, 2.25) {$A_2$};
							\node[ColorBlind3] at (-0.375, 3.75) {$A_3$};
						\end{tikzpicture}
					};
				
					\node[fill=my-blue-verylight, rounded corners=5pt] at (10, -2.5) {
						\begin{tikzpicture}[rounded corners=0]
							\draw[fill=ColorBlind2!20!white] (3.75, 1.5) rectangle (4.875, 3);
							\node at (4.3125, 3.3) {$p_2$};
							\node[ColorBlind2] at (3.375, 2.25) {$A_2$};
						\end{tikzpicture}
					};
					
					\node[fill=my-blue-verylight, rounded corners=5pt] at (10.2, -5) {
						\begin{tikzpicture}[rounded corners=0]
							\draw[fill=ColorBlind2!20!white] (3.375, 1.5) rectangle (3.75, 3);
							\draw[fill=ColorBlind2!20!white] (3.75, 1.5) rectangle (4.875, 3);
							\node at (3.5625, 3.3) {$p_1$};
							\node at (4.3125, 3.3) {$p_2$};
							\node[ColorBlind2] at (3, 2.25) {$A_2$};
						\end{tikzpicture}
					};
				\end{tikzpicture}
			}
		\end{center}
	\end{illustration}
	
	\subsubsection{Extended Justified Representation for Approval Ballots}
	
	Having defined cohesive groups, we are now ready to introduce concepts based on justified representation for approval ballots. Note that they are all parameterised by a satisfaction function. We start with Strong-EJR.
	
	\begin{definition}[Strong-EJR for Approval Ballots]
		Given an instance $I = \tuple{\projSet, c, b}$, a profile $\profile$ of approval ballots, and a satisfaction function $\satisfaction$, a budget allocation $\pi \in \allocSet(I)$ is said to satisfy \emph{strong extended justified representation for $\satisfaction$} (Strong-EJR[$\satisfaction$]) if for all $P \subseteq \projSet$ and all $P$-cohesive groups $N$, we have $\satisfaction_i(\pi) \geq \satisfaction_i(P)$ for all agents $i \in N$.
	\end{definition}
	
	\noindent As for cardinal ballots, Strong-EJR[$\satisfaction$] is quite appealing, but not satisfiable in general.
	
	\begin{proposition}
		For any satisfaction function $\satisfaction$, there exists an instance $I$ such that no budget allocation $\pi \in \allocSet(I)$ satisfies Strong-EJR[$\satisfaction$].\footnote{We are not aware of this result existing in the literature. The proof is rather simple. It relies on a counter example using three projects $p_1$, $p_2$, and $p_3$, all of cost 1. The budget limit is 2. There are four agents with the following ballots: Agent 1 approves only of $p_1$. Agent 2 approves only of $p_2$. Agent 3 approves only of $p_3$. Agent 4 approves of $p_1$, $p_2$ and $p_3$. Recall that we assume for any satisfaction function $\satisfaction$ that $\satisfaction(P) = 0$ if and only if $P = \emptyset$. Therefore, the only way to satisfy Strong-EJR[$\satisfaction$] is to select $p_1$, $p_2$, and $p_3$, which is not possible with $b = 2$.}
	\end{proposition}
	
	EJR can also be adapted quite naturally to this setting and can be shown to be always satisfiable in exponential time (using some variant of the greedy cohesive rule).
	
	\begin{definition}[Extended Justified Representation for Approval Ballots]
		Given an instance $I = \tuple{\projSet, c, b}$, a profile $\profile$ of approval ballots, and a satisfaction function $\satisfaction$, a budget allocation $\pi \in \allocSet(I)$ is said to satisfy \emph{extended justified representation for $\satisfaction$} (EJR[$\satisfaction$]) if for all $P \subseteq \projSet$ and all $P$-cohesive groups $N$, there exists $i \in N$ such that $\satisfaction_i(\pi) \geq \satisfaction_i(P)$.
	\end{definition}
	
	\begin{theorem}[\citealt{BFLMP23}]
		For every instance $I$ and every satisfaction function $\satisfaction$, there exists a budget allocation $\pi \in \allocSet(I)$ that satisfies EJR[$\satisfaction$].
	\end{theorem}
	
	Unfortunately, for large classes of satisfaction functions, it is not possible to satisfy EJR in polynomial time.
	
	\begin{theorem}[\citealt{BFLMP23}]
		\label{thm:EJR_Approval_HardToCompute}
		Let $I$ be an instance and $\satisfaction$ be a satisfaction function such that for all $P, P' \subseteq \projSet$ such that $c(P) < c(P')$ we have $\satisfaction(P) < \satisfaction(P')$. Then, there is no algorithm running in strongly polynomial time that computes, given an instance $I$ and a profile $\profile$ of approval ballots, a budget allocation $\pi \in \allocSet(I)$ satisfying EJR-[$\satisfaction$] unless $\complexP = \complexNP$, even if there is only one voter.
	\end{theorem}
	
	\noindent It is important to note that $\cardSatisfaction$ is not captured by the above statement, and indeed, budget allocations satisfying EJR[$\cardSatisfaction$] can always be computed in polynomial time using $\mesSat{\cardSatisfaction}$ \citep{PPS21, LCG22}. This is because for $\cardSatisfaction$, EJR[$\cardSatisfaction$] and EJR-1[$\cardSatisfaction$] coincide (the latter is defined below).
	
	For the same reasons as when we were considering cardinal ballots, checking whether a budget allocation satisfies EJR or not is a \complexcoNP{} problem.
	
	\medskip
		
	EJR-1 can also be adapted quite naturally to the approval setting. Remember that \citet{PPS21} proved that \mes{} always returns a budget allocation satisfying EJR-1. Since additive satisfaction functions can be interpreted as cardinal ballots, one can always compute an EJR-1[$\satisfaction$] budget allocations, for an additive satisfaction function $\satisfaction$, by running \mes{} with the cardinal ballots corresponding to $\satisfaction$. In the approval setting, we can go further and get the same result for EJR up to \emph{any} project.
	
	\begin{definition}[Extended Justified Representation up to Any Project for Approval Ballots]
		Given an instance $I = \tuple{\projSet, c, b}$, a profile $\profile$ of approval ballots, and a satisfaction function $\satisfaction$, a budget allocation $\pi \in \allocSet(I)$ is said to satisfy \emph{extended justified representation up to any project for $\satisfaction$} (EJR-X[$\satisfaction$]) if for all $P \subseteq \projSet$ and all $P$-cohesive groups $N$, there exists $i \in N$ such that for all $p^\star \in P \setminus \pi$ we have $\satisfaction_i(\pi \cup \{p^\star\}) > \satisfaction_i(P)$.
	\end{definition}
	
	\noindent EJR-X is a strengthening of EJR-1 that uses a universal quantifier on the project that bounds the distance between the justified and the actual satisfaction of an agent, instead of an existential one.
	
	One of the main results from \citet{BFLMP23} is that for a natural class of satisfaction functions, namely decreasing normalised satisfaction functions, the outcome of \mesSat{$\satisfaction$} always satisfies EJR-X[$\satisfaction$].

	\begin{definition}[DNS Function]
		\label{def:dns}
		We say a satisfaction function $\satisfaction$ has weakly \emph{decreasing normalised satisfaction} (DNS) if for all projects $p, p' \in \projSet$ with $c(p) \leq c(p')$, we have:
		\[\satisfaction(p) \leq \satisfaction(p') \qquad \text{and} \qquad \frac{\satisfaction(p)}{c(p)} \geq \frac{\satisfaction(p')}{c(p')}.\]
		In this case, we call $\satisfaction$ a DNS function.
	\end{definition}

	\noindent DNS functions ensure that more expensive projects are (weakly) better than cheaper ones; and that more expensive projects do not provide more satisfaction per cost than cheaper ones. Crucially, $\costSatisfaction$ and $\cardSatisfaction$ are DNS functions.
		
	\begin{theorem}[\citealt{BFLMP23}]
		\label{thm:EJR-X_Approval_MES}
		Let $\satisfaction$ be a DNS function. Then, for any instance $I$ and profile $\profile$ of approval ballots, $\mesSat{\satisfaction}(I, \profile)$ satisfies EJR-X[$\satisfaction$].
	\end{theorem}

	Before moving on to PJR for approval ballots, let us touch on another topic. \citet{FVMG22}  study the consequences of satisfying EJR on measures of social welfare and representation.\footnote{\citet{FVMG22} also perform a similar analysis for specific rules. However, these rules are not part of the standard set of rules we study in this paper.} Specifically, they provide bounds on the social welfare and representation guarantees of rules satisfying EJR[$\cardSatisfaction$]. In other words, they compare the maximally achievable social welfare with respect to $\cardSatisfaction$ and $\satisfaction^{CC}$ to the social welfare achieved by rules satisfying EJR.\footnote{Let us also mention that \citet{LaSk20} studied the same questions in the multiwinner voting setting.}
	
	\begin{theorem}[\citealt{FVMG22}]
		Let $\pbRule$ be a PB rule that satisfies EJR[$\cardSatisfaction$]. Then for any instance $I = \tuple{\projSet, c, b}$ and profile $\profile$ of approval ballots, we have:
		\[\frac{c_{\min}}{n \cdot b} \left\lfloor\frac{b}{c_{\max}}\right\rfloor \leq \frac{\displaystyle \sum_{i \in \agentSet} \cardSatisfaction_i(\pbRule(I, \profile))}{\displaystyle \max_{\pi \in \allocSet(I)} \sum_{i \in \agentSet} \cardSatisfaction_i(\pi)},\]
		where $c_{\min} = \min_{p \in \projSet} c(p)$ and $c_{\max} = \max_{p \in \projSet} c(p)$. Moreover, there exists an instance $I$ and a profile $\profile$ of approval ballots such that:
		\[\frac{\displaystyle \sum_{i \in \agentSet} \cardSatisfaction_i(\pbRule(I, \profile))}{\displaystyle \max_{\pi \in \allocSet(I)} \sum_{i \in \agentSet} \cardSatisfaction_i(\pi)} \leq \frac{4}{\sqrt{n}} - \frac{1}{n}.\]
		
		In addition, for any instance $I = \tuple{\projSet, c, b}$ and profile $\profile$ of approval ballots, we have:
		\[\frac{1}{n} \leq \frac{\displaystyle \sum_{i \in \agentSet} \satisfaction^{CC}(\pbRule(I, \profile))}{\displaystyle \max_{\pi \in \allocSet(I)} \sum_{i \in \agentSet} \satisfaction^{CC}(\pi)}.\]
		Moreover, there exists an instance $I$ and a profile $\profile$ of approval ballots satisfying $n \geq \nicefrac{b}{c_{\min}}$ such that:
		\[\frac{\displaystyle \sum_{i \in \agentSet} \satisfaction^{CC}(\pbRule(I, \profile))}{\displaystyle \max_{\pi \in \allocSet(I)} \sum_{i \in \agentSet} \satisfaction^{CC}(\pi)} \leq \frac{1}{n - 1}.\]
	\end{theorem}

	\subsubsection{Proportional Justified Representation for Approval Ballots}
	
	Let us now move on to PJR. Three main sets of authors have adapted PJR in the context of PB with approval ballots: \citet{ALT18}, \citet{LCG22} and \citet{BFLMP23}.
	
	We first provide the definition of PJR as stated by \citet{BFLMP23}.
	
	\begin{definition}[Proportional Justified Representation for Approval Ballots]
		Given an instance $I = \tuple{\projSet, c, b}$, a profile $\profile$ of approval ballots, and a satisfaction function $\satisfaction$, a budget allocation $\pi \in \allocSet(I)$ is said to satisfy \emph{proportional justified representation for $\satisfaction$} (PJR[$\satisfaction$]) if for all $P \subseteq \projSet$ and all $P$-cohesive groups $N$, we have:
		\[\satisfaction\left(\bigcup_{i \in N} \{p \in \pi \mid A_i(p) = 1\}\right) \geq \satisfaction(P).\]
	\end{definition}
	
	\noindent Similar adaptions of PJR to the PB setting have also been studied. PJR[$\costSatisfaction$] is equivalent to the BPJR-L property introduced by \citet{ALT18}.\footnote{Note that the definition of BPJR-L proposed by \citet{ALT18} looks more involved than PJR[$\costSatisfaction$] as they do not use the notion of cohesive groups. Close inspection should convince the reader that these two definitions are equivalent.} \citet{ALT18} also defined variants of (B)PJR based on the relative budget, which will be discussed in Section~\ref{subsec:Relativ-Budget}. Finally, PJR[$\cardSatisfaction$] has been introduced by \citet{LCG22}.
	
	For now, let us focus on PJR[$\satisfaction$]. It should be clear that for any satisfaction function $\satisfaction$, EJR[$\satisfaction$] implies PJR[$\satisfaction$]. Thus, for any instance $I$ and profile $\profile$ of approval ballots, there exists a budget allocation satisfying PJR[$\satisfaction$], however for a large class of satisfaction functions, it cannot be computed in polynomial time (see Theorem \ref{thm:EJR_Approval_HardToCompute} for the exact condition on the satisfaction function). Finally checking whether PJR[$\satisfaction$] is satisfied is \complexcoNP{}-complete for any $\satisfaction$ that is neutral with respects to projects with the same cost, and that already holds in the unit-cost setting.
	
	As we did for EJR, we can then introduce PJR-X. 
	
	\begin{definition}[Proportional Justified Representation up to Any Project for Approval Ballots]
		Given an instance $I = \tuple{\projSet, c, b}$, a profile $\profile$ of approval ballots, and a satisfaction function $\satisfaction$, a budget allocation $\pi \in \allocSet(I)$ is said to satisfy \emph{proportional justified representation up to any project for $\satisfaction$} (PJR-X[$\satisfaction$]) if for all $P \subseteq \projSet$ and all $P$-cohesive groups $N$ and any $p^\star \in P \setminus \pi$, we have:
		\[\satisfaction\left(\{p^\star\} \cup \bigcup_{i \in N} \{p \in \pi \mid A_i(p) = 1\}\right) > \satisfaction(P).\]
	\end{definition}

	Remember that we know for any DNS function $\satisfaction$ (Definition~\ref{def:dns}), that EJR-X[$\satisfaction$] can be satisfied (Theorem \ref{thm:EJR-X_Approval_MES}). Since PJR-X[$\satisfaction$] is implied by EJR-X[$\satisfaction$], this result also applies to PJR-X[$\satisfaction$]. \citet{BFLMP23} actually prove something stronger: PJR-X[$\satisfaction$] can be satisfied \emph{simultaneously} for every DNS function $\satisfaction$.
	
	\begin{theorem}[\citealt{BFLMP23}]
		Let $I$ be an instance and $\profile$ a profile. Then, $\seqPhragmen(I, \profile)$, $\maximinSupport(I, \profile)$ and $\mesSat{\cardSatisfaction}(I, \profile)$ satisfy PJR-X[$\satisfaction$] for all DNS functions $\satisfaction$ simultaneously.
	\end{theorem}
	
	\noindent Interestingly, \citet{BFLMP23} actually proved that this result holds for all rules satisfying a certain strengthening of priceability, as we will see later on (in Section~\ref{subsubsec:Priceability_PJR}).
	
	This result is rather far-reaching given its generality. Note that it generalises the result of \citet{LCG22} who prove that \seqPhragmen{} satisfies PJR-1[$\cardSatisfaction$]. It also generalises the result of \citet{ALT18} that \maximinSupport{} satisfies a property called Local-BPJR-L[$\costSatisfaction$] as explained below.
	
	Finally, note that this result cannot be generalised to EJR-X, as \citet{BFLMP23} show that there are instances where EJR-1$[\costSatisfaction]$ and EJR-1$[\cardSatisfaction]$ are incompatible.
	
	\medskip
	
	Before \citet{BFLMP23} introduced their definition of PJR parameterised by a satisfaction function, \citet{ALT18} defined PJR for PB with approval ballots. As we have mentioned before, they introduced an axiom called BPJR-L---that is equivalent to PJR[$\costSatisfaction$]---and proved that budget allocations satisfying it could not be found in polynomial time (unless $\complexP = \complexNP$). Due to this observation, they introduced Local-BPJR-L, a weakening of PJR[$\costSatisfaction$]. Let us provide the definition of this axiom. Note that we use here the definition of \citet{BFLMP23} who extended it to work with arbitrary satisfaction functions. The original definition of \citet{ALT18} would correspond to Local-BPJR-L[$\costSatisfaction$].
	
	\begin{definition}[Local Budget Proportional Justified Representation for the Budget Limit]
		Given an instance $I = \tuple{\projSet, c, b}$, a profile $\profile$ of approval ballots, and a satisfaction function $\satisfaction$, a budget allocation $\pi \in \allocSet(I)$ is said to satisfy \emph{Local-BPJR-L[$\satisfaction$]} if for all $P \subseteq \projSet$ and all $P$-cohesive groups $N$, it is the case that for every $P^\star \subseteq P$ such that $\{p \in \pi \mid \exists i \in N, A_i(p) = 1\} \subsetneq P^\star$ we have:
		\[P^\star \notin \argmax_{\substack{P' \subseteq \{p \in \projSet \mid \forall i \in N, A_i(p) = 1\}\\c(P') \leq c(P)}} \satisfaction(P').\]
	\end{definition}
	
	One of the main results of \citet{ALT18} is that \maximinSupport{} satisfies Local-BPJR-L[$\costSatisfaction$]. Later on, \citet{BFLMP23} further explored the relationship between different properties and proved that any budget allocation satisfying PJR-X[$\satisfaction$] also satisfies Local-BPJR-L[$\satisfaction$] (so \seqPhragmen{}, \maximinSupport{} and \mesSat{$\costSatisfaction$} satisfy Local-BPJR-L[$\costSatisfaction$]). In addition they showed that in the unit-cost case, Local-BPJR-L does not coincide with PJR, while PJR-X does.
	
	It is also worth mentioning that \citet{ALT18} also introduced another axiom called \emph{Strong-BPJR-L}. It is satisfied by a budget allocation $\pi$ if for every $\ell \in [1, b]$, and for every group of voters $N$ that controls $\ell$ units of budget, \textit{i.e.}, $\nicefrac{|N|}{n} \cdot b \geq \ell$, and that unanimously approve projects of total cost more than $\ell$, \textit{i.e}, $c(\{p \in \projSet \mid \forall i \in N, A_i(p) = 1\}) \geq \ell$, we have $c\left(\bigcup_{i \in N} \{p \in \pi \mid A_i(p) = 1\}\right) \geq \ell$. Because of the indivisibility of the projects, this axiom cannot always be satisfied. Note that this definition implicitly uses the satisfaction function $\costSatisfaction$ as the groups of voter claiming $\ell$ units of budget need to enjoy collectively a cost-satisfaction of at least $\ell$. Because of this limited applicability, we chose not to focus on this notion. Note that Strong-BPJR-L is a strengthening of PJR[$\costSatisfaction$] (which is equivalent to BPJR-L) as the condition on the group of agents $N$ is weaker. 
	
	\subsubsection{Illustration of Justified Representation Concepts with Cardinal Ballots}
	
	We conclude the section on approval ballots by illustrating the concepts we have introduced on an example.
	
	\begin{illustration}[Justified Representation Concepts for Approval Ballots]
		Consider again the instance $I$ and profile $\profile$ we described when illustration cohesive groups for approval ballots.
		\begin{center}
			\begin{tabular}{cccccc}
				\toprule
				& $p_1$ & $p_2$ & $p_3$ & $p_4$ & $p_5$ \\
				Cost & 1 & 3 & 8 & 9 & 10\\
				\midrule
				$A_1$ & \cmark & \xmark & \cmark & \cmark & \xmark \\
				$A_2$ & \cmark & \cmark & \xmark & \cmark & \cmark \\
				$A_3$ & \cmark & \xmark & \xmark & \cmark & \xmark \\
				\midrule
				\multicolumn{6}{c}{$b = 12$} \\
				\bottomrule
			\end{tabular}
		\end{center}
		We will review, one after the other the different fairness criteria we introduced above. We only consider the cost satisfaction function, $\costSatisfaction$. 
		
		\medskip
		
		Let us start with Strong-EJR[$\costSatisfaction$]. The only budget allocation satisfying it is $\{p_1, p_4\}$. Since $\agentSet$ is $\{p_1, p_4\}$-cohesive, all agents deserve a satisfaction of 10 according to Strong-EJR[$\costSatisfaction$]. This is only achieved by the budget allocation $\{p_1, p_4\}$. One can then check that other cohesive groups are also satisfied.
		
		\medskip
		
		We move on to EJR[$\costSatisfaction$] now. Three budget allocations satisfy it: $\{p_1, p_4\}$, $\{p_1, p_5\}$ and $\{p_2, p_4\}$. Interpreting the structure of the cohesive groups in EJR[$\costSatisfaction$] terms, we know that: $(i)$ agents 1 and 3 should have satisfaction at least 1; $(ii)$ agent 2 should have satisfaction at least 4; $(iii)$ one agent should have satisfaction at least 10. Thus if $p_4$ is selected, additionally selecting either $p_1$ or $p_2$ will satisfy these constraints. Thus $\{p_1, p_4\}$ and $\{p_2, p_4\}$ satisfy EJR[$\costSatisfaction$]. If $p_4$ is not selected, then neither $p_2$ nor $p_3$ can be selected (no agent would achieve satisfaction 10 or more). The only possibility to provide one agent with satisfaction at least 10 is thus by selecting $p_5$ and then $p_1$ to satisfy the other constraints. Overall, $\{p_1, p_5\}$ satisfies EJR[$\costSatisfaction$].
		
		Consider now EJR-X[$\costSatisfaction$]. We claim that $\{p_1\}$ satisfies EJR-X[$\costSatisfaction$]. The budget allocation $\{p_1\}$ fails EJR[$\costSatisfaction$] for two reasons: no agent receives satisfaction at least 10 and agent 2 does not reach satisfaction 4. For $\agentSet$ that is $\{p_1, p_4\}$-cohesive we can see that adding any project from $\{p_1, p_4\} \setminus \{p_1\} = \{p_4\}$ would provide all agents with satisfaction 10. This group is thus satisfied according to EJR-X[$\costSatisfaction$]. Consider now $\{2\}$ that is $\{p_1, p_2\}$-cohesive. Since adding $p_2$ would give agent 2 satisfaction 4, this is also satisfied. Overall, $\{p_1\}$ satisfies EJR-X[$\costSatisfaction$]. Other budget allocations satisfying EJR-X[$\costSatisfaction$] are presented below.
		
		We now investigate EJR-1[$\costSatisfaction$]. Consider the budget allocation $\{p_2\}$. We claim that it fails EJR-X[$\costSatisfaction$] but satisfies EJR-1[$\costSatisfaction$]. It indeed fails EJR-X[$\costSatisfaction$] as the $\{p_1, p_4\}$-cohesive group $\agentSet$ is not satisfied: adding $p_1$ does not provide any agent with satisfaction at least 10. On the other hand, adding $p_4$ provides agent 2 satisfaction 12. There exists thus a project that can be added for this group to reach the requirement imposed by EJR[$\costSatisfaction$], thus $\{p_2\}$ satisfies EJR-1[$\costSatisfaction$].
		
		\medskip
		
		We now move on to PJR[$\costSatisfaction$]. Consider the budget allocation $\{p_1, p_2, p_3\}$. It fails EJR[$\costSatisfaction$] since no agent has satisfaction of 10, though $\agentSet$ is $\{p_1, p_4\}$-cohesive. In the case of PJR, the group is considered as a single agent that would have submitted the union of the ballots of the members of the group  as an approval ballot. Following that logic, the group satisfaction of $\agentSet$ for $\{p_1, p_2, p_3\}$ is $1 + 3 + 8 = 11$, higher than the satisfaction received from $\{p_1, p_4\}$. The $\{p_1, p_4\}$-cohesive group $\agentSet$ is thus satisfied according to PJR[$\costSatisfaction$]. It is easy to check that other cohesive groups also are.
		
		We do not illustrate PJR-X[$\costSatisfaction$] and PJR-1[$\costSatisfaction$] as they are similar in spirit as EJR-X[$\costSatisfaction$] and EJR-1[$\costSatisfaction$].
		
		\medskip
		
		Let us conclude by presenting all the budget allocations satisfying the different fairness criteria for $I$ and $\profile$.
		
		\begin{center}
			\resizebox{\linewidth}{!}{
				\begin{tabular}{rl}
					\toprule
					\begin{tabular}{@{}c@{}}Criteria \\ for $\costSatisfaction$\end{tabular} & Budget Allocations Satisfying the Fairness Criteria\\
					\midrule
					Strong-EJR & $\{p_1, p_4\}$ \\
					EJR & $\{p_1, p_4\}, \{p_1, p_5\}, \{p_2, p_4\}$ \\
					EJR-X & $\{p_1, p_4\}, \{p_1, p_5\}, \{p_2, p_4\}, \{p_1\}, \{p_4\}, \{p_5\}, \{p_1, p_3\}, \{p_1, p_2\}, \{p_1, p_2, p_3\}$ \\
					EJR-1 & $\{p_1, p_4\}, \{p_1, p_5\}, \{p_2, p_4\}, \{p_1\}, \{p_4\}, \{p_5\}, \{p_1, p_3\}, \{p_1, p_2\}, \{p_1, p_2, p_3\}, \{p_2\}, \{p_2, p_3\}$ \\
					PJR & $\{p_1, p_4\}, \{p_1, p_5\}, \{p_2, p_4\}, \{p_1, p_2, p_3\}$ \\
					PJR-X & $\{p_1, p_4\}, \{p_1, p_5\}, \{p_2, p_4\}, \{p_1, p_2, p_3\}, \{p_1\}, \{p_4\}, \{p_5\}, \{p_1, p_3\}, \{p_1, p_2\}, \{p_2, p_3\}$ \\
					PJR-1 & $\{p_1, p_4\}, \{p_1, p_5\}, \{p_2, p_4\},  \{p_1, p_2, p_3\}, \{p_1\}, \{p_4\}, \{p_5\}, \{p_1, p_3\}, \{p_1, p_2\}, \{p_2, p_3\}, \{p_2\}$ \\
					\bottomrule
				\end{tabular}
			}
		\end{center}
	\end{illustration}
	
	\section{The Core}
	\label{subsec:Core}
	
	Intuitively, EJR guarantees that in every cohesive group there is at least one voter that receives as much satisfaction as the group could guarantee each member if the group could spend their part of the budget as they wished. We now introduce a property that is similar in spirit, called the \emph{core}, though it does not rely on cohesive groups.
	
	\subsection{The Core with Cardinal Ballots} We start by providing the definition of the core. Note that it was first introduce by \citet{FGM16} for PB with divisible projects. The definition below, though adapted to the indivisible PB setting, is very similar.
	
	\begin{definition}[The Core of PB with Cardinal Ballots]
		Given an instance $I = \tuple{\projSet, c, b}$ and a profile $\profile$ of cardinal ballots, a budget allocation $\pi \in \allocSet(I)$ is \emph{in the core} of $I$ if for every group of voters $N \subseteq \agentSet$ and subset of projects $P \subseteq \projSet$ such that $\nicefrac{|N|}{n} \geq \nicefrac{c(P)}{b}$, there exists a voter $i^\star \in N$ with:
		\[\sum_{p \in \pi}A_{i^\star}(p) \geq \sum_{p \in P}A_{i^\star}(p).\]
	\end{definition}
	
	\noindent The core can be seen as a kind of stability condition which guarantees that no groups of agents can ``deviate'' by taking their part of the budget to fund a set of projects $P$ that gives each agent in the group a higher satisfaction than $\pi$. The core of PB is inspired by the concept of the core in cooperative game theory \citep{FGM16}, but there is no direct technical link \citep[see][for formal link in the multi-winner voting setting]{ABCEFW17}.
	
	Interestingly, EJR can be viewed as a restriction of the core where only cohesive groups are allowed to deviate. Therefore, the core can be seen as a generalisation of EJR to arbitrary groups of agents.
	
	\medskip
	
	It is known that there are instances where no budget allocation is in the core. In this case, we say that the core of the instance is \emph{empty}. \citet{PPS21} present an instance with cardinal ballots in the unit-cost setting for which no feasible budget allocation is in the core. They strengthen a first counter-example provided by \citet{FMS18} without the unit-cost assumption.\footnote{This counterexample is described in the appendix on endowment-based core of \citet{FMS18}, available at \href{https://arxiv.org/abs/1805.03164}{arxiv.org/abs/1805.03164}.}
	
	\begin{proposition}[\citealt{PPS21}]
		There exists an instance $I$ with unit costs and profile $\profile$ of cardinal ballots such that no budget allocation $\pi \in \allocSet(I)$ is in the core, even if for every agent $i \in \agentSet$ and project $p \in \projSet$ we have $A_i(p) \in \{0, 1, 2\}$.
	\end{proposition}

	\begin{illustration}[Core for Cardinal Ballots]
		Consider the instance $I$ and the profile $\profile$ of cardinal ballots as depicted below.
		\begin{center}
			\begin{tabular}{ccccc}
				\toprule
				& $p_1$ & $p_2$ & $p_3$ & $p_4$ \\
				Cost & 1 & 3 & 5 & 5 \\
				\midrule
				$A_1$ & 2 & 3 & 1 & 2 \\
				$A_2$ & 2 & 0 & 2 & 4 \\
				$A_3$ & 1 & 4 & 0 & 3 \\
				\midrule
				\multicolumn{5}{c}{$b = 6$} \\
				\bottomrule
			\end{tabular}
		\end{center}
		Without providing any details, for $I$ and $\profile$, the budget allocation $\{p_1, p_4\}$ satisfies Strong-EJR (and EJR) and the budget allocation $\{p_1, p_2\}$ satisfies EJR but not Strong-EJR. The reason $\{p_1, p_2\}$ fails Strong-EJR is that $N = \{1, 2, 3\}$ is $(\alpha_{N, \profile}^{\min}, \{p_1, p_4\})$-cohesive and thus all agents need to have satisfaction at least 3, which is not the case in $\{p_1, p_2\}$. 
		
		Interestingly, $\{p_1, p_2\}$ is in the core of $I$ and $\profile$. Indeed, it is clear that no single-agent deviation is profitable: an agent on their own can only afford $p_1$, which is already selected. Similarly, any deviation of a group of 2 agents cannot be profitable as two agents can at most afford $p_1$ and $p_2$. We are left with checking deviations by all agents. They could deviate to either $\{p_3\}$ or $\{p_4\}$, but none of them provide agent 3 with strictly more satisfaction that they have in $\{p_1, p_2\}$.
		
		To get a better understanding of the core, consider the budget allocation $\{p_1, p_4\}$ now. It satisfies EJR---a weakening of the core---but we claim that it is not in the core of $I$ and $\profile$. Indeed the group $N = \{1, 3\}$ controls 4 units of budget and can thus afford $\{p_1, p_2\}$ on their own, which provides them with strictly more satisfaction that $\{p_1, p_4\}$. The deviation is not allowed by EJR, as according to EJR the group satisfaction of $N$ for $\{p_1, p_2\}$ is only $1 + 3$, which is less than the group satisfaction of $N$ for $\{p_1, p_4\}$ ($1 + 2$).
	\end{illustration}
	
	\subsection{Approximating the Core with Cardinal Ballots} We now know that the core can be empty. This raises the question whether it is always possible to find budget allocations that are close to the core. We will present some recent answers to this question below.
	
	\medskip
	
	We start with a multiplicative approximation to the core as defined by \citet{PPS21}. This approximates the core by bounding the satisfaction the agents would enjoy when deviating.
	
	\begin{definition}[The $\alpha$-sat Approximate Core of PB with Cardinal Ballots]
		Given an instance $I = \tuple{\projSet, c, b}$, a profile $\profile$ of cardinal ballots, and a scalar $\alpha \geq 1$, a budget allocation $\pi \in \allocSet(I)$ is \emph{in the $\alpha$-sat approximate core} of $I$ if for every group of voters $N \subseteq \agentSet$ and subset of projects $P \subseteq \projSet$ such that $\nicefrac{|N|}{n} \geq \nicefrac{c(P)}{b}$, there exists a voter $i^\star \in N$ and a project $p^\star \in \projSet$ with:
		\[\sum_{p \in \pi \cup \{p^\star\}}A_{i^\star}(p) \geq \frac{\sum_{p \in P}A_{i^\star}(p)}{\alpha}.\]
	\end{definition}
	
	\noindent Note that the above is actually an additive and multiplicative approximation of the core as an extra project is also added. This follows from the known impossibility of a (simply) multiplicative approximation of the core \citep{FMS18, CJMW20, MSWW22}.

	Using the above definition of an approximation of the core, \citet{PPS21} showed that \mes{} is never too far from the core.
	
	\begin{theorem}[\citealt{PPS21}]
		Given an instance $I = \tuple{\projSet, c, b}$ and a profile $\profile$ of cardinal ballots, let $u_{\max}$ and $u_{\min}$ be the highest and lowest possible satisfaction of a voter, defined as:
		\[u_{\min} = \min_{i \in \agentSet} \min_{\substack{\pi \in \allocSet \\ \exists p \in \pi, A_i(p) > 0}} \sum_{p \in \pi} A_i(p) \qquad \text{and} \qquad u_{\max} = \max_{i \in \agentSet} \max_{\pi \in \allocSet(I)} \sum_{p \in \pi} A_i(p).\]
		Then, $\mes(I, \profile)$ is in the $\alpha$-sat approximate core of $I$ for $\alpha = 4\log(2 \cdot \nicefrac{u_{\max}}{u_{\min}})$.
	\end{theorem}

	The previous result shows that the $\bigO(\log(|\allocSet(I)|))$-sat approximate core is always non-empty for any instance $I$. Moreover, it also implies that a suitable budget allocation can be found in polynomial time. \citet{MSWW22} extend this result by showing that the $\bigO(1)$-sat approximate core is always non-empty, however it is unknown if the corresponding budget allocation can be computed in polynomial time.
	
	\begin{theorem}[\citealt{MSWW22}]
		For every instance $I$ and profile $\profile$ of cardinal ballots, the $9.27$-sat approximate core is always non-empty.
	\end{theorem}

	\noindent This result is obtained by some rather intricate rounding of fractional budget allocations. Note that it also allows \citet{MSWW22} to obtain results for non-additive cardinal ballots. These results are out of the scope of this survey.
	
	\medskip
	
	Let us now delve into a second type of approximation of the core that has been introduced: entitlement approximation. The idea is that deviations of coalitions of voters would not be possible if we were to scale down their entitlement (which is equal to $\nicefrac{b}{n}$ in the definition of the core). We provide the definition of \citet{JMW20} below.
	
	\begin{definition}[The $\alpha$-entitlement approximate core of PB with Cardinal Ballots]
		Given an instance $I = \tuple{\projSet, c, b}$, a profile $\profile$ of cardinal ballots, and a scalar $\alpha \geq 1$, a budget allocation $\pi \in \allocSet(I)$ is \emph{in the $\alpha$-entitlement approximate core} of $I$ if for every group of voters $N \subseteq \agentSet$ and subset of projects $P \subseteq \projSet$ such that $\nicefrac{|N|}{n} \geq \alpha \cdot \nicefrac{c(P)}{b}$, there exists a voter $i^\star \in N$ with:
		\[\sum_{p \in \pi}A_{i^\star}(p) \geq \sum_{p \in P}A_{i^\star}(p).\]
	\end{definition}

	By suitable rounding of lotteries over budget allocation, \citet{JMW20} were able to show that the $\bigO(1)$-entitlement approximate core is always non-empty. 

	\begin{theorem}[\citealt{JMW20}]
		For every instance $I$ and profile $\profile$ of cardinal ballots, the $32$-entitlement approximate core is always non-empty.
	\end{theorem}

	Using the above definition of approximate core, \citet{MSW22} studied the problem of \emph{core auditing} in PB. This is the computational problem that seeks, given an instance $I$, a profile $\profile$ of cardinal ballots and a budget allocation $\pi \in \allocSet(I)$, the smallest $\alpha$ such that $\pi$ is \emph{not} in the $\alpha$-entitlement approximate core. For this problem, \citet{MSW22} prove different hardness results, including hardness of approximation, and also provide a logarithmic approximation algorithm. 
	
	\subsection{The Core with Approval Ballots}
		
	Let us now turn to approval ballots, where the core is defined as follows.
	
	\begin{definition}[The Core of PB with Approval Ballots]
		Given an instance $I = \tuple{\projSet, c, b}$, a profile $\profile$ of approval ballots and a satisfaction function $\satisfaction$, a budget allocation $\pi \in \allocSet(I)$ is \emph{in the core[$\satisfaction$]} of $I$ for $\satisfaction$ if for every group of voters $N \subseteq \agentSet$ and subset of projects $P \subseteq \projSet$ such that $\nicefrac{|N|}{n} \geq \nicefrac{c(P)}{b}$, there exists a voter $i^\star \in N$ with:
		\[\satisfaction_{i^\star}(\pi) \geq \satisfaction_{i^\star}(P).\]
	\end{definition}

	Recently, \citet{Maly23} showed that core[$\satisfaction$] of an approval-based PB instance can be empty for $\costSatisfaction$, $\satisfaction^{\share}$ and satisfaction functions based on diminishing returns. The question still remains open for $\cardSatisfaction$, though even for $\cardSatisfaction$, we know that none of the established rules satisfies the core. Moreover, observe that the non-emptiness of the core remains one of the main open problems in the literature on multi-winner voting \citep{LaSk23}.

	\section{Priceability}
	\label{subsec:Priceability}
	
	The next property on our agenda is \emph{priceability}. The idea is that voters have access to a virtual currency, and if, by following simple rules, they can use their money to fund a given budget allocation, then the latter will be called \emph{priceable}. All voters receive the same amount of virtual currency initially. In that sense, priceability is a proportionality requirement as the power to influence the outcome is shared equally among the voters. It can also be seen as an explainability requirement: a priceable budget allocation is an outcome that could have been obtained if the process had been run as a market.
	
	The initial definition of priceability---in the context of multi-winner voting---is due to \citet{PeSk20}. We present below the adaptation of this definition to the context of PB proposed by \citet{PPS21} for PB with cardinal ballots.\footnote{Note that we changed the terminology to avoid using the terms ``budget'' and ``price'', which can be confused with the basic elements of an instance. This avoids sentences such as ``$\pi$ is priceable for a budget $B \geq b$''.}

	\begin{definition}[Priceability for Cardinal Ballots]
		Given an instance $I = \tuple{\projSet, c, b}$ and a profile $\profile$ of cardinal ballots, a budget allocation $\pi$ satisfies \emph{priceability}, or is \emph{priceable}, if there exists an entitlement $\alpha \in \Rplus$ and a collection $(\gamma_i)_{i \in \agentSet}$ of contribution functions $\gamma_i: \projSet \to [0, \alpha]$ such that all of the following conditions are satisfied.
		\begin{itemize}
			\item[\textbf{C1}:] If $\gamma_i(p) > 0$ then $A_i(p) > 0$ for all $p \in \projSet$ and $i \in \agentSet$: agents only contribute to projects they derive satisfaction from.
			\item[\textbf{C2}:] If $\gamma_i(p) > 0$ then $p \in \pi$ for all $p \in \projSet$ and $i \in \agentSet$: only projects in $\pi$ receive contributions.
			\item[\textbf{C3}:] $\sum_{p \in \projSet} \gamma_i(p) \leq \alpha$ for all $i \in \agentSet$: no agent contributes more than their entitlement $\alpha$.
			\item[\textbf{C4}:] $\sum_{i \in \agentSet} \gamma_i(p) = c(p) $ for all $p \in \pi$: the projects in $\pi$ are receiving sufficient contribution to be funded.
			\item[\textbf{C5}:] $\sum_{i \in \agentSet \mid A_i(p) > 0} \left(\alpha - \sum_{p' \in \projSet} \gamma_i(p')\right) \leq c(p)$ for all $p \in \projSet \setminus \pi$: no group of agents supporting an unselected project $p$ is left with more than $c(p)$.
		\end{itemize}
	
		\noindent The pair $\tuple{\alpha, (\gamma_i)_{i \in \agentSet}}$ is called a \emph{price system}.
 	\end{definition}

	\noindent Note that it would be more natural to have a strict inequality in (C5), \textit{i.e.}, to guarantee that no group of agents has enough money left over to afford a project for which each member of the group has positive utility. Unfortunately, this would be impossible to satisfy as it is sometimes necessary to do some tie-breaking between equally popular projects.
	
	Moreover, in the definition of priceability we only distinguish between assigning a zero score to a project, or a strictly positive score. Therefore, the definition does not change whether cardinal or simply approval ballots are used. Note that this definition of priceability requires the underlying assumption that satisfaction is strictly monotonic.
	
	\paragraph{Priceable Rules.} Given the similarities between the definition of priceability and that of \mes{}, it will not surprise anyone that the latter always returns priceable budget allocations.
	Maybe more surprisingly, this also is the case for sequential Phragmén and the maximin support rules.
	
	\begin{proposition}[\citealt{PPS21}]
		For every instance $I$ and profile $\profile$ of cardinal ballots, $\mes(I, \profile)$ is priceable.
	\end{proposition}

	\begin{proposition}[\citealt{LCG22}]
		For every instance $I$ and profile $\profile$ of approval ballots, $\seqPhragmen(I, \profile)$ is priceable.
	\end{proposition}
	
	\begin{proposition}[\citealt{BFLMP23}]
		For every instance $I$ and profile $\profile$ of approval ballots, $\maximinSupport(I, \profile)$ is priceable.
	\end{proposition}

	\paragraph{Priceability and PJR.}
	\label{subsubsec:Priceability_PJR}
	
	In the context of multi-winner voting, links have been drawn between PJR and pricebility \citep{PeSk20}. \citet{BFLMP23} extend this result to PB with approval ballots. They show that priceability implies PJR-X$[\costSatisfaction]$. More importantly, they show that a stronger notion of priceability implies PJR-X$[\satisfaction]$ for all DNS functions $\satisfaction$ (see Definition~\ref{def:dns}). 
	
	\begin{theorem}[\citealt{BFLMP23}]
		\label{thm:Priceability_PJR_Appoval}
		For every instance $I = \tuple{\projSet, c, b}$ and profile $\profile$ of approval ballots, consider a budget allocation $\pi \in \allocSet(I)$ that is priceable for a price system $\tuple{\alpha, (\gamma_i)_{i \in \agentSet}}$ such that $\alpha > b$ and that also satisfies the following extra condition:
		\begin{itemize}
			\item[\textbf{C6}:] $\sum_{i \in \agentSet \mid A_i(p) > 0} \gamma_i(p') \leq c(p)$ for all $p \in \projSet \setminus \pi$ and $p' \in \pi$: no group of agents can save money by jointly moving their contributions to a project that they all support.
		\end{itemize}
		Then, $\pi$ satisfies PJR-X[$\satisfaction$] for every DNS function $\satisfaction$.
	\end{theorem}

	\noindent In particular, \citet{BFLMP23} show that \mesSat{$\cardSatisfaction$}, \seqPhragmen{} and \maximinSupport{} provide budget allocations that are priceable for their extended notion of priceability. 

	\section{Proportionality in Ordinal PB}
	\label{subsec:Proportionality_Ordinal}
	
	Until now, we have focused on cardinal ballots. In the following we consider ordinal ballots and proportionality requirements for such ballots.
	
	\citet{AzLe21} is the main reference for this case. In their work, they generalise proportionality concepts for multi-winner voting with strict ordinal ballots, to the setting of PB with weak ordinal ballots. These concepts are all based on the idea of solid coalitions, the counterpart of cohesive groups when agents submit ordinal ballots.
	
	\begin{definition}[Solid Coalition]
		Let $I = \tuple{\projSet, c, b}$ be an instance and $\profile = (\succsim_i)_{i \in \agentSet}$ a profile of weak ordinal ballots. Given a subset of projects $P \subseteq \projSet$, a group of voters $N \subseteq \projSet$ is a \emph{$P$-solid coalition} if for all voters $i \in N$ and projects $p \in P$, we have $p \succsim_i p'$ for all $p' \in \projSet \setminus P$.
	\end{definition}
	
	\noindent A group of voters $N$ is thus a $P$-solid coalition if they all prefer the projects in $P$ to the ones outside of $P$.
	
	Equipped with solid coalitions, \citet{AzLe21} define two incomparable generalisations of the \emph{proportionality for solid coalitions} \citep{Dumm84}. Before defining them, we introduce new notation. Interpret a weak order $\succsim$ over $\projSet$ as a vector of indifference classes $\succsim = (P_1, P_2, \ldots)$ such that all projects in $P_j$ are preferred to the ones in $P_{j + 1} \cup P_{j + 2} \cup \cdots$. Then, let $\mathit{top}(\succsim, k)$, for $k \in \mathbb{N}$ be defined as $\mathit{top}(\succsim, k) = P_1 \cup \cdots \cup P_{j^\star} \cup P_{j^\star + 1}$ where $j^\star \in \mathbb{N}_{\geq 0}$ is the largest number such that $|\bigcup_{j = 1}^{j^\star} P_j| < k$. Thus, $\mathit{top}(\succsim, k)$ is the smallest subset of projects of size at least $k$ for which such that all projects in $\projSet \setminus \mathit{top}(\succsim, k)$ are strictly worse than all projects in $\mathit{top}(\succsim, k)$ according to $\succsim$.
	
	The first property is comparative proportionality for solid coalitions.
	
	\begin{definition}[Comparative Proportionality for Solid Coalitions]
		Given an instance $I = \tuple{\projSet, c, b}$ and profile $\profile = (\succsim_i)_{i \in \agentSet}$ of weak ordinal ballots, a budget allocation $\pi \in \allocSet(I)$ is said to satisfy \emph{comparative proportionality for solid coalitions} (CPSC) if for every $P \subseteq \projSet$, there is no $P$-solid coalition $N \subseteq \agentSet$ for which there exists $P' \subseteq P$ such that:
		\[c\left(\{p \in \pi \mid \exists i \in N \text{ such that } p \in \mathit{top}(\succsim_i, |P|)\}\right) < c(P') \leq \frac{|N|}{n} \cdot b.\]
	\end{definition}
	
	\noindent Here, the satisfaction of a $P$-solid coalition $N \subseteq \agentSet$ is measured as the total cost of the selected projects that appear in the top-$|P|$ projects of at least one agent in $N$. Based on that, a budget allocation $\pi$ satisfies CPSC if for every $P$-solid coalition $N$ there is no subset of projects $P' \subseteq P$ that $N$ can afford with their share of the budget and that has a higher group satisfaction for $N$ than the group satisfaction of $N$ in $\pi$.
	
	Note that this definition could be rewritten using a concept like approval-based satisfaction functions to account for other ways of measuring the satisfaction than the cost of the projects.
	
	\begin{definition}[Inclusion Proportionality for Solid Coalitions]
		Given an instance $I = \tuple{\projSet, c, b}$ and a profile $\profile = (\succsim_i)_{i \in \agentSet}$ of weak ordinal ballots, a budget allocation $\pi \in \allocSet(I)$ is said to satisfy \emph{inclusion proportionality for solid coalitions} (IPSC) if for every $P \subseteq \projSet$, there is no $P$-solid coalition $N \subseteq \agentSet$ for which there exists $p^\star \in P \setminus \{p \in \pi \mid \exists i \in N \text{ such that } p \in \mathit{top}(\succsim_i, |P|)\}$ such that:
		\[c\left(\{p \in \pi \mid \exists i \in N \text{ such that } p \in \mathit{top}(\succsim_i, |P|)\}\right) + c(p^\star) \leq \frac{|N|}{n} \cdot b.\]
	\end{definition}
	
	\noindent Overall, $\pi$ satisfies IPSC if for all $P$-solid coalition $N$, there is no project $p^\star$ that appears in the top-$|P|$ projects of an agent in $N$ that $N$ could afford in addition to all the projects from $\pi$ that appear in the top-$|P|$ projects of an agent in $N$.
	
	\medskip
	
	\citet{AzLe21} show that it is not always possible to find budget allocations satisfying CPSC, but that we can always find budget allocations satisfying IPSC in polynomial time.
	
	\begin{theorem}[\citealt{AzLe21}]
		There exist an instance $I$ and a profile $\profile$ of weak ordinal ballots such that no $\pi \in \allocSet(I)$ satisfies CPSC.
		
		For every instance $I$ and a profile $\profile$ of weak ordinal ballots there exists $\pi \in \allocSet(I)$ that satisfies IPSC. Such a budget allocation can be found in polynomial time.
	\end{theorem}
	
	To conclude, note that \citet{PPS21} introduce a version of \mes{} working with strict ordinal ballots that they link to the framework of \citet{AzLe21}. In particular, they show that it satisfies PSC, a weakening of the properties we defined above.
	
	\section{Other Fairness Requirements}
	\label{subsec:Other_Proportionality_Axioms}
	
	In the following section, we go through other fairness requirements that have been introduced in the literature. Since these are properties that have received less attention, we will go a bit faster on them.
	
	\subsection{Full Justified Representation}
	\label{subsubsec:Full_Justified_Representation}
			
	The first axiom we discuss is \emph{full justified representation}. \citet{PPS21} proposed this strengthening of EJR, which is the strongest axiom based on justified representation that we know can always be satisfied. It strengthens EJR by relaxing the cohesiveness requirement.
	
	\begin{definition}[Full Justified Representation for Cardinal Ballots]
		Let $I = \tuple{\projSet, c, b}$ be an instance and $\profile$ a profile of cardinal ballots. A group of voters $N \subseteq \agentSet$ is \emph{weakly $(\beta, P)$-cohesive} for a scalar $\beta \in \mathbb{R}$ and a subset of projects $P \subseteq \projSet$ if $\nicefrac{|N|}{n} \cdot b \geq c(P)$ and $\sum_{p \in P} A_i(p) \geq \beta$ for every $i \in N$.
		
		Given $I$ and $\profile$, a budget allocation $\pi \in \allocSet(I)$ satisfies \emph{full justified representation} (FJR) if for all $P \subseteq \projSet$, all $\beta \in \mathbb{R}$ and all weakly $(\beta, P)$-cohesive groups $N$, there exists $i \in N$ such that:
		\[\sum_{p \in \pi} A_i(p) \geq \beta.\]
	\end{definition}

	Using a greedy cohesive rule, \citet{PPS21} have been able to show that we can always find a budget allocation satisfying FJR. This rule is however rather artificial. It is an open problem whether there is a polynomial time rule that satisfies FJR.

	\begin{proposition}[\citealt{PPS21}]
		For any instance $I$ and profile $\profile$ of cardinal ballots, there exists a budget allocation $\pi \in \allocSet(I)$ that satisfies FJR.
	\end{proposition}
	
	\noindent Interestingly, this applies even for cardinal ballots over budget allocations, as long as they are monotone.

	FJR can be adapted to the world of PB with approval ballots. The definition is provided below.
	
	\begin{definition}[Full Justified Representation for Approval Ballots]
		Let $I = \tuple{\projSet, c, b}$ be an instance, $\profile$ a profile of approval ballots and $\satisfaction$ a satisfaction function. A budget allocation $\pi \in \allocSet(I)$ satisfies \emph{full justified representation for $\satisfaction$} (FJR[$\satisfaction$]) if for every group of voters $N \subseteq \agentSet$ and subset of projects $P \subseteq \projSet$ such that $\nicefrac{|N|}{n} \cdot b \geq c(P)$, there exists $i \in N$ for whom:
		\[\satisfaction_i(\pi) \geq \satisfaction_i(P).\]
	\end{definition}

	Because \citet{PPS21} prove that FJR can be satisfied even for monotonic cardinal ballots over budget allocations, FJR[$\satisfaction$] can be satisfied for all $\satisfaction$.

	\subsection{Justified Representation Without Cohesiveness}

	 Motivated by the observation that large cohesive groups seldom appear in real world elections, \citet{BP23} introduced a new variant of justified representation for multiwinner voting that does not rely on cohesive groups. Instead, the axiom formalizes the idea that a group of underrepresented voters can claim to deserve an unelected candidate they all approve, even if they cannot agree on any other candidate. An additional advantage of this approach is that the satisfaction of the resulting axioms can be verified in polynomial time. \citet{BP23} extended their notion to the (approval-based) PB setting. They define their robust strengthening of EJR as follows:

	\begin{definition}[EJR+]
	Given an instance $I = \tuple{\projSet, c, b}$ and a profile $\profile$ of approval ballots, a budget allocation $\pi \in \allocSet(I)$ is said to satisfy \emph{extended justified representation plus up to any project for $\costSatisfaction$} (EJR+ up to any project for $\costSatisfaction$) if for every group $N\subseteq \agentSet$ and every project $p \in \projSet\setminus \pi$ such that $A_i(p) = 1$ for all $i \in N$, there exists a $i \in N$ such that
\[\costSatisfaction_i(\pi) + \costSatisfaction_i(p) > \frac{|N|b}{n}.\]
	\end{definition}

	Additionally, they show that MES[$\costSatisfaction$] satisfies EJR+ up to any project for $\costSatisfaction$. Observe that they did not define a version of EJR+ for any other satisfaction function and it remains open how EJR+ can be generalized to other satisfaction functions or even cardinal utilities and whether the resulting axiom would be satisfiable.

	\subsection{Variants with Relative Budgets}
	\label{subsec:Relativ-Budget}
	
	Most of the proportionality requirements we introduced heavily rely on the budget limit $b$. This is particularly true for the axioms based on justified representation. \citet{ALT18} suggest to work on properties that are independent of the budget limit and only defined in terms of the cost of the budget allocation under consideration. 
	
	They revisit their adaptations of PJR for PB by changing the notion of cohesive group, making it dependent on the cost $c(\pi)$ of the budget allocation $\pi$ under consideration instead of $b$. All of these new concepts are weaker than the standard ones. They are also all satisfiable, simply by using $\pi = \emptyset$ (note that because of how we organised the elements in our definition for cohesive groups---definitions~\ref{def:Cohesive_CardinalBallots} and~\ref{def:Cohesive_ApprovalBallots}---this does not lead to any division by 0).
	
	\subsection{Laminar Proportionality}
	
	The next property we want to mention is \emph{laminar proportionality}. It is a proportionality requirement that only applies to specific instances, the \emph{laminar} ones. These instances are very well-structured in a way that makes it obvious which outcomes are proportional. Laminar proportionality requires the outcome to be proportional with respect to this structure.
	
	This property was defined for PB by \citet{LCG22}, building on \citet{PeSk20}. They show that rules that are laminar proportional in the multi-winner setting (namely \mes{} and \seqPhragmen{}) cease to be on PB instances.
	
	\subsection{Proportionality for Solid Coalitions with Approval Ballots}
	
	In Section~\ref{subsec:Proportionality_Ordinal} we have defined two axioms for proportionality with weak ordinal ballots. Approval ballots can be seen as a special case of weak ordinal ballots where all ballots have at most two indifference classes. Following this observation, \citet{AzLe21} provide definitions of IPSC and CPSC for approval ballots. We give these definitions below. Observe that they are both closely related to PJR.
	
	\begin{definition}[CPSC with Approval Ballots]
		Given an instance $I = \tuple{\projSet, c, b}$ and a profile $\profile$ of approval ballots, a budget allocation $\pi \in \allocSet(I)$ is said to satisfy \emph{CPSC} if the following two conditions hold:
		\begin{itemize}
			\item $\pi$ satisfies PJR[$\costSatisfaction$],
			\item $\pi$ is of maximal cost: $\pi \in \argmax_{\pi' \in \allocSet(I)} (\pi')$.
		\end{itemize}
	\end{definition}

	\begin{definition}[IPSC with Approval Ballots]
		Given an instance $I = \tuple{\projSet, c, b}$ and a profile $\profile$ of approval ballots, a budget allocation $\pi \in \allocSet(I)$ is said to satisfy \emph{IPSC} if the following two conditions hold:
		\begin{itemize}
			\item for all sets of voters $N \subseteq \agentSet$ such that $c(\bigcup_{i \in N} \{p \in \pi \mid A_i(p) = 1\}) < \nicefrac{|N|}{n} \cdot b$ and for all $p \in \bigcap_{i \in N} \{p \in \projSet \setminus \pi \mid A_i(p) = 1\}$ we have:
			\[c(p) + c\left(\bigcup_{i \in N} \{p \in \pi \mid A_i(p) = 1\}\right) > \nicefrac{|N|}{n} \cdot b,\]
			\item $\pi$ is exhaustive.
		\end{itemize}
	\end{definition}
	
	\noindent The first bullet point of the above definition closely resembles PJR-X[$\costSatisfaction$]. One can actually prove that IPSC implies PJR-X[$\costSatisfaction$]. Indeed, if a budget allocation $\pi$ fails PJR-X[$\costSatisfaction$], then the $P$-cohesive $N$ witnessing this violation would also be a witness of the violation of the first bullet point of the definition of IPSC.
	
	\medskip

	It should be quite clear from the definition that CPSC is still not satisfiable with approval ballots. IPSC is, since it already was with generic weak ordinal ballots.
	
	\subsection{Proportionality with Cumulative Ballots}
	
	Among the different types of cardinal ballots we defined, there is one for which we still have not discussed proportionality requirements: \emph{cumulative ballots}. Now is the time to do so. The only study on cumulative ballots has been conducted by \citet{SSST20}. Among others, they study proportionality axioms for this setting. We present what they call \emph{proportional representation}.
	
	\begin{definition}[Proportional Representation with Cumulative Ballots]
		Given an instance $I = \tuple{\projSet, c, b}$ and a profile $\profile$ of cumulative ballots, a budget allocation $\pi \in \allocSet(I)$ is said to satisfy \emph{proportional representation} if for every $\ell \in \{1, \ldots, b\}$, every group of agents $N \subseteq \agentSet$ with $\nicefrac{|N|}{n} \cdot b \geq \ell$ and every subset of projects $P \subseteq \projSet$ with $c(P) \leq \ell$, it holds that if for all $i \in N$ and $p \in P$, we have $A_i(p) > 0$, and for all $i' \in \agentSet \setminus N$ and $p' \in \projSet \setminus P$ we have $A_{i'}(p') = 0$, then we must have $P \subseteq \pi$.
	\end{definition}

	\citet{SSST20} also introduce a weaker and a stronger variant of the above. They prove that all of them are satisfiable.
	
	\subsection{Equality of Resources} 
	
	In the context of PB with approval ballots, one needs to go through the concept of satisfaction functions to define proportionality requirements. Based on the observation that no satisfaction function can ever be satisfactory, \citet{MREL23} suggest to define fairness criteria not in terms of satisfaction but in terms of the resources spent on an agent, the so called \emph{equality of resources}. They use the concept of \emph{share} \citep{LMR21} to measure the amount of resources spent on an agent and aim at providing every agent with their \emph{fair share}.
	
	\begin{definition}[Fair Share]
		Given an instance $I = \tuple{\projSet, c, b}$ and a profile $\profile$ of approval, $\pi \in \allocSet(I)$ is said to satisfy \emph{fair share} if for every agent $i \in \agentSet$ we have:
		\[\sum_{p \in \pi} A_i(p) \cdot \frac{c(p)}{|\{i' \in \agentSet \mid A_{i'}(p) = 1\}|} \geq \min\left\{ \frac{b}{n}, \sum_{\substack{p \in \projSet\\A_i(p) = 1}} \frac{c(p)}{|\{i' \in \agentSet \mid A_{i'}(p) = 1\}|}\right\}.\]
	\end{definition}

	This requirement is not satisfiable in general, which motivated \citet{MREL23} to introduce several weakenings, either based on direct relaxations (``up-to-one project'' or ``local'' variants), or on cohesive groups. Most notably, they prove that a variant of \mes{} provides good fair share guarantees both theoretically and empirically.

	\subsection{Fair Lotteries for PB}

	Many of the difficulties in achieving fair outcomes in voting stem from the fact that we are mostly focusing on \emph{discrete} outcomes, i.e., that candidates are either selected or not. Thus, we cannot satisfy voters to any arbitrary, fractional degree. One way to address this problem is to consider probabilistic voting rules, which give every candidate a certain probability of being selected. In expectation, or ex-ante, voters can, therefore, have arbitrary levels of satisfaction. This idea has been successfully applied in different settings, see, e.g., \citet{BMS05} and \citet{CJMW20}. One specific challenge of using fair lotteries in PB is, that as PB project costs are heterogeneous, the amount of election budget spent may differ ex-ante and ex-post. \citet{ALS+24} therefore introduce the notion of an outcome being approximately within the budget, or \emph{budget balanced up to one project}:
	
	\begin{definition}[Budget Balanced Up To One Project (BB1)]
		An outcome $\pi$ is said to be \emph{budget balanced up to one project} if either
		\begin{itemize}
			\item $c(\pi) \leq b$ and there exists some project $p \in \projSet \backslash \pi$ such that $c(\pi \cup \{p\}) \geq b$, or
			\item $c(\pi) \geq b$ and there exists some project $p \in \pi$ such that $c(\pi \backslash \{p\}) \leq b$. 
		\end{itemize}
	\end{definition}

	They use this notion towards accomplishing minimal representation guarantees for individual voters. They define several ex-ante fairness concepts for \emph{fractional outcomes}, where projects are represented by the fraction of which they are funded by in said outcome. One of the introduced ex-ante fairness concepts is \emph{strong unanimous fair share (UFS)}, which guarantees strong fair share of utility for groups of voters with identical preferences.
	
	While \citet{ALS+24} show that their developed ex-ante fairness guarantees are not compatible with the usual ex-post concepts based on justified representation, they provide a randomised algorithm satisfying both Strong UFS, FJR and ex-post BB1 for PB with binary utilities.
	
	\section{Fairness in Extended Settings}
	\label{subsec:Fairness_Extended}
	
	We now mention some papers that have studied fairness in PB beyond the standard model. This section overlaps in some way with Chapter \ref{chap:Extensions}, though we focus on fairness requirements here and discuss the work only briefly.
	
	\begin{itemize}
		\item In their study of PB with multiple resources, \citet{MSRE22} introduced several proportionality axioms and studied whether they could be satisfied by some load-balancing mechanisms.
		
		\item When studying PB with uncertainty on the cost of the projects, \citet{BBL22} investigated the link between properties specific to their setting and justified representation axioms such as PJR[$\costSatisfaction$] (or BPJR-L) and EJR.
		
		\item \citet{LMR21} introduce a fairness theory for long-term PB where several instances are considered. They present several fairness concepts for their setting and study under which conditions they can be satisfied.
		
		\item In a model in which the budget is endogenous, \citet{AzGa21} studied versions of the core and of a simple proportionality axiom, investigating which welfare-maximising rule satisfy them.
	\end{itemize}

	\section{Taxonomies of Proportionality in PB}
	\label{subsec:Taxonomies_Proportionality}
	
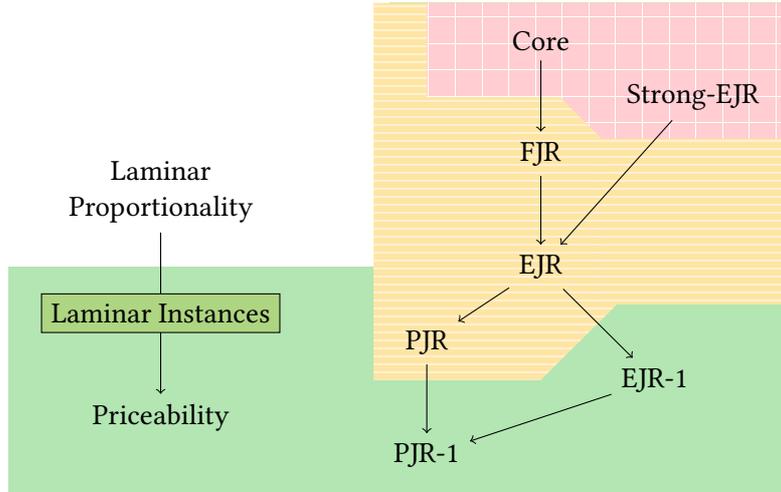
	\begin{figure}
		\begin{center}
			\begin{tikzpicture}
				
				\fill[my-green-verylight] (-7, 0.5) rectangle (3.25, -6);
				\fill[white] (-7,  0.5) rectangle (-2, -3);
				\fill[my-orange-verylight] (3.25, 0.5) -- (-2.2, 0.5) -- (-2.2, -4.5) -- (0, -4.5) -- (1, -3.5) -- (3.25, -3.5) -- cycle;
				\fill[pattern={Lines[line width = 0.25pt, distance = 3pt, yshift = 1.5pt]}, pattern color=white] (3.25, 0.5) -- (-2.2, 0.5) -- (-2.2, -4.5) -- (0, -4.5) -- (1, -3.5) -- (3.25, -3.5) -- cycle;
				\fill[my-red-verylight] (3.25, 0.5) -- (-1.5, 0.5) -- (-1.5, -0.75) -- (0.25, -0.75) -- (0.8, -1.3) -- (3.25, -1.3) -- cycle;
				\fill[pattern={Hatch[line width = 0.25pt, distance = 10pt, yshift = 1.7pt, angle=45]}, pattern color=white] (3.25, 0.5) -- (-1.5, 0.5) -- (-1.5, -0.75) -- (0.25, -0.75) -- (0.8, -1.3) -- (3.25, -1.3) -- cycle;
				
				\node (core) at (0, 0) {Core};
				
				\node (FJR) at (0, -1.5) {FJR};
				\path[->] (core) edge (FJR);
				
				\node (Strong-EJR) at (2, -0.75) {Strong-EJR};
				
				\node (EJR) at (0, -3) {EJR};
				\path[->] (FJR) edge (EJR);
				\path[->] (Strong-EJR) edge (EJR);
				
				\node (EJR-1) at (1.5, -4.5) {EJR-1};
				\path[->] (EJR) edge (EJR-1);
				
				\node (PJR) at (-1.5, -4) {PJR};
				\path[->] (EJR) edge (PJR);
				
				\node (PJR-1) at (-1.5, -5.5) {PJR-1};
				\path[->] (PJR) edge (PJR-1);
				\path[->] (EJR-1) edge (PJR-1);
				
				\node[align = center] (LP) at (-5, -2) {Laminar\\Proportionality};
				
				\node (priceability) at (-5, -5) {Priceability};
				\path[->] (LP) edge node[draw, fill = my-grass-verylight] {Laminar Instances} (priceability);
			\end{tikzpicture}
		\end{center}
		
		\resizebox{!}{0.8em}{
			\begin{tikzpicture}
				\fill[my-red-verylight] (0, 0) rectangle (2, 1);
				\draw[pattern={Hatch[line width = 0.25pt, distance = 10pt, xshift = 4pt, yshift = 5pt, angle=45]}, pattern color=white] (0, 0) rectangle (2, 1);
			\end{tikzpicture}
		} These concepts cannot always be satisfied.
		
		\resizebox{!}{0.8em}{
			\begin{tikzpicture}
				\fill[my-orange-verylight] (0, 0) rectangle (2, 1);
				\draw[pattern={Lines[line width = 0.25pt, distance = 3pt, yshift = .5pt]}, pattern color=white] (0, 0) rectangle (2, 1);
			\end{tikzpicture}
		} These concepts can always be satisfied, but finding a suitable budget allocation cannot be done in polynomial time unless $\complexP = \complexNP$.
		
		\resizebox{!}{0.8em}{
			\begin{tikzpicture}
				\draw[fill = my-green-verylight] (0, 0) rectangle (2, 1);
			\end{tikzpicture}
		} These concepts can always be satisfied, and a suitable budget allocation can be found in polynomial time.
		
		\resizebox{!}{0.8em}{
			\begin{tikzpicture}
				\draw[fill = white] (0, 0) rectangle (2, 1);
			\end{tikzpicture}
		} Laminar Proportionality is always satisfiable, and the computational complexity of finding a budget allocation satisfying it is unknown.
	
		\caption{Taxonomy of the proportionality requirements for PB with cardinal ballots. An arrow between two concepts means that any budget allocation satisfying one also satisfies the other. All missing arrows are known to be missing.\\[0.4em]
			\footnotesize
			Most of this picture is based on \citet{LCG22}, who showed: the absence of an arrow between either the core, EJR or PJR and priceability; the link between laminar proportionality and priceability (only for laminar instances); the absence of arrows between laminar proportionality and either PJR, EJR or the core. The link between FJR and EJR is due to \citet{PPS21}. For the satisfiability of the concepts, see Table~\ref{tab:Summary_FairnessRulesPreCriteria_CardinalBallots}.}
		\label{fig:Taxonomy_Cardinal_Ballots}
	\end{figure}

	Throughout this chapter we have introduced a significant number of properties related to proportionality in PB. In an attempt to clarify the relationship between these properties, we draw several taxonomies. The taxonomy for cardinal ballots can be found in Figure \ref{fig:Taxonomy_Cardinal_Ballots}. Figure \ref{fig:Taxonomy_Approval_Ballots} presents the taxonomy for approval ballots. All the details are available in the figures. We also summarise which rules satisfy which axioms, in Table \ref{tab:Summary_FairnessRulesPreCriteria_CardinalBallots} for cardinal ballots, and in Table \ref{tab:Summary_FairnessRulesPreCriteria_ApprovalBallots} for approval ballots.
	
	\begin{table}
		\begin{center}
			\begin{tabularx}{\linewidth}{rlX}
				\toprule
				\multicolumn{3}{c}{\textbf{Cardinal Ballots}} \\
				\midrule
				\textbf{Core} & None & \citet{PPS21} \\
				\midrule
				\textbf{FJR} & Greedy cohesive rule & \citet{PPS21} \\
				\midrule
				\textbf{Strong-EJR} & None & \\
				\textbf{EJR} & Greedy cohesive rule & \citet{PPS21} \\
				\textbf{EJR-1} & \mes{} & \citet{PPS21} \\
				\midrule
				\textbf{PJR} & Greedy cohesive rule & \\
				\textbf{PJR-1} & \mes{} & \citet{LCG22} \\
				\midrule
				\begin{tabular}{@{}r@{}} \textbf{Laminar} \\ \textbf{Proportionality} \end{tabular} & ? & \citet{LCG22} \\
				\midrule
				\textbf{Priceability} & \mes{} & \citet{PPS21} \\
				\bottomrule
			\end{tabularx}
		\end{center}
		\caption{Rules satisfying each of the fairness property we introduced for generic cardinal ballots}
		\label{tab:Summary_FairnessRulesPreCriteria_CardinalBallots}
	\end{table}

	\begin{figure}
		\begin{center}
			\begin{tikzpicture}
				
				\fill[my-green-verylight] (-11.75, 0.5) rectangle (2.25, -8.25);
				\fill[my-lime-verylight] (-0.35, -6) -- (-0.35, -4.4) -- (-2.25, -4.4) -- (-3, -5.25) -- (-5, -5.25) -- (-6, -6.45) -- (-6, -7.1) -- (-3.75, -7.1) -- cycle;
				\fill[pattern=vertical lines, pattern color=white] (-0.35, -6) -- (-0.35, -4.4) -- (-2.25, -4.4) -- (-3, -5.25) -- (-5, -5.25) -- (-6, -6.45) -- (-6, -7.1) -- (-3.75, -7.1) -- cycle;
				\fill[my-orange-verylight] (2.25, 0.5) -- (-2, 0.5) -- (-2, -3.25) -- (-4.25, -4.25) -- (-4.25, -5.25) -- (-3, -5.25) -- (-2.25, -4.4) -- (-0.35, -4.4) -- (-0.35, -3.3) -- (2.25, -3.3) -- cycle; 
				\fill[pattern={Lines[line width = 0.25pt, distance = 3pt, yshift = .5pt]}, pattern color=white] (2.25, 0.5) -- (-2, 0.5) -- (-2, -3.25) -- (-4.25, -4.25) -- (-4.25, -5.25) -- (-3, -5.25) -- (-2.25, -4.4) -- (-0.35, -4.4) -- (-0.35, -3.3) -- (2.25, -3.3) -- cycle; 
				\fill[my-orange-verylight] (-11.75,-2.75) rectangle (-8.75, -4.5); 
				\fill[pattern={Lines[line width = 0.25pt, distance = 3pt, yshift = .5pt]}, pattern color=white] (-11.75,-2.75) rectangle (-8.75, -4.5); 
				\fill[my-red-verylight] (-11.75, 0.5) -- (-11.75, -2.75) -- (-8, -2.75) -- (-7, -1.25) -- (2.25, -1.25) -- (2.25, 0.5) -- cycle;
				\fill[pattern={Hatch[line width = 0.25pt, distance = 10pt, xshift = 4pt, yshift = 5pt, angle=45]}, pattern color=white] (-11.75, 0.5) -- (-11.75, -2.75) -- (-8, -2.75) -- (-7, -1.25) -- (2.25, -1.25) -- (2.25, 0.5) -- cycle; 
				
				\node (core) at (-1.25, 0) {Core[$\satisfaction$]};
				
				\node (FJR) at (-1.25, -1.5) {FJR[$\satisfaction$]};
				\path[->] (core) edge (FJR);
				
				\node (Strong-EJR) at (-3.25, -0.75) {Strong-EJR[$\satisfaction$]};
				
				\node (EJR) at (-1.25, -3.75) {EJR[$\satisfaction$]};
				\path[->] (FJR) edge (EJR);
				\path[->] (Strong-EJR) edge (EJR);
				
				\node (EJR-X) at (-1.25, -5.75) {EJR-X[$\satisfaction$]};
				\path[->] (EJR) edge node[draw, fill=my-grass-verylight] {Incr.\ $\satisfaction$} (EJR-X);
				
				\node (EJR-1) at (-1.25, -7) {EJR-1[$\satisfaction$]};
				\path[->] (EJR-X) edge (EJR-1);

				\node (EJR+) at (0.9, -1.75) {EJR+[$\costSatisfaction$]};
				\path[->] (EJR+) edge node[draw, fill=my-grass-verylight] {$\costSatisfaction$} (EJR);

				\node (EJR+-X) at (0.9, -4.25) {EJR+-X[$\costSatisfaction$]};
				\path[->] (EJR+) edge (EJR+-X);
				\path[->] (EJR+-X) edge[bend left = 45] node[draw, fill=my-grass-verylight] {$\costSatisfaction$} (EJR-X);
				
				\node (PJR) at (-3.25, -4.5) {PJR[$\satisfaction$]};
				\path[->] (EJR) edge[bend right = 15] (PJR);
				
				\node (PJR-X) at (-5, -6.75) {PJR-X[$\satisfaction$]};
				\path[->] (PJR) edge node[draw, fill=my-grass-verylight] {Incr.\ $\satisfaction$} (PJR-X);
				\path[->] (EJR-X) edge (PJR-X);
				
				\node (PJR-1) at (-3.25, -7.75) {PJR-1[$\satisfaction$]};
				\path[->] (PJR-X) edge (PJR-1);
				\path[->] (EJR-1) edge (PJR-1);
				
				\node (PJR-X-cost) at (-9, -6.375) {PJR-X[$\costSatisfaction$]};
				
				\node (Local-BPJR-L) at (-9, -7.75) {Local-BPJR-L[$\satisfaction$]};
				\path[->] (PJR-X-cost) edge (Local-BPJR-L);
				\path[->] (PJR-X) edge[bend right = 15] (Local-BPJR-L);
				
				\node[align = center] (PJR-cost) at (-10.25, -3.625) {PJR[$\costSatisfaction$]\\equiv. BPJR-L};
				\path[->] (PJR-cost) edge (PJR-X-cost);
				
				\node (Strong-BPJR-L) at (-9, -1.5) {Strong-BPJR-L};
				\path[->] (Strong-BPJR-L) edge (PJR-cost);
				
				\node (IPSC) at (-9, -5) {IPSC};
				\path[->] (IPSC) edge (PJR-X-cost);
				
				\node (CPSC) at (-11, -2) {CPSC};
				\path[->] (CPSC) edge (PJR-cost);
				
				\node[align = center] (c6-priceability) at (-5.75, -2) {Priceability with \\ \textbf{C6} and $\alpha > b$};
				\path[->] (c6-priceability) edge node[draw, fill=my-grass-verylight] {DNS $\satisfaction$} (PJR-X);
				
				\node[align = center] (strict-priceability) at (-7.5, -3.625) {Priceability\\with $\alpha > b$};
				\path[->] (c6-priceability) edge (strict-priceability);
				
				\node[align = center] (priceability) at (-6.5, -5.8) {Priceability};
				\path[->] (strict-priceability) edge (priceability);
				\path[->] (strict-priceability) edge (PJR-X-cost);
				
			\end{tikzpicture}
		\end{center}

		\resizebox{!}{0.8em}{
			\begin{tikzpicture}
				\fill[my-red-verylight] (0, 0) rectangle (2, 1);
				\draw[pattern={Hatch[line width = 0.25pt, distance = 10pt, xshift = 4pt, yshift = 5pt, angle=45]}, pattern color=white] (0, 0) rectangle (2, 1);
			\end{tikzpicture}
		} These concepts cannot always be satisfied.
		
		\resizebox{!}{0.8em}{
			\begin{tikzpicture}
				\fill[my-orange-verylight] (0, 0) rectangle (2, 1);
				\draw[pattern={Lines[line width = 0.25pt, distance = 3pt, yshift = .5pt]}, pattern color=white] (0, 0) rectangle (2, 1);
			\end{tikzpicture}
		} These concepts can always be satisfied, but finding a suitable budget allocation cannot be done in polynomial time unless $\complexP = \complexNP$.
		
		\resizebox{!}{0.8em}{
			\begin{tikzpicture}
				\fill[my-lime-light] (0, 0) rectangle (2, 1);
				\draw[pattern=vertical lines, pattern color=white] (0, 0) rectangle (2, 1);
			\end{tikzpicture}
		} These concepts can always be satisfied, and a suitable budget allocation can be found in polynomial time when $\satisfaction$ is a DNS function.
	
		\resizebox{!}{0.8em}{
			\begin{tikzpicture}
				\draw[fill = my-green-verylight] (0, 0) rectangle (2, 1);
			\end{tikzpicture}
		} These concepts can always be satisfied, and a suitable budget allocation can be found in polynomial time when $\satisfaction$ is additive (for the concepts depending on $\satisfaction$).
		
		
		\vspace{0.75em}
		
		\footnotesize
		\textbf{Incr.\ $\satisfaction$:} the link only applies for satisfaction functions that are strictly increasing, \textit{i.e.}, such that for all $P \subseteq \projSet$ and $P' \subsetneq P$, we have $\satisfaction(P') < \satisfaction(P)$.
		
		\textbf{DNS $\satisfaction$:} the link only applies for DNS functions, see Definition~\ref{def:dns}.
		
		\textbf{PJR[$\costSatisfaction$] equiv. BPJR-L:} these two concepts are equivalent.
		
		\textbf{Priceability with $\alpha > b$:} priceable for a price system $\tuple{\alpha, (\gamma_i)_{i \in \agentSet}}$ where $\alpha > b$.
		
		\textbf{Priceability with \textbf{C6} and $\alpha > b$:} see Theorem \ref{thm:Priceability_PJR_Appoval}.
		
		\caption{Taxonomy of the proportionality requirements for PB with approval ballots where $\satisfaction$ is an arbitrary satisfaction function. An arrow between two concepts means that any budget allocation satisfying one also satisfies the other. Some arrows are only valid for some satisfaction functions; the conditions are indicated on the arrows. All missing arrows are known to be missing.\\[0.4em]
			\footnotesize
			The links between EJR, PJR, Local-BPJR-L and priceability concepts are due to \citet{BFLMP23}. The link from Strong-BPJR-L and BPJR-L is due to \citet{ALT18}. The link between CPSC and PJR[$\costSatisfaction$] is due to \citet{AzLe21}. The one between IPSC and PJR-X[$\costSatisfaction$] has never been published. The absence of arrows between the core, EJR and priceability is due to \citet{LCG22}. The link between FJR and EJR is due to \citet{PPS21}. All links including forms of EJR+ are due to \citet{BP23}. For the satisfiability of the concepts, see Table \ref{tab:Summary_FairnessRulesPreCriteria_ApprovalBallots}.}
		\label{fig:Taxonomy_Approval_Ballots}
	\end{figure}
	
	\begin{table}
		\begin{center}
			\renewcommand\tabularxcolumn[1]{p{#1}}
			\begin{tabularx}{\linewidth}{rX@{}}
				\toprule
				\multicolumn{2}{c}{\textbf{Approval Ballots}} \\
				
				\midrule
				\textbf{Core[$\satisfaction$]} & %
				{%
					\renewcommand\tabularxcolumn[1]{m{#1}}
					\begin{tabularx}{\linewidth}{Xc}
						? &
					\end{tabularx}
				} \\
				
				\midrule
				\textbf{FJR[$\satisfaction$]} & %
				{%
					\renewcommand\tabularxcolumn[1]{m{#1}}
					\begin{tabularx}{\linewidth}{Xc}
						$\blacktriangleright$ for any $\satisfaction$, a greedy cohesive rule & [1]
					\end{tabularx}
				} \\
			
				\midrule
				\textbf{Strong-EJR[$\satisfaction$]} & %
				{%
					\renewcommand\tabularxcolumn[1]{m{#1}}
					\begin{tabularx}{\linewidth}{Xc}
						$\blacktriangleright$ None &
					\end{tabularx}
				} \\
				
				\midrule
				\textbf{EJR[$\satisfaction$]} & %
				{%
					\renewcommand\tabularxcolumn[1]{m{#1}}
					\begin{tabularx}{\linewidth}{Xc}
						$\blacktriangleright$ A greedy cohesive rule for any $\satisfaction$ & [1] \\
						$\blacktriangleright$ \mesSat{$\cardSatisfaction$} for $\satisfaction = \cardSatisfaction$ & [1]
					\end{tabularx}
				} \\
				
				\midrule
				\textbf{EJR-X[$\satisfaction$]} & %
				{%
					\renewcommand\tabularxcolumn[1]{m{#1}}
					\begin{tabularx}{\linewidth}{Xc}
						$\blacktriangleright$ for any $\satisfaction$, a greedy cohesive rule & [1] \\
						$\blacktriangleright$ for any DNS function $\satisfaction$, \mesSat{$\satisfaction$} & [2]
					\end{tabularx}
				} \\
				
				\midrule
				\textbf{EJR-1[$\satisfaction$]} & %
				{%
					\renewcommand\tabularxcolumn[1]{m{#1}}
					\begin{tabularx}{\linewidth}{Xc}
						$\blacktriangleright$ for any $\satisfaction$, a greedy cohesive rule & [1] \\
						$\blacktriangleright$ for any additive $\satisfaction$, \mesSat{$\satisfaction$} & [1] 
					\end{tabularx}
				} \\
			
				\midrule
				\textbf{PJR[$\satisfaction$]} & %
				{%
					\renewcommand\tabularxcolumn[1]{m{#1}}
					\begin{tabularx}{\linewidth}{Xc}
						 $\blacktriangleright$ for any $\satisfaction$, a greedy cohesive rule & [1] \\
					\end{tabularx}
				} \\
				
				\midrule
				\textbf{PJR-X[$\satisfaction$]} & %
				{%
					\renewcommand\tabularxcolumn[1]{m{#1}}
					\begin{tabularx}{\linewidth}{Xc}
						$\blacktriangleright$ for any $\satisfaction$, a greedy cohesive rule & [1] \\
						$\blacktriangleright$ for any DNS function $\satisfaction$, MES[$\satisfaction$], \seqPhragmen{}, and \newline \maximinSupport{} & [2]
					\end{tabularx}
				} \\
			
				\midrule
				\textbf{CPSC} & %
				{%
					\renewcommand\tabularxcolumn[1]{m{#1}}
					\begin{tabularx}{\linewidth}{Xc}
						$\blacktriangleright$ None & [3]
					\end{tabularx}
				} \\
				
				\midrule
				\textbf{IPSC} & %
				{%
					\renewcommand\tabularxcolumn[1]{m{#1}}
					\begin{tabularx}{\linewidth}{Xc}
						$\blacktriangleright$ The expanding approval rule & [3] \\
					\end{tabularx}
				} \\
				
				\midrule
				\textbf{Local-BPJR-L[$\satisfaction$]} & %
				{%
					\renewcommand\tabularxcolumn[1]{m{#1}}
					\begin{tabularx}{\linewidth}{Xc}
						$\blacktriangleright$ \mesSat{$\satisfaction$}, \seqPhragmen{}, and \maximinSupport{} & [2, 4]
					\end{tabularx}
				} \\
				
				\midrule
				\textbf{Strong-BPJR-L} & %
				{%
					\renewcommand\tabularxcolumn[1]{m{#1}}
					\begin{tabularx}{\linewidth}{Xc}
						$\blacktriangleright$ None & [4]
					\end{tabularx}
				} \\
				
				\midrule
				\textbf{Priceability} & %
				{%
					\renewcommand\tabularxcolumn[1]{m{#1}}
					\begin{tabularx}{\linewidth}{Xc}
						$\blacktriangleright$ \mesSat{$\satisfaction$}, \seqPhragmen{}, and \maximinSupport{} & [1, 2]
					\end{tabularx}
				} \\
				
				\midrule
				\begin{tabular}{@{}r@{}} \textbf{Priceability} \\ \textbf{with $\alpha > b$} \end{tabular} &%
				{%
					\renewcommand\tabularxcolumn[1]{m{#1}}
					\begin{tabularx}{\linewidth}{Xc}
						$\blacktriangleright$ \mesSat{$\satisfaction$}, \seqPhragmen{}, and \maximinSupport{} & [1, 2]
					\end{tabularx}
				} \\
				
				\midrule
				\begin{tabular}{@{}r@{}} \textbf{Priceability} \\ \textbf{with \textbf{C6} and $\alpha > b$} \end{tabular} & %
				{%
					\renewcommand\tabularxcolumn[1]{m{#1}}
					\begin{tabularx}{\linewidth}{Xc}
						$\blacktriangleright$ \mesSat{$\cardSatisfaction$}, \seqPhragmen{}, and \maximinSupport{} & [2]
					\end{tabularx}
				} \\ 
				\bottomrule
			\end{tabularx}
		\end{center}
		
		[1] \citet{PPS21}
		
		[2] \citet{BFLMP23}
		
		[3] \citet{AzLe21}
		
		[4] \citet{ALT18}
		
		\caption{Rules satisfying each of the fairness properties we introduced for approval ballots. $\satisfaction$ is an arbitrary satisfaction function.}
		\label{tab:Summary_FairnessRulesPreCriteria_ApprovalBallots}
	\end{table}

	\chapter{Axiomatic Analysis}
	\label{chap:Axiomatic}
	
	Fairness requirements are the most studied properties in the literature on PB but are not the only ones. In the following, we review other axioms that have been introduced.
	
	Our analysis will start with a discussion around exhaustiveness (Section \ref{subsec:Exhaustiveness}) and a presentation of the monotonicity axioms that have been introduced for PB (Section \ref{subsec:Monotonicty_Axioms}). From there, we will move on to axioms relating to strategic behaviour of the agents (Section \ref{subsec:StrategyProofness}). We will conclude this section by our usual discussion of the concepts that exist in the literature but do not fit in earlier sections (Section \ref{subsec:Other_Axioms}).
	
	\section{Exhaustiveness}
	\label{subsec:Exhaustiveness}
	
	\begin{table}
		\centering
		\begin{tabular}{ccc}
			\toprule
			& \multicolumn{2}{c}{\textbf{Exhaustiveness}} \\
			& General Instances & Unit-Cost Instances \\
			\midrule
			\maxWelCard{} & \cmark & \cmark \\
			\greWelCard{} & \cmark & \cmark \\
			\maxWelCost{} & \cmark & \cmark \\
			\greWelCost{} & \cmark & \cmark \\
			\midrule
			\seqPhragmen{} & \xmark & \cmark \\
			\maximinSupport{} & \xmark & \cmark \\
			\mes{} & \xmark & \xmark \\
			\bottomrule
		\end{tabular}
		\caption{Satisfaction of exhaustiveness for different rules.}
		\label{tab:summary_exhaustiveness}
	\end{table}
	
	Let us start with \emph{exhaustiveness}, an efficiency requirement that states that the budget should not be underused. It is sometimes considered a standard requirement that should be enforced by default. However, as we will see, it is incompatible with some other axioms, notably priceability. Note that \citet{TaFa19} introduced an axiom called \emph{budget monotonicity} that is equivalent to exhaustiveness for resolute rules and very similar to it for irresolute rules; the name exhaustiveness is due to \citet{ALT18}.
	
	Let us first introduce the idea of exhaustiveness.
	
	\begin{definition}[Exhaustiveness]\label{def:exhaustive}
		Given an instance $I = \tuple{\projSet, c, b}$, a feasible budget allocation $\pi \in \allocSet(I)$ is said to be \emph{exhaustive} if there are no project $p \in \projSet \setminus \pi$ such that $c(\pi \cup \{p\}) \leq b$.
	\end{definition}
	
	Table \ref{tab:summary_exhaustiveness} summarizes which of the usual rules satisfy exhaustiveness. The results for the welfare-maximising and greedy rules are straightforward. Interestingly, the fact that \seqPhragmen, \maximinSupport{} and \mes{} fail exhaustiveness is due the fact that they are priceable. Indeed, the two requirements are incompatible.
	
	\begin{proposition}[\citealt{PPS21}]
		There exists an instance $I = \tuple{\projSet, c, b}$ and a profile $\profile$, such that there is no budget allocation $\pi \in \allocSet(I)$ which is both priceable and exhaustive, even though there are feasible budget allocations that are priceable, and others that are exhaustive.
	\end{proposition}

	Since exhaustiveness is sometimes considered to be a must, \citet{PPS21} proposed several ways to obtain exhaustive outcomes when using non-exhaustive rules.
	\begin{itemize}
		\item \textbf{Completion via Exhaustive Rule}: This technique consists of completing the original outcome of the rule by applying another rule, which is exhaustive, on the reduced instance where the selected projects have been removed and the budget reduced accordingly. Typically, one could use a greedy selection procedure or an exhaustive variant of \seqPhragmen{}.
		
		\item \textbf{Exhaustion by Variation of the Budget Limit}: Using this technique, the rule is run several times for different values of the budget limit until finding an outcome that is feasible and exhaustive for the initial budget. Typically, the budget limit is increased by one unit per voter at each iteration and the final outcome is the first exhaustive one that is found, or the first one for which increasing the budget one more time would lead to an outcome that is not feasible for the original budget limit.
		
		Note that this technique does not guarantee that the outcome will be exhaustive (notably because when used with \mes{}, the outcome would still be priceable). Moreover, this is not necessarily a ``completion technique'' since many rules are not limit monotonic (see Section~\ref{subsec:Monotonicty_Axioms}), so the final outcome does not need to be a superset of the initial outcome.
		
		\item \textbf{Exhaustion by Perturbation of the Ballots:} This final technique modifies the profile slightly so that the outcome is guaranteed to be exhaustive. Which perturbation mechanism should be used depends on the rule. For instance, for \mes{} with cardinal ballots, it is know that if every voter reports a \emph{strictly} positive score for all the projects, then the outcome of \mes{} is exhaustive. Therefore, one could apply \mes{} on the modified profile in which all 0 scores have been replaced by an arbitrary small value.
	\end{itemize}

	For MES specifically, \citet{KE23} introduced the \emph{Adaptive Method of Equal Shares (AMES)}, which, when given an MES outcome, utilises this to greedily reach a stable one. In an effort to avoid redundant computation, this method is what they call budget-adaptive: When updating budget $b$ to $b'$, the computation of $\pi'$ is computationally inexpensive by leveraging similarities between $\pi$ and $\pi'$. In experiments on real-world data, the authors observe that small budget increases usually lead to only small changes in the election outcome. Furthermore, they show that an AMES outcome can be proven to satisfy EJR in polynomial time. 
	
	\section{Monotonicity Requirements}
	\label{subsec:Monotonicty_Axioms}
	
	\citet{TaFa19} introduced several monotonicity axioms for PB that represent to this date the largest corpus of axioms that has been proposed (if we disregard proportionality requirements). All of these axioms regard the behaviour of PB rules in dynamic environments: when the cost function changes, when the set of projects changes, \textit{etc.} Hence, they can also be interpreted as robustness requirements: they enforce that the outcome does not change much with small variations of the instance. We will define these axioms in the following and present what is known about them.
	
	\medskip
	
	The first axiom we consider constrains the behaviour of the rule when the cost function changes.
	
	\begin{definition}[Discount monotonicity]
		\label{def:discountMono}
		A PB rule $\pbRule$ is said to be discount-monotonic if, for any two PB instances $I = \tuple{\projSet, c, b}$ and $I' = \tuple{\projSet, c', b}$ such that for some distinguished project $p^\star \in \projSet$, we have $c(p^\star) > c'(p^\star)$, and $c(p)=c'(p)$ for all $p \in \projSet \setminus \{p^\star\}$, it is the case that $p^\star \in \pbRule(I,\profile)$ implies $p^\star \in \pbRule(I',\profile)$ for all profiles~$\profile$.
	\end{definition}
	
	\noindent Thus, a rule is discount monotonic if whenever the price of a selected project $p$ decreases, the rule would still select project $p$.

	\medskip

	The second axiom, inspired by \emph{committee monotonicity} in the multi-winner voting literature \citep{LaSk23}, investigates the behaviour of the rule when the budget limit changes.
	
	\begin{definition}[Limit monotonicity]
		A PB rule $\pbRule$ is said to be limit-monotonic if, for any two PB instances $I = \tuple{\projSet, c, b}$ and $I' = \tuple{\projSet, c, b'}$ with $b < b'$ and $c(p) \leq  b$ for all projects $p \in \projSet$, it is the case that $\pbRule(I,\profile) \subseteq \pbRule(I',\profile)$ for all profiles~$\profile$.
	\end{definition}

	\noindent Thus, a rule is limit monotonic if it selects a superset of the original set of selected projects when the budget limit increases.
	
	\medskip
	
	The next two axioms concern cases where the project set changes, with some projects being either merged or split. Note that these axioms have only been considered for approval ballots. Since generalising them to arbitrary cardinal ballots is not straightforward, we focus on approval profiles here.
	
	Given a PB instance $I = \tuple{\projSet, c, b}$ and a profile $\profile$ of approval ballots, we say that the instance $I' = \tuple{\projSet', c', b}$ and the profile $\profile'$ of approval ballots are \emph{the result of splitting project $p^\star \in \projSet$ into $P^\star \subseteq \projSet'$} (with $P^\star \cap \projSet = \emptyset$, \textit{i.e.}, $P^\star$ is a set of new projects), if the following conditions are satisfied:
	\begin{itemize}
		\item The project $p^\star$ is replaced by $P^\star$ in the set of projects: $\projSet' = (\projSet \setminus \{p^\star\}) \cup P^\star$;
		\item The total cost of $P^\star$ is that of $p^\star$, \textit{i.e.}, $c'(P^\star) = c(p^\star)$; and for all $p \in P^\star$, it is the case that $c'(p) > 0$;
		\item The cost of every other project is as in $I$: $c'(p) = c(p)$ for all projects $p \in \projSet' \setminus P^\star$;
		\item The project $p^\star$ is replaced by $P^\star$ in the approval ballots containing it: for every $i \in \agentSet$ with $A_i(p^\star) = 0$, we have $A'_i = A_i$, and for every $i \in \agentSet$ with $A_i(p^\star) = 1$, we have $A_i'(p) = 1$ for all $p \in P^\star$, and $A'_i(p) = A_i(p)$ for all $p \in \projSet' \setminus P^\star$.
	\end{itemize} 
	We also say that $I$ and $\profile$ are \emph{the result of merging $P^\star$ into $p^\star$} given $I'$ and $\profile'$.
	
	\begin{definition}[Splitting monotonicity]
		\label{def:splittingMono}
		A PB rule~$\pbRule$ is said to be splitting-monotonic if, for any two  PB instances $I = \tuple{\projSet, c, b}$ and $I' = \tuple{\projSet', c', b}$ with corresponding profiles of approval ballots $\profile$ and $\profile'$ and any project $p \in \pbRule(I, \profile)$ such that $I'$ and $\profile'$ are the result of splitting project~$p$ into a subset of projects~$P$ given $I$ and $\profile$, it is the case that $\pbRule(I',\profile') \cap P \neq \emptyset$.
	\end{definition}
	
	\begin{definition}[Merging monotonicity]
		\label{def:mergingMono}
		A PB rule~$\pbRule$ is said to be merging-monotonic if, for any two PB instances $I = \tuple{\projSet, c, b}$ and $I' = \tuple{\projSet', c', b}$ with corresponding profiles of approval ballots $\profile$ and $\profile'$, and any subset of projects $P \subseteq \pbRule(I, \profile)$ such that $I'$ and $\profile'$ are the result of merging project set~$P$ into project~$p$ given $I$ and $\profile$, it is the case that $p \in \pbRule(I',\profile')$.
	\end{definition}

	\noindent These two axioms thus require the rule to also apply the splitting and merging operations on its outcome. Note that for splitting monotonicity, a stronger version of it would require \emph{all} the smaller projects to be selected (instead of only one).
	
	\medskip

	\begin{table}
		\centering
		\begin{tabular}{ccccc}
			\toprule
			& \multicolumn{4}{c}{\textbf{Monotonicity}} \\
			& Limit & Discount & Splitting & Merging \\
			\midrule
			\maxWelCard{} & \xmark & \cmark & \cmark & \xmark \\
			\greWelCard{} & \xmark & \cmark & \cmark & \xmark \\
			\midrule
			\maxWelCost{} & \xmark & \xmark & \cmark & \cmark \\
			\greWelCost{} & \xmark & \xmark & \xmark & \cmark \\
			\midrule
			\mes & \xmark & & & \\
			\bottomrule
		\end{tabular}
		\caption{
			Summary of the results concerning the monotonicity axioms for rules used with approval ballots.\\[0.4em]
			\footnotesize The results for \maxWelCard{}, \greWelCard{}, \maxWelCost{} and \greWelCost{} are due to \citet{TaFa19}. Note that their proofs contained several mistakes, corrected in part by \citet{BBS20}. Specifically, the proof that \greWelCard{} fails merging monotonicity is wrong, but the results still holds (though it is solely based on the use of tie-breaking rules that apply differently before and after merging projects). \mes{} fails limit montonicity as it already did on unit-cost instances \citep{LaSk23}.
		}
		\label{tab:summary_monotonicity}
	\end{table}
	
	We present in Table \ref{tab:summary_monotonicity} what is known about the standard PB rules regarding those axioms. The relevant references are provided there. Observe that it is not known which monotonicity axioms are satisfied by \seqPhragmen, \maximinSupport{} and \mes. One exception is that we know that \mes{} cannot satisfy limit monotonicity, as it does not satisfy committee monotonicity, the equivalent of limit monotonicity for unit-cost instances \citep{LaSk23}. 
	
	\medskip
	
	The definitions we provided above concern \emph{resolute} PB rules, that is, rules that always output a single budget allocation. They have been extended to \emph{irresolute} rules---that can return more than one budget allocation---in two different general ways. For a given irresolute rule $\pbRule$:
	\begin{itemize}
		\item \citet{BBS20} \citep[and subsequently][]{SBY22} extend the monotonicity axioms in an \emph{existential} fashion: for a given instance $I$ and profile $\profile$, and for every budget allocation $\pi \in \pbRule(I, \profile)$ that satisfy a specific pre-condition, it must be the case that for every suitable $I'$ and $\profile'$, there exist a budget allocation $\pi' \in \pbRule(I', \profile')$ satisfying the specific post-condition;
		\item \citet{REH20} extend them in a \emph{universal} fashion: for a given instance $I$ and profile $\profile$ if every budget allocation $\pi \in \pbRule(I, \profile)$ satisfies a specific pre-condition, it must be the case that for every suitable $I'$ and $\profile'$ and every $\pi' \in \pbRule(I', \profile')$, the required post-condition is satisfied.
	\end{itemize}
	
	\section{Strategy-Proofness}
	\label{subsec:StrategyProofness}
	
	The next class of requirements we consider is that of incentive compatibility axioms. These axioms are concerned with preventing agents from engaging in strategic behaviour.
	
	Let us first discuss the concept of \emph{strategy-proofness}. Intuitively speaking, it states that no agent should be able to obtain a better outcome by reporting a ballot that is different from their true preferences. To define it, we thus need a way of comparing outcomes from the point of view of the agents. When using cardinal ballots, we will assume that the ballot represents the agents' utility for the projects. For approval ballots, we will use the notion of satisfaction function as the measure of utility.\footnote{Note here that we are indeed discussing utilities and not satisfaction levels since we are considering behaviours that the agents engage in themselves, according to their private information.}
	
	\begin{definition}[Strategy-Proofness for Cardinal Ballots]
		A PB rule $\pbRule$ is said to be \emph{strategy-proof} if for every instance $I$ and profile $\profile$ of cardinal ballots, and  for every agent $i \in \agentSet$, there exists no cardinal ballot $A_i'$ such that for the profile $\profile' = (A_1$, $\ldots$, $A_{i - 1}$, $A_i'$, $A_{i + 1}$, $\ldots$, $A_n)$ we have:
		\[\sum_{p \in \pbRule(I, \profile')} A_i(p) > \sum_{p \in \pbRule(I, \profile)} A_i(p).\]
	\end{definition}
	
	\noindent Observe that the satisfaction of the manipulating agent $i$ with the output under the new profile $\profile'$ is computed with regards to the initial ballot $A_i$.
	
	\begin{definition}[Strategy-Proofness for Approval Ballots]
		Given a satisfaction function $\satisfaction$, a PB rule $\pbRule$ is said to be \emph{strategy-proof for $\satisfaction$} if for every instance $I$ and profile $\profile$ of approval ballots, for every agent $i \in \agentSet$, there exists no approval ballot $A_i'$ such that for the profile $\profile' = (A_1, \ldots, A_{i - 1}, A_i', A_{i + 1}, \ldots, A_n)$ we have:
		\[\satisfaction(\pbRule(I, \profile') \cap A_i) > \satisfaction(\pbRule(I, \profile) \cap A_i).\]
	\end{definition}

	It is already known from multi-winner voting, {i.e.}, when instances have unit costs, that strategy-proofness is incompatible with very weak notions of proportionality \citep{Peters18, Peters19}. This result obviously also applies to general PB instances.
	
	\begin{theorem}[\citealt{Peters18}]
		A PB rule $\pbRule$ is said to be \emph{weakly proportional on unit-cost instances} if for every unit-cost instance $I$ and profile $\profile$ of cardinal ballots such that for all voters $i, i' \in \agentSet$ either $\{p \in \projSet \mid A_i(p) > 0\} = \{p \in \projSet \mid A_{i'}(p) > 0\}$, or these two sets do not intersect (meaning that $\profile$ is a party-list profile), then for any project $p \in \projSet$ such that $|\{i \in \agentSet \mid A_i(p) > 0\}| \geq \nicefrac{n}{b}$ we have $p \in \pbRule(I, \profile)$.
		
		There is no rule that satisfies simultaneously weak proportionality on unit-cost instances and strategy-proofness.
	\end{theorem}

	\noindent Note that in the actual statement of \citet{Peters18, Peters19}, an additional efficiency requirement is needed. This is because in the multi-winner voting setting, one \emph{has} to ensure that a rule selects the required number of candidates (\textit{i.e.}, the rule has to be exhaustive). Since this constraint is lifted in the PB setting, there is no need for such an additional axiom.
	
	It should also be noted that the proportionality axiom defined in the above statement is particularly weak and is known to be implied by all kinds of other requirements \citep{Peters18}, including all the ones introduced in Section \ref{subsec:Justified_Representation}. In particular, this implies that rules such as \seqPhragmen{}, \maximinSupport{} or \mes{} are not strategy-proof.

	This result has been replicated in the multi-resource PB case, for suitable adaptations of the axioms \citep{MSRE22}. Moreover, it also applies to irresolute rules \citep{KVVBE20}.
	
	\medskip
	
	It is known that with unit-cost instances, welfare-maximising rules such as \greWelCost{} (which is equivalent to \maxWelCost{} on unit-cost instances) are strategy-proof. When moving to general PB instances, we can show that \greWelCost{} is only \emph{approximately} strategy-proof. We provide below the definition of \citet{GKSA19}, which weakens strategy-proofness in a ``up-to-one'' fashion.
	
	\begin{definition}[Approximate Strategy-Proofness for Approval Ballots]
		Given a satisfaction function $\satisfaction$, a PB rule $\pbRule$ is said to be \emph{approximately strategy-proof for $\satisfaction$} if for every instance $I$ and profile $\profile$ of approval ballots, for every agent $i \in \agentSet$, there exists no approval ballot $A_i'$ such that for the profile $\profile' = (A_1, \ldots, A_{i - 1}, A_i', A_{i + 1}, \ldots, A_n)$, for all $p \in \projSet$ we have:
		\[\satisfaction(\pbRule(I, \profile') \cap A_i) > \satisfaction((\pbRule(I, \profile) \cap A_i) \cup \{p\}).\]
	\end{definition}

	\begin{proposition}[\citealt{GKSA19}]
		$\greWelCost{}$ is approximately strategy-proof for $\costSatisfaction$.
	\end{proposition}

	\noindent Note that the result by \citet{GKSA19} uses knapsack ballots. This is not required when projects are indivisible.\footnote{Let us sketch the proof, originally devised by Ulle Endriss. For any given $I = \tuple{\projSet, c, b}$, consider $I' = \tuple{\projSet', c', b}$, where projects in $\projSet$ have been split into sets of subprojects, each of cost~1. $I'$ is thus a unit-cost instance. We can transform any given profile $\profile$ of approval ballots in the same manner to obtain a profile $\profile'$ of approval ballots. Now, it is clear that the approval scores of the projects in $\profile'$ are equal to those of the projects in $\projSet$ they come from in $\profile$. Assume that the tie-breaking rule is extended in a consistent way from projects in $\projSet$ to projects in $\projSet'$. Then we know that there exists at most one project $p \in \projSet$ such that $\greWelCost(I', \profile')$ contains a proper subset of its corresponding subprojects. Let $\pi' \subseteq \projSet$ be the budget allocation that includes any project in $\projSet$ for which at least one corresponding subproject is in $\greWelCost(I', \profile')$. We thus have  $\greWelCost(I, \profile) \cup \{p\} = \pi'$. Since $\greWelCost$ is strategy-proof over unit-cost instances \citep{Peters18}, no agent can reach a better budget allocation than $\pi'$ by strategising, when considering the satisfaction function $\costSatisfaction$.} It is also worth noting that this result does not hold for $\cardSatisfaction$.
	
	Interestingly, exact welfare maximising rules such as \maxWelCard{} or \maxWelCost{} fail even approximate strategy-proofness on PB instances, for large sets of satisfaction functions. This can come as a surprise since they are strategy-proof on unit-cost instances. Note that this also holds if ballot are knapsack ballots. We present below an example for \maxWelCost.
	
	\begin{illustration}[\maxWelCost Strategy-Proofness]
		Consider the instance $I = \tuple{\projSet, c, b}$ with $\projSet = \{p_1, \ldots p_5\}$, the cost are such that $c(p_1) = 6$, $c(p_2) = 3$ and $c(p_3) = c(p_4) = c(p_5) = 1$, and the budget limit is $b = 6$.
		
		Assume that three agents are involved in the process for whom the truthful ballots are to approve of $p_1$ for agent~1; $p_2$ for agent~2; and $p_3$, $p_4$ and $p_5$ for agent~3. If ties are broken lexicographically, the outcome of \maxWelCost{} would then be $\pi = \{p_1\}$. Note that agent~3 has satisfaction 0 for $\pi$. Now, if agent~3 were to approve of $p_2, p_3, p_4$ and $p_5$ instead, the outcome would be $\pi' = \{p_2, p_3, p_4, p_5\}$. Is it clear that for any satisfaction function that is strictly monotonic\footnote{A satisfaction function $\satisfaction$ is strictly monotonic if for all $P \subseteq \projSet$ and $P' \subsetneq P$, we have $\satisfaction(P') < \satisfaction(P)$.} and for every project $p \in \projSet$, agent~3 prefers $\pi'$ over $\pi \cup \{p\}$.
	\end{illustration}

	Finally, for a slightly differing budgeting model motivated by PB, \citet{WM23} introduced a strategy-proof VCG-like mechanism.
	
	\section{Other Axioms}
	\label{subsec:Other_Axioms}
	
	Let us conclude by mentioning some other axioms and axiomatic directions that have been followed in the context of PB.
	
	In their study on maximin PB with approval ballots, \citet{SBY22} adapt several axioms from the multi-winner literature to the context of PB with irresolute rules. These axioms are the \emph{narrow-top criterion} (an adaptation of unanimity) and \emph{clone-proofness} (the outcome of a rule remains the same if projects are cloned). They also introduce a new axiom called \emph{maximal coverage} stating that no redundant project should ever be selected unless it is not possible to cover more voters, where a voter is \emph{covered} if at least one of their approved projects has been selected, and a project is \emph{redundant} if removing it does not change the set of covered voters. Note that this axiom can be seen as a fairness requirement.
	
	Following a more typical social choice route, \citet{CGN22} initiated the axiomatic characterisation of PB rules, focusing on \greWelCost{} for now.
	
	Finally, it is also worth mentioning that \citet{GKSA19} provided the first analysis of PB rules in terms of epistemic criteria (being a maximum likelihood estimator) to date, another branch of the axiomatic approach \citep{ElSl16, Piva19}. The epistemic analysis of PB with approval ballots has then been the focus of \citet{ReEn23}.
	
	\chapter{Algorithmic Considerations}
	\label{chap:Algorithmic}
	
	Another large part of the literature focuses on the algorithmic aspects of PB. This usually concerns computing outcomes of PB rules and the exact complexity of welfare maximisation under different models.
	
	We will discuss these different aspects, focusing first on outcome determination (Section \ref{subsec:OutcomeDetermination}), then on the complexity of welfare maximisation (Section \ref{subsec:AlgorithmicSocialWelfare}), and finally on the other algorithmic problems that have been studied (Section \ref{subsec:OtherAlgorithmicProblems}).
	
	\section{Outcome Determination of Standard PB Rules}
	\label{subsec:OutcomeDetermination}
	
	The main focus of the algorithmic perspective on social choice is to assess the computational complexity of computing ``good'' outcomes. With all that has been presented so far, we already know a lot about the quality of the outcome of the standard PB rules. The last step is thus to assess how hard it is to compute said outcomes.
	
	\medskip
	
	Formally speaking, this is the problem of computing the outcome of a given rule, the so-called \emph{outcome determination problem}. We present below one version of this problem for a given resolute PB rule $\pbRule$.

	\begin{center}
		\begin{tabular}{rl}
			\toprule
			\multicolumn{2}{c}{\textsc{OutcomeDetermination($\pbRule$)}} \\
			\midrule
			\textbf{Input:} & An instance $I = \tuple{\projSet, c, b}$, a profile $\profile$, and a project $p \in \projSet$. \\
			\textbf{Question:} & Is $p \in \pbRule(I, \profile)$? \\
			\bottomrule
		\end{tabular}
	\end{center}

	\noindent Note that this definition only makes sense for \emph{resolute} PB rules. Other formulations are also possible, for example as a function problem. 
	
	The complexity of the winner determination problem for irresolute PB rules has not been considered in the literature yet and it is not immediately clear how the outcome determination problem should be formulated. One natural idea would be to define the problem as checking whether a project is \emph{always} selected, or whether it is \emph{sometimes} selected.
	
	\medskip

	It should be more or less clear that the outcome determination problem can be efficiently solved for most of the rules that we have focused on, at least in the resolute case. The definitions of \greWelCard{}, \greWelCost{} and \seqPhragmen{} should make it somewhat obvious that computing their outcomes can be done efficiently. For \maximinSupport{}, \citet{ALT18} presents a linear program allowing to compute efficiently an optimum load distribution at each round. Finally, \citet{PPS21} discuss how to efficiently compute outcomes of \mes{}.
	
	The only rules whose outcomes cannot be computed efficiently are the ones that relate to exact welfare maximisation. Indeed, maximising the social welfare is usually hard, as we shall see next.
	
	\section{Maximising Social Welfare}
	\label{subsec:AlgorithmicSocialWelfare}
	
	Let us now turn to the computational problem of maximising measures of social welfare. 
	
	First, we introduce the different notions of social welfare that have been studied in the literature. Note that throughout this section, we will work with cardinal ballots. We also repeat the definition of \utilSW{} so that the reader does not need to go back to Section \ref{subsec:Welfare_Rules}.
	
	\begin{itemize}
		\item \textbf{Utilitarian Social Welfare:} Given an instance $I$ and a profile $\profile$ of cardinal ballots, the \emph{utilitarian social welfare} achieved by a budget allocation $\pi$ is defined as:
		\[\utilSW{}(I, \profile, \pi) = \sum_{i \in \agentSet} \sum_{p \in \pi} A_i(p).\]
		This is the most common definition of social welfare simply considering the sum of the satisfactions of the individuals. A budget allocation maximising \utilSW{} selects the items that are individually best, \textit{i.e.}, it ignores any interactions between the projects.
		
		\item \textbf{Chamberlin-Courant Social Welfare:} Given an instance $I$ and a profile $\profile$ of cardinal ballots, the \emph{Chamberlin-Courant social welfare} achieved by a budget allocation $\pi$ is defined as:
		\[\ccSW{}(I, \profile, \pi) = \sum_{i \in \agentSet} \max_{p \in \pi} A_i(p).\]
		The Chamberlin-Courant social welfare assumes that agents only consider one project from each budget allocation, namely the one that leads to the highest satisfaction. Maximising \ccSW{} corresponds thus to aim for a budget allocation that represents as many voters as possible.
		
		Note that \ccSW{} has been studied by \citet{Laru21} under the name \emph{Rawlsian social welfare}.
				
		\item \textbf{Egalitarian Social Welfare:} Given an instance $I$ and a profile $\profile$ of cardinal ballots, the \emph{egalitarian social welfare} achieved by a budget allocation $\pi$ is defined as:
		\[\egalSW{}(I, \profile, \pi) = \min_{i \in \agentSet} \sum_{p \in \pi} A_i(p).\]
		The egalitarian social welfare assumes that the welfare of a society is the satisfaction of its most dissatisfied member. Maximising \egalSW{} hence means maximising the satisfaction of the worst-off voter.
		
		\egalSW{} is studied by \citet{SBY22} under the name \emph{maximin PB}.
		
		\item \textbf{Nash Social Welfare:} Given an instance $I$ and a profile $\profile$ of cardinal ballots, the \emph{Nash social welfare} achieved by a budget allocation $\pi$ is defined as:
		\[\nashSW{}(I, \profile, \pi) = \prod_{i \in \agentSet} \sum_{p \in \pi} A_i(p).\]
		The Nash social welfare measure can be seen as a compromise between utilitarian and egalitarian social welfare. By maximising \nashSW{}, one aims to find a fair budget allocation \citep{FSTW19}.
		
		Note that maximising \nashSW{} is equivalent to maximising the sum of the logarithms of the satisfactions of the agents.
	\end{itemize}

	The typical computational problem is then to determine whether there is a budget allocation that provides at least a certain amount of satisfaction according to a specific measures of welfare. \citet{FSTW19} studied this problem for \utilSW{}, \nashSW{} and \ccSW{}. \citet{SBY22} considered the case of \egalSW{}, in the context of approval ballots with $\costSatisfaction$. \citet{TaFa19} focused on \utilSW{} with approval ballots and several satisfaction functions. We summarise the main findings in Table~\ref{tab:Complexity_WelfareMaximisation}.
	
	\medskip	
	
	Welfare maximisation problems have also been studied for many of the variations of the standard model that have been introduced. We just mention them here and refer the reader to Chapter~\ref{chap:Extensions} for more details. \citet{HKPP21} studied welfare maximisation in a model in which projects are grouped into districts. Similarly, \citet{JSTZ21} and \citet{PKL21} investigated different social welfare maximisation when projects are grouped in categories. \citet{JST20} looked into social welfare for non-additive satisfaction functions. Social welfare has been studied in multi-resource PB \citep{MSRE22}, when the cost is dependent on the number of users of the projects \citep{LuBo11}, and when the budget is endogenous \citep{AzGa21, AGPSV22, CLM22}.
	
	\begin{table}
		\centering
		\renewcommand\tabularxcolumn[1]{p{#1}}
		\begin{tabularx}{\linewidth}{rX@{}}
			\toprule
			
			\textbf{\utilSW{}} & 
			{
				\renewcommand\tabularxcolumn[1]{m{#1}}
				\begin{tabularx}{\linewidth}{p{10em}X}
					Weakly \complexNP-complete & $\blacktriangleright$ Even with one voter \\
					\midrule
					Pseudopoly.\ solvable & \\
					\midrule
					Poly.\ solvable & $\blacktriangleright$ If $A_i(p) \in \{0, 1\}$ for all $i \in \agentSet$ and $p \in \projSet$ \\
				\end{tabularx}
			} \\
		
			\midrule
			
			\textbf{\nashSW{}} & 
			{
				\renewcommand\tabularxcolumn[1]{m{#1}}
				\begin{tabularx}{\linewidth}{p{10em}X}
					Strongly \complexNP-complete & $\blacktriangleright$ Even with one voter \newline $\blacktriangleright$ Even with two voters and unit-cost instances \newline $\blacktriangleright$ Even with unit-cost instances and $A_i(p) \in \{0, 1\}$ for all $i \in \agentSet$ and $p \in \projSet$ \\
					\midrule
					$\mathtt{W}$[1]-hard & $\blacktriangleright$ Parameterised by the budget limit $b$, even with unit-cost instances and $A_i(p) \in \{0, 1\}$ for all $i \in \agentSet$ and $p \in \projSet$ \newline $\blacktriangleright$ Parameterised by the budget limit $b$ and the number of voters $n$, even with unit-cost instances and unary encoding \newline $\blacktriangleright$ Even with single-peaked or single crossing profiles \\
					\midrule
					$\mathtt{XP}$ & $\blacktriangleright$ Parameterised by the number of voters $n$ \\
					\midrule
					\complexFPT & $\blacktriangleright$ Parameterised by the number of voters $n$ and $\max_{i \in \agentSet} |\{\sum_{p \in \pi} A_i(p) \mid \pi \in \allocSet(I)\}|$ \\
				\end{tabularx}
			} \\
			
			\midrule
			
			\textbf{\ccSW{}} & 
			{
				\renewcommand\tabularxcolumn[1]{m{#1}}
				\begin{tabularx}{\linewidth}{p{10em}X}
					Pseudopoly.\ solvable & $\blacktriangleright$ For single-peaked and single-crossing profiles \\
					\midrule
					Strongly \complexNP-complete & $\blacktriangleright$ Even for binary valuations, i.e., ballots with only two different values, and unit-cost instances \\
					\midrule
					\complexFPT & $\blacktriangleright$ Parameterised by the number of voters and $\sum_{i \in \agentSet} \sum_{p \in \projSet} A_i(p)$ \\
					\midrule
					$\mathtt{W}$[2]-hard & $\blacktriangleright$ Parameterised by the budget limit $b$ \\
				\end{tabularx}
			} \\
		
			\midrule
			
			\textbf{\egalSW{}} & 
			{
				\renewcommand\tabularxcolumn[1]{m{#1}}
					\begin{tabularx}{\linewidth}{p{10em}X}
					Strongly \complexNP-complete & $\blacktriangleright$ Even if $A_i(p) \in \{0, c(p)\}$ for all $i \in \agentSet$ and $p \in \projSet$ \\
					\midrule
					Pseudopoly.\ solvable & $\blacktriangleright$ When $A_i(p) \in \{0, c(p)\}$ for all $i \in \agentSet$ and $p \in \projSet$ and the number of distinct ballots is constant \\
					\midrule
					Poly.\ solvable & $\blacktriangleright$ When $A_i(p) \in \{0, c(p)\}$ for all $i \in \agentSet$ and $p \in \projSet$, the number of distinct ballots is constant, and $\frac{\max_{p \in \projSet} c(p)}{\mathit{GCD}\{c(p) \mid p \in\projSet\}}$ is constant.
				\end{tabularx}
			} \\
			\bottomrule
		\end{tabularx}
		\caption{
			Computational complexity of the decision problem corresponding to the maximisation of different types of social welfare. For a given measure of welfare $\textsc{SW}$, the exact decision problem that is considered is the following: given an instance $I = \tuple{\projSet, c, b}$, a profile $\profile$ of cardinal ballots, and a target $x \in \mathbb{Q}_{\geq 0}$, is there a budget allocation $\pi \in \allocSet(I)$ such that $\textsc{SW}(I, \profile, \pi) \geq x$?\\[0.4em]
			\footnotesize Statements for \utilSW{} follow immediately from the literature on the knapsack problem \citep{KPP04} as explained by \citet{TaFa19}. The results for \nashSW{} and \ccSW{} are due to \citet{FSTW19}. \ccSW{} with approval ballots was studied by \citet{TaFa19}. \citet{SBY22} studied \egalSW{}. \\[0.4em]
			\footnotesize Note that $A_i(p) \in \{0, 1\}$ for all $i \in \agentSet$ and $p \in \projSet$ simulates approval ballots with the satisfaction function $\cardSatisfaction$, and $A_i(p) \in \{0, c(p)\}$ for all $i \in \agentSet$ and $p \in \projSet$ simulates approval ballots used with $\costSatisfaction$.
		}
		\label{tab:Complexity_WelfareMaximisation}
	\end{table}
	
	\section{Other Algorithmic Problems}
	\label{subsec:OtherAlgorithmicProblems}
	
	Participatory budgeting offers other avenues for studies focusing on the computational complexity of related problems.
	
	For instance, \citet{BBH21} study the computational complexity of control in PB instances with approval ballots. Control problems are problems of the form ``Can the decision maker achieve certain objectives by changing certain parameters of the instance?''. More specifically, \citet{BBH21} studied two types of control for \greWelCard{}, \greWelCost{}, \maxWelCard{}, and \maxWelCost{} when the decision maker can decide on the price of a project, or on the budget limit. Under \emph{constructive control}, the decision maker aims at forcing the selection of a given project, while under \emph{destructive control}, they aim at preventing a given project from being selected.

	To better understand the robustness of PB outcomes, \citet{BFJK23} studied the computational complexity of what they call the \normalfont\textsc{\#Flip-Bribery} problem. The latter is intended to model the effect of adding random noise to an approval-based PB instance, and asks for the number of ways in which a given number of approvals can be flipped to result in a specific project being included in the outcome. \citet{BFJK23} show that \normalfont\textsc{\#Flip-Bribery} is intractable even for simple PB rules. Following this result, they used \href{http://pabulib.org}{pabulib.org} data to empirically reason about the extent to which more simple, greedy PB rules are more robust than proportional ones, and to identify three groups of non-robust projects.

	\chapter{Variations and Extensions of the Standard Model}
	\label{chap:Extensions}
	
	The literature we reviewed so far studied what could be called the standard model of PB. Beyond that, a myriad of variations of the model have been introduced. In the following we delve into these variations, in an order based on their (assumed) end-goal. We will first look into variations of the standard model that aim at capturing real-life PB processes more accurately (Section \ref{subsec:AccurateModel}). Our focus will then shift to models that propose extensions of the standard model (Section \ref{subsec:EnrichingModel}).
	
	\section{Towards More Accurate Models of PB}
	\label{subsec:AccurateModel}
	
	A large chunk of the variations of the standard model have been introduced with the aim of better capturing real-life PB processes. Among others, the repetitive aspect of PB, its multistage implementation, and its geographical constraints have been studied.
	
	\subsection{End-to-End Model for PB}
	
	We start with the integration of the several stages of a PB process in the analysis. As detailed in the introduction already, a typical PB process has several stages including two during which the citizens are consulted: for the exploration of the projects to consider, and for the vote on which of the selected proposals should actually be implemented. This two stage model has been formalised and studied by \citet{REH21}. They focus on two specific aspects: shortlisting rules, and strategic behaviours in this integrated model.
	
	\citet{REH21} approach the first stage of the process---the \emph{shortlisting stage}, during which agents submit proposals for projects that could be considered throughout the process---as a multi-winner election for which there is no specific requirements regarding the size, or cost, of the outcome. They define and analyse several \emph{shortlisting rules} that could be used to create the shortlist based on different ideas: diversity in the voters represented, diversity in the essence of the proposals, and limited size of the shortlist to reduce the cognitive burden induced by the voting stage.
	
	In addition to the first stage, \citet{REH21} also study their end-to-end model in its entirety. They focus on its strategic aspects and try to answer questions of the type: can an agent improve their satisfaction with the final outcome by not proposing a project in the first stage? Their findings indicate that, unsurprisingly, it is difficult to prevent strategic behaviours.
	
	\subsection{Local versus Global Processes}
	
	Real-life PB processes tend to be implemented at the scale of a municipality. It is very common for the municipality to actually implement several local PB processes, one for each district for instance, instead of one general process. This is the case in \textit{e.g.} Amsterdam \citep{amspb}, and to some extent in Paris \citep{parispb}\footnote{In Paris, the PB process combines local and global aspects: voters can vote on the projects for their district together with some Paris-wide projects.}. Motivated by this observation \citet{HKPP21} investigate the effect of the local versus global implementation of PB processes.
	
	In their study, \citet{HKPP21} introduce a model of  \emph{district-based PB}. Each project belongs to a specific district and contributes a fixed additive amount to the welfare of its district. In addition, there is a budget limit for each district. A budget allocation is called \emph{district-fair} if it provides each district at least as much social welfare as they could achieve with their share of the budget limit. The authors then consider the problem of selecting a global budget allocation that is district-fair.
	
	\citet{HKPP21} show that it is computationally hard to maximise social welfare under district-fairness constraints. In addition, they show that one can, in polynomial time, find probabilistic outcomes that maximise the global social welfare while being almost district-fair in expectation. Finally, they show that by slightly overspending (by a factor $1.647 + \epsilon$, with $\epsilon > 0$), one can find budget allocations that maximise the global social welfare while providing ``district-fairness up to one project'' to each district in polynomial time.
	
	\subsection{Temporal Aspects of PB} PB processes are rarely single-shot instances as they often span several years, one PB process being organised each year. Based on that insight, \citet{LMR21} introduced a model for long-term PB based on the \emph{perpetual voting} framework \citep{Lack20}.
	
	In their work, \citet{LMR21} introduce what they call a fairness theory for long-term PB. They assume that agents are partitioned into types, and they try to achieve fairness for the types over time. They study three fairness requirements based on $\costSatisfaction$, $\mathit{relsat}$ and $\share$: Enforcing that all types enjoy the same welfare, that all types converge towards equal welfare if the instance would be infinite, or that the welfare across types is distributed optimally (according to the Gini-coefficient). Each of these fairness concepts is analysed in terms of whether they are satisfiable or not. Their findings suggest that it is difficult to provide such fairness guarantees.
	
	\section{Enriching the Standard Model}
	\label{subsec:EnrichingModel}
	
	The works we have presented above aimed at capturing real-life PB processes more accurately. In the following, we will review works that aim at enriching the standard model by proposing extension of the model that could improve the PB process, but are, to the best of our knowledge, currently not widely implemented in practice.
	
	\subsection{Additional Distributional Constraints}
	
	We first focus on a strand of the literature that deals with incorporating additional constraints into the standard model. These constraints are usually distributional ones that affect which projects can be selected. They can model the fact that some projects are \emph{incompatible}, or that some projects have \emph{positive interactions} for instance. We will see several examples in the following.
	
	\medskip
	
	Developing a very general framework for this task, \citet{REH20} demonstrate how to encode PB problems into \emph{judgment aggregation}, a very expressive framework for constrained aggregation \citep{Endr16}. Their framework allows for the addition of any additional constraint that can be expressed in propositional logic. They study the computational cost of such an approach, and show that as long as the constraints can be efficiently encoded in some compact logical representations, the computational overhead is not too large. They also provide an axiomatic analysis \citep[following][]{TaFa19} of some rules that can be used in this context.
	
	A similar general approach was also considered by \citet{FMS18}, though in a context more general than PB. They provide a framework of public decision making with matroid, matching, and packing constraints, allowing for great flexibility on what can be modelled. Note that packing constraints correspond to what we call budget constraints.
	
	\medskip
	
	In addition to this, several papers focus on specific constraints that can be implemented in PB.
	
	\begin{itemize}
		\item \textbf{Dependency constraints:} \citet{REH20} study how to include dependency constraints in their framework described above. By dependencies, they mean that the implementation of some projects is dependent on the status of some others.
		
		\item \textbf{Categorical constraints:} These constraints model the idea that projects are grouped into categories and that additional constraints apply regarding which of the projects can be selected within each category.
		
		Still within their framework, \citet{REH20} introduce \emph{quota constraints} that enforce some lower and upper quota to be satisfied for each category. They provide two example of such quotas: on the number of selected projects from a category, or on the total cost.
		
		\citet{JSTZ21} also study what \citet{REH20} called cost quota constraint, and what they refer to as \emph{PB with project groups}. They focus on the computational aspects of finding a feasible budget allocation maximising the social welfare, and they provide an in-depth analysis of this extended PB setting: Parameterised complexity analysis, and approximability and inapproximability results. In particular, they provide efficient algorithms to maximise or to approximate the social welfare when the number of categories is small; while proving hardness for arbitrary number of categories.
		
		\citet{PKL21} investigate the computational complexity of selecting group-fair knapsacks. This problem is equivalent to selecting budget allocations maximising the utilitarian social welfare in PB instances with categories over the projects, and upper and lower quotas on the categories. The quotas are expressed either in terms of number of selected projects per category, or contribution to the social welfare per category. They prove hardness results, and provide intricate dynamic programming algorithms that compute approximate solutions.
		
		Quotas on the number of projects selected per category have also been considered by \citet{CLM22} in a model with endogenous funding.
		
		Note that both \citet{REH20} and \citet{CLM22} do not assume categories to be disjoint while \citet{JSTZ21} and \citet{PKL21} do.
	\end{itemize}

	Let us finally mention that when studying PB with multidimensional costs, \citet{MSRE22} show how to encode distributional constraints simply by using extra resources. They discuss dependency constraints, categorical constraints (upper quota on the cost of a category), and incompatibility constraints (categorical constraints with quotas on the upper number of projects selected in a category).
	
	\subsection{Interaction Between Projects}
	
	An assumption that is almost always made is that projects are independent. We have seen above how to incorporate distributional constraints challenging that assumption at the level of which budget allocations are admissible or not. In a similar spirit, \citet{JST20} challenge the independence assumption from the perspective of the voters, assuming that the satisfaction of the voters is not \emph{additive}, \textit{i.e.}, can be more, or less, than the sum of its parts.
	
	Specifically, \citet{JST20} assume that there is an \emph{interaction structure} partitioning the projects into categories. The utility of the voters is defined as the sum of their satisfaction for each category, the latter being an increasing, but potentially non-linear, function of the number of approved and selected projects from within the category. This model enables the study of substitution or complementarity effects between the projects from the perspective of the voters.
	
	In addition to their conceptual contribution, \citet{JST20} present a computational analysis of welfare-maximising problems in this setting. They provide a mixture of hardness results and (fixed parameter) efficient algorithms. They also identify restrictions of the ballots submitted by the voters, defined with respect to a specific interaction structure, for which the computational problems become tractable.
	
	Note that in the work of \citet{JST20}, the interaction structure is given and fixed for all voters. In subsequent work, \citet{JTB21} analysed how to obtain such an interaction structure based on several partitions of the projects submitted by the agents. The focus is computational there as well.

	Also based on the assumption that project preferences are connected, \citet{LT23} propose a recommendation system for PB, which asks the voters to only fill out partial ballots and, based on this preference information, completes their ballots for them. In their findings, the outcome based on these artificially completed ballots differs only slightly from the outcome based on full preference information.

	\subsection{Enriched Cost Functions}
	
	Another typical assumption that is made is to assume that the cost of the projects is fixed and expressed in only one dimension. Both of these aspects of the cost function have been challenged by different authors.
	
	In one of the first papers on a model at the time not yet called participatory budgeting, \citet{LuBo11} consider the problem of selecting multiple costly alternatives under a given budget constraint. Their model is slightly different from the standard one for PB as they aim at modelling recommendation systems. In particular,  selected alternatives are assigned to some agents. What is more interesting for us here is that they assume that the cost of a project is composed of a fixed part and of a variable part. Specifically, the cost of a project is an affine function of the number of agents assigned to that project. 
	
	The assumption that costs are unidimensional has also been lifted. In their framework developed to include additional constraints in PB (see above), \citet{REH20} assume that the costs are expressed over several dimensions. More interestingly, \citet{MSRE22} focus on analysing the effect of multidimensional costs. They extend the standard model for PB, assuming that the costs of the projects are expressed in terms of several resources. In this setting, they define and study proportionality requirements, incentive compatibility axioms, and their interactions. They also touch on the computational aspect of maximising social welfare.

	Finally, \citet{Sreedurga23IJCAI} introduces a model where each project has a set of permissible costs, with voters giving a lower and upper bound for each project. \citet{Sreedurga23IJCAI} introduces several PB rules for this setting and studies their computational complexity and their axiomatic properties.
	
	\subsection{Uncertainty in PB}
	
	In practice there is a lot of uncertainty around the actual implementation of the projects. It is for instance rarely possible to assess the cost of the projects exactly, let alone their completion time. \citet{BBL22} initiated the study of PB under uncertainty about the projects.
	
	In their model, \citet{BBL22} assume that the costs of the projects are uncertain. For each project, its cost is described as a probability distribution over a specific interval. Projects are associated with a completion time and the actual cost of a project is revealed only once the project has been completed. They consider online mechanisms that select the projects to be funded in a dynamic fashion. Within this framework, they provide a series of impossibility results showing that no online mechanism can be at the same time punctual (finishes within the given time bound), not too risky (the probability of exceeding the budget is never too high, or the excess is never too high), and exhaustive (the budget is not underused). They also adapt the justified representation axioms to this setting, showing that an adaptation of \mes{} provides some fairness guarantees here.
	
	\subsection{PB with Endogenous Funding}
	
	The standard PB model assumes that the budget is provided by the organising entity (a municipality for instance). Several authors have proposed alternative models in which the voters can actually contribute their own funds to help implement some projects.
	
	In a model in which voters submit cardinal ballots over the projects, \citet{CLM22} introduce the idea that voters can also submit monetary contributions to specific projects, thus reducing the amount of public money needed to select the projects. They investigate suitable aggregation methods for this framework. The risk with donation is that some voters could have too much influence on the final outcome. Therefore, the authors focus on devising rules for which the satisfaction of no voter decreases when taking into account donations, compared to the case where the donations are ignored. They provide several such rules, and study their merits regarding some donation-specific monotonicity requirements. They conclude their analysis by studying the computational complexity of outcome determination problems, and the problem of finding an optimal donation policy for the voters.

	Based upon this, \citet{WWWJ23} explore the idea of donations to projects under approval preferences specifically, concluding that while the positive results of \citet{CLM22} can be applied, not all negative results in the general setting remain negative in the approval setting. Furthermore, they extend the study to what they call proportional greedy rules under donations.
	
	Moving further away from PB, \citet{AzGa21} propose a setting in which there is no exogenous fund, instead, each agent joins the process with a given personal budget that will be used to fund the projects. Agents submit approval ballots and a rule in this setting determines, given an approval profile and the personal budget of the agents, a set of projects to be funded and the monetary contribution of each individual to the selected projects. This model is slightly different from PB in the sense that it is not about the allocation of public funds. It is nevertheless a framework studying aggregation problems when selecting costly alternatives. They introduce and study several axioms dealing with efficiency (Pareto-optimality), and fairness (core and proportionality). Finally, they investigate several welfare-maximisation rules---based on utilitarian, egalitarian, or Nash social welfare---in terms of these axioms.
	
	\citet{AGPSV22} study a similar model except that agents submit cardinal ballots instead of approval ones, and that they have quasi-linear utilities (that depend on the money they spend). The authors focus on the computational aspects of maximising the utilitarian social welfare subject to some participation requirements (that guarantees the agents not to contribute more than they receive), showing both computational hardness and inapproximability of the problem.
	
	\subsection{Weighted PB}
	
	The fact that projects have different costs in PB can be interpreted as them having different weights. In their study of PB with ordinal ballots, \citet{AzLe21} make a symmetrical assumption that the voters have different weights. Their analysis does not really focus on this assumption however, and little is known about what its impact is in general.

	\subsection{Consensus through Reinforcement Learning}

	\citet{MP23} propose a PB model which utilises multi-agent reinforcement learning to reach what they call a consensus in a PB election. They argue for a PB implementation where voters systematically deliberate with each other. Policy makers assist in filtering out infeasible projects during the first stage. Researchers are available to aid in the best choice of vote aggregation methods. \citet{MP23} reason that data from real-world elections prove that consensus is thus reachable in practice, however, their approach remains unused to this day.
	
	\chapter{Empirical Analysis of Participatory Budgeting}
	
	So far, we only focused on the formal analysis that has been developed for PB. However, a growing share of the literature approaches PB from an empirical point of view. This includes \textit{e.g.} analysing real-life PB data, investigating the behaviours of PB rules on real-life and synthetic data, testing the relevance of some concepts in lab experiments.
	
	\section{Tools}
	
	Several tools have been developed in recent years to investigate PB concepts empirically. They are mainly concerned with numerical simulations.
	
	One of the most important tool is the website \href{http://pabulib.org/}{PaBuLib.org}, a library of real-life PB data \citep{SST20, FFP+23}. The website hosts a large number of PB elections encoded in the \texttt{.pb} format. This format is used to encode all the information related to a PB election: general information about the election, details of all the projects, and a description of the voters' ballots. All ballot formats are represented on the website, approval ballots being the most frequent.
	
	To work with PB data, one can use the \texttt{pabutools}, a Python library providing all the necessities to handle PB elections. It has initially been developed by Piotr Skowron and Grzegorz Pierczyński. Simon Rey and Markus Utke later on completely re-wrote it. The package is hosted on PyPI, so can be installed and used easily. (Almost) all of the PB rules described in this survey are implemented, together with a long list of analysis tools. It also includes a parser for \texttt{.pb} files, as well as a visualisation tool for election outcomes. Most examples and illustrations from the survey have been generated using this package for instance. We provide some useful links:

	\begin{itemize}
		\item The GitHub repository: \href{https://github.com/COMSOC-community/pabutools}{github.com/COMSOC-community/pabutools};
		\item The full documentation:  \href{https://COMSOC-community.github.io/pabutools/}{pbvoting.COMSOC-community.io/pabutools}.
	\end{itemize}
	
	

	
	
	
	
	
	
	
	

	\section{Comparing Voting Rule Perception}

	\citet{YHP+24} asked 180 student participants, after voting in a fictional PB setting, to give their opinion on two differing outcomes, one achieved with \greWelCost{} and the other with \mes{}. Towards the participants, the used rules were being anonymised, and they were then asked to mark their satisfaction and perceived fairness of each outcome, as well as the perceived trustworthiness of the voting rules. A clear correlation between the overall satisfaction achieved by the outcome and the perceived proportionality representation could be observed. \mes{} also achieved higher fairness ratings independent of a participant's personal utlity-based satisfaction. In the following, both rules were explained. While \mes{} was seen as more complex than \greWelCost{}, it was also viewed by participants as being more trustworthy and fairer. The authors therefore argue in favour of pre-hoc explanations. 
	
	\chapter{Beyond the Social Choice Take on Participatory Budgeting}
	\label{chap:Beyond}
	
	The focus of this survey, as its title suggests, is the (computational) social choice literature on PB. Nevertheless, some related topics are worth presenting. First, we briefly discuss some frameworks from the social choice literature that are related to PB (Section~\ref{sec:RelatedFrameworks}), and then we take a more general look at PB and how it is implemented in practice (Section~\ref{sec:PBPractice}).

	\section{Related Frameworks and Fields}
	\label{sec:RelatedFrameworks}
	
	In this section, we present several frameworks that relate to PB in some ways. We do not provide much detail about them but merely give pointers for the interested reader.
	
	\subsection{Multi-Winner Voting}
	
	The most obviously related framework, as we have mentioned several times already, is multi-winner voting. It is a special case of PB---where instances have unit costs and the budget allocation is required to be exhaustive---and has been extensively studied for many years, way before PB became a topic of interest. A recent book by \citet{LaSk23} presents a large part of that literature for approval ballots and provides many relevant references. A good starting point for multi-winner voting beyond approval ballots is the chapter by \citet{FSST17}. Other relevant pointers have already been included in the different sections above.
	
	\subsection{Collective Optimisation Problems}
	
	As we have observed, PB can be seen as a collective variant of the knapsack problem \citep[see, \textit{e.g.},][]{FSTW19}. The idea of looking at collective variants of optimisation problems is a growing field into which PB fits nicely \citep{BCGLN21}. Other optimisation problems for which their collective variants have been studied include finding spanning trees or scheduling jobs on machines \citep{DKP09, DKP11, BCELP16b, PRS18}.
	
	\subsection{Divisible Participatory Budgeting}
	
	Throughout this paper, we only focused on the case of indivisible PB where the projects are either fully funded or not at all. Relaxing this assumption by allowing projects to receive any amount of funding leads to the world of divisible PB. This framework has sometimes been called \emph{portioning} where a given public resource has to be shared among different divisible projects. Its study dates back to \citet{BMS05} and has since then received substantial attention. Perspectives that have been considered include welfare maximisation \citep{GKSA19, MPS20}, fairness guarantees \citep{FGM16, CCP22, AACKLP23} and strategic behaviours \citep{ABM19, FPPV21, BBPS21}. This setting is also closely related to \emph{probabilistic social choice} \citep{Bran18}.
	
	\subsection{Fair Allocation}
	
	Additionally, PB relates to the literature on fair allocation \citep{Roth15, BCELP16} and more specifically on the \emph{fair allocation of public goods} \citep{CFS17} where the allocated items can impact several agents (they are not privately owned as is assumed in the typical fair division literature). This framework can be seen as an unconstrained version of PB as there needs not be a budget constraint. Note that \citet{FMS18} consider the same model but with constraints on the outcome, though not necessarily budget ones.
	
	\section{PB in Practice}
	\label{sec:PBPractice}
	
	So far we have taken a very theoretical look at PB. However, theoretical analysis should be grounded in some observable facts. In this section we provide pointers to real-life PB processes for the interested researcher.
	
	\medskip

	For a more empirical analysis of PB, we can look to political scientists. The seminal paper on the topic is probably that of \citet{Caba04} who describes the first PB processes in Brazil. Subsequently, \citet{SHR08} analysed how PB was adapted from Brazil to Europe; \citet{Wamp12} analysed the core principles of PB; and several books have been published, presenting the relevant literature and the recent developments regarding PB \citep{Shah07, Oliv17, VNS21, WaGo22}. Finally, several books giving an overview over the different forms of PB processes around the world have been published in recent years, such as the books from \citet{Dias18}, \citet{DEJ19} or \citet{WNT21}.
	
	\medskip
	
	A lot of papers we have gone through also contain experimental studies they performed. Getting access to data about real-life PB processes is of critical importance here. Thankfully, the website \href{http://pabulib.org/}{PaBuLib.org} \citep{SST20} provides a lot of such data.
	
	\medskip
	
	Finally, Dominik Peters compiled a \href{https://en.wikipedia.org/wiki/List_of_participatory_budgeting_votes}{list}\footnote{In case the clickable link does not work: \href{https://en.wikipedia.org/wiki/List_of_participatory_budgeting_votes}{wikipedia.org/wiki/List\_of\_participatory\_budgeting\_votes}.} of PB instances that includes many interesting parameters for the social choice scientist: number of votes and projects, budget limit, ballot format, \textit{etc.} Let us provide some interesting facts from that list below. Plenty more are to be found out by going through the list. 
	\begin{itemize}
		\item Sometimes the budget is increased after the vote to afford more projects, occasionally by up to 250\% (Montreal 2021, Toulouse 2019, Gdynia 2021, Cambridge 2015-2021).
		\item Sometimes participation is incentivised by giving a bonus to districts with high turnout or to individual voters (Rome 2019, Gdynia 2016-2018, Kraków 2019).
		\item Sometimes there is a minimum requirement for projects to be selected, for instance, a project has to receive at least 200 ``points'' to be selected (Gdańsk 2021).
		\item Sometimes unused funds are transferred to the next year (Gdańsk 2014, 2018).
		\item Sometimes projects are partially funded by individual donors (Gdynia 2021).
	\end{itemize}
	
	\chapter{Conclusion}
	\label{chap:Conclusion}
	
	We have now presented almost the entire research that has been conducted by social choice scientist on the topic of PB. This line of research is still quite young and there are many things that can be explored further. We conclude this survey by presenting several directions we believe are worth exploring for PB.
	
	\begin{itemize}
		\item \textbf{Investigating more expressive ballots.} As we have seen, PB with approval ballots is the most studied framework for PB. It seems to us that these ballots are unfortunately not expressive enough for a framework in which alternatives have different cost. In particular, it is problematic that the meaning of not approving a project is unclear. One ballot format that has not received much attention but that we find appealing is the cumulative ballot format. It would also be interesting to initiate a study of PB processes where ``negative'' opinions can be submitted (with explicit disapproval for instance).
		
		\item \textbf{Extending the literature on fairness.} Even though the literature on fairness in PB is already quite extensive, there are still several interesting directions to pursue.
		\begin{itemize}
			\item Knowing whether the core of PB with approval ballots is always non-empty or not remains open for $\cardSatisfaction$. So far, it is known that for $\cardSatisfaction$ none of the established rules satisfies the core. 
			\item Among the proportionality axioms that we know can always be satisfied, FJR is one of the strongest. However, we don't know any \emph{natural} rule that satisfies it. Devising such a rule, therefore, is an important open problem.
			\item The existing analysis of the \emph{price of fairness} in PB \citep{FVMG22} is still at a preliminary stage and there is a lot of room for improvement.
			\item Linking to our first point, when considering approval ballots, it is particularly interesting to find results that apply to whole classes of satisfaction functions rather than a single one. This would mitigate the criticism that no satisfaction function on its own is fully convincing.
		\end{itemize} 
		
		\item \textbf{Deepening the axiomatic analysis.} The corpus of axioms that have been introduced in the literature on PB is still rather slim. We believe that there is a crucial need to develop that side of the literature to have other means to compare rules than fairness guarantees. Investigating how to adapt the characterisation results from the multi-winner voting literature \citep{SFS19, LaSk21} could be an interesting starting point (though potentially rather technical).
		
		\item \textbf{Devising a theory of explainable PB.} Taking the risk to be called trend-followers, we believe that there is room for explaining outcomes of voting procedures to the citizen. PB makes no exception here. The outcome of PB rules could be explained in a counterfactual fashion: ``How much cheaper should the project have been to be selected? How many more supporters?...'' A theory of explainable PB could also take the form of more principled, axiomatically-guided, approaches \citep[see, \textit{e.g.},][]{Proc19, BoEn20}.
		
		\item \textbf{Developing a Python library.} Several papers presents simulation results using similar approaches, rules, \textit{etc.}, meaning that a large set of authors must have the exact same code somewhere on their laptop. Some of this code is available, see e.g., \href{https://github.com/Grzesiek3713/pabutools}{Pabutools}. It would be nice to gather that in a unified Python package \citep[inspired by the \texttt{abcvoting} package][]{abcvoting,LRK23}. This would make the results based on simulations more reliable as they would be less susceptible to having errors in the code. It would also help with reproducibility of said results. Finally, it would be a great tool for the adoption of newly developed PB rules: verified and approved code would be available.
		
		\item \textbf{Looking beyond the voting stage.} As we mentioned in the introduction, PB is usually a longer process that has several steps. Despite this, the social choice literature on PB has almost exclusively focused on the voting stage, with the end-to-end model of \citet{REH21} being the only exception. We believe that there are many interesting questions which can be answered with social choice methods that arise from a more holistic view of the PB process. This includes, for example, which incentives voters have when planning a project that they want to propose, \textit{i.e.}, whether it is beneficial to make a project as cheap as possible or to merge two similar projects.
		
		\item \textbf{Stepping outside the Western world.} It is worth pointing out that the (computational) social choice literature so far has almost exclusively used PB processes in the Western world as examples. However, there is a large diversity in the actual implementation of PB around the world. For example in Western countries, generally only a small percentage of a municipalities budget is allocated to PB and most projects funded through PB are small ``quality of life'' improvements that are not essential to the functioning of the city. In contrast, for example in early implementations of PB in Brazil, significant parts of a city's budget was spend through PB and many projects funded through PB addressed crucial parts of life, such as access to basic health care \citep{Caba04}. For a systematic analysis of the differences between PB processes in different parts of the world, we refer to the book by \citet{WNT21}. Taking a more global perspective on PB could open interesting new research directions.
	\end{itemize}

	\bibliographystyle{ACM-Reference-Format}
	\bibliography{PB}


\begin{thebibliography}{127}


\ifx \showCODEN    \undefined \def \showCODEN     #1{\unskip}     \fi
\ifx \showDOI      \undefined \def \showDOI       #1{#1}\fi
\ifx \showISBNx    \undefined \def \showISBNx     #1{\unskip}     \fi
\ifx \showISBNxiii \undefined \def \showISBNxiii  #1{\unskip}     \fi
\ifx \showISSN     \undefined \def \showISSN      #1{\unskip}     \fi
\ifx \showLCCN     \undefined \def \showLCCN      #1{\unskip}     \fi
\ifx \shownote     \undefined \def \shownote      #1{#1}          \fi
\ifx \showarticletitle \undefined \def \showarticletitle #1{#1}   \fi
\ifx \showURL      \undefined \def \showURL       {\relax}        \fi
\providecommand\bibfield[2]{#2}
\providecommand\bibinfo[2]{#2}
\providecommand\natexlab[1]{#1}
\providecommand\showeprint[2][]{arXiv:#2}

\bibitem[\protect\citeauthoryear{Abers}{Abers}{2000}]%
        {Aber00}
\bibfield{author}{\bibinfo{person}{Rebecca Abers}.}
  \bibinfo{year}{2000}\natexlab{}.
\newblock \bibinfo{booktitle}{\emph{Inventing Local Democracy: Grassroots
  Politics in {B}razil}}.
\newblock \bibinfo{publisher}{Lynne Rienner Publishers}.
\newblock


\bibitem[\protect\citeauthoryear{Airiau, Aziz, Caragiannis, Kruger, Lang, and
  Peters}{Airiau, Aziz, Caragiannis, Kruger, Lang and Peters}{2023}]%
        {AACKLP23}
\bibfield{author}{\bibinfo{person}{St{\'e}phane Airiau}, \bibinfo{person}{Haris
  Aziz}, \bibinfo{person}{Ioannis Caragiannis}, \bibinfo{person}{Justin
  Kruger}, \bibinfo{person}{J{\'e}r{\^o}me Lang}, {and}
  \bibinfo{person}{Dominik Peters}.} \bibinfo{year}{2023}\natexlab{}.
\newblock \showarticletitle{Portioning Using Ordinal Preferences: {F}airness
  and Efficiency}.
\newblock \bibinfo{journal}{\emph{Artificial Intelligence}}
  \bibinfo{volume}{314} (\bibinfo{year}{2023}), \bibinfo{pages}{103809}.
\newblock


\bibitem[\protect\citeauthoryear{Arrow, Sen, and Suzumura}{Arrow, Sen and
  Suzumura}{2002}]%
        {ASS02}
\bibfield{editor}{\bibinfo{person}{Kenneth~J. Arrow}, \bibinfo{person}{Amartya
  Sen}, {and} \bibinfo{person}{Kotaro Suzumura}} (Eds.).
  \bibinfo{year}{2002}\natexlab{}.
\newblock \bibinfo{booktitle}{\emph{Handbook of Social Choice and Welfare}}.
  Vol.~\bibinfo{volume}{1}.
\newblock \bibinfo{publisher}{North-Holland}.
\newblock


\bibitem[\protect\citeauthoryear{Arrow, Sen, and Suzumura}{Arrow, Sen and
  Suzumura}{2011}]%
        {ASS11}
\bibfield{editor}{\bibinfo{person}{Kenneth~J. Arrow}, \bibinfo{person}{Amartya
  Sen}, {and} \bibinfo{person}{Kotaro Suzumura}} (Eds.).
  \bibinfo{year}{2011}\natexlab{}.
\newblock \bibinfo{booktitle}{\emph{Handbook of Social Choice and Welfare}}.
  Vol.~\bibinfo{volume}{2}.
\newblock \bibinfo{publisher}{North-Holland}.
\newblock


\bibitem[\protect\citeauthoryear{Aziz, Bogomolnaia, and Moulin}{Aziz,
  Bogomolnaia and Moulin}{2019}]%
        {ABM19}
\bibfield{author}{\bibinfo{person}{Haris Aziz}, \bibinfo{person}{Anna
  Bogomolnaia}, {and} \bibinfo{person}{Herv{\'e} Moulin}.}
  \bibinfo{year}{2019}\natexlab{}.
\newblock \showarticletitle{Fair Mixing: the Case of Dichotomous Preferences}.
  In \bibinfo{booktitle}{\emph{Proceedings of the 20th ACM Conference on
  Economics and Computation (ACM-EC)}}. \bibinfo{pages}{753--781}.
\newblock


\bibitem[\protect\citeauthoryear{Aziz, Brill, Conitzer, Elkind, Freeman, and
  Walsh}{Aziz, Brill, Conitzer, Elkind, Freeman and Walsh}{2017}]%
        {ABCEFW17}
\bibfield{author}{\bibinfo{person}{Haris Aziz}, \bibinfo{person}{Markus Brill},
  \bibinfo{person}{Vincent Conitzer}, \bibinfo{person}{Edith Elkind},
  \bibinfo{person}{Rupert Freeman}, {and} \bibinfo{person}{Toby Walsh}.}
  \bibinfo{year}{2017}\natexlab{}.
\newblock \showarticletitle{Justified Representation in Approval-Based
  Committee Voting}.
\newblock \bibinfo{journal}{\emph{Social Choice and Welfare}}
  \bibinfo{volume}{48}, \bibinfo{number}{2} (\bibinfo{year}{2017}),
  \bibinfo{pages}{461--485}.
\newblock


\bibitem[\protect\citeauthoryear{Aziz, Elkind, Huang, Lackner,
  S{\'a}nchez-Fern{\'a}ndez, and Skowron}{Aziz, Elkind, Huang, Lackner,
  S{\'a}nchez-Fern{\'a}ndez and Skowron}{2018}]%
        {AEHLSS18}
\bibfield{author}{\bibinfo{person}{Haris Aziz}, \bibinfo{person}{Edith Elkind},
  \bibinfo{person}{Shenwei Huang}, \bibinfo{person}{Martin Lackner},
  \bibinfo{person}{Luis S{\'a}nchez-Fern{\'a}ndez}, {and}
  \bibinfo{person}{Piotr Skowron}.} \bibinfo{year}{2018}\natexlab{}.
\newblock \showarticletitle{On the Complexity of Extended and Proportional
  Justified Representation}. In \bibinfo{booktitle}{\emph{Proceedings of the
  32rd AAAI Conference on Artificial Intelligence (AAAI)}}.
\newblock


\bibitem[\protect\citeauthoryear{Aziz and Ganguly}{Aziz and Ganguly}{2021}]%
        {AzGa21}
\bibfield{author}{\bibinfo{person}{Haris Aziz} {and} \bibinfo{person}{Aditya
  Ganguly}.} \bibinfo{year}{2021}\natexlab{}.
\newblock \showarticletitle{Participatory Funding Coordination: Model, Axioms
  and Rules}. In \bibinfo{booktitle}{\emph{Proceedings of the 7th International
  Conference on Algorithmic Decision Theory (ADT)}}.
\newblock


\bibitem[\protect\citeauthoryear{Aziz, Gujar, Padala, Suzuki, and Vollen}{Aziz,
  Gujar, Padala, Suzuki and Vollen}{2022}]%
        {AGPSV22}
\bibfield{author}{\bibinfo{person}{Haris Aziz}, \bibinfo{person}{Sujit Gujar},
  \bibinfo{person}{Manisha Padala}, \bibinfo{person}{Mashbat Suzuki}, {and}
  \bibinfo{person}{Jeremy Vollen}.} \bibinfo{year}{2022}\natexlab{}.
\newblock \showarticletitle{Coordinating Monetary Contributions in
  Participatory Budgeting}.
\newblock \bibinfo{journal}{\emph{arXiv preprint arXiv:2206.05966}}
  (\bibinfo{year}{2022}).
\newblock


\bibitem[\protect\citeauthoryear{Aziz and Lee}{Aziz and Lee}{2021}]%
        {AzLe21}
\bibfield{author}{\bibinfo{person}{Haris Aziz} {and} \bibinfo{person}{Barton~E.
  Lee}.} \bibinfo{year}{2021}\natexlab{}.
\newblock \showarticletitle{Proportionally Representative Participatory
  Budgeting with Ordinal Preferences}. In \bibinfo{booktitle}{\emph{Proceedings
  of the 35th AAAI Conference on Artificial Intelligence (AAAI)}}.
\newblock


\bibitem[\protect\citeauthoryear{Aziz, Lee, and Talmon}{Aziz, Lee and
  Talmon}{2018}]%
        {ALT18}
\bibfield{author}{\bibinfo{person}{Haris Aziz}, \bibinfo{person}{Barton~E.
  Lee}, {and} \bibinfo{person}{Nimrod Talmon}.}
  \bibinfo{year}{2018}\natexlab{}.
\newblock \showarticletitle{Proportionally Representative Participatory
  Budgeting: Axioms and Algorithms}. In \bibinfo{booktitle}{\emph{Proceedings
  of the 17th International Conference on Autonomous Agents and Multiagent
  Systems (AAMAS)}}.
\newblock


\bibitem[\protect\citeauthoryear{Aziz, Lu, Suzuki, Vollen, and Walsh}{Aziz, Lu,
  Suzuki, Vollen and Walsh}{2024}]%
        {ALS+24}
\bibfield{author}{\bibinfo{person}{Haris Aziz}, \bibinfo{person}{Xinhang Lu},
  \bibinfo{person}{Mashbat Suzuki}, \bibinfo{person}{Jeremy Vollen}, {and}
  \bibinfo{person}{Toby Walsh}.} \bibinfo{year}{2024}\natexlab{}.
\newblock \bibinfo{title}{Fair Lotteries for Participatory Budgeting}.
\newblock
\newblock
\urldef\tempurl%
\url{https://arxiv.org/abs/2404.05198}
\showURL{%
\tempurl}


\bibitem[\protect\citeauthoryear{Aziz and Shah}{Aziz and Shah}{2021}]%
        {AzSh21}
\bibfield{author}{\bibinfo{person}{Haris Aziz} {and} \bibinfo{person}{Nisarg
  Shah}.} \bibinfo{year}{2021}\natexlab{}.
\newblock \showarticletitle{Participatory Budgeting: Models and Approaches}.
\newblock In \bibinfo{booktitle}{\emph{Pathways between Social Science and
  Computational Social Science: Theories, Methods and Interpretations}}.
  \bibinfo{publisher}{Springer-Verlag}.
\newblock


\bibitem[\protect\citeauthoryear{Baumeister, Boes, and Hillebrand}{Baumeister,
  Boes and Hillebrand}{2021}]%
        {BBH21}
\bibfield{author}{\bibinfo{person}{Dorothea Baumeister}, \bibinfo{person}{Linus
  Boes}, {and} \bibinfo{person}{Johanna Hillebrand}.}
  \bibinfo{year}{2021}\natexlab{}.
\newblock \showarticletitle{Complexity of Manipulative Interference in
  Participatory Budgeting}. In \bibinfo{booktitle}{\emph{Proceedings of the 7th
  International Conference on Algorithmic Decision Theory (ADT)}}.
  \bibinfo{pages}{424--439}.
\newblock


\bibitem[\protect\citeauthoryear{Baumeister, Boes, and
  Lau{\ss}mann}{Baumeister, Boes and Lau{\ss}mann}{2022}]%
        {BBL22}
\bibfield{author}{\bibinfo{person}{Dorothea Baumeister}, \bibinfo{person}{Linus
  Boes}, {and} \bibinfo{person}{Christian Lau{\ss}mann}.}
  \bibinfo{year}{2022}\natexlab{}.
\newblock \showarticletitle{Time-Constrained Participatory Budgeting Under
  Uncertain Project Costs}. In \bibinfo{booktitle}{\emph{Proceedings of the
  31st International Joint Conference on Artificial Intelligence (IJCAI)}}.
\newblock


\bibitem[\protect\citeauthoryear{Baumeister, Boes, and Seeger}{Baumeister, Boes
  and Seeger}{2020}]%
        {BBS20}
\bibfield{author}{\bibinfo{person}{Dorothea Baumeister}, \bibinfo{person}{Linus
  Boes}, {and} \bibinfo{person}{Tessa Seeger}.}
  \bibinfo{year}{2020}\natexlab{}.
\newblock \showarticletitle{Irresolute Approval-based Budgeting}. In
  \bibinfo{booktitle}{\emph{Proceedings of the 19th International Conference on
  Autonomous Agents and Multiagent Systems (AAMAS)}}.
  \bibinfo{pages}{1774--1776}.
\newblock


\bibitem[\protect\citeauthoryear{Baumeister, Faliszewski, Lang, and
  Rothe}{Baumeister, Faliszewski, Lang and Rothe}{2012}]%
        {BFLR12}
\bibfield{author}{\bibinfo{person}{Dorothea Baumeister}, \bibinfo{person}{Piotr
  Faliszewski}, \bibinfo{person}{J{\'{e}}r{\^{o}}me Lang}, {and}
  \bibinfo{person}{J{\"{o}}rg Rothe}.} \bibinfo{year}{2012}\natexlab{}.
\newblock \showarticletitle{Campaigns for lazy voters: truncated ballots}. In
  \bibinfo{booktitle}{\emph{Proceedings of the 11th International Conference on
  Autonomous Agents and Multiagent Systems (AAMAS)}}.
  \bibinfo{pages}{577--584}.
\newblock


\bibitem[\protect\citeauthoryear{Beikverdi, Tehrani, and Shahanaghi}{Beikverdi,
  Tehrani and Shahanaghi}{2024}]%
        {BTS24}
\bibfield{author}{\bibinfo{person}{Majid Beikverdi},
  \bibinfo{person}{Nasim~Ghanbar Tehrani}, {and} \bibinfo{person}{Kamran
  Shahanaghi}.} \bibinfo{year}{2024}\natexlab{}.
\newblock \showarticletitle{A Bi-level model for district-fairness
  participatory budgeting: Decomposition methods and application}.
\newblock \bibinfo{journal}{\emph{Eur. J. Oper. Res.}} (\bibinfo{year}{2024}).
\newblock


\bibitem[\protect\citeauthoryear{Benadè, Nath, Procaccia, and Shah}{Benadè,
  Nath, Procaccia and Shah}{2021}]%
        {BNPS21}
\bibfield{author}{\bibinfo{person}{Gerdus Benadè}, \bibinfo{person}{Swaprava
  Nath}, \bibinfo{person}{Ariel~D. Procaccia}, {and} \bibinfo{person}{Nisarg
  Shah}.} \bibinfo{year}{2021}\natexlab{}.
\newblock \showarticletitle{Preference Elicitation for Participatory
  Budgeting}.
\newblock \bibinfo{journal}{\emph{Management Science}} \bibinfo{volume}{67},
  \bibinfo{number}{5} (\bibinfo{year}{2021}), \bibinfo{pages}{2813--2827}.
\newblock


\bibitem[\protect\citeauthoryear{Bendor, Diermeier, Siegel, and Ting}{Bendor,
  Diermeier, Siegel and Ting}{2011}]%
        {BDST11}
\bibfield{author}{\bibinfo{person}{Jonathan Bendor}, \bibinfo{person}{Daniel
  Diermeier}, \bibinfo{person}{David~A Siegel}, {and} \bibinfo{person}{Michael
  Ting}.} \bibinfo{year}{2011}\natexlab{}.
\newblock \bibinfo{booktitle}{\emph{A Behavioral Theory of Elections}}.
\newblock \bibinfo{publisher}{Princeton University Press}.
\newblock


\bibitem[\protect\citeauthoryear{Blum, Cucker, Shub, and Smale}{Blum, Cucker,
  Shub and Smale}{1998}]%
        {BCSS98}
\bibfield{author}{\bibinfo{person}{Lenore Blum}, \bibinfo{person}{Felipe
  Cucker}, \bibinfo{person}{Michael Shub}, {and} \bibinfo{person}{Steve
  Smale}.} \bibinfo{year}{1998}\natexlab{}.
\newblock \bibinfo{booktitle}{\emph{Complexity and Real Computation}}.
\newblock \bibinfo{publisher}{Springer-Verlag}.
\newblock


\bibitem[\protect\citeauthoryear{Boehmer, Faliszewski, Janeczko, and
  Kaczmarczyk}{Boehmer, Faliszewski, Janeczko and Kaczmarczyk}{2023}]%
        {BFJK23}
\bibfield{author}{\bibinfo{person}{Niclas Boehmer}, \bibinfo{person}{Piotr
  Faliszewski}, \bibinfo{person}{Lukasz Janeczko}, {and}
  \bibinfo{person}{Andrzej Kaczmarczyk}.} \bibinfo{year}{2023}\natexlab{}.
\newblock \showarticletitle{Robustness of Participatory Budgeting Outcomes:
  Complexity and Experiments}. In \bibinfo{booktitle}{\emph{Proceedings of the
  16th International Symposium on Algorithmic Game Theory (SAGT)}}.
  \bibinfo{pages}{161--178}.
\newblock


\bibitem[\protect\citeauthoryear{Boes, Colley, Grandi, Lang, and Novaro}{Boes,
  Colley, Grandi, Lang and Novaro}{2021}]%
        {BCGLN21}
\bibfield{author}{\bibinfo{person}{Linus Boes}, \bibinfo{person}{Rachael
  Colley}, \bibinfo{person}{Umberto Grandi}, \bibinfo{person}{Jerome Lang},
  {and} \bibinfo{person}{Arianna Novaro}.} \bibinfo{year}{2021}\natexlab{}.
\newblock \showarticletitle{Collective Discrete Optimisation as Judgment
  Aggregation}.
\newblock \bibinfo{journal}{\emph{arXiv preprint arXiv:2112.00574}}
  (\bibinfo{year}{2021}).
\newblock


\bibitem[\protect\citeauthoryear{Bogomolnaia, Moulin, and Stong}{Bogomolnaia,
  Moulin and Stong}{2005}]%
        {BMS05}
\bibfield{author}{\bibinfo{person}{Anna Bogomolnaia},
  \bibinfo{person}{Herv{\'e} Moulin}, {and} \bibinfo{person}{Richard Stong}.}
  \bibinfo{year}{2005}\natexlab{}.
\newblock \showarticletitle{Collective Choice Under Dichotomous Preferences}.
\newblock \bibinfo{journal}{\emph{Journal of Economic Theory}}
  \bibinfo{volume}{122}, \bibinfo{number}{2} (\bibinfo{year}{2005}),
  \bibinfo{pages}{165--184}.
\newblock


\bibitem[\protect\citeauthoryear{Boixel and Endriss}{Boixel and
  Endriss}{2020}]%
        {BoEn20}
\bibfield{author}{\bibinfo{person}{Arthur Boixel} {and} \bibinfo{person}{Ulle
  Endriss}.} \bibinfo{year}{2020}\natexlab{}.
\newblock \showarticletitle{Automated Justification of Collective Decisions via
  Constraint Solving}. In \bibinfo{booktitle}{\emph{Proceedings of the 19th
  International Conference on Autonomous Agents and Multiagent Systems
  (AAMAS)}}. \bibinfo{pages}{168--176}.
\newblock


\bibitem[\protect\citeauthoryear{Brandl, Brandt, Peters, and Stricker}{Brandl,
  Brandt, Peters and Stricker}{2021}]%
        {BBPS21}
\bibfield{author}{\bibinfo{person}{Florian Brandl}, \bibinfo{person}{Felix
  Brandt}, \bibinfo{person}{Dominik Peters}, {and} \bibinfo{person}{Christian
  Stricker}.} \bibinfo{year}{2021}\natexlab{}.
\newblock \showarticletitle{Distribution Rules Under Dichotomous Preferences:
  Two Out of Three Ain't Bad}. In \bibinfo{booktitle}{\emph{Proceedings of the
  22nd ACM Conference on Economics and Computation (ACM-EC)}}.
  \bibinfo{pages}{158--179}.
\newblock


\bibitem[\protect\citeauthoryear{Brandt}{Brandt}{2018}]%
        {Bran18}
\bibfield{author}{\bibinfo{person}{Felix Brandt}.}
  \bibinfo{year}{2018}\natexlab{}.
\newblock \showarticletitle{Collective Choice Lotteries: {D}ealing with
  Randomization in Economic Design}.
\newblock In \bibinfo{booktitle}{\emph{The Future of Economic Design}},
  \bibfield{editor}{\bibinfo{person}{Jean-François Laslier},
  \bibinfo{person}{Hervé Moulin}, \bibinfo{person}{Remzi Sanver}, {and}
  \bibinfo{person}{William~S. Zwicker}} (Eds.).
  \bibinfo{publisher}{Springer-Verlag}.
\newblock


\bibitem[\protect\citeauthoryear{Brandt, Conitzer, Endriss, Lang, and
  Procaccia}{Brandt, Conitzer, Endriss, Lang and Procaccia}{2016a}]%
        {BCELP16}
\bibfield{editor}{\bibinfo{person}{Felix Brandt}, \bibinfo{person}{Vincent
  Conitzer}, \bibinfo{person}{Ulle Endriss}, \bibinfo{person}{J{\'e}r{\^o}me
  Lang}, {and} \bibinfo{person}{Ariel~D. Procaccia}} (Eds.).
  \bibinfo{year}{2016}\natexlab{a}.
\newblock \bibinfo{booktitle}{\emph{Handbook of Computational Social Choice}}.
\newblock \bibinfo{publisher}{Cambridge University Press}.
\newblock


\bibitem[\protect\citeauthoryear{Brandt, Conitzer, Endriss, Lang, and
  Procaccia}{Brandt, Conitzer, Endriss, Lang and Procaccia}{2016b}]%
        {BCELP16b}
\bibfield{author}{\bibinfo{person}{Felix Brandt}, \bibinfo{person}{Vincent
  Conitzer}, \bibinfo{person}{Ulle Endriss}, \bibinfo{person}{Jérôme Lang},
  {and} \bibinfo{person}{Ariel~D. Procaccia}.}
  \bibinfo{year}{2016}\natexlab{b}.
\newblock \showarticletitle{Introduction to Computational Social Choice}.
\newblock In \bibinfo{booktitle}{\emph{Handbook of Computational Social
  Choice}}, \bibfield{editor}{\bibinfo{person}{Felix Brandt},
  \bibinfo{person}{Vincent Conitzer}, \bibinfo{person}{Ulle Endriss},
  \bibinfo{person}{Jérôme Lang}, {and} \bibinfo{person}{Ariel~D. Procaccia}}
  (Eds.). \bibinfo{publisher}{Cambridge University Press}, Chapter~1.
\newblock


\bibitem[\protect\citeauthoryear{Bredereck, Faliszewski, Kaczmarczyk, and
  Niedermeier}{Bredereck, Faliszewski, Kaczmarczyk and Niedermeier}{2019}]%
        {BFKN19}
\bibfield{author}{\bibinfo{person}{Robert Bredereck}, \bibinfo{person}{Piotr
  Faliszewski}, \bibinfo{person}{Andrzej Kaczmarczyk}, {and}
  \bibinfo{person}{Rolf Niedermeier}.} \bibinfo{year}{2019}\natexlab{}.
\newblock \showarticletitle{An Experimental View on Committees Providing
  Justified Representation}. In \bibinfo{booktitle}{\emph{Proceedings of the
  28th International Joint Conference on Artificial Intelligence (IJCAI)}}.
  \bibinfo{pages}{109--115}.
\newblock


\bibitem[\protect\citeauthoryear{Brill, Forster, Lackner, Maly, and
  Peters}{Brill, Forster, Lackner, Maly and Peters}{2023}]%
        {BFLMP23}
\bibfield{author}{\bibinfo{person}{Markus Brill}, \bibinfo{person}{Stefan
  Forster}, \bibinfo{person}{Martin Lackner}, \bibinfo{person}{Jan Maly}, {and}
  \bibinfo{person}{Jannik Peters}.} \bibinfo{year}{2023}\natexlab{}.
\newblock \showarticletitle{Proportionality in Approval-Based Participatory
  Budgeting}. In \bibinfo{booktitle}{\emph{Proceedings of the 37th AAAI
  Conference on Artificial Intelligence (AAAI)}}.
\newblock


\bibitem[\protect\citeauthoryear{Brill, Freeman, Janson, and Lackner}{Brill,
  Freeman, Janson and Lackner}{2017}]%
        {BFJL17}
\bibfield{author}{\bibinfo{person}{Markus Brill}, \bibinfo{person}{Rupert
  Freeman}, \bibinfo{person}{Svante Janson}, {and} \bibinfo{person}{Martin
  Lackner}.} \bibinfo{year}{2017}\natexlab{}.
\newblock \showarticletitle{Phragm{\'e}n's Voting Methods and Justified
  Representation}. In \bibinfo{booktitle}{\emph{Proceedings of the 31st AAAI
  Conference on Artificial Intelligence (AAAI)}}. \bibinfo{pages}{406--413}.
\newblock


\bibitem[\protect\citeauthoryear{Brill and Peters}{Brill and Peters}{2023}]%
        {BP23}
\bibfield{author}{\bibinfo{person}{Markus Brill} {and} \bibinfo{person}{Jannik
  Peters}.} \bibinfo{year}{2023}\natexlab{}.
\newblock \showarticletitle{Robust and Verifiable Proportionality Axioms for
  Multiwinner Voting}.
\newblock \bibinfo{journal}{\emph{CoRR}}  \bibinfo{volume}{abs/2302.01989}
  (\bibinfo{year}{2023}).
\newblock


\bibitem[\protect\citeauthoryear{Cabannes}{Cabannes}{2004}]%
        {Caba04}
\bibfield{author}{\bibinfo{person}{Yves Cabannes}.}
  \bibinfo{year}{2004}\natexlab{}.
\newblock \showarticletitle{Participatory budgeting: A significant contribution
  to participatory democracy}.
\newblock \bibinfo{journal}{\emph{Environment and Urbanization}}
  \bibinfo{volume}{16}, \bibinfo{number}{1} (\bibinfo{year}{2004}),
  \bibinfo{pages}{27--46}.
\newblock


\bibitem[\protect\citeauthoryear{Caragiannis, Christodoulou, and
  Protopapas}{Caragiannis, Christodoulou and Protopapas}{2022}]%
        {CCP22}
\bibfield{author}{\bibinfo{person}{Ioannis Caragiannis},
  \bibinfo{person}{George Christodoulou}, {and} \bibinfo{person}{Nicos
  Protopapas}.} \bibinfo{year}{2022}\natexlab{}.
\newblock \showarticletitle{Truthful Aggregation of Budget Proposals with
  Proportionality Guarantees}. In \bibinfo{booktitle}{\emph{Proceedings of the
  36th AAAI Conference on Artificial Intelligence (AAAI)}}.
  \bibinfo{pages}{4917--4924}.
\newblock


\bibitem[\protect\citeauthoryear{Ceron, Gonzalez, and Navarro-Ramos}{Ceron,
  Gonzalez and Navarro-Ramos}{2022}]%
        {CGN22}
\bibfield{author}{\bibinfo{person}{Federica Ceron}, \bibinfo{person}{Stéphane
  Gonzalez}, {and} \bibinfo{person}{Adriana Navarro-Ramos}.}
  \bibinfo{year}{2022}\natexlab{}.
\newblock \showarticletitle{Axiomatic Characterizations of the Knapsack and
  Greedy Participatory Budgeting Methods}.
\newblock \bibinfo{journal}{\emph{Working paper}} (\bibinfo{year}{2022}).
\newblock


\bibitem[\protect\citeauthoryear{Cevallos and Stewart}{Cevallos and
  Stewart}{2021}]%
        {CeSt21}
\bibfield{author}{\bibinfo{person}{Alfonso Cevallos} {and}
  \bibinfo{person}{Alistair Stewart}.} \bibinfo{year}{2021}\natexlab{}.
\newblock \showarticletitle{A Verifiably Secure and Proportional Committee
  Election Rule}. In \bibinfo{booktitle}{\emph{Proceedings of the 3rd ACM
  Conference on Advances in Financial Technologies (AFT)}}.
  \bibinfo{pages}{29--42}.
\newblock


\bibitem[\protect\citeauthoryear{Chen, Lackner, and Maly}{Chen, Lackner and
  Maly}{2022}]%
        {CLM22}
\bibfield{author}{\bibinfo{person}{Jiehua Chen}, \bibinfo{person}{Martin
  Lackner}, {and} \bibinfo{person}{Jan Maly}.} \bibinfo{year}{2022}\natexlab{}.
\newblock \showarticletitle{Participatory Budgeting with Donations and
  Diversity Constraints}. In \bibinfo{booktitle}{\emph{Proceedings of the 36th
  AAAI Conference on Artificial Intelligence (AAAI)}}.
  \bibinfo{pages}{9323--9330}.
\newblock


\bibitem[\protect\citeauthoryear{Cheng, Jiang, Munagala, and Wang}{Cheng,
  Jiang, Munagala and Wang}{2020}]%
        {CJMW20}
\bibfield{author}{\bibinfo{person}{Yu Cheng}, \bibinfo{person}{Zhihao Jiang},
  \bibinfo{person}{Kamesh Munagala}, {and} \bibinfo{person}{Kangning Wang}.}
  \bibinfo{year}{2020}\natexlab{}.
\newblock \showarticletitle{Group Fairness in Committee Selection}.
\newblock \bibinfo{journal}{\emph{ACM Transactions on Economics and Computation
  (TEAC)}} \bibinfo{volume}{8}, \bibinfo{number}{4} (\bibinfo{year}{2020}),
  \bibinfo{pages}{1--18}.
\newblock


\bibitem[\protect\citeauthoryear{{City of Amsterdam}}{{City of
  Amsterdam}}{2022}]%
        {amspb}
\bibfield{author}{\bibinfo{person}{{City of Amsterdam}}.}
  \bibinfo{year}{2022}\natexlab{}.
\newblock \bibinfo{title}{Oost Begroot}.
\newblock
  \bibinfo{howpublished}{\href{https://www.amsterdam.nl/stadsdelen/oost/oost-begroot/}{https://www.amsterdam.nl/stadsdelen/oost/oost-begroot}}.
\newblock
\newblock
\shownote{Last accessed on March the 1st 2023.}


\bibitem[\protect\citeauthoryear{{City of Paris}}{{City of Paris}}{2020}]%
        {parispb}
\bibfield{author}{\bibinfo{person}{{City of Paris}}.}
  \bibinfo{year}{2020}\natexlab{}.
\newblock \bibinfo{title}{{Paris Budget Participatif}}.
\newblock
  \bibinfo{howpublished}{\href{https://decider.paris.fr/decider/jsp/site/Portal.jsp}{https://decider.paris.fr/decider/jsp/site/Portal.jsp}}.
\newblock
\newblock
\shownote{Last accessed on March the 1st 2023.}


\bibitem[\protect\citeauthoryear{Conitzer, Freeman, and Shah}{Conitzer, Freeman
  and Shah}{2017}]%
        {CFS17}
\bibfield{author}{\bibinfo{person}{Vincent Conitzer}, \bibinfo{person}{Rupert
  Freeman}, {and} \bibinfo{person}{Nisarg Shah}.}
  \bibinfo{year}{2017}\natexlab{}.
\newblock \showarticletitle{Fair Public Decision Making}. In
  \bibinfo{booktitle}{\emph{Proceedings of the 18th ACM Conference on Economics
  and Computation (ACM-EC)}}. \bibinfo{pages}{629--646}.
\newblock


\bibitem[\protect\citeauthoryear{Darmann, Klamler, and Pferschy}{Darmann,
  Klamler and Pferschy}{2009}]%
        {DKP09}
\bibfield{author}{\bibinfo{person}{Andreas Darmann}, \bibinfo{person}{Christian
  Klamler}, {and} \bibinfo{person}{Ulrich Pferschy}.}
  \bibinfo{year}{2009}\natexlab{}.
\newblock \showarticletitle{Maximizing the Minimum Voter Satisfaction on
  Spanning Trees}.
\newblock \bibinfo{journal}{\emph{Mathematical Social Sciences}}
  \bibinfo{volume}{58}, \bibinfo{number}{2} (\bibinfo{year}{2009}),
  \bibinfo{pages}{238--250}.
\newblock


\bibitem[\protect\citeauthoryear{Darmann, Klamler, and Pferschy}{Darmann,
  Klamler and Pferschy}{2011}]%
        {DKP11}
\bibfield{author}{\bibinfo{person}{Andreas Darmann}, \bibinfo{person}{Christian
  Klamler}, {and} \bibinfo{person}{Ulrich Pferschy}.}
  \bibinfo{year}{2011}\natexlab{}.
\newblock \showarticletitle{Finding Socially Best Spanning Trees}.
\newblock \bibinfo{journal}{\emph{Theory and Decision}} \bibinfo{volume}{70},
  \bibinfo{number}{4} (\bibinfo{year}{2011}), \bibinfo{pages}{511--527}.
\newblock


\bibitem[\protect\citeauthoryear{De~Oliveira}{De~Oliveira}{2017}]%
        {Oliv17}
\bibfield{author}{\bibinfo{person}{Osmany~Porto De~Oliveira}.}
  \bibinfo{year}{2017}\natexlab{}.
\newblock \bibinfo{booktitle}{\emph{International Policy Diffusion and
  Participatory Budgeting: Ambassadors of Participation, International
  Institutions and Transnational Networks}}.
\newblock \bibinfo{publisher}{Springer-Verlag}.
\newblock


\bibitem[\protect\citeauthoryear{De~Vries, Nemec, and
  {\v{S}}pa{\v{c}}ek}{De~Vries, Nemec and {\v{S}}pa{\v{c}}ek}{2021}]%
        {VNS21}
\bibfield{author}{\bibinfo{person}{Michiel~S. De~Vries}, \bibinfo{person}{Juraj
  Nemec}, {and} \bibinfo{person}{David {\v{S}}pa{\v{c}}ek}.}
  \bibinfo{year}{2021}\natexlab{}.
\newblock \bibinfo{booktitle}{\emph{International Trends in Participatory
  Budgeting: Between Trivial Pursuits and Best Practices}}.
\newblock \bibinfo{publisher}{Springer-Verlag}.
\newblock


\bibitem[\protect\citeauthoryear{Dhillon and Peralta}{Dhillon and
  Peralta}{2002}]%
        {Dhillon02}
\bibfield{author}{\bibinfo{person}{Amrita Dhillon} {and}
  \bibinfo{person}{Susana Peralta}.} \bibinfo{year}{2002}\natexlab{}.
\newblock \showarticletitle{Economic Theories of Voter Turnout}.
\newblock \bibinfo{journal}{\emph{The Economic Journal}} \bibinfo{volume}{112},
  \bibinfo{number}{480} (\bibinfo{year}{2002}), \bibinfo{pages}{F332--F352}.
\newblock


\bibitem[\protect\citeauthoryear{Dias}{Dias}{2018}]%
        {Dias18}
\bibfield{editor}{\bibinfo{person}{Nelson Dias}} (Ed.).
  \bibinfo{year}{2018}\natexlab{}.
\newblock \bibinfo{booktitle}{\emph{Hope for democracy: 30 years of
  participatory budgeting}}.
\newblock \bibinfo{publisher}{Epopeia Records and Oficina},
  \bibinfo{address}{Vila Ruiva and Faro}.
\newblock


\bibitem[\protect\citeauthoryear{Dias, Enr{\'i}quez, and J{\'u}lio}{Dias,
  Enr{\'i}quez and J{\'u}lio}{2019}]%
        {DEJ19}
\bibfield{editor}{\bibinfo{person}{Nelson Dias}, \bibinfo{person}{Sahsil
  Enr{\'i}quez}, {and} \bibinfo{person}{Simone J{\'u}lio}} (Eds.).
  \bibinfo{year}{2019}\natexlab{}.
\newblock \bibinfo{booktitle}{\emph{The Participatory Budgeting World Atlas}}.
\newblock \bibinfo{publisher}{Epopee Records: Officinal Coordination}.
\newblock


\bibitem[\protect\citeauthoryear{Dietrich and List}{Dietrich and List}{2007}]%
        {DiLi07}
\bibfield{author}{\bibinfo{person}{Franz Dietrich} {and}
  \bibinfo{person}{Christian List}.} \bibinfo{year}{2007}\natexlab{}.
\newblock \showarticletitle{Strategy-Proof Judgment Aggregation}.
\newblock \bibinfo{journal}{\emph{Economics \& Philosophy}}
  \bibinfo{volume}{23}, \bibinfo{number}{3} (\bibinfo{year}{2007}),
  \bibinfo{pages}{269--300}.
\newblock


\bibitem[\protect\citeauthoryear{Dummett}{Dummett}{1984}]%
        {Dumm84}
\bibfield{author}{\bibinfo{person}{Michael Dummett}.}
  \bibinfo{year}{1984}\natexlab{}.
\newblock \bibinfo{booktitle}{\emph{Voting Procedures}}.
\newblock \bibinfo{publisher}{Oxford University Press}.
\newblock


\bibitem[\protect\citeauthoryear{Elkind and Slinko}{Elkind and Slinko}{2016}]%
        {ElSl16}
\bibfield{author}{\bibinfo{person}{Edith Elkind} {and} \bibinfo{person}{Arkadii
  Slinko}.} \bibinfo{year}{2016}\natexlab{}.
\newblock \showarticletitle{Rationalizations of Voting Rules}.
\newblock In \bibinfo{booktitle}{\emph{Handbook of Computational Social
  Choice}}, \bibfield{editor}{\bibinfo{person}{Felix Brandt},
  \bibinfo{person}{Vincent Conitzer}, \bibinfo{person}{Ulle Endriss},
  \bibinfo{person}{Jérôme Lang}, {and} \bibinfo{person}{Ariel~D. Procaccia}}
  (Eds.). \bibinfo{publisher}{Cambridge University Press}, Chapter~8.
\newblock


\bibitem[\protect\citeauthoryear{Endriss}{Endriss}{2016}]%
        {Endr16}
\bibfield{author}{\bibinfo{person}{Ulle Endriss}.}
  \bibinfo{year}{2016}\natexlab{}.
\newblock \showarticletitle{Judgment Aggregation}.
\newblock In \bibinfo{booktitle}{\emph{Handbook of Computational Social
  Choice}}, \bibfield{editor}{\bibinfo{person}{Felix Brandt},
  \bibinfo{person}{Vincent Conitzer}, \bibinfo{person}{Ulle Endriss},
  \bibinfo{person}{Jérôme Lang}, {and} \bibinfo{person}{Ariel~D. Procaccia}}
  (Eds.). \bibinfo{publisher}{Cambridge University Press},
  \bibinfo{address}{New York}, Chapter~17, \bibinfo{pages}{399--426}.
\newblock


\bibitem[\protect\citeauthoryear{Fain, Goel, and Munagala}{Fain, Goel and
  Munagala}{2016}]%
        {FGM16}
\bibfield{author}{\bibinfo{person}{Brandon Fain}, \bibinfo{person}{Ashish
  Goel}, {and} \bibinfo{person}{Kamesh Munagala}.}
  \bibinfo{year}{2016}\natexlab{}.
\newblock \showarticletitle{The Core of the Participatory Budgeting Problem}.
  In \bibinfo{booktitle}{\emph{Proceedings of the 12th International Workshop
  on Internet and Network Economics (WINE)}}.
\newblock


\bibitem[\protect\citeauthoryear{Fain, Munagala, and Shah}{Fain, Munagala and
  Shah}{2018}]%
        {FMS18}
\bibfield{author}{\bibinfo{person}{Brandon Fain}, \bibinfo{person}{Kamesh
  Munagala}, {and} \bibinfo{person}{Nisarg Shah}.}
  \bibinfo{year}{2018}\natexlab{}.
\newblock \showarticletitle{Fair Allocation of Indivisible Public Goods}. In
  \bibinfo{booktitle}{\emph{Proceedings of the 19th ACM Conference on Economics
  and Computation (ACM-EC)}}. \bibinfo{pages}{575--592}.
\newblock


\bibitem[\protect\citeauthoryear{Fairstein, Benadè, and Gal}{Fairstein,
  Benadè and Gal}{2023}]%
        {FBG23}
\bibfield{author}{\bibinfo{person}{Roy Fairstein}, \bibinfo{person}{Gerdus
  Benadè}, {and} \bibinfo{person}{Kobi Gal}.} \bibinfo{year}{2023}\natexlab{}.
\newblock \showarticletitle{Participatory Budgeting Design for the Real World}.
  In \bibinfo{booktitle}{\emph{Proceedings of the 37th AAAI Conference on
  Artificial Intelligence (AAAI)}}.
\newblock


\bibitem[\protect\citeauthoryear{Fairstein, Vilenchik, Meir, and
  Gal}{Fairstein, Vilenchik, Meir and Gal}{2022}]%
        {FVMG22}
\bibfield{author}{\bibinfo{person}{Roy Fairstein}, \bibinfo{person}{Dan
  Vilenchik}, \bibinfo{person}{Reshef Meir}, {and} \bibinfo{person}{Kobi Gal}.}
  \bibinfo{year}{2022}\natexlab{}.
\newblock \showarticletitle{Welfare vs.\ Representation in Participatory
  Budgeting}. In \bibinfo{booktitle}{\emph{Proceedings of the 21st
  International Conference on Autonomous Agents and Multiagent Systems
  (AAMAS)}}.
\newblock


\bibitem[\protect\citeauthoryear{Faliszewski, Flis, Peters, Pierczy{\'n}ski,
  Skowron, Stolicki, Szufa, and Talmon}{Faliszewski, Flis, Peters,
  Pierczy{\'n}ski, Skowron, Stolicki, Szufa and Talmon}{2023}]%
        {FFP+23}
\bibfield{author}{\bibinfo{person}{Piotr Faliszewski},
  \bibinfo{person}{Jaros{\l}aw Flis}, \bibinfo{person}{Dominik Peters},
  \bibinfo{person}{Grzegorz Pierczy{\'n}ski}, \bibinfo{person}{Piotr Skowron},
  \bibinfo{person}{Dariusz Stolicki}, \bibinfo{person}{Stanis{\l}aw Szufa},
  {and} \bibinfo{person}{Nimrod Talmon}.} \bibinfo{year}{2023}\natexlab{}.
\newblock \showarticletitle{Participatory Budgeting: Data, Tools, and
  Analysis}. In \bibinfo{booktitle}{\emph{Proceedings of the 32th International
  Joint Conference on Artificial Intelligence (IJCAI)}}.
  \bibinfo{pages}{2667--2674}.
\newblock


\bibitem[\protect\citeauthoryear{Faliszewski, Skowron, Slinko, and
  Talmon}{Faliszewski, Skowron, Slinko and Talmon}{2017}]%
        {FSST17}
\bibfield{author}{\bibinfo{person}{Piotr Faliszewski}, \bibinfo{person}{Piotr
  Skowron}, \bibinfo{person}{Arkadii Slinko}, {and} \bibinfo{person}{Nimrod
  Talmon}.} \bibinfo{year}{2017}\natexlab{}.
\newblock \showarticletitle{Multiwinner Voting: {A} New Challenge for Social
  Choice Theory}.
\newblock In \bibinfo{booktitle}{\emph{Trends in Computational Social Choice}},
  \bibfield{editor}{\bibinfo{person}{Ulle Endriss}} (Ed.).
  \bibinfo{publisher}{AI Access}.
\newblock


\bibitem[\protect\citeauthoryear{Fluschnik, Skowron, Triphaus, and
  Wilker}{Fluschnik, Skowron, Triphaus and Wilker}{2019}]%
        {FSTW19}
\bibfield{author}{\bibinfo{person}{Till Fluschnik}, \bibinfo{person}{Piotr
  Skowron}, \bibinfo{person}{Mervin Triphaus}, {and} \bibinfo{person}{Kai
  Wilker}.} \bibinfo{year}{2019}\natexlab{}.
\newblock \showarticletitle{Fair Knapsack}. In
  \bibinfo{booktitle}{\emph{Proceedings of the 33rd AAAI Conference on
  Artificial Intelligence (AAAI)}}.
\newblock


\bibitem[\protect\citeauthoryear{Freeman, Pennock, Peters, and
  Vaughan}{Freeman, Pennock, Peters and Vaughan}{2021}]%
        {FPPV21}
\bibfield{author}{\bibinfo{person}{Rupert Freeman}, \bibinfo{person}{David~M.
  Pennock}, \bibinfo{person}{Dominik Peters}, {and}
  \bibinfo{person}{Jennifer~Wortman Vaughan}.} \bibinfo{year}{2021}\natexlab{}.
\newblock \showarticletitle{Truthful Aggregation of Budget Proposals}.
\newblock \bibinfo{journal}{\emph{Journal of Economic Theory}}
  \bibinfo{volume}{193} (\bibinfo{year}{2021}), \bibinfo{pages}{105234}.
\newblock


\bibitem[\protect\citeauthoryear{Gelauff and Goel}{Gelauff and Goel}{2024}]%
        {GG24}
\bibfield{author}{\bibinfo{person}{Lodewijk Gelauff} {and}
  \bibinfo{person}{Ashish Goel}.} \bibinfo{year}{2024}\natexlab{}.
\newblock \showarticletitle{Rank, Pack, or Approve: Voting Methods in
  Participatory Budgeting}. In \bibinfo{booktitle}{\emph{Proceedings of the
  18th International Conference on Web and Social Media (ICWSM)}}.
  \bibinfo{pages}{448--461}.
\newblock


\bibitem[\protect\citeauthoryear{Gibbard}{Gibbard}{1973}]%
        {Gibb73}
\bibfield{author}{\bibinfo{person}{Allan Gibbard}.}
  \bibinfo{year}{1973}\natexlab{}.
\newblock \showarticletitle{Manipulation of Voting Schemes: a General Result}.
\newblock \bibinfo{journal}{\emph{Econometrica}} (\bibinfo{year}{1973}),
  \bibinfo{pages}{587--601}.
\newblock


\bibitem[\protect\citeauthoryear{Goel, Krishnaswamy, Sakshuwong, and
  Aitamurto}{Goel, Krishnaswamy, Sakshuwong and Aitamurto}{2019}]%
        {GKSA19}
\bibfield{author}{\bibinfo{person}{Ashish Goel}, \bibinfo{person}{Anilesh~K.
  Krishnaswamy}, \bibinfo{person}{Sukolsak Sakshuwong}, {and}
  \bibinfo{person}{Tanja Aitamurto}.} \bibinfo{year}{2019}\natexlab{}.
\newblock \showarticletitle{Knapsack Voting for Participatory Budgeting}.
\newblock \bibinfo{journal}{\emph{{ACM} Transactions on Economics and
  Computation}} \bibinfo{volume}{7}, \bibinfo{number}{2}
  (\bibinfo{year}{2019}), \bibinfo{pages}{8:1--8:27}.
\newblock


\bibitem[\protect\citeauthoryear{Hansson}{Hansson}{2001}]%
        {hansson01}
\bibfield{author}{\bibinfo{person}{Sven~Ove Hansson}.}
  \bibinfo{year}{2001}\natexlab{}.
\newblock \bibinfo{booktitle}{\emph{The Structure of Values and Norms}}.
\newblock \bibinfo{publisher}{Cambridge University Press}.
\newblock


\bibitem[\protect\citeauthoryear{Hershkowitz, Kahng, Peters, and
  Procaccia}{Hershkowitz, Kahng, Peters and Procaccia}{2021}]%
        {HKPP21}
\bibfield{author}{\bibinfo{person}{D.~Ellis Hershkowitz},
  \bibinfo{person}{Anson Kahng}, \bibinfo{person}{Dominik Peters}, {and}
  \bibinfo{person}{Ariel~D. Procaccia}.} \bibinfo{year}{2021}\natexlab{}.
\newblock \showarticletitle{District-Fair Participatory Budgeting}. In
  \bibinfo{booktitle}{\emph{Proceedings of the 35th AAAI Conference on
  Artificial Intelligence (AAAI)}}.
\newblock


\bibitem[\protect\citeauthoryear{Jain, Sornat, and Talmon}{Jain, Sornat and
  Talmon}{2020}]%
        {JST20}
\bibfield{author}{\bibinfo{person}{Pallavi Jain}, \bibinfo{person}{Krzysztof
  Sornat}, {and} \bibinfo{person}{Nimrod Talmon}.}
  \bibinfo{year}{2020}\natexlab{}.
\newblock \showarticletitle{Participatory Budgeting with Project Interactions}.
  In \bibinfo{booktitle}{\emph{Proceedings of the 29th International Joint
  Conference on Artificial Intelligence (IJCAI)}}. \bibinfo{pages}{386--392}.
\newblock


\bibitem[\protect\citeauthoryear{Jain, Sornat, Talmon, and Zehavi}{Jain,
  Sornat, Talmon and Zehavi}{2021}]%
        {JSTZ21}
\bibfield{author}{\bibinfo{person}{Pallavi Jain}, \bibinfo{person}{Krzysztof
  Sornat}, \bibinfo{person}{Nimrod Talmon}, {and} \bibinfo{person}{Meirav
  Zehavi}.} \bibinfo{year}{2021}\natexlab{}.
\newblock \showarticletitle{Participatory Budgeting with Project Groups}. In
  \bibinfo{booktitle}{\emph{Proceedings of the 30th International Joint
  Conference on Artificial Intelligence (IJCAI)}}.
\newblock


\bibitem[\protect\citeauthoryear{Jain, Talmon, and Bulteau}{Jain, Talmon and
  Bulteau}{2021}]%
        {JTB21}
\bibfield{author}{\bibinfo{person}{Pallavi Jain}, \bibinfo{person}{Nimrod
  Talmon}, {and} \bibinfo{person}{Laurent Bulteau}.}
  \bibinfo{year}{2021}\natexlab{}.
\newblock \showarticletitle{Partition Aggregation for Participatory Budgeting}.
  In \bibinfo{booktitle}{\emph{Proceedings of the 20th International Conference
  on Autonomous Agents and Multiagent Systems (AAMAS)}}.
  \bibinfo{pages}{665–673}.
\newblock


\bibitem[\protect\citeauthoryear{Janson}{Janson}{2016}]%
        {Jans16}
\bibfield{author}{\bibinfo{person}{Svante Janson}.}
  \bibinfo{year}{2016}\natexlab{}.
\newblock \showarticletitle{{P}hragmén’s and {T}hiele’s election methods}.
\newblock \bibinfo{journal}{\emph{arXiv preprint arXiv:1611.08826}}
  (\bibinfo{year}{2016}).
\newblock


\bibitem[\protect\citeauthoryear{Jiang, Munagala, and Wang}{Jiang, Munagala and
  Wang}{2020}]%
        {JMW20}
\bibfield{author}{\bibinfo{person}{Zhihao Jiang}, \bibinfo{person}{Kamesh
  Munagala}, {and} \bibinfo{person}{Kangning Wang}.}
  \bibinfo{year}{2020}\natexlab{}.
\newblock \showarticletitle{Approximately Stable Committee Selection}. In
  \bibinfo{booktitle}{\emph{Proceedings of the 52nd Annual ACM Symposium on
  Theory of Computing (STOC)}}. \bibinfo{pages}{463--472}.
\newblock


\bibitem[\protect\citeauthoryear{Kellerer, Pferschy, and Pisinger}{Kellerer,
  Pferschy and Pisinger}{2004}]%
        {KPP04}
\bibfield{author}{\bibinfo{person}{Hans Kellerer}, \bibinfo{person}{Ulrich
  Pferschy}, {and} \bibinfo{person}{David Pisinger}.}
  \bibinfo{year}{2004}\natexlab{}.
\newblock \bibinfo{booktitle}{\emph{Knapsack Problems}}.
\newblock \bibinfo{publisher}{Springer-Verlag}.
\newblock


\bibitem[\protect\citeauthoryear{Kluiving, de~Vries, Vrijbergen, Boixel, and
  Endriss}{Kluiving, de~Vries, Vrijbergen, Boixel and Endriss}{2020}]%
        {KVVBE20}
\bibfield{author}{\bibinfo{person}{Boas Kluiving}, \bibinfo{person}{Adriaan de
  Vries}, \bibinfo{person}{Pepijn Vrijbergen}, \bibinfo{person}{Arthur Boixel},
  {and} \bibinfo{person}{Ulle Endriss}.} \bibinfo{year}{2020}\natexlab{}.
\newblock \showarticletitle{Analysing Irresolute Multiwinner Voting Rules with
  Approval Ballots via SAT Solving}. In \bibinfo{booktitle}{\emph{Proceedings
  of the 24th European Conference on Multi-Agent Systems (EUMAS)}}.
\newblock


\bibitem[\protect\citeauthoryear{Kraiczy and Elkind}{Kraiczy and
  Elkind}{2023}]%
        {KE23}
\bibfield{author}{\bibinfo{person}{Sonja Kraiczy} {and} \bibinfo{person}{Edith
  Elkind}.} \bibinfo{year}{2023}\natexlab{}.
\newblock \showarticletitle{An Adaptive and Verifiably Proportional Method for
  Participatory Budgeting}.
\newblock \bibinfo{journal}{\emph{CoRR}}  \bibinfo{volume}{abs/2310.10215}
  (\bibinfo{year}{2023}).
\newblock


\bibitem[\protect\citeauthoryear{Lackner}{Lackner}{2020}]%
        {Lack20}
\bibfield{author}{\bibinfo{person}{Martin Lackner}.}
  \bibinfo{year}{2020}\natexlab{}.
\newblock \showarticletitle{Perpetual Voting: Fairness in Long-Term Decision
  Making}. In \bibinfo{booktitle}{\emph{Proceedings of the 24th AAAI Conference
  on Artificial Intelligence (AAAI)}}. \bibinfo{pages}{2103--2110}.
\newblock


\bibitem[\protect\citeauthoryear{Lackner, Maly, and Rey}{Lackner, Maly and
  Rey}{2021}]%
        {LMR21}
\bibfield{author}{\bibinfo{person}{Martin Lackner}, \bibinfo{person}{Jan Maly},
  {and} \bibinfo{person}{Simon Rey}.} \bibinfo{year}{2021}\natexlab{}.
\newblock \showarticletitle{Fairness in Long-Term Participatory Budgeting}. In
  \bibinfo{booktitle}{\emph{Proceedings of the 30th International Joint
  Conference on Artificial Intelligence (IJCAI)}}.
\newblock


\bibitem[\protect\citeauthoryear{Lackner, Regner, and Krenn}{Lackner, Regner
  and Krenn}{2023}]%
        {LRK23}
\bibfield{author}{\bibinfo{person}{Martin Lackner}, \bibinfo{person}{Peter
  Regner}, {and} \bibinfo{person}{Benjamin Krenn}.}
  \bibinfo{year}{2023}\natexlab{}.
\newblock \showarticletitle{\texttt{abcvoting}: {A} {P}ython Package for
  Approval-Based Multi-Winner Voting Rules}.
\newblock \bibinfo{journal}{\emph{Journal of Open Source Software}}
  \bibinfo{volume}{8}, \bibinfo{number}{81} (\bibinfo{year}{2023}),
  \bibinfo{pages}{4880}.
\newblock


\bibitem[\protect\citeauthoryear{Lackner, Regner, Krenn, Cela, Kompauer,
  Lackner, Szufa, and Forster}{Lackner, Regner, Krenn, Cela, Kompauer, Lackner,
  Szufa and Forster}{2021}]%
        {abcvoting}
\bibfield{author}{\bibinfo{person}{Martin Lackner}, \bibinfo{person}{Peter
  Regner}, \bibinfo{person}{Benjamin Krenn}, \bibinfo{person}{Elvi Cela},
  \bibinfo{person}{Jonas Kompauer}, \bibinfo{person}{Florian Lackner},
  \bibinfo{person}{Stanis\l{}aw Szufa}, {and} \bibinfo{person}{Stefan~Schlomo
  Forster}.} \bibinfo{year}{2021}\natexlab{}.
\newblock \bibinfo{title}{{abcvoting: A Python library of approval-based
  committee voting rules}}.
\newblock
\newblock
\urldef\tempurl%
\url{https://doi.org/10.5281/zenodo.3904466}
\showDOI{\tempurl}
\newblock
\shownote{Current version: \url{https://github.com/martinlackner/abcvoting}.}


\bibitem[\protect\citeauthoryear{Lackner and Skowron}{Lackner and
  Skowron}{2020}]%
        {LaSk20}
\bibfield{author}{\bibinfo{person}{Martin Lackner} {and} \bibinfo{person}{Piotr
  Skowron}.} \bibinfo{year}{2020}\natexlab{}.
\newblock \showarticletitle{Utilitarian Welfare and Representation Guarantees
  of Approval-Based Multiwinner Rules}.
\newblock \bibinfo{journal}{\emph{Artificial Intelligence}}
  \bibinfo{volume}{288} (\bibinfo{year}{2020}), \bibinfo{pages}{103366}.
\newblock


\bibitem[\protect\citeauthoryear{Lackner and Skowron}{Lackner and
  Skowron}{2021}]%
        {LaSk21}
\bibfield{author}{\bibinfo{person}{Martin Lackner} {and} \bibinfo{person}{Piotr
  Skowron}.} \bibinfo{year}{2021}\natexlab{}.
\newblock \showarticletitle{Consistent Approval-Based Multi-Winner Rules}.
\newblock \bibinfo{journal}{\emph{Journal of Economic Theory}}
  \bibinfo{volume}{192} (\bibinfo{year}{2021}), \bibinfo{pages}{105173}.
\newblock


\bibitem[\protect\citeauthoryear{Lackner and Skowron}{Lackner and
  Skowron}{2023}]%
        {LaSk23}
\bibfield{author}{\bibinfo{person}{Martin Lackner} {and} \bibinfo{person}{Piotr
  Skowron}.} \bibinfo{year}{2023}\natexlab{}.
\newblock \bibinfo{booktitle}{\emph{Multi-Winner Voting with Approval
  Preferences}}.
\newblock \bibinfo{publisher}{Springer-Verlag}.
\newblock


\bibitem[\protect\citeauthoryear{Laruelle}{Laruelle}{2021}]%
        {Laru21}
\bibfield{author}{\bibinfo{person}{Annick Laruelle}.}
  \bibinfo{year}{2021}\natexlab{}.
\newblock \showarticletitle{Voting to Select Projects in Participatory
  Budgeting}.
\newblock \bibinfo{journal}{\emph{European Journal of Operational Research}}
  \bibinfo{volume}{288}, \bibinfo{number}{2} (\bibinfo{year}{2021}),
  \bibinfo{pages}{598--604}.
\newblock


\bibitem[\protect\citeauthoryear{Leibiker and Talmon}{Leibiker and
  Talmon}{2023}]%
        {LT23}
\bibfield{author}{\bibinfo{person}{Gil Leibiker} {and} \bibinfo{person}{Nimrod
  Talmon}.} \bibinfo{year}{2023}\natexlab{}.
\newblock \showarticletitle{A Recommendation System for Participatory
  Budgeting}. In \bibinfo{booktitle}{\emph{International Conference on
  Autonomous Agents and Multiagent Systems (AAMAS). Workshop on Optimization
  and Learning in Multiagent Systems}}.
\newblock


\bibitem[\protect\citeauthoryear{Lewin}{Lewin}{1996}]%
        {Lewin96}
\bibfield{author}{\bibinfo{person}{Shira~B. Lewin}.}
  \bibinfo{year}{1996}\natexlab{}.
\newblock \showarticletitle{Economics and Psychology: Lessons for Our Own Day
  From the Early Twentieth Century}.
\newblock \bibinfo{journal}{\emph{Journal of Economic Literature}}
  \bibinfo{volume}{34}, \bibinfo{number}{3} (\bibinfo{year}{1996}),
  \bibinfo{pages}{1293--1323}.
\newblock


\bibitem[\protect\citeauthoryear{Los, Christoff, and Grossi}{Los, Christoff and
  Grossi}{2022}]%
        {LCG22}
\bibfield{author}{\bibinfo{person}{Maaike Los}, \bibinfo{person}{Zo{\'e}
  Christoff}, {and} \bibinfo{person}{Davide Grossi}.}
  \bibinfo{year}{2022}\natexlab{}.
\newblock \showarticletitle{Proportional Budget Allocations: {T}owards a
  Systematization}. In \bibinfo{booktitle}{\emph{Proceedings of the 31st
  International Joint Conference on Artificial Intelligence (IJCAI)}}.
\newblock


\bibitem[\protect\citeauthoryear{Lu and Boutilier}{Lu and Boutilier}{2011}]%
        {LuBo11}
\bibfield{author}{\bibinfo{person}{Tyler Lu} {and} \bibinfo{person}{Craig
  Boutilier}.} \bibinfo{year}{2011}\natexlab{}.
\newblock \showarticletitle{Budgeted Social Choice: From Consensus to
  Personalized Decision Making}. In \bibinfo{booktitle}{\emph{Proceedings of
  the 22nd International Joint Conference on Artificial Intelligence (IJCAI)}}.
  \bibinfo{pages}{280--286}.
\newblock


\bibitem[\protect\citeauthoryear{Majumdar and Pournaras}{Majumdar and
  Pournaras}{2023}]%
        {MP23}
\bibfield{author}{\bibinfo{person}{Srijoni Majumdar} {and}
  \bibinfo{person}{Evangelos Pournaras}.} \bibinfo{year}{2023}\natexlab{}.
\newblock \showarticletitle{Consensus-Based Participatory Budgeting for
  Legitimacy: Decision Support via Multi-agent Reinforcement Learning}. In
  \bibinfo{booktitle}{\emph{Proceedings of the 9th International Conference on
  Machine Learning, Optimization and Data Science (LOD)}}.
  \bibinfo{pages}{1--14}.
\newblock


\bibitem[\protect\citeauthoryear{Maly}{Maly}{2023}]%
        {Maly23}
\bibfield{author}{\bibinfo{person}{Jan Maly}.} \bibinfo{year}{2023}\natexlab{}.
\newblock \bibinfo{title}{The core of an approval-based PB instance can be
  empty for nearly all cost-based satisfaction functions and for the share}.
\newblock
\newblock
\showeprint[arxiv]{2311.06132}


\bibitem[\protect\citeauthoryear{Maly, Rey, Endriss, and Lackner}{Maly, Rey,
  Endriss and Lackner}{2023}]%
        {MREL23}
\bibfield{author}{\bibinfo{person}{Jan Maly}, \bibinfo{person}{Simon Rey},
  \bibinfo{person}{Ulle Endriss}, {and} \bibinfo{person}{Martin Lackner}.}
  \bibinfo{year}{2023}\natexlab{}.
\newblock \showarticletitle{Fairness in Participatory Budgeting via Equality of
  Resources}. In \bibinfo{booktitle}{\emph{Proceedings of the 22th
  International Conference on Autonomous Agents and Multiagent Systems
  (AAMAS)}}.
\newblock


\bibitem[\protect\citeauthoryear{Meir}{Meir}{2018}]%
        {Meir18}
\bibfield{author}{\bibinfo{person}{Reshef Meir}.}
  \bibinfo{year}{2018}\natexlab{}.
\newblock \bibinfo{booktitle}{\emph{Strategic Voting}}.
\newblock \bibinfo{publisher}{Morgan \& Claypool Publishers}.
\newblock


\bibitem[\protect\citeauthoryear{Michorzewski, Peters, and
  Skowron}{Michorzewski, Peters and Skowron}{2020}]%
        {MPS20}
\bibfield{author}{\bibinfo{person}{Marcin Michorzewski},
  \bibinfo{person}{Dominik Peters}, {and} \bibinfo{person}{Piotr Skowron}.}
  \bibinfo{year}{2020}\natexlab{}.
\newblock \showarticletitle{Price of Fairness in Budget Division and
  Probabilistic Social Choice}. In \bibinfo{booktitle}{\emph{Proceedings of the
  34th AAAI Conference on Artificial Intelligence (AAAI)}}.
  \bibinfo{pages}{2184--2191}.
\newblock


\bibitem[\protect\citeauthoryear{Motamed, Soeteman, Rey, and Endriss}{Motamed,
  Soeteman, Rey and Endriss}{2022}]%
        {MSRE22}
\bibfield{author}{\bibinfo{person}{Nima Motamed}, \bibinfo{person}{Arie
  Soeteman}, \bibinfo{person}{Simon Rey}, {and} \bibinfo{person}{Ulle
  Endriss}.} \bibinfo{year}{2022}\natexlab{}.
\newblock \showarticletitle{Participatory Budgeting with Multiple Resources}.
  In \bibinfo{booktitle}{\emph{Proceedings of the 19th European Conference on
  Multi-Agent Systems (EUMAS)}}.
\newblock


\bibitem[\protect\citeauthoryear{Munagala, Shen, and Wang}{Munagala, Shen and
  Wang}{2022}]%
        {MSW22}
\bibfield{author}{\bibinfo{person}{Kamesh Munagala}, \bibinfo{person}{Yiheng
  Shen}, {and} \bibinfo{person}{Kangning Wang}.}
  \bibinfo{year}{2022}\natexlab{}.
\newblock \showarticletitle{Auditing for Core Stability in Participatory
  Budgeting}. In \bibinfo{booktitle}{\emph{Proceedings of the 18th
  International Workshop on Internet and Network Economics (WINE)}}.
  \bibinfo{pages}{292--310}.
\newblock


\bibitem[\protect\citeauthoryear{Munagala, Shen, Wang, and Wang}{Munagala,
  Shen, Wang and Wang}{2022}]%
        {MSWW22}
\bibfield{author}{\bibinfo{person}{Kamesh Munagala}, \bibinfo{person}{Yiheng
  Shen}, \bibinfo{person}{Kangning Wang}, {and} \bibinfo{person}{Zhiyi Wang}.}
  \bibinfo{year}{2022}\natexlab{}.
\newblock \showarticletitle{Approximate Core for Committee Selection via
  Multilinear Extension and Market Clearing}. In
  \bibinfo{booktitle}{\emph{Proceedings of the SODA22 Annual ACM-SIAM Symposium
  on Discrete Algorithms (SODA)}}. \bibinfo{pages}{2229--2252}.
\newblock


\bibitem[\protect\citeauthoryear{Pascual, Rzadca, and Skowron}{Pascual, Rzadca
  and Skowron}{2018}]%
        {PRS18}
\bibfield{author}{\bibinfo{person}{Fanny Pascual}, \bibinfo{person}{Krzysztof
  Rzadca}, {and} \bibinfo{person}{Piotr Skowron}.}
  \bibinfo{year}{2018}\natexlab{}.
\newblock \showarticletitle{Collective Schedules: {S}cheduling Meets
  Computational Social Choice}. In \bibinfo{booktitle}{\emph{Proceedings of the
  17th International Conference on Autonomous Agents and Multiagent Systems
  (AAMAS)}}. \bibinfo{pages}{667--675}.
\newblock


\bibitem[\protect\citeauthoryear{Patel, Khan, and Louis}{Patel, Khan and
  Louis}{2021}]%
        {PKL21}
\bibfield{author}{\bibinfo{person}{Deval Patel}, \bibinfo{person}{Arindam
  Khan}, {and} \bibinfo{person}{Anand Louis}.} \bibinfo{year}{2021}\natexlab{}.
\newblock \showarticletitle{Group Fairness for Knapsack Problems}. In
  \bibinfo{booktitle}{\emph{Proceedings of the 20th International Conference on
  Autonomous Agents and Multiagent Systems (AAMAS)}}.
\newblock


\bibitem[\protect\citeauthoryear{Peters}{Peters}{2018}]%
        {Peters18}
\bibfield{author}{\bibinfo{person}{Dominik Peters}.}
  \bibinfo{year}{2018}\natexlab{}.
\newblock \showarticletitle{Proportionality and Strategyproofness in
  Multiwinner Elections}. In \bibinfo{booktitle}{\emph{Proceedings of the 17th
  International Conference on Autonomous Agents and Multiagent Systems
  (AAMAS)}}. \bibinfo{pages}{1549--1557}.
\newblock


\bibitem[\protect\citeauthoryear{Peters}{Peters}{2019}]%
        {Peters19}
\bibfield{author}{\bibinfo{person}{Domink Peters}.}
  \bibinfo{year}{2019}\natexlab{}.
\newblock \emph{\bibinfo{title}{Fair Division of the Commons}}.
\newblock {DPhil} Thesis. \bibinfo{school}{University of Oxford}.
\newblock


\bibitem[\protect\citeauthoryear{Peters, Pierczyński, and Skowron}{Peters,
  Pierczyński and Skowron}{2021}]%
        {PPS21}
\bibfield{author}{\bibinfo{person}{Dominik Peters}, \bibinfo{person}{Grzegorz
  Pierczyński}, {and} \bibinfo{person}{Piotr Skowron}.}
  \bibinfo{year}{2021}\natexlab{}.
\newblock \showarticletitle{Proportional Participatory Budgeting with Additive
  Utilities}. In \bibinfo{booktitle}{\emph{Proceedings of the 35th Annual
  Conference on Neural Information Processing Systems (NeurIPS)}}.
\newblock


\bibitem[\protect\citeauthoryear{Peters and Skowron}{Peters and
  Skowron}{2020}]%
        {PeSk20}
\bibfield{author}{\bibinfo{person}{Dominik Peters} {and} \bibinfo{person}{Piotr
  Skowron}.} \bibinfo{year}{2020}\natexlab{}.
\newblock \showarticletitle{Proportionality and the Limits of Welfarism}. In
  \bibinfo{booktitle}{\emph{Proceedings of the 21st ACM Conference on Economics
  and Computation (ACM-EC)}}.
\newblock


\bibitem[\protect\citeauthoryear{Pivato}{Pivato}{2019}]%
        {Piva19}
\bibfield{author}{\bibinfo{person}{Marcus Pivato}.}
  \bibinfo{year}{2019}\natexlab{}.
\newblock \showarticletitle{Realizing Epistemic Democracy}.
\newblock In \bibinfo{booktitle}{\emph{The Future of Economic Design}},
  \bibfield{editor}{\bibinfo{person}{Jean-François Laslier},
  \bibinfo{person}{Hervé Moulin}, \bibinfo{person}{M.~Remzi Sanver}, {and}
  \bibinfo{person}{William~S. Zwicker}} (Eds.).
  \bibinfo{publisher}{Springer-Verlag}, \bibinfo{pages}{103--112}.
\newblock


\bibitem[\protect\citeauthoryear{Procaccia}{Procaccia}{2019}]%
        {Proc19}
\bibfield{author}{\bibinfo{person}{Ariel~D. Procaccia}.}
  \bibinfo{year}{2019}\natexlab{}.
\newblock \showarticletitle{Axioms Should Explain Solutions}.
\newblock In \bibinfo{booktitle}{\emph{The Future of Economic Design}},
  \bibfield{editor}{\bibinfo{person}{Jean-François Laslier},
  \bibinfo{person}{Hervé Moulin}, \bibinfo{person}{Remzi Sanver}, {and}
  \bibinfo{person}{William~S. Zwicker}} (Eds.).
  \bibinfo{publisher}{Springer-Verlag}.
\newblock


\bibitem[\protect\citeauthoryear{Procaccia and Rosenschein}{Procaccia and
  Rosenschein}{2006}]%
        {PrRo06}
\bibfield{author}{\bibinfo{person}{Ariel~D. Procaccia} {and}
  \bibinfo{person}{Jeffrey~S. Rosenschein}.} \bibinfo{year}{2006}\natexlab{}.
\newblock \showarticletitle{The Distortion of Cardinal Preferences in Voting}.
  In \bibinfo{booktitle}{\emph{Proceedings of the International Workshop on
  Cooperative Information Agents X (CIA)}}. \bibinfo{pages}{317--331}.
\newblock


\bibitem[\protect\citeauthoryear{Rey and Endriss}{Rey and Endriss}{2023}]%
        {ReEn23}
\bibfield{author}{\bibinfo{person}{Simon Rey} {and} \bibinfo{person}{Ulle
  Endriss}.} \bibinfo{year}{2023}\natexlab{}.
\newblock \showarticletitle{Epistemic Selection of Costly Alternatives: The
  Case of Participatory Budgeting}.
\newblock \bibinfo{journal}{\emph{arXiv preprint arXiv:2304.10940}}
  (\bibinfo{year}{2023}).
\newblock


\bibitem[\protect\citeauthoryear{Rey, Endriss, and de~Haan}{Rey, Endriss and
  de~Haan}{2020}]%
        {REH20}
\bibfield{author}{\bibinfo{person}{Simon Rey}, \bibinfo{person}{Ulle Endriss},
  {and} \bibinfo{person}{Ronald de Haan}.} \bibinfo{year}{2020}\natexlab{}.
\newblock \showarticletitle{Designing Participatory Budgeting Mechanisms
  Grounded in Judgment Aggregation}. In \bibinfo{booktitle}{\emph{Proceedings
  of the 17th International Conference on Principles of Knowledge
  Representation and Reasoning (KR)}}.
\newblock


\bibitem[\protect\citeauthoryear{Rey, Endriss, and de~Haan}{Rey, Endriss and
  de~Haan}{2021}]%
        {REH21}
\bibfield{author}{\bibinfo{person}{Simon Rey}, \bibinfo{person}{Ulle Endriss},
  {and} \bibinfo{person}{Ronald de Haan}.} \bibinfo{year}{2021}\natexlab{}.
\newblock \showarticletitle{Shortlisting Rules and Incentives in an End-to-End
  Model for Participatory Budgeting}. In \bibinfo{booktitle}{\emph{Proceedings
  of the 30th International Joint Conference on Artificial Intelligence
  (IJCAI)}}.
\newblock


\bibitem[\protect\citeauthoryear{Rothe}{Rothe}{2015}]%
        {Roth15}
\bibfield{editor}{\bibinfo{person}{J{\"o}rg Rothe}} (Ed.).
  \bibinfo{year}{2015}\natexlab{}.
\newblock \bibinfo{booktitle}{\emph{Economics and Computation}}.
\newblock \bibinfo{publisher}{Springer-Verlag}.
\newblock


\bibitem[\protect\citeauthoryear{S{\'a}nchez-Fern{\'a}ndez, Elkind, Lackner,
  Fern{\'a}ndez, Fisteus, Val, and Skowron}{S{\'a}nchez-Fern{\'a}ndez, Elkind,
  Lackner, Fern{\'a}ndez, Fisteus, Val and Skowron}{2017}]%
        {SELFFVS17}
\bibfield{author}{\bibinfo{person}{Luis S{\'a}nchez-Fern{\'a}ndez},
  \bibinfo{person}{Edith Elkind}, \bibinfo{person}{Martin Lackner},
  \bibinfo{person}{Norberto Fern{\'a}ndez}, \bibinfo{person}{Jes{\'u}s
  Fisteus}, \bibinfo{person}{Pablo~Basanta Val}, {and} \bibinfo{person}{Piotr
  Skowron}.} \bibinfo{year}{2017}\natexlab{}.
\newblock \showarticletitle{Proportional justified representation}. In
  \bibinfo{booktitle}{\emph{Proceedings of the 31st AAAI Conference on
  Artificial Intelligence (AAAI)}}. \bibinfo{pages}{670--676}.
\newblock


\bibitem[\protect\citeauthoryear{S{\'a}nchez-Fern{\'a}ndez,
  Fern{\'a}ndez-Garc{\'\i}a, Fisteus, and Brill}{S{\'a}nchez-Fern{\'a}ndez,
  Fern{\'a}ndez-Garc{\'\i}a, Fisteus and Brill}{2016}]%
        {SFFB16}
\bibfield{author}{\bibinfo{person}{Luis S{\'a}nchez-Fern{\'a}ndez},
  \bibinfo{person}{Norberto Fern{\'a}ndez-Garc{\'\i}a},
  \bibinfo{person}{Jes{\'u}s~A Fisteus}, {and} \bibinfo{person}{Markus Brill}.}
  \bibinfo{year}{2016}\natexlab{}.
\newblock \showarticletitle{Fully Open Extensions to the {D}'{H}ondt Method}.
\newblock \bibinfo{journal}{\emph{arXiv preprint arXiv:1609.05370}}
  (\bibinfo{year}{2016}).
\newblock


\bibitem[\protect\citeauthoryear{S{\'a}nchez-Fern{\'a}ndez,
  Fern{\'a}ndez-Garc{\'\i}a, Fisteus, and Brill}{S{\'a}nchez-Fern{\'a}ndez,
  Fern{\'a}ndez-Garc{\'\i}a, Fisteus and Brill}{2022}]%
        {SFFB22}
\bibfield{author}{\bibinfo{person}{Luis S{\'a}nchez-Fern{\'a}ndez},
  \bibinfo{person}{Norberto Fern{\'a}ndez-Garc{\'\i}a},
  \bibinfo{person}{Jes{\'u}s~A Fisteus}, {and} \bibinfo{person}{Markus Brill}.}
  \bibinfo{year}{2022}\natexlab{}.
\newblock \showarticletitle{The Maximin Support Method: An Extension of the
  {D}'{H}ondt Method to Approval-Based Multiwinner Elections}.
\newblock \bibinfo{journal}{\emph{Mathematical Programming}}
  (\bibinfo{year}{2022}).
\newblock


\bibitem[\protect\citeauthoryear{Satterthwaite}{Satterthwaite}{1975}]%
        {Satt75}
\bibfield{author}{\bibinfo{person}{Mark~Allen Satterthwaite}.}
  \bibinfo{year}{1975}\natexlab{}.
\newblock \showarticletitle{Strategy-Proofness and {A}rrow's Conditions:
  {E}xistence and Correspondence Theorems for Voting Procedures and Social
  Welfare Functions}.
\newblock \bibinfo{journal}{\emph{Journal of Economic Theory}}
  \bibinfo{volume}{10}, \bibinfo{number}{2} (\bibinfo{year}{1975}),
  \bibinfo{pages}{187--217}.
\newblock


\bibitem[\protect\citeauthoryear{Shah}{Shah}{2007}]%
        {Shah07}
\bibfield{editor}{\bibinfo{person}{Anwar Shah}} (Ed.).
  \bibinfo{year}{2007}\natexlab{}.
\newblock \bibinfo{booktitle}{\emph{Participatory budgeting}}.
\newblock \bibinfo{publisher}{The World Bank}.
\newblock


\bibitem[\protect\citeauthoryear{Sintomer, Herzberg, and R{\"o}cke}{Sintomer,
  Herzberg and R{\"o}cke}{2008}]%
        {SHR08}
\bibfield{author}{\bibinfo{person}{Yves Sintomer}, \bibinfo{person}{Carsten
  Herzberg}, {and} \bibinfo{person}{Anja R{\"o}cke}.}
  \bibinfo{year}{2008}\natexlab{}.
\newblock \showarticletitle{Participatory Budgeting in Europe: {P}otentials and
  Challenges}.
\newblock \bibinfo{journal}{\emph{International journal of urban and regional
  research}} \bibinfo{volume}{32}, \bibinfo{number}{1} (\bibinfo{year}{2008}),
  \bibinfo{pages}{164--178}.
\newblock


\bibitem[\protect\citeauthoryear{Skowron, Faliszewski, and Slinko}{Skowron,
  Faliszewski and Slinko}{2019}]%
        {SFS19}
\bibfield{author}{\bibinfo{person}{Piotr Skowron}, \bibinfo{person}{Piotr
  Faliszewski}, {and} \bibinfo{person}{Arkadii Slinko}.}
  \bibinfo{year}{2019}\natexlab{}.
\newblock \showarticletitle{Axiomatic Characterization of Committee Scoring
  Rules}.
\newblock \bibinfo{journal}{\emph{Journal of Economic Theory}}
  \bibinfo{volume}{180} (\bibinfo{year}{2019}), \bibinfo{pages}{244--273}.
\newblock


\bibitem[\protect\citeauthoryear{Skowron, Slinko, Szufa, and Talmon}{Skowron,
  Slinko, Szufa and Talmon}{2020}]%
        {SSST20}
\bibfield{author}{\bibinfo{person}{Piotr Skowron}, \bibinfo{person}{Arkadii
  Slinko}, \bibinfo{person}{Stanis{\l}aw Szufa}, {and} \bibinfo{person}{Nimrod
  Talmon}.} \bibinfo{year}{2020}\natexlab{}.
\newblock \showarticletitle{Participatory Budgeting with Cumulative Votes}.
\newblock \bibinfo{journal}{\emph{arXiv preprint arXiv:2009.02690}}
  (\bibinfo{year}{2020}).
\newblock


\bibitem[\protect\citeauthoryear{Sreedurga}{Sreedurga}{2023}]%
        {Sreedurga23IJCAI}
\bibfield{author}{\bibinfo{person}{Gogulapati Sreedurga}.}
  \bibinfo{year}{2023}\natexlab{}.
\newblock \showarticletitle{Participatory Budgeting with Multiple Degrees of
  Projects and Ranged Approval Votes}. In \bibinfo{booktitle}{\emph{Proceedings
  of the 32nd International Joint Conference on Artificial Intelligence
  (IJCAI)}}. \bibinfo{pages}{2870--2877}.
\newblock


\bibitem[\protect\citeauthoryear{Sreedurga, Bhardwaj, and Narahari}{Sreedurga,
  Bhardwaj and Narahari}{2022}]%
        {SBY22}
\bibfield{author}{\bibinfo{person}{Gogulapati Sreedurga},
  \bibinfo{person}{Mayank~Ratan Bhardwaj}, {and} \bibinfo{person}{Y.
  Narahari}.} \bibinfo{year}{2022}\natexlab{}.
\newblock \showarticletitle{Maxmin Participatory Budgeting}. In
  \bibinfo{booktitle}{\emph{Proceedings of the 31st International Joint
  Conference on Artificial Intelligence (IJCAI)}}.
\newblock


\bibitem[\protect\citeauthoryear{Sreedurga and Narahari}{Sreedurga and
  Narahari}{2022}]%
        {SrNa22}
\bibfield{author}{\bibinfo{person}{Gogulapati Sreedurga} {and}
  \bibinfo{person}{Yadati Narahari}.} \bibinfo{year}{2022}\natexlab{}.
\newblock \showarticletitle{Indivisible Participatory Budgeting under Weak
  Rankings}.
\newblock \bibinfo{journal}{\emph{arXiv preprint arXiv:2207.07981}}
  (\bibinfo{year}{2022}).
\newblock


\bibitem[\protect\citeauthoryear{Stolicki, Szufa, and Talmon}{Stolicki, Szufa
  and Talmon}{2020}]%
        {SST20}
\bibfield{author}{\bibinfo{person}{Dariusz Stolicki},
  \bibinfo{person}{Stanis{\l}aw Szufa}, {and} \bibinfo{person}{Nimrod Talmon}.}
  \bibinfo{year}{2020}\natexlab{}.
\newblock \showarticletitle{Pabulib: {A} Participatory Budgeting Library}.
\newblock \bibinfo{journal}{\emph{arXiv preprint arXiv:2012.06539}}
  (\bibinfo{year}{2020}).
\newblock


\bibitem[\protect\citeauthoryear{Talmon and Faliszewski}{Talmon and
  Faliszewski}{2019}]%
        {TaFa19}
\bibfield{author}{\bibinfo{person}{Nimrod Talmon} {and} \bibinfo{person}{Piotr
  Faliszewski}.} \bibinfo{year}{2019}\natexlab{}.
\newblock \showarticletitle{A Framework for Approval-Based Budgeting Methods}.
  In \bibinfo{booktitle}{\emph{Proceedings of the 33rd AAAI Conference on
  Artificial Intelligence (AAAI)}}.
\newblock


\bibitem[\protect\citeauthoryear{Wagner and Meir}{Wagner and Meir}{2023}]%
        {WM23}
\bibfield{author}{\bibinfo{person}{Jonathan Wagner} {and}
  \bibinfo{person}{Reshef Meir}.} \bibinfo{year}{2023}\natexlab{}.
\newblock \showarticletitle{Strategy-Proof Budgeting via a VCG-Like Mechanism}.
  In \bibinfo{booktitle}{\emph{Proceedings of the 16th International Symposium
  on Algorithmic Game Theory (SAGT)}}, Vol.~\bibinfo{volume}{14238}.
  \bibinfo{pages}{401--418}.
\newblock


\bibitem[\protect\citeauthoryear{Wampler}{Wampler}{2000}]%
        {Wamp00}
\bibfield{author}{\bibinfo{person}{Brian Wampler}.}
  \bibinfo{year}{2000}\natexlab{}.
\newblock \showarticletitle{A Guide to Participatory Budgeting}.
\newblock \bibinfo{journal}{\emph{Third conference of the International Budget
  Project}} (\bibinfo{year}{2000}).
\newblock


\bibitem[\protect\citeauthoryear{Wampler}{Wampler}{2012}]%
        {Wamp12}
\bibfield{author}{\bibinfo{person}{Brian Wampler}.}
  \bibinfo{year}{2012}\natexlab{}.
\newblock \showarticletitle{Participatory Budgeting: Core Principles and Key
  Impacts}.
\newblock \bibinfo{journal}{\emph{Journal of Public Deliberation}}
  (\bibinfo{year}{2012}).
\newblock


\bibitem[\protect\citeauthoryear{Wampler and Goldfrank}{Wampler and
  Goldfrank}{2022}]%
        {WaGo22}
\bibfield{author}{\bibinfo{person}{Brian Wampler} {and}
  \bibinfo{person}{Benjamin Goldfrank}.} \bibinfo{year}{2022}\natexlab{}.
\newblock \bibinfo{booktitle}{\emph{The Rise, Spread, and Decline of
  {B}razil’s Participatory Budgeting: the Arc of a Democratic Innovation}}.
\newblock \bibinfo{publisher}{Springer-Verlag}.
\newblock


\bibitem[\protect\citeauthoryear{Wampler, McNulty, and Touchton}{Wampler,
  McNulty and Touchton}{2021}]%
        {WNT21}
\bibfield{author}{\bibinfo{person}{Brian Wampler}, \bibinfo{person}{Stephanie
  McNulty}, {and} \bibinfo{person}{Michael Touchton}.}
  \bibinfo{year}{2021}\natexlab{}.
\newblock \bibinfo{booktitle}{\emph{Participatory Budgeting in Global
  Perspective}}.
\newblock \bibinfo{publisher}{Oxford University Press}.
\newblock


\bibitem[\protect\citeauthoryear{Wang, Wang, Wang, and Jia}{Wang, Wang, Wang
  and Jia}{2023}]%
        {WWWJ23}
\bibfield{author}{\bibinfo{person}{Shiwen Wang}, \bibinfo{person}{Chenhao
  Wang}, \bibinfo{person}{Tian Wang}, {and} \bibinfo{person}{Weijia Jia}.}
  \bibinfo{year}{2023}\natexlab{}.
\newblock \showarticletitle{Approval-Based Participatory Budgeting with
  Donations}. In \bibinfo{booktitle}{\emph{Proceedings of the 29th
  International Computing and Combinatorics Conference (COCOON)}}.
  \bibinfo{pages}{404--416}.
\newblock


\bibitem[\protect\citeauthoryear{Yang, Hausladen, Peters, Pournaras, Fricker,
  and Helbing}{Yang, Hausladen, Peters, Pournaras, Fricker and Helbing}{2024}]%
        {YHP+24}
\bibfield{author}{\bibinfo{person}{Joshua~C Yang}, \bibinfo{person}{Carina~I
  Hausladen}, \bibinfo{person}{Dominik Peters}, \bibinfo{person}{Evangelos
  Pournaras}, \bibinfo{person}{Regula~Haenggli Fricker}, {and}
  \bibinfo{person}{Dirk Helbing}.} \bibinfo{year}{2024}\natexlab{}.
\newblock \showarticletitle{Designing Digital Voting Systems for Citizens:
  Achieving Fairness and Legitimacy in Digital Participatory Budgeting}.
\newblock \bibinfo{journal}{\emph{arXiv preprint arXiv:2310.03501}}
  (\bibinfo{year}{2024}).
\newblock


\end{thebibliography}
	
\end{document}